(64.3 k words inclusive; Table 11;Figure 41; Ref 563; Updated 12/4/2007 5:18 PM)

Being updated at: www.ntu.edu.sg/home/ecqsun/rtf/mesomechanics.doc

# The thermo-mechanical behavior of low-dimensional materials

Chang Q. Sun

School of Electrical and Electronic Engineering, Nanyang Technological University, Singapore 639798, Singapore

E-mail: ecqsun@ntu.edu.sg

URL: www.ntu.edu.sg/home/ecqsun/

Abstract

With the miniaturization of a solid, surface and quantum effects become increasingly important. As a result, low-dimensional materials manifest unusual features in particular in the energetic and mechanical performance. Consistently atomistic understanding of the mechanism behind the intriguing behavior of low-dimensional systems including monatomic chains, hollow tubes, surface skins, nanocavities, nanowires, and nanograins has long been a high challenge. This article reports recent progress in this regard. A survey is presented and then is followed by analytical approaches in terms of local bond average (LBA) from the perspective of bonding energetics and its functional dependence on external stimuli of coordination environment and temperature change. It is shown that the measurable quantities of a specimen can be functionally correlated to the identities of the representative bonds and their responses to the external stimuli. It is understood that the shortened and strengthened bonds between the under-coordinated atoms and the associated local strain and energy trapping dictate *intrinsically* the mechanical behavior of systems with large portion of under-coordinated atoms. The thermal softening of a substance arises from thermally-induced bond expansion and lattice vibration that weakens the bonds through the internal energy increase. The competition between the energy-density-gain and the residual atomic cohesive-energy in the relaxed surface skin determines *intrinsically* the mechanical performance of a mesoscopic specimen, whereas competition between the activation and inhibition of atomic dislocations dominates *extrinsically* the yield strength of the specimen in plastic deformation. Therefore, the mechanical behavior of a specimen depends intrinsically on its shape and size, the nature of the bond involved, surface and interface conditions, and the temperature of operations. Solutions have enabled reproduction of the observations such as the bond strain limit in monatomic chains, size and temperature dependence of elasticity and extensibility, temperature dependence of surface tension, and the inverse Hall-Petch relation in nanograins with derivative of quantitative information about the bonding identities and the factors dominating the strongest grain sizes. Further extension of the approaches to junction interfaces, liquid surfaces, defects and impurities, chemically adsorbed systems, amorphous states, and substances under other stimuli such as pressure and electric field would contribute to knowledge of such systems and could lead to the development of even more fascinating and rewarding substances.

**Keywords:** nanostructures; bond contraction; monatomic chain; nanotubes; thermal stability; surface energy, surface tension; mechanical strength; extensibility; Hall-Petch relationship; atomic vacancy; nanocavity; atomistic modeling.

**Nomenclature**

| | |
|---|---|
| $\omega$ | The angular frequency |
| $\mu$ | Atomic magnetic momentum |
| $\beta$ | Compressibility/extensibility |
| $\eta_1$ | Specific heat per coordinate |
| $\eta_2$ | Thermal energy per coordinate for evaporating a molten atom |
| $\theta_D$ | Debye temperature |
| $\gamma_d$ | Energy density gain |
| $\gamma_f$ | Atomic residual cohesive energy |
| $\alpha_i(T)$ | Temperature dependent thermal expansion coefficient (TEC) |
| $\gamma_{ij}$ | Atomic portion in the i th atomic shell over the entire solid of size $D_j$ |
| $\gamma_{qj}$ | Gruneisen mode parameter |
| $\gamma_s$ | Surface tension or surface energy |
| AFAM | Atomic force acoustic microscopy |
| AFM | Atomic force microscopy |
| BBB | Bond-band-barrier |
| BOLS | Bond order-length-strength |
| $C_i(z_i)$ | CN dependent bond contraction coefficient |
| CN(z) | Coordination number |
| $c_v(T/\theta_D)$ | Debye specific heat |
| DFT | Density functional theory |
| DOS | Density-of-state |
| $E_B$ | Atomic cohesive energy /vacancy formation energy |
| $E_b$ | Cohesive energy per bond |
| $E_F$ | Fermi energy |
| EXAFS | Extended X-ray absorption fine structure spectroscopy |
| GB | Grain boundary |
| GIXR | Grazing incidence X-ray reflectivity |
| IHPR | Inverse Hall-Petch relationship |
| $k_B$ | Boltzmann constant |
| $K_j$ | Dimensionless form of the radius of a sphere or the thickness of a plate |
| LBA | Local bond average |
| MC | Monatomic chain |
| MD | Molecular dynamics |
| MWCNT | Multi-walled carbon nanotube |
| NW | Nanowire |
| P | Stress |
| $Q(K_j)$ | Measurable quantity of a nanosolid |
| SAW | Surface acoustic wave |
| SEM | Scanning electron microscopy |
| STM/S | Scanning tunneling microscopy/spectroscopy |
| SWCNT | Single-walled carbon nanotube |
| T-BOLS | Temperature dependent Bond-order-length-strength correlation |
| $T_C$ | Critical/Curie temperature |
| TEM | Transition electronic spectroscopy |
| $T_m$ | Melting point |
| $U(T/\theta_D)$ | Atomic vibration/internal energy |
| UHV | Ultra-high-vacuum |
| UPS | Ultraviolet photoelectron spectroscopy |
| $v_i$ | Atomic volume of the specific ith atom |
| VLEED | Very-low-energy electron diffraction |
| XANES | X-ray absorption near edge spectroscopy |
| XPD/S | X-ray photoelectron diffraction/spectroscopy |
| Y | Young's modulus |

# I Introduction

## 1.1 Scope

The report starts in Section 1 with a brief overview on the mechanically unusual behavior of the mesoscopic systems including monatomic chains (MCs), nanotubes (NTs), nanowires (NWs), solid and liquid surface skins, nanocavities, and solids in nanometer and micrometer regimes. A brief summary will be given outlining existing questions and emerging problems that are challenging for consistent understanding with analytical expressions. Section 2 presents theory considerations of the local bond average (LBA) and the often-overlooked effect of bond broken on the behavior of the remaining bonds of the under-coordinated atoms that follows the recently developed bond-order-length-strength (BOLS) correlation mechanism.[1] In order to deal with the size and temperature dependence of the mechanical properties of the mesoscopic systems, the BOLS correlation has been extended to include temperature, as the thermal stimulus affects directly the length and strength of the bond through thermal expansion and thermal vibration. Sections 3-9 extend the analytical solutions to the functional dependence of mechanical properties in various situations on the bonding identities and temperature dependence. Solutions have been applied to typical situations on the size and temperature dependence of the intrinsic behaviors of elasticity and extensibility and the plastic deformation where extrinsic factors may dominate. Wherever possible, existing modeling considerations and understandings from various perspectives are comparatively discussed. The survey and the LBA analysis showed consistently that the shortened and strengthened bonds of the less-coordinated atoms dictate the behavior of a mesoscopic substance, being quite different from that of the isolated constituent atoms or that of its bulky counterpart. Agreement between predictions and experimental observations on the mechanical properties of monatomic chains, solid and liquid surface skins, nanotubes, nanowires, nanocavities, and the inverse Hall-Petch relationship (IHPR) has been realized with improved understanding of the commonly intrinsic origin behind the observations from the perspective of LBA in the cases of broken bond dominance. Chemisorption induced surface stress and its effect on the surface tension of metallic liquid are discussed as well from the perspective of charge repopulation and polarization upon adsorbate bond making. The LBA treatment in terms of bonding energetics has led to quantitative information about bonding identities and an improved understanding of the factors governing the intrinsically mechanical performance of mesoscopic systems. Artifacts are emphasized very important in the indentation test of plastic deformation and hence the observed inverse Hall-Petch relationship is suggested to arise from the competition between the intrinsic and the extrinsic contributions. The negatively curved surface in nanocavities is naturally the same to the flat surface or the positively curved surface of a nanosolid in determining the mechanical properties and thermal stabilities of the porous structures. Atomic vacancies of nanocavities play dual roles in determining the mechanical properties. The broken bond induced strain and trapping surrounding the defects serve as centers inhibiting motion of atomic dislocations yet the pores provide sites initialing structure failure in plastic deformation. The interface bond strain and the associated pinning or trapping and the bond nature alteration upon alloy or compound formation should be responsible for the mechanical strengthening of the twin grains and interface mixing. Section 10 presents a discussion on the attainments and limitations of the present approaches. The prospects of further extension of the developed approaches to atomic defects, impurities, adsorbed surfaces, liquid surfaces, junction interfaces, and systems under other stimuli such as pressure, electronic and magnetic fields are briefly addressed. Some open problems and continuing challenges in particular for the plastic deformation are also highlighted in the last section.

## 1.2 Overview

### 1.2.1 Fundamentals

As the bridge between the atomistic and macroscopic scales, the mesoscopic systems have attracted tremendous interest in recent years because of their intriguing properties from a basic scientific viewpoint, as well as from their great potential in upcoming technological applications such as nanomechanoelectronic devices.[2] The significance of a mesoscopic specimen is the tunability in physical properties compared with the corresponding more bulky samples of which the portion of the under-coordinated surface atoms is negligible. The coordination deficiency makes the mesoscopic systems differ substantially from the isolated atoms of their constituent elements or the corresponding bulk counterparts in performance. Because of the reduction of mean atomic coordination numbers (CNs), mesoscopic systems display novel mechanical, thermal, acoustic, optical, electronic, dielectric, and magnetic properties.[1,3,4,5] Unfortunately, the unusual behavior of a nanostructure goes beyond the expectation and description of the classical theories in terms of the continuum medium mechanics and the statistic thermodynamics. The quantities such as the Young's modulus and the extensibility of a solid remain no longer constant but change with the solid size. In general, the mechanical properties of a solid vary with the temperature and pressure of operation. Thermal softening and pressure hardening are very common. It is fascinating that the new degree of the freedom of size and its combination with temperature or pressure not only offer us opportunities to tune the physical properties of a nanosolid but also allow us to gain information that may be beyond the scope of conventional approaches.

In dealing with the mechanical behavior of the mesoscopic systems, the following concepts could be of importance:

- Surface energetics can be categorized as follows: (i) excessive energy stored per unit area of the surface skin of a certain thickness, (ii) residual cohesive energy per discrete atom at the surface upon bond breaking and, (iii) the conventional definition of surface energy that refers to energy consumed (loss) for making a unit area of surface.
- Surface stress (P) is the change of surface energy with respect to surface strain, corresponding to the first order differential of the binding energy with respect to volume. Surface stress, being the same in dimension (in unit of $J/m^3$ or $N/m^2$) to surface energy and hardness (H), reflects intrinsically the internal energy response to volume change at a given temperature. Hardness is the ability of one material to resist being scratched or dented by another in plastic deformation. The stress often applies to elastic regime while the hardness or flow stress applies to plastic deformation where creeps, grain glide, dislocation movements, and strain gradient work hardening are competitively involved.[6,7]

- Elastic bulk modulus is the second order differential of the binding energy with respect to volume strain and is proportional to the sum of binding energy per unit volume from the perspective of dimensionality. Differing from the definition of bulk modulus, Young's modulus gives the elastic response of a material to an applied uni-axial stress and is therefore directionally dependent on the orientation of the defect structure and/or crystal. Young's modulus represents the stiffness of the material that correlates with the atomic structure. The Y value also relates to other quantities such as Debye temperature, sound velocity, specific heat at constant volume, and the thermal conductivity of a substance.
- Compressibility (β, also called extensibility) is theoretically proportional to the inverse of modulus. The stiffness of a specimen refers to its elastic strength that is the product of Young's modulus and the thickness of the specimen; the toughness of a specimen refers to its plastic strength that involves activation and inhibition of atomic dislocation, bond unfolding, grain gliding and work hardening during deformation. A specimen that is stiff may not be tough, and vise versa, although both the elastic and the plastic strength are in principle proportional to the binding energy density.
- Surface tension referring to the surface energy of a liquid phase is one of the important physical quantities that control the growth of a material on a substrate as well as different phenomena, such as coalescence, melting, evaporation, phase transition, crystal growth, and so on. Temperature dependence of the surface tension gives profound information in particular for the surfaces with adsorbed molecules or with multi components for alloying or compound forming.
- Critical temperatures ($T_C$) represent the thermal stability of a specimen such as solid-liquid, liquid-vapor, or ferromagnetic, ferroelectric, and superconductive phase transitions, or glass transition in amorphous states.

From an experimental point of view, the values of the bulk modulus B and pressure P can be measured by equilibrating the external mechanical stimulus to the responses of the interatomic bonding of the solid,

$$P = \frac{F}{A} = \frac{E}{V} \propto -\left.\frac{\partial u(r)}{\partial V}\right|_{r=d}; \quad B = \frac{P}{\Delta V/V} = V\left.\frac{\partial P}{\partial V}\right|_{r=d} \propto V\left.\frac{\partial^2 u(r)}{\partial V^2}\right|_{r=d}; \beta = \frac{\partial V}{V\partial P} = B^{-1}$$

$$[P] \propto [B] \propto [Y] \propto \frac{E_b}{d^3}$$

**(1)**

where the parameters P, F, V, and A correspond to stress, force, volume, and the area on which the F is acting. The function u(r) is a pairing potential for two atoms and r is the atomic distance. $E_b$ is the cohesive energy per bond at equilibrium atomic distance d. We used the proportional relation herewith because we will pursue the relative change of the P and B with respect to the given bulk values. According to the LBA (more details in section 2.1) approach,[8] the elasticity and stress can be related to the bond length (volume) and the bond energy of the representative bonds, as the nature and the total number of bonds for a given specimen do not change during measurement unless phase transition occurs.

From an atomistic and dimensional point of view, the terms of B, Y, and P as well as the surface tension and surface energy are the same in dimension ($Jm^{-3}$) because they are all intrinsically proportional to the sum of bond energy per unit volume. The numerical expressions in Eq (1) apply in principle to any substance in any phase and processes including elastic, plastic, recoverable, or non-recoverable deformation without contribution from extrinsic artifacts. The fact that the hardness for various carbon materials, silicon, and SiC[9] varies linearly with elastic modulus may provide evidence for this concept. . Nanoindentation revealed that the hardness and modulus of Ni films are linearly dependence.[10] However, counter examples may be observed such as polycrystalline metals whose Young's modulus is essentially independent of grain size, but whose hardness varies following the inverse Hall-Petch relation (IHPR)[11] because the involvement of artifacts as extrinsic factors in the contact measurement for the latter. Artifacts such as purity, strain rate, load scale and direction, will come into play as an addition. The artifacts are unavoidable in some measurement methods such as nanoindentation or the Vikers microhardness test in particular nanograins. It will be shown in section 7.5 that contributions of such artifacts are more significant to nanograins than to the surfaces of thin films. Therefore, the measured results are a collection of intrinsic and extrinsic changes of the mechanical performance, which make it difficult to discriminate intrinsic information from extrinsic contributions to the mechanical behavior of the mesoscopic systems.

Many techniques have been developed to measure the Young's modulus and the stress of the mesoscopic systems.[12] Besides the traditional Vickers microhardness test, techniques mostly used for nanostructures are tensile test using an atomic force microscope (AFM) cantilever, a nanotensile tester, a transmission electron microscopy (TEM) based tensile tester, an AFM nanoindenter, an AFM three-point bending tester, an AFM wire free-end displacement tester, an AFM elastic-plastic indentation tester, and a nanoindentation tester. Surface acoustic waves (SAWs), ultrasonic waves, atomic force acoustic microscopy (AFAM), and electric-field induced oscillations in AFM and in TEM, are also used. Comparatively, the methods of SAWs, ultrasonic waves, field-induced oscillations, and AFAM could minimize the artifacts because of their nondestructive nature though these techniques collect statistic information from responses of all the chemical bonds involved. A comprehensive review on the *in situ* microscopic measurement of the elastic modulus of nanowire, and nanotubes has been given by Han et al.[13]

### 1.2.2 Challenges

A huge experimental database has been generated in mesoscopic mechanics in past decades. As the mesoscopic mechanics is an emerging field of study, fundamental progress is lagging far behind the experimental exploitations. Many questions and challenges are still open for discussion. A few typical samples regarding the mechanical puzzles of the mesoscopic systems are summarized in the following.

(i) Size dependence of elasticity and strength

Measurements have revealed that the elasticity and the strength of a nanosolid changes with solid size exhibiting three seemingly conflicting trends, as summarized in Table 1. Sophisticated theoretical models have been developed from various perspectives to explain the intriguing mechanical performance of the mesoscopic systems. For instance, the elastic response and mechanical strength of nanostructures have been attributed to the nonlinear effects,[14] surface reconstruction[15] and relaxation,[16] surface stress or surface tension,[17,18] surface strain,[16] dislocation starvation,[19,20] the stochastic of dislocation source lengths,[21] mismatch stress,[22] and stronger bonds of the less-coordinated atoms.[23] Large-scale atomistic simulations[24,25] of the plastic deformation of nanocrystalline materials suggest that both the inter- and intra-granular deformation processes under uni-axial tensile and nano-indentation are leading conditions. In the scoping study, [26,27] various parametric effects on the stress state and kinematics have been quantified. The considered parameters include crystal orientation (single slip, double slip, quadruple slip, octal slip), temperature, applied strain rate, specimen size, specimen aspect ratio size, deformation path (compression, tension, shear, and torsion), and material (Ni, Al, and Cu). Although the thermodynamic force (stress) varies at different size scales, the kinematics of deformation is found to be very similar based on atomistic simulations, finite element simulations, and physical experiments. Atomistic simulations, that inherently include extreme strain rates and size scales, give results that agree with the phenomenological attributes of plasticity observed in macro scale experiments. These include strain rate dependence of the flow stress into a rate independent regime, approximate Schmidt type behavior; size scale dependence on the flow stress, and kinematic behavior of large deformation plasticity. However, an atomistic understanding and analytical expressions in terms of the intrinsically key factors for the size and temperature dependence of the strength and extensibility of nanostructures are yet lacking though a recent molecular dynamics (MD) simulation[28] suggests that surface atoms play an important yet unclear role in mesomechanics.

Table 1 A summary of the experimentally observed changes in elastic modulus or mechanical strength at a surface or for a nanosolid upon size reduction with respect to the bulk values.

| Observed trends | Specimens and references | Methods used |
|---|---|---|
| Hardening | TiCrN,[29] AlGaN,[30] a-C, and a-C:N surface[31,32] | Nanoindentation |
| | Ni,[33,34] Ag, Ni, Cu, Al $\alpha_2$-TiAl and $\gamma$-TiAl surfaces[35] | |
| | Au and Ag films[36] | |
| | TiC, ZrC, and HfC surfaces[37] | |
| | Nanograined steel[38] | AFM |
| | ZnO nanobelts[39,40,41] wires[42], and surface[43,44] | |
| | Ag wires[45] | SAW |
| | Poly(L-lactic acid) (PLLA) fibers[46] | DFT |
| | SiTiN fibers[47] | AFM |
| | Au-Au bond[48] | |
| | SWCNT, MWCNT, and SiC wire[49] | |
| | CNT spun fibers[50] | Nanoindentation |
| | Ag and Pd wires,[51] | |
| | GaN wires[52] | |
| Softening | Ni surface[50] | AFAM |
| | Polymer surface[53] | AFM |
| | polystyrene surface[54] | |
| | ZnO nanobelts/wires[55,56,57] | AFM three-point bending |
| | Cr[58] and Si[59] nanocantilevers | AFM |
| | ZnS nanofilaments[60] | Force deflection spectroscopy |
| Retention or irregularity | ZnS nanobelts hardness increase yet elasticity decreases.[61] | AFM |
| | Au wires[62] | |
| | $SiO_2$ wires[63] | |
| | Silver nanowires[64] | Nanoindenter |
| | 20-80 nm Ge wires[65] | |

(ii) Surface energetics, strength, and thermal stability

Normally, the surface skin of an inorganic solid is harder at temperatures far below the melting point ($T_m$) but the hard skin melts more easily in a shell-resolved manner.[66,67] However, surface hardening does not occur so often for specimens with lower $T_m$ values. In comparison, the surface of a liquid solidifies first associated with lattice contraction and crystallization in the outermost atomic layers.[68] Although a critical-depth mechanism[69,70] has been developed to ascribe the surface hardening as a surface effect, strain gradient work hardening, and non-dislocation mechanisms of deformation,[71] these phenomena appear beyond the scope of classical theory considerations in terms of entropy, enthalpy, or free energy. An atomic scale understanding of the origin of surface tension and its temperature and adsorbate dependence is highly desirable.[72,73] Furthermore, conventional definition of surface energetics with involvement of classical statistic thermodynamics and continuum medium considerations for larger scales may need revision because the discrete quantized nature in mesoscopic scales becomes dominant.

(iii) Thermally induced softening

Thermally-induced softening of a substance has been widely seen for substances disregarding the shape and size. Generally, when the testing temperature is raised, the compressibility/extensibility of the solid increases rendering the mechanical strength as observed in the cases of nanograined Al[74] and diamond films.[75] At higher temperatures, the bending stiffness and the apparent Young's modulus of the diamond beams are drastically reduced to one third of the initial value before fracture. The flexural strength and the modulus of the hydrosilylated and condensated curable silicone resins also decrease when the testing temperature is raised.[76] The yield strength of Mg nanosolid[77] of a given size drops when the temperature is increased. An atomic-scale simulation[78] suggests that the material becomes softer in both the elastic and the plastic regimes as the operating temperature is raised. When measured at 200 °C, the strength of the 300-nm-sized Cu nanograins is lowered by 15% and the ductility increases substantially.[79] The biaxial Young's modulus of Si(111) and Si(100) was measured to drop linearly when the T is increased. [80,81] The Young's modulus of the TiN/$Mo_xC$ multilayer films drops when the temperature is raised from 100 to 400 °C though the modulus increases with the decrease of the modulation period.[82] An MD investigation suggested that the longitudinal Young's modulus and the shear modulus for both the armchair and the zigzag nanotubes change in different trends over the temperature range of 300 - 1200 K. The Y value drops while the shear modulus increases as the temperature is increased.[83]

Detailed examinations have shown that the measured Y values drop nonlinearly at very low temperatures and then follow a linear relation at higher temperatures.[8,84] When the operation temperature is increased from room temperature to 400 °C, the ductility of the ultrafine-grained $FeCo_2V$ samples of 100-290 nm size increases from 3 - 13% to 22% associated with strength reduction.[85] Superplasticity of individual single-walled CNTs has been observed at elavated temperatures.[86] The ductility of a nanosolid increases exponentially with temperature up to almost infinity at $T_m$. An analytical expression for the thermally driven softening and the enhanced ductility is yet lacking.

(iv) Monatomic chain forming and breaking

A monatomic chain (MC) is an ideal prototype for mechanical testing as no processes of bond unfolding or atomic glide dislocating are involved in the deformation. It is intriguing that at 4.2 K and under UHV conditions, the breaking length of an Au-Au bond in the Au-MC is measured to be 0.23 nm, being 20% shorter than the equilibrium Au-Au distance of 0.29 nm in the bulk. [87] At room temperature, the breaking limits were measured to vary from 0.29 to 0.48 nm.[88] However, the controllable formation of a MC of other metals is rare. Theoretical reproduction of the scattered data of measurement, in particular, the extreme values of 0.23 and 0.48 nm, have been hardly possible although the mechanisms of fuzzy imaging,[89] atomistic impurity mediation,[2] and charge mediation[90] in the state-of-the-art computational approaches have allowed some progress to be made.

(v) Stiffness and stability of nanotubes and nanowires

Carbon nanotubes and compound nanobeams such as ZnO and SiC nanowires exhibit extremely high strength yet relatively lower thermal stability compared to their bulk counterparts. The elastic modulus of the single-walled CNT (SWCNT) was measured to have values varying from 0.5 to 5.5 TPa depending on the presumption of the wall thickness of the CNT. The Young's modulus of the multi-walled CNTs (MWCNTs) drops with the inverse number of walls (or wall thickness) and it is less sensitive to the outermost radius of the MWCNTs if the wall thickness remains unchanged.[91] Atoms in the open edge of a SWCNT coalesce at 1593 K and a ~280% extensibility of the CNT occurs at ~2000 K. Under the flash of an ordinary camera, the SWCNT burns under the ambient conditions. The mechanisms behind the Y value enhancement, the $T_m$ suppression, and the high-temperature superplasticity of the CNTs are still puzzling. The uncertainty in the wall thickness and the Young's modulus of the C-C bond in the SWCNTs have long been issues of challenge, although the atoms that surround defects or are located at the tip ends or at the surface are expected to play some unusual, and yet unclear, roles in dominating the mechanical and thermal responses of the CNTs.[51] A consistent insight into the mechanism behind these observations from the perspective of under-coordination is necessary.

(vi) Inverse Hall-Petch relation

Under a tensile or compressive stress, the hardness or the plastic flow stress of crystals in the size range of 100 nm or higher is subject to the classical Hall-Petch relation (HPR).[11] The hardness increases linearly with the inverse square root of solid size. With further size reduction, the mechanical strength of the solid continues increasing but deviates from the initially linear HPR until a critical size of the strongest hardness, of the order of 10 nm. At the critical grain size, the slope of the IHPR curve will change from positive to negative and then the nanosolid turns to be softer. Although there is a growing body of experimental evidence pointing to the unusual IHPR deformation in the nanometer regime, the underlying atomistic mechanisms behind are yet unclear.[11,51] As pointed out by Kumar et al[27] and Mayrhofer et al, [92] the physical origin of the IHPR transition has been a long-standing puzzle and the factors that dominate the critical size at which the HPR transits are far from clear. The HPR-IHPR transition seems to be a topic of endless discussions because of the competition between the intrinsic contributions and the extrinsic artifacts that involve the activation and prohibition of grain boundary sliding due to the difficulty of partial dislocation movement.

(vii) Vacancy and nanocavity induced hardening and melting

It is expected that atomic vacancies reduce the number of chemical bonds and hence the strength of a porous material. However, the hardness of a specimen does not follow this simple picture of coordination counting. Vacancies not only act as pinning centers inhibiting dislocation motion and thus enhancing the mechanical strength within a certain concentration but also provide sites initiating structure failure. An introduction of a limited amount of atomic vacancies or nanocavities could indeed enhance the mechanical strength of the porous specimen. Atomic vacancies or discretely distributed nanometer-sized cavities could not only enhance the mechanical strength of the specimen, but also cause a substantial depression of the temperature of melting. Metallic foams of 40~60 % mass density are several times harder yet lighter compared with the standard materials. However, excessive amount of cavities or large pores are detrimental to the mechanical strength of the specimen. For instance, the Young's modulus of a defected nanotube is reduced gradually with each atomic defect and the plastic strength of the nanotubes is catastrophically influenced by the existence of just a few atomic defects.[93] Understanding the discrepancy between expectations and observations of cavity-induced hardening and melting is also a challenge.

(viii) Adsorbate-induced surface stress

The surface tension of a liquid drops linearly then the measuring temperature is raised. However, contamination or adsorption may change the slope of the temperature coefficient. Adsorbate bond making at a solid surface could alter the surface stress in various ways. For instance, hydrogen addition could embrittle the metals and C addition usually induces compressive stress at the surface. Even on the same surface, different adsorbents such as C, N, O, S, and CO, result in different kinds of stresses. The adsorbate-induced stress may change its sign with the coverage of the specific adsorbate. A specific adsorbate may induce different kinds of stress at different faces of the same material. An understanding of the adsorbate - induced surface stress and its correlation to the adsorbate-induced slope inflection of the T-dependent surface tension from the perspective of charge polarization and repopulation upon bond making is necessary.

(ix) Interface and nanocomposite

Mechanical strengthening of a material can occur by using multilayer formation of different kinds of compounds or by composite formation with nanostructured fillers such as carbon nanotubes, fibers, or clays inserted into the polymer matrix. The roles of the interface mixing, the dissociated interface skins, or the hard infillers are yet unclear in the process of nanocomposite reinforcement.

(x) Limitations of the classical and quantum approximations

The physical properties of a macroscopic system can be well described using classical approaches in terms of the Gibbs free energy or the continuum medium mechanics, for instances, that relate the detectable quantities directly to the external stimulus such as the temperature T and entropy S, pressure P and volume V, surface area A and surface energy $\gamma$, chemical potential $\mu_I$ and composition $n_i$, chrge q and electric field, magnetic momentum $\mu_B$ and magnetic field, etc, without needing consideration of atomistic origin:

$$G(T, P, A, n_i, E, B,..) = \Omega(ST, VP, \gamma A, \mu_i n_i, qE, \mu_B B,...)$$

At the atomic scale, quantum effect becomes dominant and the physical properties of a small object can be reliably optimized in computations by solving the Schrödinger equations for the behavior of electrons or the Newtonian motion of equations for the atoms with a sum of averaged interatomic potentials as key factors to the single body systems:

$$H_i \quad = T_i + v_i(r) + V_{crystal}(r + R_{ij})$$

$$F \quad = -Grad\left[V_{crystal}(r + R_{ij})\right] = Mr''$$

Where H is the Hamiltonian and F is the force. $T_i$ is the kinetic energy, $v_i$ the intraatomic potential and $V_{Crystal}$ is the periordic crystal potential. $R_{ij}$ is the atomic distance. M is the massofthe atom.

However, for a small system at the manometer regime, both the classical and the quantum approaches encountered some difficulties. For instances, the statistics is conducted over a large N number of atoms with a standard deviation that is proportional to $N^{-1/2}$. Quantities such as the entropy, the volume, the surface energy, and the chemical potential of a particular element remain no longer constant but change with the solid size. Quantum approaches are facing the boundary-condition problems that are the core of nanoscience. The broken-bond induced surface trapping and the associated charge, energy, and mass densification in the surface skin play a role of significance.[1] In fact, the real system is atomic site anisotropic, kinetic, with strongly localizd features. Describing the effect of skin trapping using an average of interatomic potentials and under the periodic or free boundary conditions in quantum mechanical approaches may be too ideal. Therefore, as a complement to the classical and quantum theories, a set of analytical expressions from the perspective of local bond average for the size, temperature, and bond nature dependence of the intrinsic mechanical properties of a specimen are necessary.

## 1.3 Objectives

The main objectives of this report are as following:

(i) To survey recent progress in experimental and theoretical observations on the size and temperature dependences of the elastic and plastic deformation of mesoscopic systems including atomic chains, nanotubes, liquid and solid surface skins, nanocavities, nanograins, and nanocomposites to provide interested readers with the latest and comprehensive information.

(ii) To develop analytical solutions for predicatively reconciling the size, temperature, and bond nature dependence of the intrinsic mechanical behavior of systems from atomic chain to macro specimen as mentioned above from the perspective of bond formation, dissociation, relaxation, and vibration by extending the recently developed BOLS correlation[1] to the temperature domain. It could be possible to employ the approach of local bond average to connect the macroscopic properties of a specimen to the atomistic factors (e.g., bond nature, bond order, bond length, and bond strength) of the representative local bonds. We need to establish the functional dependence of the detectable quantities on the bonding identities and the response of the bonding identities to external stimulus such as coordination environment (BOLS effect), temperature (thermal expansion and vibration), and stress field (deformation and deformation energy). A combination of the LBA approach and the BOLS correlation could provide complementary to the classical approaches.

(iii) To gain atomistic insight into the origin of the size and temperature dependence of the mesoscopic mechanics in a "bottom up" way. Deeper insight into the consequences of bond making and breaking and a grasp of the factors controlling bond making and bond breaking are necessary to improve works in other fields such as nano and microelectronics, catalytic electronics, and biotechnology.

(iv) To find factors dominating the mechanical performance of the mesoscopic systems and the interdependence of various quantities from the perspective of bond making and bond breaking to provide guideline for nanomechanical device design. Discriminating the intrinsic contributions from the extrinsic artifacts involved in the indentation test may allow us to understand the correlation between the elastic and the plastic deformations. Importantly, besides the performance and its origin, we need to know the trends and the limitations of the changes.

(v) To elucidate information such as single bond energy, maximum strain, in particular, the length, strength, extensibility, breaking limit, specific heat, and melting point of the single bond in monatomic chains and CNTs, by matching the theoretical predictions to the experimental observations. The new degree-of-freedom of size not only allows one to tune the physical properties of a specimen but also provides us with an opportunity to gain information such as the energy levels of an isolated atom[94,95] and the vibration frequency of an isolated dimer. [96,97] A combination of the degrees of freedom of size and temperature may allow us to gain more information such the cohesive energy per bond in various systems. All these quantities are of elemental importance to surface and materials sciences.

## II Principles

### 2.1 Local bond average (LBA)

A bulk solid is formed by numerous atoms with bonds connected one to another. The involvement of interatomic bonding distinguishes the solid in performance from the isolated constituent atoms. For a given specimen whether it is crystal, non-crystal, or with defects or impurities being involved, the nature and the total number of bonds do not change under the external stimulus unless phase transition occurs. However, the length and strength of all the involved bonds will response to the applied stimulus. If the functional dependence of a detectable quantity on the bonding identities is established, one would readily be able to know the performance of the entire specimen under external stimulus by focusing on the response of the length and strength of the representative bonds at different sites or their average.

The LBA approach is substantially the same to the volume partitioning approximation implemented by Delph and co-workers[98,99] who have opened a way in improving the calculation of such local quantities through volume partitioning of the problem showing that volume averaging is the correct way to obtain physically meaningful stress and elastic properties of complex microstructures. They examined the problem of deriving the absolute values of local stress and associated elastic constants for a region containing a fixed number of atoms that is part of a large body. By means of an expansion of the local interatomic potential energy, they derived expressions for the local second Piola-Kirchoff stress. Using this approach, they have obtained good agreement with the suitably averaged continuum solutions in the far-field regime.

The LBA approach may represent the true situations of measurements and theoretical computations that collect statistic information from large number of atoms or bonds. Further more, compared with the measurement and computation, the LBA could discriminate the behavior of local bonds at different sites. Furnished with the LBA approximation, the difficulties encountered by classical and quantum approaches in particular for the small object could be readily solved.[8]

Unlike the volume partition approximation, the LBA approach seeks for the relative change of a quantity with the applied stimulus to the known bulk value. The LBA approach focuses merely on the performance of the local representative bonds disregarding the number of bonds in the given specimen. The presence of broken bonds, defects, impurities, or the non-crystallinity will affect the reference values of concern rather the nature of observations. Contribution from long-order interaction or the high-order coordinates can be simplified by folding them into the bonds of the specific atom to the nearest neighbors.

### 2.2 The BOLS correlation

The core ideas of the BOLS correlation mechanism are the following: [1]

(i) If one bond breaks, the neighboring ones become shorter and stiffer. Consequently, local strain and trapping are formed immediately nearby the potential barrier at site of the broken bond.

(ii) The local strain and trapping localizes and densifies charge, energy, and mass at the trapping site, which contribute not only to the local change of the static properties but also to dynamic behavior such as transport dynamics of electrons, phonons, and photons in the specimen.

(iii) The shortened and strengthened bonds and the associated trappings originate the tunability of a mesoscopic system that contains varied portion of the under-coordinated atoms of the entire specimen.

Figure 1 illustrates the BOLS correlation mechanism using a typical pair potential function. The shortened atomic distance at equilibrium and the deepened potential well represent the length and strength of a bond between the under-coordinated atoms. The BOLS is an extension of the "atomic CN – atomic size" correlation mechanism of Goldschmidt,[100] Pauling,[101] and Feibelman[102] to energy domain. The "atomic CN – atomic size" holds disregarding the nature of the specific chemical bond, sample dimension, or the structural phase of the substance.[103] The depressed potential well should serve as a trap for the under-coordinated atoms exhibiting the pinning effect[104] to inhibit atomic glide dislocations and hence enhances the mechanical strength locally.

The localization and densification of charge and energy modifies the atomic cohesive energy, electro-affinity, Hamiltonian, Young's modulus, and work function, which dictates the unusual tunability of mesoscopic systems in thermal stability, lattice dynamics, photonic performance, electronic structures, magnetism, dielectrics, and chemical reactivity because of the dominance of interaction between the under-coordinated atoms. Potential well depression also contributes to the transport dynamics of phonons, electrons, and photons because of the additional trapping sites near the edges.[105,106] Densification and localization of electrons with lowered binding energy in the traps have been observed as defect states,[107] chain end states,[108,109] terrace edge states,[110,111,112] and surface states.[113,114,115] Therefore, as a rule, the broken bond, or the dangling bond, may not itself contribute directly to the performance of a substance but its effect on the remaining ones of the under-coordinated atoms is indeed profoundly significant.

Recent development shows that bond contraction occurs also at an interface, liquid surface, and sites surrounding impurities. For instance, a 10% contraction of spacing between the first and second atomic surface layers has been detected in the liquid phase of Sn.[68] For the three alkalis, Li, Na, and K, first principle calculations suggested that there is a 7% contraction of the first interlayer spacing.[116] Liquid surface layering is quite often for metals.[117] For example, x-ray reflectivity measurement in the surface normal structure of the Hg liquid-vapor interface reveals that the layering amplitude decays into the bulk with a characteristic length of 0.3-0.35 nm. The layering for Ga is as high as 0.6 nm, being attributed to the formation of directional bonds that prevent the close packing achieved by the nearly free electron metals.[118] A three-wave Bragg-surface x-ray diffraction[119] revealed that the Au/GaAs(001) interface bonds change by −49% along the surface normal [001] direction, and a −27% and a 2% change in the $[\bar{1}\,\bar{1}\,0]$ and the $[1\,\bar{1}\,0]$ in-plane directions, respectively. Therefore, it is practical to extend the BOLS correlation mechanism to systems such as point defects, liquid surfaces, and junction interfaces as well as amorphous glass in which the bond-order loss may take place randomly. A detailed description of the BOLS correlation and its applications to the size dependence of nanostructures has been presented in Ref [1].

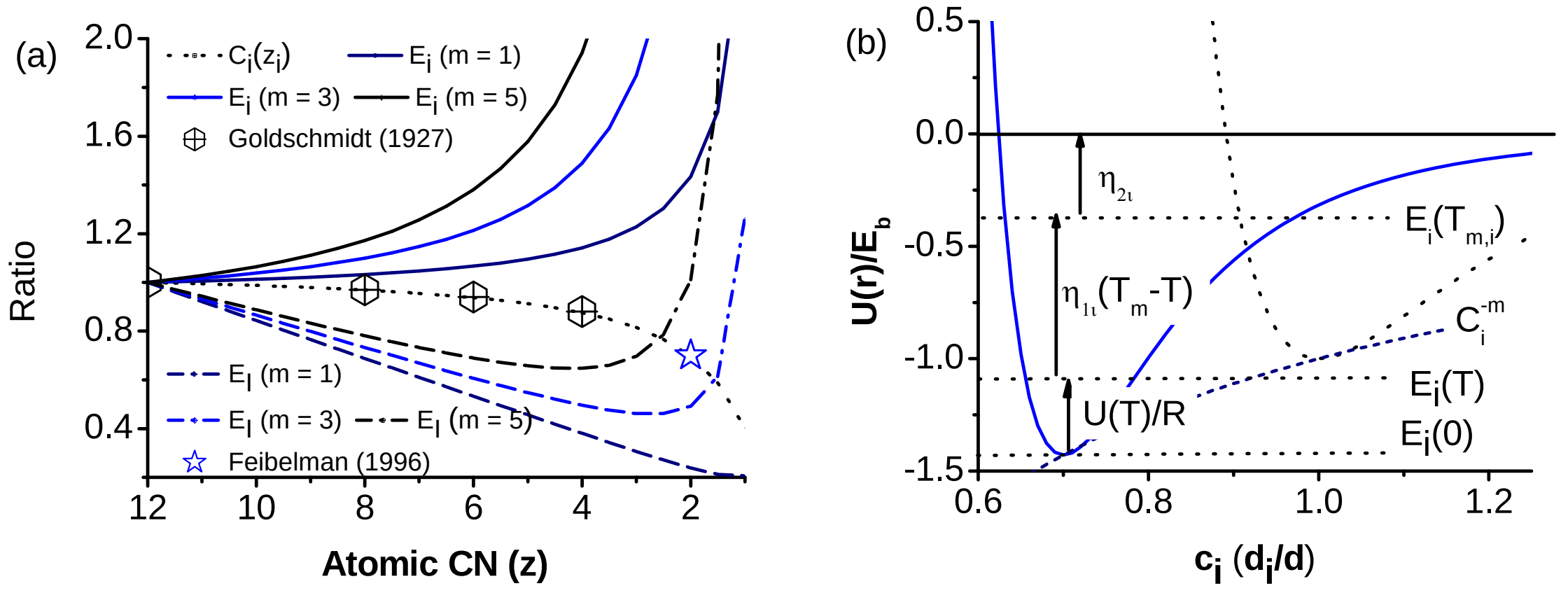

Figure 1 Illustration of the BOLS correlation. (a) CN dependence of the normalized bond length $d_i/d = C_i(z_i)$, and the CN and bond nature dependence of bond energy $E_i/E_b = C_i^{-m}$ and the atomic cohesive energy $E_I/E_B = z_{ib}C_i^{-m}$. The scattered data are from Goldschmidt and Feibelman [1]. (b) Atomic CN imperfection causes the remaining bonds of the under-coordinated atom to contract from one unit (in d) to $C_i$ and to increase the cohesive energy per coordinate from one unit (in $E_b$) to $C_i^{-m}$. Separation between the $E_i(T)$ and the $E_i(0)$ is the thermal vibration or internal energy, $U(T/\theta_D) = \int_0^T \eta_{1i}(t)dt \cong \eta_{1i}T$ . Separation between $E_i(T_{m,i})$ and $E_i(T)$, $\int_T^{T_{mi}} \eta_{1i}(t)dt \cong \eta_{1i}(T_{mi} - T)$ corresponds to the energy required for melting, which dominates the extensibility and the yield strength in plastic deformation. The separation between $E_i(T)$ and $E_i = 0$ (at evaporation) corresponds to energy for thermal or mechanical rupture of the pairing bond and the elastic modulus. $T_{m,i}$ is the melting point, which is proportional to atomic cohesive energy, $E_{C,i}$. $\eta_{1i}$ is the specific heat per bond and $\eta_{2i}$ is $1/z_i$ fold energy required for evaporating an atom in the molten state [1]. Link

The analytical form of the BOLS correlation is given as follows:[1]

$$
\begin{cases}
C_i(z_i) & = d_i/d = 2\{1+\exp[(12-z_i)/(8z_i)]\}^{-1} & (bond-contraction-coefficient) \\
E_i & = C_i^{-m}E_b & \begin{pmatrix} Single-bond-energy \\ Trapping-potential-well-depression \end{pmatrix} \\
E_I & = z_iE_i = z_{ib}C_i^{-m}E_b & (Atomic-cohesive-energy)
\end{cases}
\tag{2}
$$

Subscript i and b denote an atom in the ith atomic layer and in the bulk, respectively. The i is counted from the outermost atomic layer to the centre of the solid up to three, as no bond order loss occurs for i > 3. The $C_i$, being the bond contraction coefficient, varies with the effective atomic $CN(z_i)$. The $d_i$ is referred to the bond without specification of direction. The off-plane bond contraction is obvious yet the in-plane bond contraction will cause surface reconstruction with even cases of atom missing. The index m, however, is an indicator for bond nature of a specific material, which is not freely adjustable. Previous studies have revealed that, for elemental metals, m = 1; for alloys or compounds, m ~ 4; and for carbon and silicon, the m has been optimized to be 2.56[120] and 4.88,[121] respectively. For compounds, the m value should change in a continuous way that is to be clear by fitting a quantity change with both sample size and composition change. The m value has been found to increase when the $z_i$ is smaller than three for the IIIa and IVa elements.[122] The term, $z_{ib}C_i^{-m} = E_{C,i}/E_{C,b} = z_i/z_b \times E_i/E_b = 1+\Delta_i$, is the dimensionless form of atomic cohesive energy being normalized by the bulk value with $\Delta_i$ being the perturbation to atomic cohesive energy. Figure 2(a) illustrates the CN dependence of bond length and bond energy. The scattered symbols are from Goldschmidt and Feibelman.

The $z_i$ varies with the curvature of a solid surface (-) or a nanovoid surface (+):

$$
\begin{aligned}
z_1 &= \begin{cases} 4(1 \pm 0.75/K_j) & curved-surface \\ 4 & flat-surface \end{cases}; \\
z_2 &= z_1 + 2 \\
z_3 &= 12
\end{aligned}
\tag{3}
$$

The term $K_j = R_j/d$ is the dimensionless form of size, which corresponds to the number of $K_j$ atoms, with mean diameter or bond length d, lined up along the radius $R_j$ of a spherical-like nanosolid or cross the thickness of a thin film. For a spherical dot, the curvature is positive; for a spherical hollow, the curvature takes negative values, and $z_1 = 4 \times (1 + 0.75/K_j)$ with $K_j$ here being the radius of the hollow sphere.

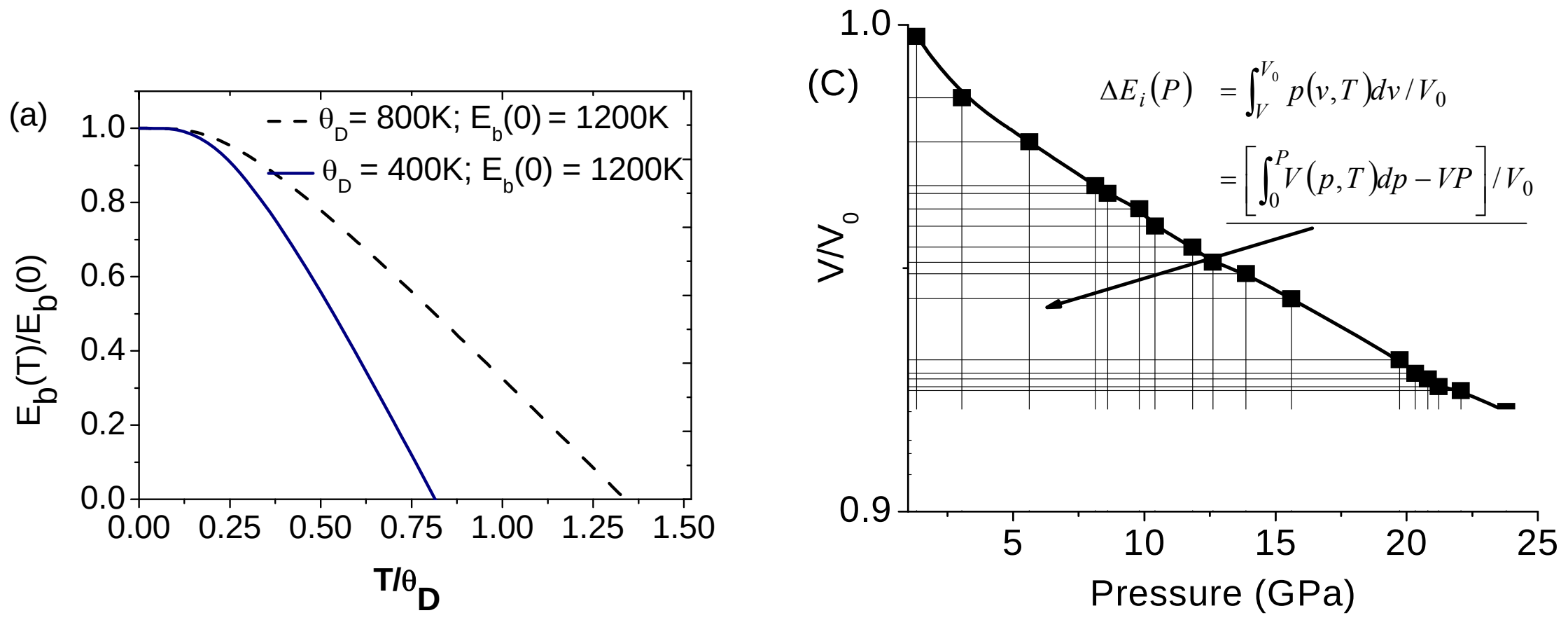

Figure 2 (a) Temperature dependence of bond energy $E_b(T)/E_b(0) = 1 - U(T/\theta_D)/E_b(0)$ with $U(T/\theta_D)$ being the thermal or internal vibration energy per coordinate in the bulk of Debye approximation. $E_b(T)/E_b(0)$ depends linearly on temperature at $T > \theta_D/3$ and the slope changes with the $E_b(0)$ values. (b) Pressure dependence of volume change and pressure induced cohesive energy elevation. The grid area in the typical V-P profile contributes to the cohesive energy elevation of the entire body.[123] link

2.3 Temperature dependent BOLS

2.3.1 Low temperature approximation

When the testing temperature is raised or a mechanical stimulus such as a tensile stress field is applied, the length and the strength of the representative bonds will change. If the effects of atomic CN deficiency, temperature, and stress come into play simultaneously, the length, $d(z_i, T, P)$, and the net energy, $E_i(z_i, T, P)$, of the representative bond will change accordingly. Therefore, based on the LBA approach, we can extend the BOLS correlation to temperature and pressure domains, leading to the T-BOLS correlation,

$$\begin{aligned} d_i(z_i,T,P) &= d_b(z,0,0)\times C_i \times\left(1+\int_0^T \alpha_i(t)dt\right)\times\left(1+\int_0^P \beta_i(p)dp\right) \\ E_i(z_i,T,P) &= E_b(z,0,0)\times C_i^{-m} - \int_0^T \eta_{1i}(t)dt + \left[\int_0^p v_i(z_i,t,p)dp - v_i p\right] \end{aligned} \quad \mathbf{(4)}$$

where $\alpha_i(t)$ is the temperature dependent thermal expansion coefficient (TEC). $\beta_i = -\partial v/(v\partial p)$ is the compressibility ($p < 0$, compressive stress) or extensibility ($p > 0$ tensile stress) that is proportional to the inverse of elastic bulk modulus. $\eta_{1i}(t)$ is the T-dependent specific heat per bond that is assumed to follow Debye approximation, $c_v(T/\theta_D)$, for a $z_i$-coordinated atom. Generally, the thermal measurement is conducted under constant pressure the $\eta_{1i}(t)$ should be related to the $C_p$. However, only a few percent difference between the $C_p$ and $C_v$.[124] Figure 2a illustrates the temperature and Debye temperature dependence of the inner bond energy. In the V-P profile, as illustrated in Figure 2b, only the gridded part $\left[\int_0^p v_i(z_i,t,p)dp - V_i P\right]/V_0$ contributes to the cohesive energy of the entire body.[123] For the single bond, the atomic volume, $v_i(z_i, t, p)$, is replaced by the bond length, $d_i(z_i, t, p)$, and the p is replaced by the force, f. Evidence[125] shows that the $\beta_i(p)$ remains constant at constant temperature within the regime of elastic deformation and then the integration $\int_0^P \beta_i(p)dp = \beta_i P$ can be simplified, unless phase transition occurs[126,127] or the thermal effect is involved.

The $\alpha_i(t)$ depends non-linearly on temperature in the low temperature range and the $\alpha_i(t)$ increases with the feature sizes of nanostructures.[128,129] From the perspective of LBA, the $\alpha_i(t)$ for the representative bond can be derived from the differential of the thermal expansion relation,[130]

$L = L_0\left(1 + \int_0^T \alpha_i(t)dt\right)$, and,

$$\alpha_i(t) \cong \frac{dL}{L_0 dt} = \frac{1}{L_0}\left(\frac{\partial L}{\partial u}\right)\frac{du}{dt} \propto -\frac{\eta_{1i}(t)}{L_0 F(r)} = A(r)\eta_{1i}(t) \tag{5}$$

with $(\partial L/\partial u) = -F^{-1}$ is the inverse of interatomic force in dimension at a non-equilibrium atomic distance. The value $A(r) = (-L_0F(r))^{-1}$ can be obtained by matching theory to measurement. Therefore, the derived thermal expansion coefficient follows approximately the trend of Debye specific heat and increases with the CN reduction, being consistent with observations.[131,132]

According to a sophisticated modeling,[133] the lattice thermal expansion of a cubic crystal could be expressed in terms of the Gruneisen mode parameter, $\gamma_{qj}$, and lattice vibration in the frequency of $\omega_{qj}$,

$$\begin{aligned}
\frac{\Delta d}{d_0} = \alpha T &= \frac{\hbar}{3BV}\sum_{qj}\gamma_{qj}\omega_{qj}\left[n_B(\omega_{qj}) + \frac{1}{2}\right] \\
&\propto \begin{cases} \dfrac{2kT}{BV_C} < \gamma_{qj} > & (T > \theta_D) \\ \int_0^{\omega_D} < \gamma_{qj} > \omega^3\left\{[\exp(\hbar\omega/kT) - 1]^{-1} + 1/2\right\}d\omega & (else) \end{cases} \\
\gamma_{qj} &= -\frac{\partial Ln(\omega_{qj})}{\partial Ln(V)}
\end{aligned} \tag{6}$$

where B is the bulk modulus and V the volume. $V_C$ is the volume of the primary unit cell and $<\gamma_{qj}>$ the average of $\gamma_{qj}$ over all branches of the Brillouin zone. The $n_B(\omega_{qj})$ is the Bose-Einstein population function.

In fact, eqs (5) and (6) are substantially the same following the Debye or Einstein population. Normally, the $\alpha_i(t > \theta_D)$ is in the order of $10^{-(6\sim7)}$ $K^{-1}$. A reproduction of the measured thermal expansion of Si and some nitrides eq (5) is given in Figure 3. It is clear that the present form is much simpler and straightforward without the parameters of B, V, or $\gamma_{qj}$ being involved.

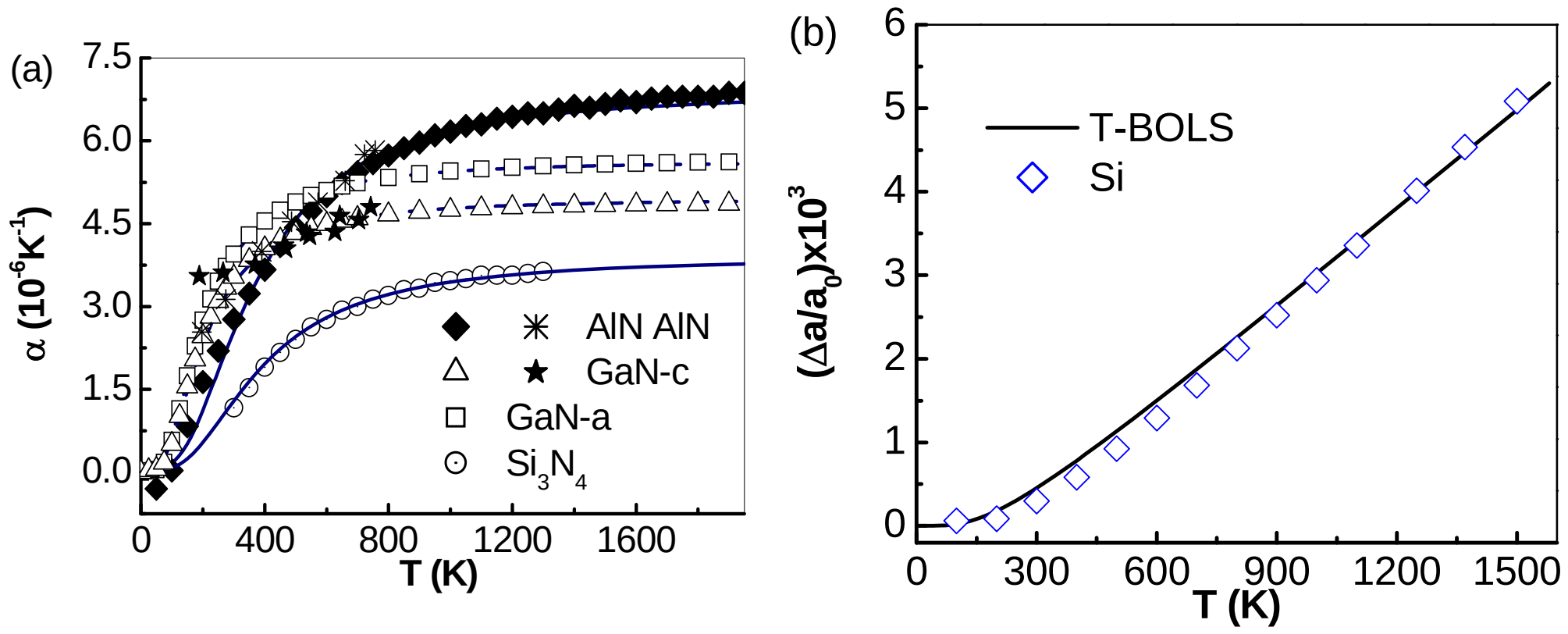

Figure 3 LBA reproduction (solid lines, Ref 130) of the temperature dependence of (a) the TECs of (a) AlN, (Ref 134) $Si_3N_4$ (Ref 134), and GaN (Ref 132, 135) and (b) thermal expansion of Si, which also agree with the model of eq (6) [133]. Link

The $\eta_{1i}(t)$ and the integration of $\eta_{1i}(t)$ with respect to T, or the conventionally termed internal or vibration energy, $U(T/\theta_D)$, follow the relation,

$$
\begin{aligned}
\eta_{1i}(t/\theta_D) &= \frac{c_v(T/\theta_D)}{z_i R} = \frac{\kappa^2}{z_i}\left(\frac{T}{\theta_D}\right)^{\kappa}\int_0^{\theta_D/T}\frac{x^{\kappa+1}\exp(x)}{(e^x-1)^2}dx \\
U_i(T/\theta_D) &= \begin{cases} = \int_0^T \eta_{1i}(t)dt & = \kappa\left(\frac{T}{\theta_D}\right)^{\kappa+1}\int_0^{\theta_D/T}\frac{x^{\kappa}dx}{e^x-1} \\ \cong \sum_{n>1}\frac{1}{n!}\left.\frac{\partial^n u(x)}{\partial x^n}\right|_{x=0} x^n & \sim \omega^2 x^2/2 + u'''(r)x^3/6 + 0\left(x^{n>4}\right) \end{cases}
\end{aligned}
\tag{7}
$$

with $x = \hbar\omega/k_BT$ = and κ the dynamic dimensionality considered in transport dynamics (for a spherical dot, $\kappa = 0$; for atomic chains and thin wires, $\kappa = 1$; for thin surface slabs, $\kappa = 2$; for large bulk, $\kappa = 3$). The $U_i(T/\theta_D)$ corresponds to the internal bond energy including all acoustic and optical modes of harmonic and anharmonic vibrations. The $U_i(T/\theta_D)$ is the amount of energy for bond weakening.

### 2.3.2 High temperature approximation

At temperature higher than the Debye temperature, the $\eta_1(T > \theta_D)$ equals unity. Therefore, the absolute value of $E_i(z_i, T, P)$ can be simplified as:

$$E_i(z_i,T,P) \underset{T>\theta_D}{=} E_i(z_i,0,0) - \eta_{1i}T + d_i(z_i,T,0)[1+\beta P/2]P$$
$$\cong \eta_{2i} + \eta_{1i}(T_{mi} - T) + d_i(z_i,T,0)[1+\beta P/2]P \quad (8)$$

$\eta_{2i}$ is the latent heat per coordinate of atomization for an atom in the molten state. At $T > \theta_D$, the integral will degenerate into the linear relation as given by Nanda,[136] $E_b(0) = \eta_2 + \eta_1 T_m$. Therefore, the magnitude of the net binding energy in eq (8) contains three parts: (i) the binding energy per bond at 0 K, $E_i(z_i, 0, 0)$; (ii) the internal or thermal energy, $U_i(T/\theta_D) = \eta_{1i}T$, and the deformation energy due to the field of stress, $d_i(z_i,T,0)[1+\beta P/2]P$. For an atom, the last two terms have to be revised to $z_i\eta_{1i}T$ and $d_i^3(z_i,T,0)[1+3\beta P/2]P$ because of the three-dimensional nature. It can be estimated that the deformation energy is rather small (at $10^{-1}$ to $10^{-2}$ eV levels) and it can be neglected in comparison to the bond energy in the order of $10^0$ eV. Figure 2(b) illustrates the temperature dependence of the reduced bond energy for an atom in the bulk. Considering the single bond, one has to divide the values by the atomic CN. The $E_b(0)$ and the $\theta_D$ are so important that they determine the slope of the $E_b(T)/E_b$-T curve and the transition point (about $\theta_D/3$) at which the $E_b(T)/E_b$ approaches to a linear temperature dependence.

It is emphasized that for the bond of an under-coordinated atom, we have a similar form to the standard case, $E_i = \eta_{2i} + \eta_{1i}T_{mi}$. Because of the relations, $T_m \propto z_b E_b$, $T_{mi} \propto z_i E_i$, and $E_i = c_i^{-m} E_b$, we have the relations of $\eta_{2i} + \eta_{1i}T_{mi} = c_i^{-m}(\eta_2 + \eta_1 T_m)$ and,

$$\frac{T_{mi}}{T_m} = z_{ib}c_i^{-m} = 1 + \Delta_i; \frac{\eta_{2i}}{\eta_2} = c_i^{-m}; \frac{\eta_{1i}}{\eta_1} = \frac{z_b}{z_i} = z_{bi} \quad (9)$$

for the quantities between the localized and the bulk standard. The $\Delta_i$ is the perturbation to atomic cohesion energy. This correlation will be used often in later discussions.

2.3.3 Factors dominating mechanical strength

As indicated in eq (1), the stress and modulus have the same dimension being proportional to the sum of binding energy per unit volume. According to the expression, mechanical elasticity strengthening only takes place once the bond energy increases and/or the bond length contracts. No other factors contribute intrinsically. Therefore, a superhard covalent crystal should be reached with the factors of higher bond density or electronic density, shorter bond length, and greater degree of bond covalency.[137]

According to Born's criterion,[138] the shear modulus disappears when a solid is in molten state. If Born's criterion holds, the latent heat of atomization, $\eta_2$, does not contribute to the elasticity. However, an elastic modulus should be present in the liquid and even in the gaseous phases because of the non-zero sound velocity in these phases. The sound velocity depends functionally on the elastic constant and the mass density [139]of the specimen in the form of $(Y/\rho)^{1/2}$. As we have demonstrated, the $\eta_2$ contributes indeed to the extensibility of atomic wires.[140] Measurements[141,142] revealed that the tensile strength of alloys drops from the bulk values to approximately zero when the temperature is approaching $T_m$. Therefore, Born's criterion may be extended to cover the extensibility and plastic yield strength.

Instead of the classical Gibbs free energy [$G(T, P, n_i, A) = U - TS + PV + \sum\mu_i n_i + \sigma A$], Helmholtz premise of free energies [$F(T, V, n_i, A) = U - TS + PV + \sum\mu_i n_i + \sigma A$], internal energy U(S, V), or enthalpy [H(S, P) = U + PV], we proposed the LBA approach in terms of bond length and bond energy and their response to atomic CN, temperature, and stress. We also use the reduced specific heat per bond, $c_v(T/\theta_D)$, without taking into account the classical statistic thermodynamic quantities such as entropy S that is suitable for a body with infinitely large number of atoms. Neither the chemical potential $\mu_i$ for the component $n_i$ nor the tension $\sigma$ of a given surface are needed in the current approach. Reproduction of the measured temperature dependence of: (i) the redshift of the Raman optical modes,[8, 97,143] (ii) surface tension,[144] (iii) elastic modulus,[8] and (iv) lattice expansion[145] may evidence the validity of the current approach of LBA for the thermally induced mechanical behavior of materials.

## 2.4 Scaling relations

### 2.4.1 LBA scaling relation

Generally, the experimentally observed size-and-shape dependence of a detectable quantity, Q, of a nanosolid follows a scaling relation based on the LBA consideration. The size-induced property changes generally with the inverse of the solid size, $K_j$:[1]

$$
\begin{aligned}
\frac{Q(K_j, T_0) - Q(\infty, T_0)}{Q(\infty, T_0)} &= \begin{cases} bK_j^{-1} & (measurement) \\ \Delta_{qj} & (theory) \end{cases} \\
\Delta_{qj} &= \sum_{i \le 3} \gamma_{ij} \left(\Delta q_i / q\right) \\
\gamma_{ij} &= \frac{N_i}{N_j} = \frac{V_i}{V_j} = \frac{\tau c_i}{K_j} \le 1
\end{aligned}
$$

For the joint effect of temperature and size, the scaling relation becomes,

$$
\frac{Q(K_j, T)}{Q(\infty, T_0)} = \frac{Q(K_j, T)}{Q(\infty, T)} \times \frac{Q(\infty, T)}{Q(\infty, T_0)} = \left(1 + \Delta_{qj}\right) \times \frac{Q(\infty, T)}{Q(\infty, T_0)} \tag{10}
$$

with $K_j$ being the dimensionless form of size, which is the number of atoms lined along the radius of the jth sphere or cross the thickness of jth thin plate. $N_i$ is the number of atoms and $V_i$ the partitioned volume of the ith atomic layer, respectively. $Q(K_j)$ and $Q(\infty)$ represent any measurable quantity for the same solid with and without consideration of the effect of broken surface bond. The q and $q_i$ correspond to the local density of Q inside the bulk and at the ith atomic site, respectively. The q(z, $d_i$, $E_i$) depends functionally on the bonding identities of bond order, length, and strength, and their coordination, bond nature, and temperature dependence. $T_0$ is the temperature of reference. The quantity $\Delta q_i = q_i - q$ generates the property change at the referred local site. The layer-counting premise in Figure 4(a) represents that atoms in the surface skins dictate the size-induced property change yet atoms in the core interior remain as they are in the bulk because of the broken bond induced nearby strain and trapping. The weighting factor, $\gamma_{ij}$, or the portion of the under-coordinated atoms of the whole solid, represents the geometrical contributions from the dimension ($K_j$) and the static dimensionality (τ) of the solid, which determines the speed of change with the shrinkage of the solid size. The static dimensionality τ represents the factor of shape such as a thin plate (τ = 1, and monatomic chain as well), a rod (τ = 2), and a spherical dot (τ = 3) of any size, which is different from the dynamic dimensionality defined in transport considerations. The $\sum_{i\le 3}\gamma_{ij}$ drops in a $K_j^{-1}$ fashion from unity to become infinitely small when the solid dimension grows from the atomic level to infinitely large. At $K_j < 3$, the performance of the surface atoms will dominate because at the smallest size the $\gamma_1$ approaches unity. At $K_j \le 0.75$, the solid will degenerate into an isolated atom. For a spherical dot at the lower end of the size limit, $K_j = 1.5$ ($K_j d = 0.43$ nm for an Au spherical dot example, or an fcc unit cell), $\gamma_{1j} = 1$, $\gamma_{2j} = \gamma_{3j} = 0$, and $z_1 = 2$, which is identical in situation to an atom in a monatomic chain despite the geometrical orientation of the two interatomic bonds. Actually, bond orientation is not involved in the current modeling consideration. Therefore, the performance of an atom in the fcc unit cell is identical to the same atom in a monatomic chain from the perspective of bond order loss. At the lower end of the size limit, the property change of a nanosolid relates directly to the behavior of the single bond, which forms the starting point of the current "bottom up" approach and covers the whole range of solid size. The variable T is separated from the size effect as all the bonds in the substance response to the temperature change.

### 2.4.2 Surface-to-volume ratio for solid and hollow systems

The difference between a positively curved surface and a negatively curved surface is nothing more than the surface-to-volume ratio, $\gamma_{ij}$, and the slight difference in the coordinating environment of the under-coordinated atoms. Considering a sphere of $K_j$ radius with (n + 1/2) spherical cavities of $L_j$ radius lined along the $K_j$ radius, as illustrated in Figure 4(b), the total number of voids is $4\pi(n+1/2)^3/3$. For the hollow sphere (n = 0) with only one void in the center, this expression needs revision. We can estimate the entire volume $V_0$ occupied by atoms is the volume of the sphere with removed voids. The sum of the volume of skins of the voids and the sphere surface is $V_i$. The $V_0$ and the $V_i$ are calculated as follows:

$$V_0 = \frac{4\pi}{3}\left[K_j^3 - \frac{4\pi}{3}\left(n+\frac{1}{2}\right)^3 L_j^3\right]$$

$$V_i = 4\pi\left[K_j^2 C_{io} + \frac{4\pi}{3}\left(n+\frac{1}{2}\right)^3 L_j^2 C_{ii}\right]$$

$C_{ii}$ and $C_{io}$ represent the bond contraction coefficient for atoms in the inner negatively curved skins of the cavities and atoms at the outer positively curved surface of the sphere, respectively. The ratio between the volume sum of the skins and the volume entirely occupied by atoms can be derived as,

$$\begin{aligned} r_{ij}(n, L_j, K_j) &= \frac{V_i}{V_0} = \frac{3}{K_j} 3 \frac{3C_{io} + 4\pi(n+1/2)^3 (L_j/K_j)^2 C_{ii}}{1 - 4\pi(n+1/2)^3 (L_j/K_j)^3} \\ &= \frac{3}{K_j}\begin{cases} \dfrac{3C_{io} + 4\pi(n+1/2)^3 (L_j/K_j)^2 C_{ii}}{3 - 4\pi(n+1/2)^3 (L_j/K_j)^3} & (Porous-sphere) \\ C_{io} & (Solid-sphere) \\ \dfrac{3C_{io} + 4\pi(L_j/K_j)^2 C_{ii}}{3 - 4\pi(L_j/K_j)^3} & (Hollow-sphere) \end{cases} \end{aligned} \quad (11)$$

The parameters of n, $L_j$ and $K_j$ are constrained by the relation: $2(L_j+1)(n+1/2) \le K_j - 2$ because of the allowed maximum number of cavities aligned along the radius $K_j$. $2(L_j+1)$ represents the diameter of the void including the one layer of surface skin and $K_j - 2$ the radius of the sphere excluding the surface skin. This expression covers situations of a solid sphere, a hollow sphere, and a sphere with uniformly distributed cavities of the same size. This relation can be extended to a solid rod, a hollow tube, and a porous nanowire as well.

With the derived $r_{ij}(n, L_j, K_j)$ relation and the given expressions for the q($z_i$, $d_i$(t), $E_i$(t)), one can readily predict the size, cavity density, and temperature dependence of the Q of a system with large fraction of under-coordinated atoms without involving hypothetic parameters.

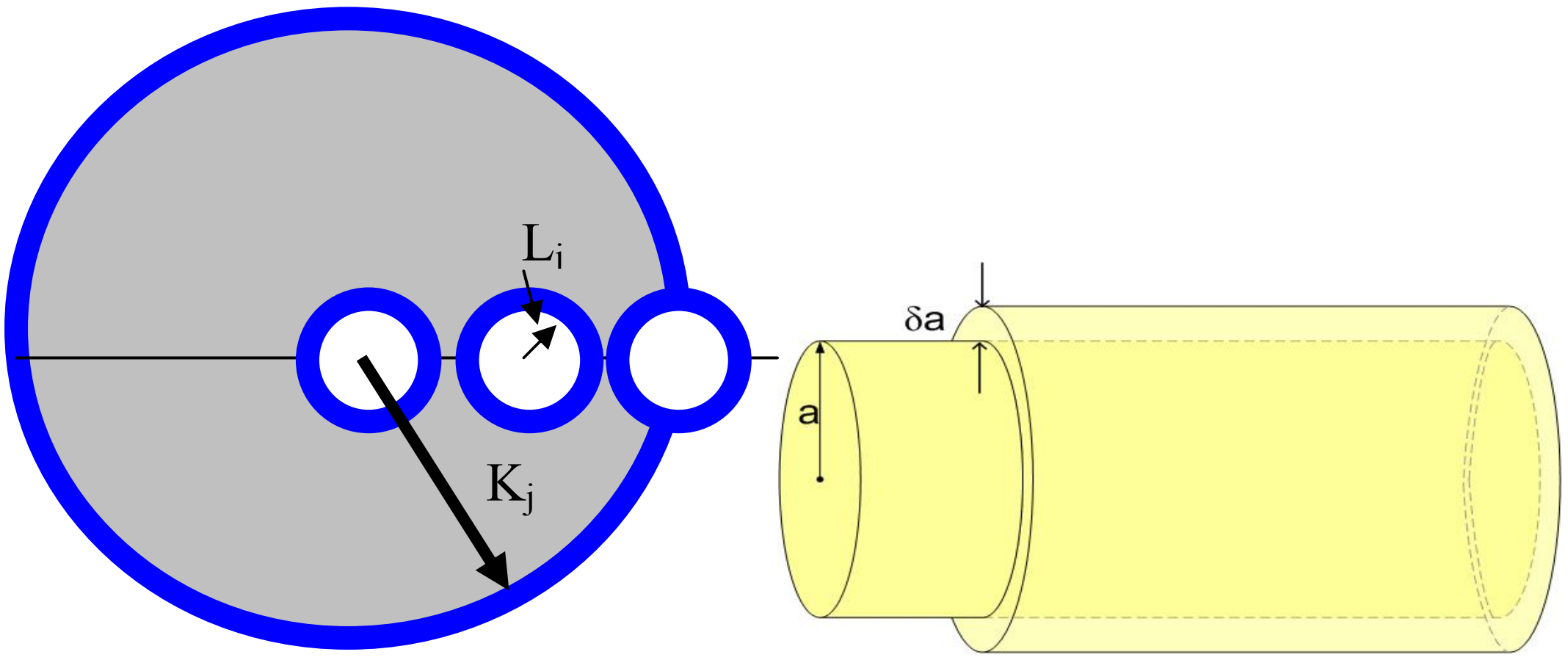

Figure 4 Schematic illustration of the surface-to-volume ratio of (a) a sphere with $4\pi(n+1/2)^3/3$ cavities and (b) the core-shell structure of a nanorods. Only atoms in the skin of $\Delta\alpha$ thick contribute to the property change yet atoms in the core region remain their bulk nature. link.

# III Liquid and solid surfaces

## 3.1 Observations

### 3.1.1 Surface energetics: classical concepts

- Surface free energy

Surface energetics including the terms of surface energy, surface free energy, surface tension, surface stress, and their correlations, plays the key and central role in surface and nanosolid sciences. Despite some confusion about these terms, the surface energetics is of great importance to a qualitative and sometimes even quantitative understanding of the microscopic and mesoscopic processes at a surface, as it links the atomistic bonding configuration at the interfacial region with its macroscopic properties, such as strength, elasticity, wettability, reactivity, diffusivity, adhesion, and so on.[146,147]

During the last few decades, tremendous attention has been paid to the energetically induced surface processes such as reconstruction, relaxation, interfacial mixing, segregation, self-organization, adsorption, and melting at the solid surfaces. Increased knowledge about the surface stress or surface energy of clean and adsorbate-covered metals has been established by the development of experimental and theoretical methods such as Raman shift, x-ray glancing diffractions, and the microcantilever sensors that allows the detection of extremely small amounts of substances in gases or liquids.[148] However, detailed knowledge about the dynamics of surface energetic identities is yet lacking,[146,147,149] in particular, the atomistic origin, temperature and adsorbate dependence, and the analytical description of the interdependence between surface energetic identities and the energetically driven phenomena and processes.

Traditionally, surface energy ($\gamma_s$), or surface free energy for a solid, is defined as the energy needed to cut a given crystal into two halves, or energy consumed (loss) in making a unit area of surface.[84] The surface energy or stress is involved in the Helmholtz or Gibbs free energy in the classical statistic thermodynamics, [150]

$$dF = -SdT - PdV + \sum \mu_i dn_i + \sigma dA \quad (Helmhotz)$$
$$dG = -SdT + VdP + \sum \mu_i dn_i + \sigma dA \quad (Gibbs)$$

with parameters of entropy (S), chemical potential ($\mu_i$) of the ith component ($n_i$), surface tension ($\sigma$), and the surface area A being involved.

Usually, un-relaxed structures at zero temperature are considered in the discussion of surface energies. The values obtained are then corrected for relaxations of the surface atoms, without mentioning reconstructions of the surface. In some cases, these corrections are thought very small, so a simplified model can be made without including relaxations. The temperature dependence involves the phonons and their modification on the surface; the vibrational effects sometimes have to be taken into account when the temperature dependence is studied.

The atomistic origin of the surface energy is usually explained as follows. A surface atom has fewer neighboring atoms and experiences fewer attractive interaction forces from its surroundings than an atom in the bulk interior. Consequently, the atoms at the surface experience a net force pointing to the bulk, or in other words, the potential energy of a surface atom is higher than that of a bulk atom because of the lowered atomic coordination. If one wants to create two new surfaces with a total area of A by cutting a solid at constant elastic strain ($\varepsilon_{ij}$) and constant temperature, one requires the surface energy,

$$\gamma_s = \left(F_S - F_0\right)_{\varepsilon_{ij},T} / A$$

where $F_S$ and $F_0$ are respectively the system free energy after and before the cut.

The simplest approach to get a rough estimate of the surface energy is to determine the number of bonds that have to be broken in order to create a unit area of surface. One can cut a crystal along a certain crystallographic plane and multiply this number of broken bonds with the energy per bond without considering the bond energy change caused by the reduction of atomic coordination. The typical approaches for the defined surface energy are comparatively summarized as follows:[147,151,152,153]

$$\gamma_s = \begin{cases} \dfrac{W_S - W_B}{20} n_d \left(n_d - 10\right) & (\text{Galanakis}) \\ \left(1 - z_s / z_b\right) E_B & (\mathit{Haiss}) \\ \left(1 - \sqrt{z_s / z_b}\right) E_B & (\text{Desjonqueres}) \\ \dfrac{\left[2 - z_s/z_b - \left(z_s/z_b\right)^{1/2}\right] + \lambda\left[2 - z'_s/z'_b - \left(z'_s/z'_b\right)^{1/2}\right]}{2 + 2\lambda} E_B & (\mathit{Jiang}) \end{cases} \tag{12}$$

- Galanakis et al[84] correlated the surface energy of some *d*-metals to the broken bond in the tight-binding approximation. $n_d$ is the number of d-electrons. $W_S$ and $W_B$ are the bandwidths for the surface and the bulk density of states, which are assumed in rectangular forms.
- Haiss et al[147] related the surface energy directly to the multiplication of the number of broken bonds and the cohesive energy per bond $E_b = E_B/z_b$ at 0 K. The $\gamma_s$ values are estimated by determining the broken bond number $z_{hkl} = z_b - z_s$ for creating a unit surface area by cutting a crystal along a certain crystallographic plane with a Miller index (hkl) where $z_s$ is the CN of a surface atom and $z_b$ the corresponding bulk atomic CN.

- A second-moment tight-binding approximation conducted by Desjonquères et al,[151] suggested that the surface energy gain is proportional to $\sqrt{z_S}$ , instead of $z_s$, because of the lowering of the occupied surface energy states or the surface-induced positive core-level shift as observed using XPS.[94] According to this approximation, the rearrangement of the electronic charge does not practically change the nature of the remaining bonds if one bond breaks. Thus, the energy needed to break a bond is independent of the surface orientation, so that the $\gamma_s$ value is proportional to the square root of the number of the nearest-neighboring bonds.
- In order to obtain a more general expression, Jiang et al[153] suggested that an average of the approximations of Haiss and Desjonquères and an extension to counting the contribution from the next nearest neighbors could be more comprehensive. The prime in the expression denotes the next-nearest neighbors of the surface atoms and λ is the total bond strength ratio between the next-nearest neighbor and the nearest one.
- In addition, Xie et al[154] derived an expression for the size dependence of surface energy of nanostructures: $\gamma_s = E_B / \pi d^2$ with d being the mean atomic radius and $E_B$ the atomic vacancy formation energy.[155]

Besides the thermodynamic considerations, the kinetic processes of lattice vibration play significant roles in the anisotropy of surface properties.[156] The amplitudes and frequencies of atomic vibrations,[96] as well as the bond lengths and strengths at a surface are different from their corresponding bulk values because of the effect of bond broken.[1] During film growth, the adatoms and atomic vacancies also contribute to the surface energy. With these contributing factors, the surface energetics becomes even more complicated.

- Surface stress

In contrast to the surface energy, which is related to energy change during the plastic deformation of a unit area of surface, the surface stress ($P_{ij}$) is related to energy changes during the elastic stretching of a pre-existing surface. Since the surface stress in general may be dependent on the direction of the stretching, the term $P_{ij}$ is a second-rank tensor and the stretching has to be expressed by the elastic strain tensor $\varepsilon_{ij}$.

$$P_{ij} = \frac{\partial \gamma_s}{\partial \varepsilon_{ij}}$$

where the $\gamma_s$ is the free energy at any strain. If $\gamma_s$ is lowered for negative strain values, the surface tends to shrink and the surface stress has a positive sign. In this situation, the stress is 'tensile'. If the surface tends to expand, the corresponding stress is 'compressive' with a negative sign.

According to Haiss,[147] the surface stress originates physically from the atomic CN reduction of surface atoms and the corresponding charge redistribution. The charge redistribution alters the nature of the chemical bonding and the equilibrium interatomic distances. Hence, the surface atoms attempt to assume their equilibrium interatomic distances and exert forces on the bulk, as long as the potential corrugation of the underlying atomic layer holds the topmost atoms in registry with the bulk lattice. The surface stress can be quantified as the sum of such forces.

It has been found[157] recently that the stress in CrN films varies with film thickness; see Figure 5, which is attributed to the sum of the tensile stress generated at the grain boundaries, compressive stress due to ion peening in deposition, and thermal stress due to the difference in thermal expansion between the coating and the substrate. The tensile part due to grain boundaries is thickness dependent. The other two contributions are thickness independent. Summation of the three components leads to a stress gradient in the coating.

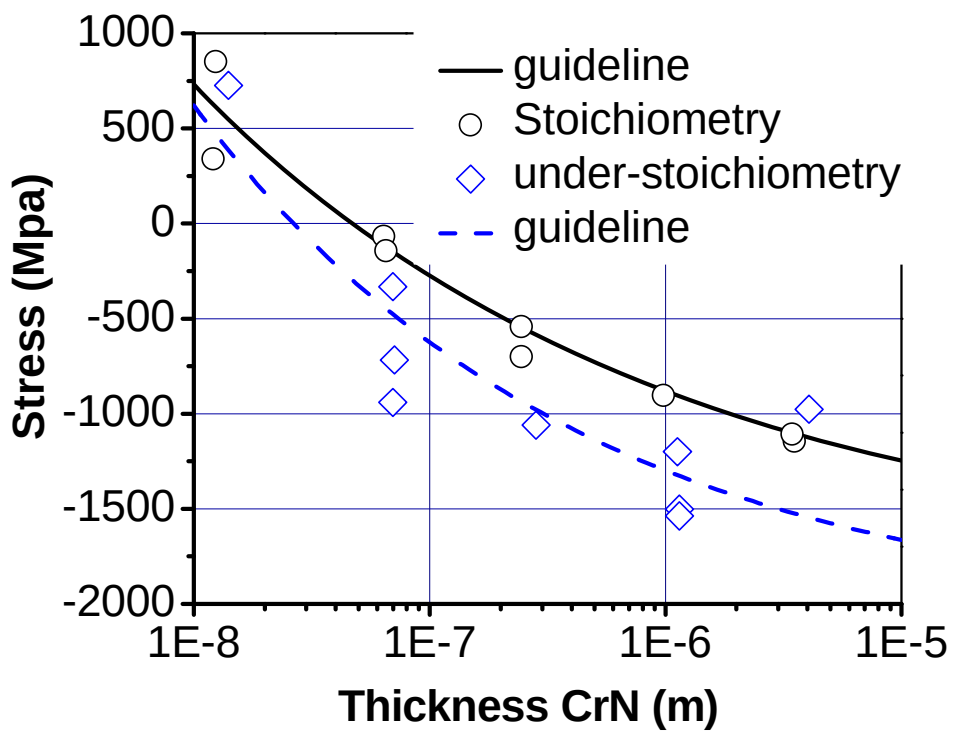

Figure 5 Thickness dependence of CrN film stress [157] showing dominance of the surface and the influence of chemical stoichiometry. Link

- Surface tension

Surface tension is expressed either as a force per unit length or as an energy per unit area of the interface between air and liquid, which results from the force of cohesion between liquid molecules or from the attraction of molecules to one another on a liquid surface. Thus, a potential barrier is created between the air and the liquid. The elastic-like force between surface molecules tends to minimize or constrict the area of the surface. In physics, surface tension is an effect within the surface layer of a liquid that causes the layer to behave as an elastic sheet with higher strength and elasticity of the liquid. However, it is unclear yet why the liquid skin is more elastic and stronger than the liquid core interior although the surface tension of an organic specimen was suggested[158] to be proportional to the product of molecular weight and the topological wiener index to a power of 3/2, $MW^{3/2}$. The topological winner index is the sum of all the shortest distance in the molecular crystallography and the $MW^{3/2}$ is treated as characterizing the moment of inertia of the rotational motion of the molecules.

### 3.1.2 Solid surface: skin hardening or softening

The surface elasticity and hardiness of a surface depend on several factors:[159] (i) the surface curvature, (ii) the nature of the bond involved and, (iii) the ratio between the operating temperature and the melting point. Experimental methods have an influence on the measured results. Normally, the surface of a solid is harder than the bulk interior at temperature far below the melting point but the hard surface melts more easily. For instances, as shown in Figure 6(a), the hardness measurements of Si surfaces with penetration depths as small as 1 nm yield $H \sim 25$ GPa, showing a drastic increase to 75 GPa with penetration depths to 5 nm compared with the bulk value of 12 GPa.[160] Similarly, the maximum hardness for nano- and micro-crystalline pure nickel films is also peaked at a penetration depth of ~5 nm.[33,34] The hardness of Ni films varies in a range from 6 to 20 GPa depending not only on the geometrical shapes (conical, Berkovich, and cube-corner) of the indenter tips but also on the strain rate in measurement. The peak position (5 nm in depth) changes with neither the shapes of the indenter tips nor the strain rates, indicating the intrinsic nature of surface hardening. The values of both the Y and the P for nitrogen-doped amorphous carbon (a-C) films[31] at the ambient temperature also show maximum hardness near the surface. The maximum hardness is 3 ~ 4 times higher than the bulk values whereas the peak positions in the hardness-depth profiles remain unchanged when the nitrogen content or film thickness changes. The hardness of TiCrN films at 5-10 nm depth[29] reaches a maximum of 50 GPa, being twice the bulk value. The same trend holds for a-C[32] and AlGaN[30] films with peaks positioned at several nanometers in depth that corresponds to the surface roughness. Surfaces of Ag, Ni, Cu, and Al thin films are 4 ± 0.5 times harder than the bulk interior and the hardness of $\alpha_2$-TiAl and $\gamma$-TiAl surfaces is ~2 times the corresponding bulk values.[35] The hardness of Ti, Zr, and Hf carbide films on silicon substrate increases from the bulk value of 18 to 45 GPa when the film thickness is decreased from 9000 to 300 nm.[37] The Young's modulus of nanograined steel was determined to increase from 218 to 270 GPa associated with a mean lattice contraction from 0.2872 to 0.2864 nm when the grain size is reduced from 700 to 100 nm.[38] These observations evidence the skin hardening effect arising from the shortened and strengthened bonds between the under-coordinated atoms. Furthermore, surface passivation with electronegative elements could alter the bond nature in the surface skins and hence the surface stress. A combination of the DFT and MD calculations[161] suggested that a surface can be softer or stiffer depending on the competition between electron redistribution and the atomic coordination on surfaces. Results show that the Young's modulus along <110> direction on (100) surface is higher than its bulk counterpart; meanwhile, it is smaller along <100> direction on (100) surface.

The tangent modulus of a CNT turf (a complex structure of intertwined nanotubes cross-linked by adhesive contact) was measured to be one order higher near the surface and the modulus drops with the penetration depth.[162] The modulus near the terrace edge is even harder than the flat surface; see Figure 6(b).

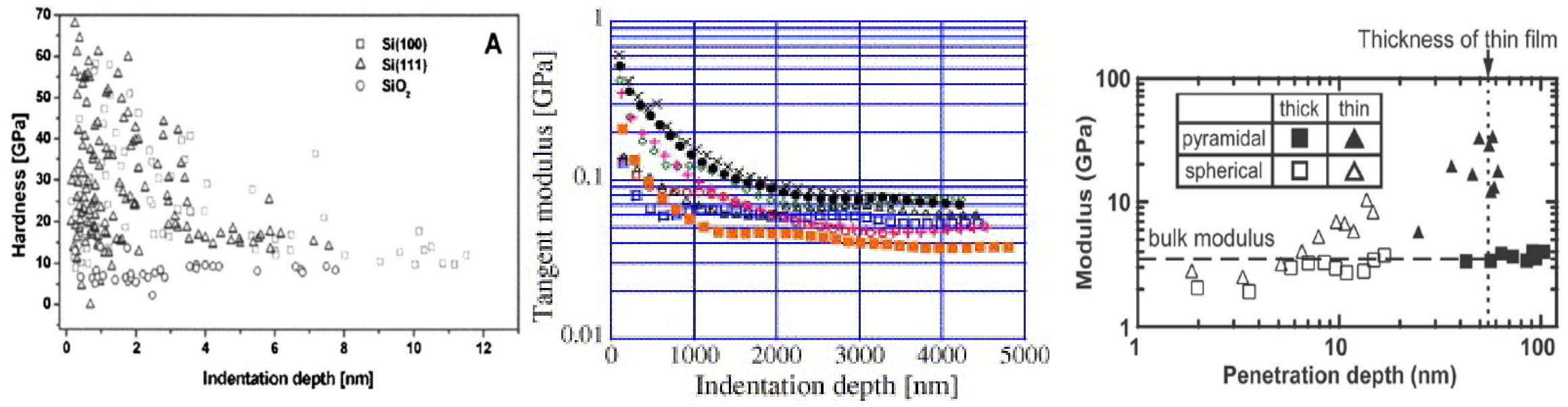

**(a)** **(b)** **(c)**

Figure 6 (a) Skin hardening of the Si(001), Si(111) and $SiO_2$ surfaces.[160] (b) the modulus-depth profile of a CNT turf showing one order higher in magnitude at the skins. Indentations are taken at different distances from the edge of the turf. The more compliant cases correspond to indentations closer to the edge [162]. (c) Skin softening of polymer surface,[54] measured by AFM indentation. Outlined and filled symbols correspond to the values estimated using the pyramidal diamond and spherical indenters, respectively for thick (squares) and thin (triangles) polymer films. Link

In contrast, skin softening occurs to some polymers or materials of lower melting points. An AFM room temperature measurement[53] has shown that the local Young's modulus of the organic thin films that can evaporate at ~150 °C decreases with particle size. The elastic modulus of polystyrene films increases with the penetration depth.[54] When the penetration depth is less than 5 nm, the elastic modulus of the films is smaller than that of the bulk and then approaches the bulk value when the depth is more than 10 nm, as shown in Figure 6(c). These observations suggest the significance of the difference between the temperature of melting and the temperature of testing to the measured values.

Ideally, an enhancement of the elasticity and hardness are readily observed from the skins of compounds, alloys, or specimens with high melting points whereas the enhancement is not seen so often for specimens with lower melting points such as Sn, Zn, Al and organic specimens unless they are chemically passivated.[163] The measured skin hardening or softening also varies with methods of detection. For instance, the modulus of nanocrystalline of 50 - 800 nm thick Ni films, measured using acoustic AFM,[164] is lower than that of the bulk, as no artifacts such as accumulation of dislocations and possibly no passivation are involved in the non-contact acoustic AFM method. SiTiN films are measured approximately 10–20% harder using nanoindentation than the values obtained using SAWs methods.[47]

Normally, deformation resistance from the effect of the pile up of dislocations, strain gradient work hardening, and from the artifacts due to indentation tip shapes, strain rates, loading scales, etc, involved in the contact mode of indentation would play a role of significance.[165] The surface smoothness has less influence on the measurement.[166] Figure 7 shows the measured hardness dependence of Cr[71], 6H-SiC[165] and Ni[10] films on the load scales in the indentation methods. The hardness of the single crystal moissanite (6H-SiC) obtained by ten-second loading parallel to the crystallographic c axis varies with the loading magnitude. If the load is 0.5 N, the derived hardness is 26 GPa whereas, loading at 29 and 50 N, the corresponding hardness values are 22.5 and 22 GPa, respectively. [165] Therefore, it is difficult to be certain of the hardness of a material because of the joint contribution from the intrinsic and extrinsic processes. However, an experimentally carefully calibration of the area function over the contact depth range prior to nanoindentation tests may improve the accuracy of the derived information.[167]

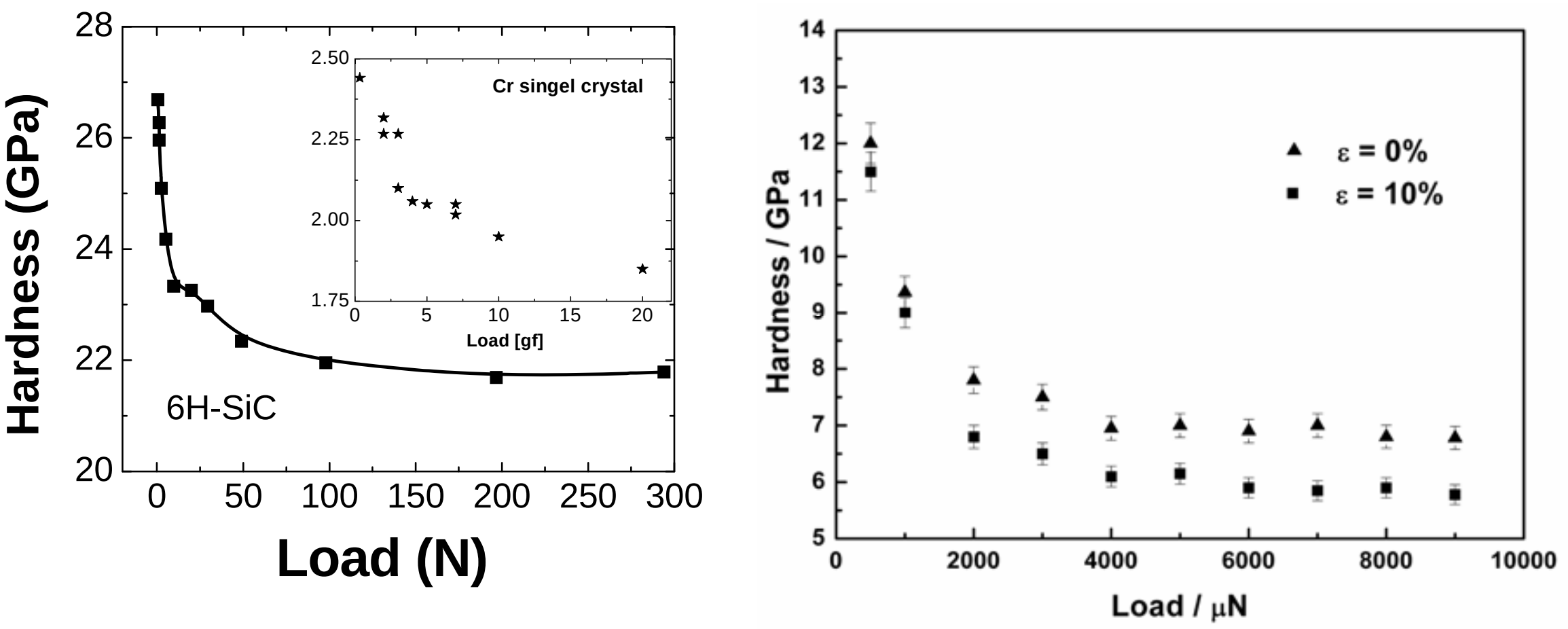

Figure 7 Load dependence of (a) the microhardness of Cr crystal (insert) [71] and 6H-SiC[165] and (b) the nanohardness of Ni films under zero and 10% tensile strain,[10] showing the general trend of the uncertain accuracy of hardness in indentation methods. link.

Hence, one could not simply tell that a specimen is harder or softer compared with the bulk standards without indicating the conditions of surface passivation, temperature of operation, and the testing methods used. Artifacts are hardly excluded in practice. What one is concerned about is the intrinsic change of the mechanical properties with temperature, solid size, and the nature of the bond involved. Therefore, we have to minimize contributions from artifacts in modeling considerations.

There have been several models on the mechanism for the observed surface skin hardening or softening. In order to understand the unusual surface mechanics, Zhang and co-workers[69,70] proposed a critical-depth mechanism stating that the apparent surface stress plays an important role in the depth-dependent hardness for various types of materials, such as metals, ceramics, and polymers and there exists a critical indentation depth. The bulk deformation predominates when the indentation depth is deeper than the critical depth; otherwise, surface deformation predominates. However, the origin of this critical depth is unknown. The trend of hardness versus indentation depth after the peak is described using the mechanism of strain gradient[168] or attributed to the contributions from surface energy and the increase of dislocation density with the depth.[169] The surface hardening has also been attributed to the surface effect, strain gradient work hardening, and non-dislocation mechanisms of deformation,[71] besides the currently proposed T-BOLS correlation.[170] The T-BOLS correlation mechanism attributes the surface hardening to the broken bond induced shortening and strengthening of the bonds between the under-coordinated atoms and the surface softening to the high ratio of $T/T_m$. Operation temperature closing to the melting point of the specimen will trigger the situation described by Born's criterion.

3.1.3 Liquid surface elasticity: adsorption and heating

The surface tension for a liquid takes on an important part in the liquid drop or gas bubble formation in a liquid. Compared with the solid skin pre-melting, a liquid surface tends to solidify at temperatures below the bulk melting point. Normally, the temperature dependence of the surface tension of a molten substance follows a linear relation to the temperature of testing:[171,172,173,174,175,176]

$$\begin{aligned} \gamma_s(T) &= \gamma_s(T_m) + \alpha_t(T_m - T) = \gamma_s(0) - \alpha_t T \\ \gamma_s(0) &= \gamma_s(T_m) + \alpha_t T_m \end{aligned} \quad (\mathrm{J/m^2}) \qquad \textbf{(13)}$$

where $\gamma_s(T_m)$ corresponds to the $\gamma_s$ value at melting; $\alpha_t$ is the thermal coefficient or the slope of variation. For pure metals or alloys, the surface tension drops linearly with the increase of temperature. However, the surface tension is very sensitive to the chemical environment, or contamination.[177,178] The surface tension of an alloy also varies with composition concentration. Typical samples in Figure 8 show the linear temperature dependence of surface tension of a CoSi alloy[179] and the inflected temperature dependence of the surface tension of a Sn surface under different ambient of oxygen.[180] With the given composition, the tension drops linearly with the increase in temperature. However, the temperature coefficient inflects from negative to positive at a certain temperature and then drops following that of the ideal situations. The turning point changes with oxygen dosage. As discussed later, the slope inflection can be envisioned as the contribution from the additionally compressive or tensile stress induced by surface adsorption to compensate or enhance the original surface tension of the liquid. For a specimen with a melting point much higher than the temperature of desorption of the specific adsorbate, such slope inflection may not be readily seen because of the thermal desorption of the electronegative additives, which occurs at temperature below 1200 K for oxygen.[181]

A huge database has been established regarding the temperature coefficient of surface tension for metals, alloys, and polymers. Table 2 tabulates the data for some typical samples and includes information derived in this work and discussed later in section 3.4.2. The temperature dependence of surface tension has provided us with an opportunity to derive information regarding atomic cohesive energy in the bulk and with possible mechanism for the adsorbate-induced surface stress. The latter could be a challenging topic of research on adsorption of various adsorbates to liquid surfaces of relatively low-$T_m$ metals.

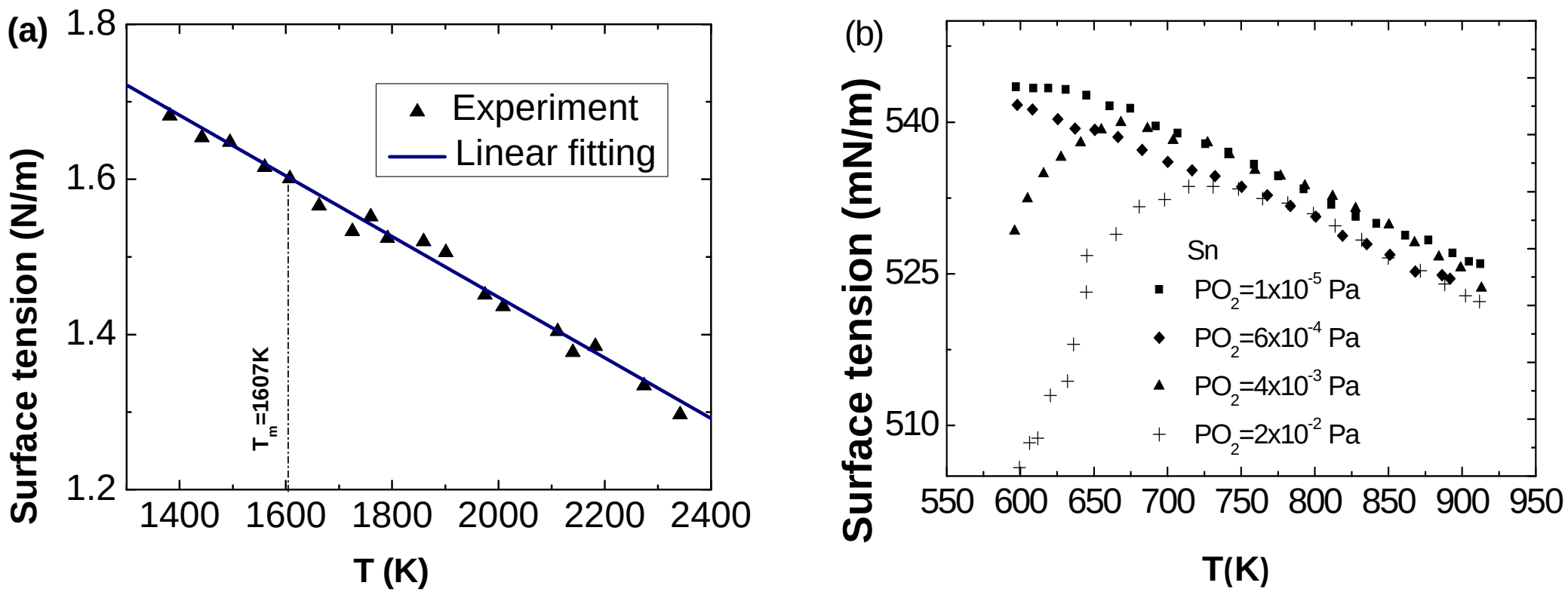

Figure 8 Temperature dependence of surface tension of (a) CoSi liquid[179] and (b) Sn liquid under different oxygen partial pressures.[180] The inflection of the coefficient can be viewed as the results of a competition between the adsorbate induced compressive stress and the broken bond induced tensile stress according to the present understanding. link.

Table 2a Information of the mean atomic cohesive energy $E_B(0)$ derived from fitting the measured T-dependent surface tension, as shown in detail later in section 3.4.2. with the measured surface tension and its temperature coefficient as listed in the first two columns. The bond energy $E_b(0)$ is available by dividing the atomic cohesive energy $E_B(0)$ with bulk coordination $z_b$ (=12) for elemental specimen. $\theta_D$ and $T_m$ are the Debye temperature and melting point as input in calculations. $\Delta T$(K) is the temperature range of testing. Scattered data for a specific substance and the deviation from the reference values show the sensitivity of the LBA derived $E_B(0)$ to the extrinsic factors such as surface contamination in comparison to the reference data.

| | $\gamma_s(T_m)$ (mJ/m$^2$) | $\alpha_t$ (mJ/(m$^2$K) | $\alpha_l$* ($10^{-6}K^{-1}$) | $\theta_D$(K) | $T_m$(K) | $\Delta T$(K) | **$E_B(0)$ (eV)** | $E_B(0)$ Ref [104] |
|---|---|---|---|---|---|---|---|---|
| **Hg[174]** | **493** | **0.2** | **60.4** | **100** | **234.32** | **273-523** | **1.47** | **0.67** |
| Si[182] | 783.5 | 0.65 | | | | 1350 -1900 | **5.33** | **4.63** |
| Si[183] | 735 | 0.074 | 2.6 | 647 | 1687 | 1457-1890 | **2.04** | |
| Si[175] | 763 | 0.219 | | | | 1690-1750 | **1.15** | |
| Ga-added Si [184] | 777 | 0.243 | | | | 1680-1760 | **2.12** | |
| B-added Si [175] | 721 | 0.098 | | | | 1690-1750 | **3.92** | |
| $Al_2O_3$[175] | 640 | 0.082 | | | | 2190-2500 | **4.39** | |
| $Al_2O_3$[179] | 550~700 | 0.082 ~0.48 | | 1045 | 2327 | | | |
| $Co_{25}Si_{75}$ [184] | 1604 | 0.4 | | | 1607 | 1384-2339 | **2.43** | |
| **Ni[173]** | **1823** | **0.46** | **1.6** | **375** | **1728** | | **3.645** | **4.44** |
| Ni[185] | 1868 | 0.22 | | | | 1300-1625 | **6.03** | |
| Ni[186] | 1854 | 0.364 | | | | | **4.02** | |
| Ni[187] | 1846 | 0.25 | | | | | **5.38** | |
| Co[188] | 1875 | 0.348 | 13 | 385 | 1768 | | **4.22** | **4.39** |
| Co[189] | 1881 | 0.34 | | | | | **4.31** | |
| Co[190] | 1887 | 0.33 | | | | | **4.42** | |
| **Co[191]** | **1930** | **0.33** | | | | **1500-2000** | **4.49** | |
| Ag[192] | 925.4 | 0.228 | 18.9 | 215 | 1234.78 | 1250-1500 | **3.12** | **2.95** |
| Ta[193] | 2150 | 0.21 | 6.3 | 225 | 3269 | 2970-3400 | **7.98** | **8.1** |
| W[194] | 2478 | 0.31 | 4.5 | 310 | 3695 | 3360-3700 | **6.90** | **8.9** |
| Ga[195] | 718.2 | 0.062 | 18 | 240 | 302.77 | 823-993 | **7.01** | **2.81** |
| In[192] | 573.5 | 0.099 | 32.1 | 129 | 429.6 | 500-1400 | **3.67** | **2.52** |
| In[195] | 546.8 | 0.082 | | | | 673-993 | **4.32** | |
| Sn[195] | 545.66 | 0.066 | 22 | 170 | 504.93 | 723-993 | **5.18** | **3.14** |
| Sn[196,192] | 547.17 | 0.065 | | | | 500-1300 | **5.26** | |
| Bi[195] | 378.9 | 0.070 | 13.4 | 120 | 544.5 | 773-873 | **3.51** | **2.18** |
| Pb[195] | 445.54 | 0.089 | 28.9 | 88 | 600.46 | 757-907 | **2.95** | **2.03** |
| $H_2O$ [197] | 75.4 | 0.162 | - | 192 | 273 | 273-373 | **0.38** | |

*Coefficient of linear thermal expansion at 25°C in $K^{-1}$ (Handbook of chemistry and physics)

Table 2b (continued) for polymers. Thermal expansion coefficients are not available and therefore not used in the practices on polymers. Experimental data are sourced from Ref [198].

| | $\gamma_s(T_m)$ (mJ/m$^2$) | $\alpha_t$ (mJ/(m$^2$K) | ΔT(°C) | $T_m$ (K) | **$E_B$ (0) (eV)** |
|---|---|---|---|---|---|
| hexadecane (C16) | 29 | 0.094 | 298-373 | 291 | **0.30** |
| PE (C2000) | 35.6 | 0.065 | 403-493 | 407 | **0.41** |
| PEO | 46.7 | 0.08 | 343-463 | 333 | **0.40** |
| PCAP | 44.4 | 0.068 | 373-398 | 333 | **0.43** |
| PEKK | 63.8 | 0.08 | 571-618 | 578 | **0.60** |
| PBT, poly (butylene terephthalate) | 59.3 | 0.08 | 493-523 | 496 | **0.54** |
| Poly (trimethylene terephthalate) | 53.8 | 0.067 | 538-562 | 496 | **0.56** |
| PET | 54.2 | 0.0646 | 513-593 | 528 | **0.59** |
| Poly (amide ester) copolymer | 60.4 | 0.08 | 433-463 | 433 | **0.52** |
| Nylon 66 | 64 | 0.115 | 543-563 | 533 | **0.47** |
| Polyamide MPMD-12 | 54.55 | 0.081 | 463-623 | 463 | **0.49** |

## 3.2 Atomistic origin of surface energetics

### 3.2.1 Motivation

By definition, the dimension of surface energy, surface stress, and surface tension is expressed in terms of energy per unit area, eV/nm$^2$; however, according to the coordinate-counting premise, the unit of the surface energy is in eV/atom. The former represents the energy density per unit area, whereas the latter reflects the energy loss per discrete atom upon a surface being made. Observations show that the surface energy in terms of $eV/nm^2$ is often higher while the surface energy in terms of eV/atom is lower in magnitude compared with the bulk values because of the difference in definition. Such inconsistency has caused long confusion about the definition of surface energetics. On the other hand, the atomistic origin, temperature dependence, and the responsibility of the surface energy either in terms of eV/atom or in terms of $eV/nm^2$ are yet far from clear. Most importantly, the effect of a broken bond on the length and strength of the remaining bonds between the under-coordinated atoms has been overlooked in the available models. The shortened and strengthened bonds are indeed crucial to surface energetics. For example, a considerable percentage contraction occurs to the first and the second layer of the diamond surface, which leads to a substantial reduction in the surface energy according to MD calculations.[199] The contraction varies from 11.2 to 56.2% depending on the surface plane and the potential used. Furthermore, the classical theories of continuum medium mechanics and the statistic thermodynamics could hardly cover the atomistic origin of the mesoscopic mechanics. In fact, the performance of a surface is dictated by the energy-gain per unit volume in the surface skin or the cohesive energy remnant per bond of the under-coordinated surface atoms instead of the energy cost for surface formation. It would be necessary to point out that the surface energy density gain arises from surface relaxation or bond contraction that cause the reconstruction according to the BOLS consideration. Meanwhile, strain causes the surface stress other than the stress induces the strain. Therefore, a complementary formulation to take into account the bond nature and temperature dependence of surface energetics is necessary.

### 3.2.2 Atomistic definition

In fact, the performance of a surface is governed by the remaining energies in the surface skin or by the residual bond energy of the discrete surface atoms instead of the energy loss upon surface formation. In order to describe the phenomena and processes at a surface effectively, we may propose the following concepts to complement the conventional term of surface energy consumed for surface formation. Firstly, the surface is envisioned as a sheet or a skin of a certain thickness (two-to-three atomic layers) rather than an ideal two-dimension sheet without thickness. Secondly, the concept of energy-density-gain ($\gamma_{ds}$) refers to the energy stored per unit volume (in the units of $eV/nm^3$) in the surface skin upon making a surface. Lastly, the concept of residual atomic-cohesive energy ($\gamma_{fs}$) represents for the cohesive energy per discrete surface atom (in the units of eV/atom) upon the broken of surface bond, which equals the multiplication of the remaining number of bonds with a bond energy that varies with the number of the bonds broken.

From the analytical expression of binding energy per unit area, $E_b d^{-2}$, and the current T-BOLS correlation, the shortened and strengthened bonds of the under-coordinated surface atoms originate the surface tension, surface stress, and surface energy. Correspondingly, we may also clarify the difference between surface stress and surface tension as the energy gradient in the surface normal and in the surface plane directions, as summarized in Table 3. The $d_i(z_i, T)$ and $E_i(z_i, T)$ follow the temperature and $z_i(K_j)$ dependence, as discussed in Section 2. From the atomistic perspective, the surface tension is no longer a second-rank tensor but a vector with x and y components because $\partial F_x / \partial y = \partial F_y / \partial x = 0$.

**Table 3**

Proposed terms of curvature $K_j(z_i)$ and T dependence of surface energetics arising from the effect of a broken bond represented by the T-BOLS correlation and their expressions, atomistic origins, and their responsibilities for the surface phenomena and processes.[144]

| Definitions | T-dependent expression | Atomistic origin | Functionality |
|---|---|---|---|
| Surface energy density ($\gamma_{ds}$) (eV/nm$^3$) | $\gamma_{ds} = \frac{\sum_{i\le 2} E_i(z_i,T)/d_i^3(z_i,T) \times d_i(z_i,T)}{\sum_{i\le 2} d_i(z_i,T)}$ | Energy gain per unit area of ($d_1$+$d_2$) thick due to the broken bond induced strain and bond strength gain | Surface stress; elasticity; Hamiltonian; Surface optics; Dielectrics; Surface trapping states; electron and photon transport dynamics; work function, etc |
| Surface atomic coherence energy ($\gamma_{fs}$) (eV/atom) | $\gamma_{fs} = \frac{\sum_{i\le 2} z_i E_i(z_i,T)}{2}$ | Binding energy remnant per discrete atom upon surface formation | Thermal stability; melting and evaporating ability; wettability; diffusivity; reactivity; acoustics; self-assembly; reconstruction. |
| Energy cost for surface formation ($\gamma_s$) (eV/atom) | $\gamma_s = z_b E_b - \gamma_{fs}$ or equivalent to the forms in Eq (12) | Traditional definition of energy loss per atom upon surface formation | |
| Surface stress (P) (eV/nm$^2$) | $\frac{\partial F_x}{\partial x} i + \frac{\partial F_y}{\partial y} j + \frac{\partial F_z}{\partial z} k$ | Force gradient in the surface | Surface relaxation, strength, hardness |
| Surface tension ($\tau$) (eV/nm$^2$) | $\frac{\partial F_x}{\partial x} i + \frac{\partial F_y}{\partial y} j$ | Force per unit length in the surface xy plane | Surface reconstruction, strength, hardness |

3.3 Analytical expressions

3.3.1 Surface energetics

According to eq (7), the proposed curvature and temperature dependence of surface energy density, $\gamma_{di}$, and the surface atomic cohesive energy, $\gamma_{fi}$, will follow the relations:

$$\gamma_{di}(z_i,T) \propto \frac{E_i(z_i,T)}{d_i^3(z_i,T)} = \frac{E_i(0) - \int_0^T \eta_{1i}(t)dt}{d_i^3(1+\alpha_i T)^3} = \frac{E_i(0) - U(T/\theta_D)}{d_i^3(1+\alpha_i T)^3}$$

$$\gamma_{fi}(z_i,T) \propto z_i E_i(z_i,T) = z_i\left[E_i(0) - \int_0^T \eta_{1i}(t)dt\right] = z_i\left[E_i(0) - U(T/\theta_D)\right]$$

(14)

The internal energy $\int_0^T \eta_{1i}(t)dt \cong \omega^2 x^2/2 + u'''(r)x^3/3 + 0(x^{n>3})$ corresponds to the thermal vibration of various modes, harmonic and anharmonic, which weakens the bond strength. The $\omega$ and $x$ represent the frequency and amplitude of vibration, respectively. The relative values of the surface energetics in the ith atomic layer to the bulk values of $\gamma_{db}$ and $\gamma_{fb}$, measured at $T_0 = 0$ K, can be derived as:

$$\begin{aligned}\frac{\gamma_{di}(z_i,m,T)}{\gamma_d(z_b,m,0)} &= \frac{d^3}{d_i^3(1+\alpha_i T)^3} \times \frac{E_i(0) - \int_0^T \eta_{1i}(t)dt}{E_b(0)} \\ &= \frac{c_i^{-(3+m)}}{(1+\alpha_i T)^3}\left(1 - \frac{\int_0^T \eta_1(t)dt}{z_{ib}c_i^{-m}E_b(0)}\right) \\ \frac{\gamma_{fi}(z_i,m,T)}{\gamma_f(z_b,m,0)} &\cong z_{ib}c_i^3(1+\alpha_i T)^3 \frac{\gamma_{di}(z_i,m,T)}{\gamma_d(z_b,m,0)}\end{aligned}$$

(15)

At T = 0 K, the surface energy density is always higher and the surface atomic cohesive energy is always lower than the corresponding bulk values. These definitions may clarify discrepancies in observations.

3.3.2 Elasticity and strength

The surface stress, or surface tension, and bulk modulus at a specific atomic site share the same dimension of the surface energy density:

$$P_i(T) = -\left.\frac{\partial u(r)}{\partial V}\right|_{r=d_i} \propto Y_i(T) \propto B_i(T) = -V\left.\frac{\partial^2 u(r)}{\partial V^2}\right|_{r=d_i} \propto \frac{E_i(T)}{d_i^3} \propto \gamma_{di}(T)$$

**(16)**

This expression demonstrates the correspondence of surface energy density to the mechanical properties.

Precisely, the B and P are the derivatives of interatomic potential. However, there exists uncertainty in choosing the exact form of $u$(r). What we are concerned with is the relative change of the B and P to the bulk values. No particular form of the u(r) is involved in the present exercise, as we need only the values of $E_i$ and $d_i$ at equilibrium positions. On the other hand, from dimensional analysis, the P and B are both proportional to the energy per unit volume. Therefore, it is reasonable to take the above approximation in the analysis. It is unnecessary and impractical to consider the differentiations at a point away from the equilibrium atomic distance. It would be sufficient to consider the equilibrium positions (bond length and energy) and the temperature induced energy change. An exact solution may be obtainable in the first principle calculations but the outcome will also be subject to the form of the *u*(r) chosen.

Eq (16) indicates that the $Y_i$ and $P_i$ enhancement depends uniquely on the shortened and the strengthened bonds. The relative values for the local $Y_i$, $P_i$ and $\gamma_{di}$ to those of the bulk measured at $T_0$= 0 K follow the same relation:

$$\left.\begin{array}{c}\dfrac{Y_i(m,z_i,T)}{Y(m,z,0)}\\ \dfrac{P_i(m,z_i,T)}{P(m,z,0)}\\ \dfrac{\gamma_{di}(m,z_i,T)}{\gamma_d(m,z,0)}\end{array}\right\}=\frac{C_i^{-(3+m)}}{(1+\alpha_i T)^3}\times\left[1-\frac{\int_0^T \eta_1(t)dt}{z_{ib}C_i^{-m}E_b(0)}\right] \tag{17}$$

3.4 Strain induced surface elasticity and stress

3.4.1 Bond nature and curvature dependence

- A flat surface

Generally, for a flat surface, $z_1 = 4$, $z_2 = 6$, and the effective $z_{i\geq 3} = 12$, and correspondingly, the bond contraction coefficient $C_i(z_i) = 2/(1+\exp((12-z_i)/(8z_i)))$ leads to $c_1 = 0.88$, $c_2 = 0.94$, and $C_{i\geq 3} = 1$. For the diamond cubic materials the effective $z_b$ is also 12 though they have a first neighbor coordination of 4 because the complex unit cell in diamond is formed by an offset of two fcc structures. Averaging the sum over the top two atomic layers, we can obtain the mean energy-density-gain per unit area of ($d_1+d_2$) thick and the remaining cohesive-energy per discrete atom in the top two atomic layers of a flat surface at very low temperatures:

$$\begin{array}{ll}\langle\gamma_{ds}\rangle=\left(\sum_{i\leq 2}\gamma_{di}d_i\right)/\left(\sum_{i=2}d_i\right), & \left(eV/nm^3\right)\\ \langle\gamma_{fs}\rangle=\left[\sum_{i\leq 2}\gamma_{fi}\right]/2, & \left(eV/atom\right)\end{array},$$

and the relative changes for a flat surface,

$$\left\langle \frac{\Delta\gamma_{ds}}{\gamma_{db}} \right\rangle = \sum_{i\le 2}\left(C_i^{-(m+2)}-1\right)\Big/\left(C_1+C_2\right)$$
$$\left\langle \frac{\Delta\gamma_{fs}}{\gamma_{fb}} \right\rangle = \sum_{i\le 2}\left(z_{ib}C_i^{-m}-1\right)\Big/2$$

**(18)**

The estimated ratios of the defined surface energy density, surface atomic cohesive energy, and the traditionally defined surface energy are compared in Table 4. Results show that $\langle \gamma_{ds} \rangle$ is always higher and $\langle \gamma_{fs} \rangle$ is always lower than the bulk values. Therefore, it is not surprising that measurements show inconsistent values because of the definitions.

**Table 4**

Predicted bond nature (m) dependence of the ratios of surface energy density, $<\gamma_{ds}>$, per unit area and surface cohesion (free energy), $<\gamma_f>$, per atom with respect to the bulk values. Subscript 1 and 2 refers to the top first and the second atomic layers in a flat surface. $<\gamma_{ss}/\gamma_{sb}>$ is approximately equal to 1-$<\gamma_{fs}/\gamma_{fb}>$.

| M | $\gamma_{d1}/\gamma_{db}$ | $\gamma_{d2}/\gamma_{db}$ | $<\Delta\gamma_{ds}/\gamma_{db}>$ | $\gamma_{f1}/\gamma_{fb}$ | $\gamma_{f2}/\gamma_{fb}$ | $<\Delta\gamma_{fs}/\gamma_{fb}>$ |
|---|---|---|---|---|---|---|
| 1 | 1.668 | 1.281 | 0.39 | 0.379 | 0.532 | -0.54 |
| 3 | 2.153 | 1.45 | 0.73 | 0.489 | 0.602 | -0.44 |
| 5 | 2.981 | 1.68 | 1.16 | 0.632 | 0.681 | -0.33 |

- A nanosolid: curvature and bond nature dependences

Considering the outermost two atomic layers of a spherical dot, it is possible to derive the T-independent $<\gamma_{ds}/\gamma_{db}>$ and $<\gamma_{fs}/\gamma_{fb}>$ as a function of the bond nature and the curvature by using, $z_1 = 4\times(1\text{-}0.75/K_j)$ and $z_2 = z_1+2$. According to the core-shell structure, the size dependent relative change of a measurable quantity $Q$ follows the scaling relation in eq (10) with the following quantities:

$$q_i = \begin{cases} \gamma_{di} & = E_i/d_i^3 \propto Y_i \\ \gamma_{fi} & = z_i E_i \propto T_{Ci} \end{cases}$$

**(19)**

Based on eq (19) we can predict the trend of bond nature and curvature dependence of the surface energetics, as shown in Figure 9 (a) and (b). It is seen that these identities are not simply a linear dependence of the curvature or solid size because of the non-linear dependence of the bond contraction coefficient on the atomic coordination. The surface energetics changes slightly with the curvature. The volume average of the energy-density-gain agrees with the observed size dependence of the Young's modulus of $ZnO_2$ nanowires.[42] The size dependence of residual cohesive-energy is consistent with the measured trends of critical temperature for evaporation $T_C$ ($T_{onset}$) of Ag, Au, and PbS nanostructures,[200] as shown in Figure 9 (c) and (d). Ideally, the $T_C$ for Au and Ag nanosolid should follow the *m* =1 curve and that of PbS follow the curve of *m* = 4. However, contamination during heating should alter the surface bond nature and energy. Therefore, it is no surprising why the measured data for Au and Ag not follow the *m* = 1 curve, instead. Measurement under ultrahigh vacuum may rectify the deviation. The predicted size dependence of remaining cohesive-energy also agrees with the measured size effect on the binding energy per atom of Ag particles,[201] and the structural phase transition temperature for Pb nanoislands on Si substrate.[202] Intensive investigation of the size-induced solid-liquid, magnetic, ferroelectric, and superconductive phase transitions of nanostructures has been reported in ref [[203]].

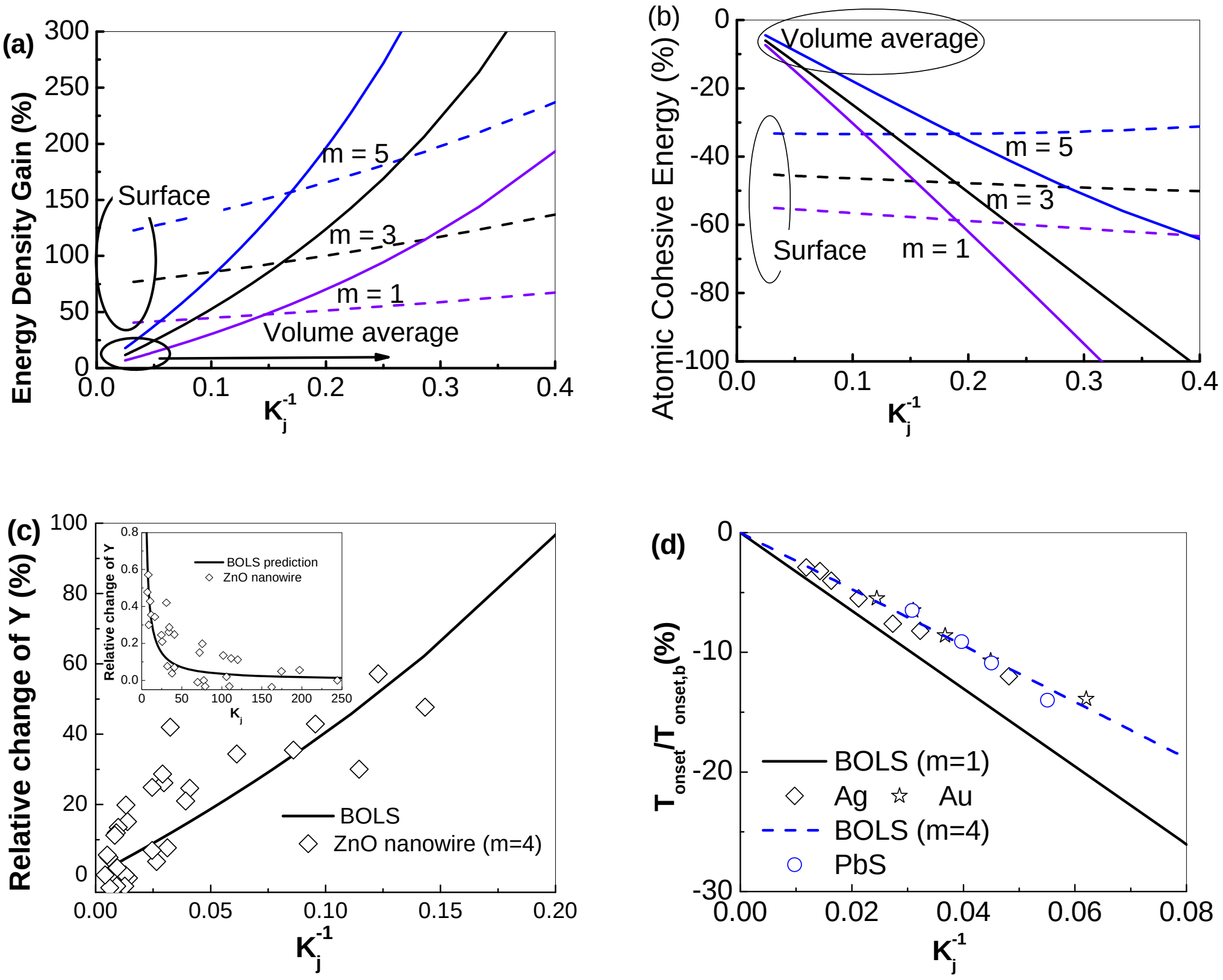

Figure 9 Prediction of curvature - induced (a) energy-density-gain per unit volume and (b) the remaining cohesive-energy per discrete atom averaged over the surface skins (surface) of two atomic layer thickness and averaged over the entire spherical solid (volume average or size dependence). The former increases whereas the latter drops with the decrease of solid size. The volume average of (a) determines the size dependence of strength and elasticity and the volume average of (b)

dictates the thermal stability. Panels (c) and (d) compare the predictions to the measured size dependent relative change of Young's modulus of ZnO nanowires [42] and the cohesive energy (evaporation temperature) for PbS, Ag and Au nanosolids [200]. Insert in (c) is the same set of data expressed in Y-$K_j$ wise [144]. Link

3.4.2 T-dependent surface tension: bond energy

The $z_i$, m, and T dependence of the mean surface energy density, $\langle \gamma_{ds} \rangle$ and mean atomic cohesive energy, $\langle \gamma_{fs} \rangle$ can be expressed as,

$$\langle \gamma_{ds} \rangle = \frac{\sum_{i\le 2} E_i(z_i,T)/d_i^3(z_i,T)\times d_i(z_i,T)}{\sum_{i\le 2} d_i(z_i,T)} \text{ (eV/nm}^3\text{)}$$

$$\langle \gamma_{fs} \rangle = \frac{\sum_{i\le 2} z_i E_i(z_i,T)}{2} \text{ (eV/atom),}$$

Under the given conditions of $\alpha T_m << 1$ and $T >> \theta_D$, the $\langle \gamma_{ds} \rangle$ can be reduced as,

$$\left\langle \frac{\gamma_{ds}(T)}{\gamma_{db}(0)} \right\rangle \overset{T>>\theta_D}{\cong} \left[ \frac{c_1^{-(m+2)} + c_2^{-(m+2)}}{c_1+c_2} - \frac{z_{b1}c_1^{-2} + z_{b2}c_2^{-2}}{c_1+c_2}\frac{\eta_1 T}{E_b(0)} \right] \frac{1}{(1+\alpha_i T)^3} \overset{\alpha T_m<<1}{\cong} \frac{c_1^{-(m+2)} + c_2^{-(m+2)}}{c_1+c_2} - \frac{z_{b1}c_1^{-2} + z_{b2}c_2^{-2}}{c_1+c_2}\frac{\eta_1 T}{E_b(0)} \quad (20)$$

To be in line with the current definition of surface energy density, we need to assume a thickness D (= $d_1$+ $d_2$) to apply to the classical expression for surface tension or stress, $\gamma_s(T)/D$. The introduction of D does not vary the reduced form of surface tension and thus we have,

$$\frac{\gamma_s(T)}{\gamma_b(0)} = \frac{\gamma_s(T_m)}{\gamma_b(0)} + \frac{\alpha_t T_m}{\gamma_b(0)}\left(1 - \frac{T}{T_m}\right) = \frac{\gamma_s(0)}{\gamma_b(0)} - \frac{\alpha_t T}{\gamma_b(0)} \quad (21)$$

Equilibrating eq (20) to eq (21) leads to an estimation of the single bond energy $E_b(0)$ and the bulk energy density, $\gamma_b(0)$/D, with the measured temperature dependence of surface tension:

$$\begin{cases} \gamma_b(0) \overset{\alpha_i T<<1}{=} \dfrac{(c_1+c_2)\times(\gamma_s(T_m)+\alpha_t T_m)}{c_1^{-(m+2)} + c_2^{-(m+2)}} = A_1(m)[\gamma_s(T_m)+\alpha_t T_m] = A_1(m)\gamma_s(0) \\ E_b(0) \overset{\alpha_i T<<1}{=} \dfrac{z_{b1}c_1^{-2} + z_{b2}c_2^{-2}}{c_1+c_2} \times \dfrac{\eta_1 \gamma_b(0)}{a_t} \end{cases} \quad \textbf{(22)}$$

$A_1(m)$ is bond nature dependent. The calculated $A_1$ (m = 1) = 0.6694 and $A_2$ = 3.4174. A precise value of $E_b(0)$ can be obtained by a careful fitting to the measured surface tension with the given Debye temperature and the coefficient of thermal expansion for a specimen. Based on this relation we can reproduce the measured temperature dependence of surface energy, surface tension, and Young's modulus (refer to section 6.3.2) with derivatives of the $E_b(0)$ that represents the mean bond energy of the entire solid. Involvement of artifacts and impurities will cause deviation of the $E_b(0)$ from the true values. Figure 10 shows the reproduction of the measured temperature dependence of surface tension of (a) liquid Hg[174] and Ni,[173] (b) Co[191] and $H_2O$, (c) Hexadecane and Polyethylene. In the calculations, the Debye temperature and the thermal expansion coefficient for the corresponding specimens are the input parameters; the $E_b(0)$ is the derived information. No other parameters are involved. Table 2 summarizes information of the estimated results for more specimens.

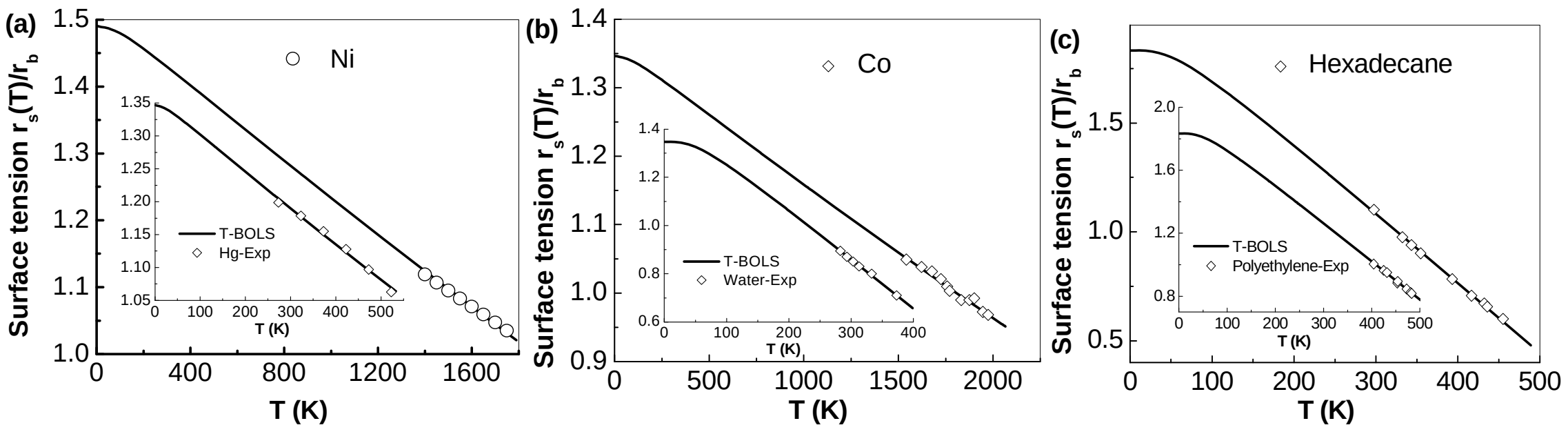

Figure 10 Estimation of $E_b(0)$ by reproducing the measured temperature dependence of surface tension for (a) Hg[174] and Ni[173] liquid, (b) Co[191] and $H_2O$[197] liquid, and (c) Hexadecane and Polyethylene,[198] The input and output are summarized in Table 2. link.

Therefore, the temperature dependence of surface tension has given us an opportunity to derive information regarding the mean atomic cohesive energy in the bulk. The accuracy of the determination is strictly subject to the measurement and that can be improved by increasing the droplet size or minimizing the curvature and by improving the sample purity [195,204] because artifacts such as surface contamination and sample purity may lead to error to the derived $E_b(0)$ value. The high sensitivity of the derived $E_B(0)$ to the chemical condition of the liquid and solid surfaces could be an advantage of the approach by offering a promising way to detect chemical reactions of a specimen by attaching other guest atoms such as biomolecules and cells to the surface of low-$T_m$ specimens.

It is noted that the adatoms or atomic vacancies during growth will affect the surface energy in a dynamic way by introducing additional traps nearby because of the bond order deficiency, which is within the BOLS expectation. As we focus here on a surface in static states, the dynamic behavior of adatoms and vacancies is not immediate concern in the present case. However, the dynamic process of adatom growth or defect/impurity formation and its influence on the surface energy would be a challenging topic for further exploitation. The present approach derives information of bond energy limiting only to elemental specimens. For compounds or alloys, we can obtain the mean value of atomic cohesive energy. The accuracy of estimation is strictly subject to the measurement. Other factors such as materials purity, defect concentration, and testing techniques may lead to the accuracy of the derived $E_b(0)$ values. Discriminating the contribution of defect concentration, surface chemical contamination, or artifacts due to experimental techniques from the intrinsically true contribution to the derived $E_b(0)$ would be even more interesting. Nevertheless, results given here and progress made insofar may demonstrate that our approach could represent the true situation of observations with the seemly simple approaches.

The O-induced inflection of the temperature coefficient of Sn surface tension,[205] as seen in Figure 8, can be attributed to the adsorbate-induced compressive stress that compensates for the surface tension, as will be further discussed in the next section.

3.4.3 Strain-induced elasticity and strength

Incorporating the nanoindentation measurement of the hardness (stress) for TiCrN[29] and Si[160] surfaces, we have the following relations,

$$\left(P_s - P\right)/P = C_i^{-(m+3)} - 1 = \begin{cases} \dfrac{(50-25)}{25} = 1 & \left(TiCrN; C_i = 0.9; m = 4\right) \\ \dfrac{(70-12)}{12} = 4.7 & \left(Si; C_i = 0.84; m = 4.88\right) \end{cases}.$$

With the above relation, we can estimate the m values and the extent of bond contraction using eq (17). For the TiCrN surface, the $C_i$ value is estimated as ~ 0.9 and the associated m = 4. The $C_i$ for Si is estimated as 0.84 with the given m = 4.88. Similarly, for amorphous carbon films[31] the $C_i$ is around 0.8±0.2 with the given m = 2.56. Predictions also agree with the theoretically calculated thickness dependence of Young's modulus for Ni, Cu, and Ag thin films at T << $T_m$.[206,207] Incorporating the maximum Y values and the lattice contraction measured by Liu et al[38] to the BOLS expression ($Y_i/Y_b = C_i^{-(m+3)}$ with m = 1) leads to a $C_i$ value of 0.95 for steel. Unfortunately, the high $P_i/P > 4$ values for Ag, Ni, Cu, Al (m = 1) thin films[35] are beyond the BOLS expectation. The exceedingly high surface hardness may result from artifacts such as surface passivation that alters the nature of the initially metallic surface bonds.

The Y-suppression of an organic specimen indicates the importance of the ($T_m$-T) contribution to the mechanical strength. The Y value drops for polymers with size, according to the empirical relation $Y = 0.014\times\ln(x) + 0.903 \pm 0.045$,[53] with x being the particle size in nm. The Y-suppression of polymers results from the low $T_m$ (~450 K) of the specimen. The ambient testing temperature is only 2/3 of the $T_m$.

### 3.5 Adsorbate-induced surface stress – bonding effect

#### 3.5.1 Observations

Atoms at the pure metal surfaces relax inwardly towards the bulk interior and reconstruct in the surface planes because of the spontaneous bond contraction and the derived tensile stress. The stress-induced inward relaxation is quite common as discussed in Ref [1] despite the discrepancy in the number of layers being involved in the relaxation. Occasionally, surface tension breaks the bonds between the surface atoms and then atoms may move away from the surface without being disturbed by external stimulus. For instances, missing rows could often form at the Au, Pt, Ag, Pd, and Rh surfaces, in particular, in the (110) plane.[208,209,210,211,212] The surface relaxation and reconstruction are consequences of bond order deficiency according to the current understanding.

However, adsorbate could induce various kinds of stresses associated with versatile patterns of relaxation and reconstruction.[181,213,214] The spacing between the first and the second atomic layers may expand if an adsorbate such as C, N, and O is inserted into space between the atomic layers even if there is contraction of bonds between the adsorbates and the host atoms.[214] For example, H, C, N, O, S, and CO adsorbates on a metal surface could change the surface stress and cause surface reconstructions because of bond making and breaking.

The adsorbate-induced stress is versatile and is capricious. One kind of adsorbate can induce different kinds of stresses at different planes of a specimen and different kinds of adsorbates can induce different stresses at a specific surface. The adsorbate-induced stress may change its sign when the adsorbate coverage changes. For example, one monolayer (ML) of oxygen on a Si(111) surface could induce -7.2 N/m compressive stress while, on O-Si(001), a tensile stress of 0.26 N/m was measured.[215] One monolayer S, O and C addition to the Ni(100) surface could induce a c(2×2) reconstruction with compressive stress of -6.6, -7.5, and -8.5 N/m, respectively.[216] The C, O, and S adsorbate-induced stresses provide driving forces for the Ni(111) surface reconstruction upon adsorption.[217] A recent STM bending bar measurement[218] suggested that the (2×1)-O phase induces compressive stress on the Cu(110) surface and the stress varies significantly with oxygen coverage. Using the cantilever vibration method, Hwang et al[219] studied the correlation between bimolecular interactions and the dynamical response of nanocantilevers in terms of the resonant frequency shift. They found that the surface stress increases linearly with the concentration of antigen, driven by the specific protein-protein interactions.

Room-temperature adsorption of CO molecules could induce exclusively a compressive stress on the Ni(111) surface. The stress is tensile on the Ni(100) at CO coverage below 0.2 ML.[220] The tensile stress passes through a maximum of 0.96 N/m at a coverage of $\theta = 0.09$ and then becomes compressive for $\theta > 0.25$ with a value of -0.54 N/m at $\theta = 0.5$, as compared in Figure 11. The latter value is small compared with the stress produced by other more strongly chemisorbed adsorbates on the same surface. The sign reversal of the CO-Ni(100) surface stress was attributed to a coverage dependent variation in the net charge transfer between the metal surface atoms and the adsorbates, involving an enhanced splitting of the bonding and the CO-2π* antibonding orbitals of neighboring molecules at higher adsorbate coverage.[220] Nitrogen addition is often more beneficial to the mechanical and corrosion properties of a metallic surface because N could induce different kinds of stress.[221,222]

A full understanding of the adsorbate-induced stress may help to explain the adsorbate-induced slope change in the temperature dependence of the surface tension such as Sn liquid, as shown in Figure 8. Charge repopulation and polarization upon chemisorption should provide forces driving the observed reconstruction and the measured stress.

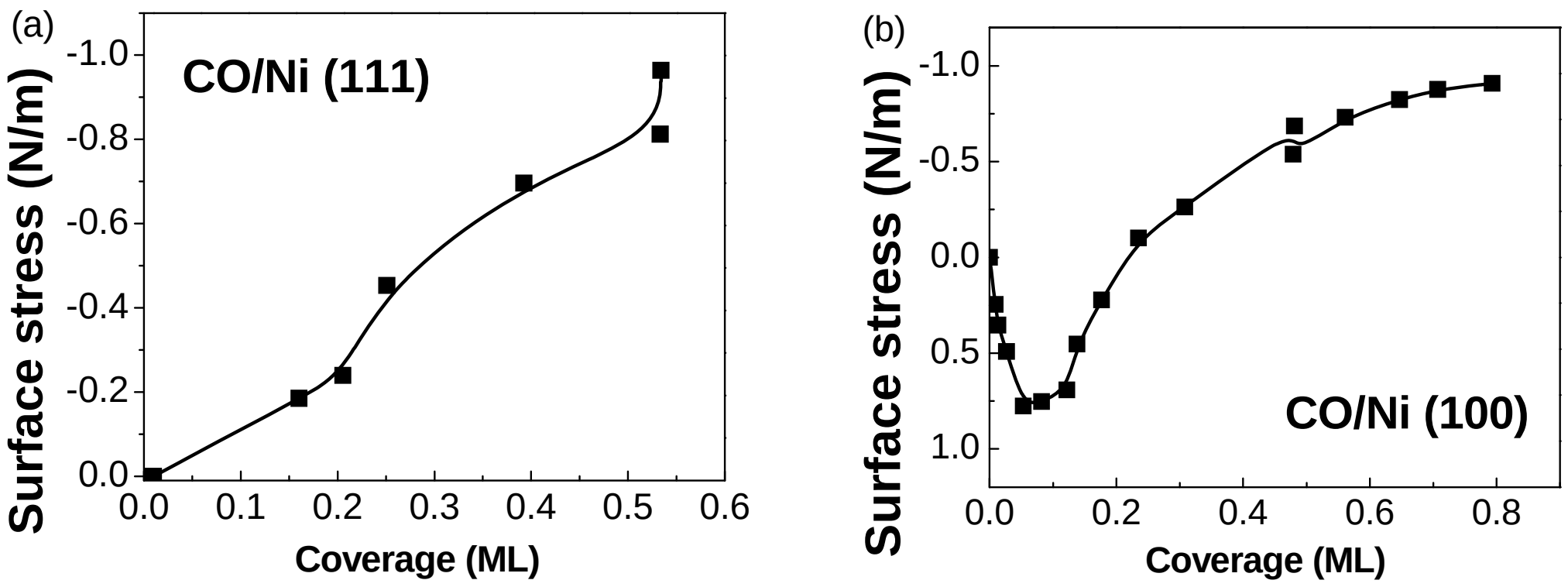

Figure 11 Coverage dependence of CO induced Ni(001) and Ni(111) surface stress.[220] CO changes the Ni(100) stress from tensile to compressive with the increase of CO coverage. Link

It has been found that hydrogen plasma sputtering could embrittle the Ti surface and causes a detrimental effect on the toughness of a Ti specimen that would otherwise behave as a ductile material when breaking.[223] Nanovoids about two nm in diameter are formed on the epitaxial GaN layers upon high-dose hydrogen implantation and subsequent annealing, while large microcracks of 150–200 nm long occurred after 1 hr annealing at 700 °C, leading to surface blistering.[224] The nanovoids also serve as precursors to the microcrack formation and the blistering process at Si surfaces as well.[225] These observations manifest the special function of the hydrogen atoms in terminating the metallic bonding networks because of the presence of $H^+$ valence. A better understanding of $H^+$ termination of bonding may explain the results of $Na^+$ and $Ca^{2+}$ induced embrittlement of Al grain boundaries. [226]

XRD and Raman measurement of the residual stresses of the Ti/TiC/diamond interfaces at different stages in diamond growth revealed that carbon addition could change the tensile stress of the Ti specimen into compressive.[227] As shown in Table 5, the tensile stress dominates in the pure Ti surface and the carbon turns the tensile surface stress of the Ti into compression upon TiC interlayer formation. The compressive stress in the TiC interlayer and the tensile stress of the Ti are reduced gradually with the thickness increases of the grown diamond films.

Table 5 Residual stresses (GPa) of Ti, TiC and diamond coatings deposited with $CH_4/H_2$ = 196: 4 at a plasma power of 1 kW.[227]

| Diamond thickness (μm) | | 3.8 | 6.4 | 8.1 | 9.2 |
|---|---|---|---|---|---|
| Diamond | XRD (311) | -3.42 | -3.19 | -2.85 | -2.41 |
| | Raman (1332 $m^{-1}$) | -3.5 | -3.1 | -3.2 | -2.1 |
| TiC interlayer | XRD (420) | -0.505 | -0.280 | -0.344 | -0.303 |
| Ti substrate | XRD (420) | 0.267 | 0.120 | 0.118 | 0.105 |

The tensile stress of pure metals can readily be attributed to the effect of the broken bond induced bond contraction and bond strength gain from the perspective of BOLS correlation. However, the mechanism for the adsorbate-induced stress, being a long-historical issue, is yet unclear, in particular, how the stress changes its sign upon increasing the dosage of adsorbate and why different adsorbates cause different kinds of stress at a given surface. Apparently, the processes of adsorbate-induced stress cannot be explained in terms of the potential wells associated with adsorption.[181]

### 3.5.2 Electronic origin: bond making

Observations[181,214] suggested that adsorbate-induced charge redistribution and polarization could be responsible for the adsorbate-induced surface stress and the corresponding reconstruction. In order to understand the atomistic and electronic origin for the adsorbate-induced surface stress, we briefly discuss here the bond making dynamics with a focus on its consequences on the evolution of bond stress during adsorbate bond making with reference to previous reports.[181,213,214,228]

- $O^{1-}$ induced stress at Cu(001) and Rh(001) surfaces

Figure 12 compares the STM images and the corresponding bond configurations for the c(2×2)-$2O^{-1}$ reconstruction at the Cu(001) and the Rh(001) surfaces. In both cases, $O^{1-}$ adsorbate occupies the fourfold hollow site. Because of the difference in lattice size ($d_{Cu}$ = 0.255 nm; $d_{Rh}$ = 0.268 nm) and the electro-negativity ($\xi_{Cu}$=2.5; $\xi_{Rh}$ = 2.2), $O^{1-}$ prefers one bond forming with a Cu atom (labeled +) on the surface and polarizes the remaining three Cu atoms nearby. A group of four $O^{1-}$ ions and a pair of $O^{1-}$ ions form two domains (depressed patches in dark) with walls consisting of the interlocked dipoles (bright protrusions). Therefore, the Cu(001)-c(2×2)-$2O^{-1}$ surface is fully occupied by the interlocked dipoles and the atoms attract one another. The surface stress tends to be tensile. Very-low-energy electron diffraction (VLEED) calculations suggested that the $O^{1-}$ forms an off-centered (by 0.02 nm) pyramid positioned 0.04 nm above the surface plane. The spacing between the top and the second atomic layers remains the value for clean Cu(001) surface with the usually measured contraction of ~10%.

However, $O^{1-}$ forms a bond with the Rh atom immediately underneath (labeled 1) and polarizes the remaining four atoms (labeled 2) in the surface. The dipoles point to the open end of the surface. The surface dipoles repel one another and, therefore, compressive stress dominates in the Rh(001)-c(2×2)-2$O^{1-}$ surface with a "radial outgoing" pattern of reconstruction. The Rh(001)-c(2×2)-2$O^{1-}$ scenario holds for the Ni(001)-c(2×2)-2$C^{1-}$ surface as well with compressive stress dominance. Therefore, the same $O^{1-}$ adsorbate induces an opposite stress on the fcc(001) surface of Cu and Rh because of the difference in the electro-negativity and the lattice size of the host elements.

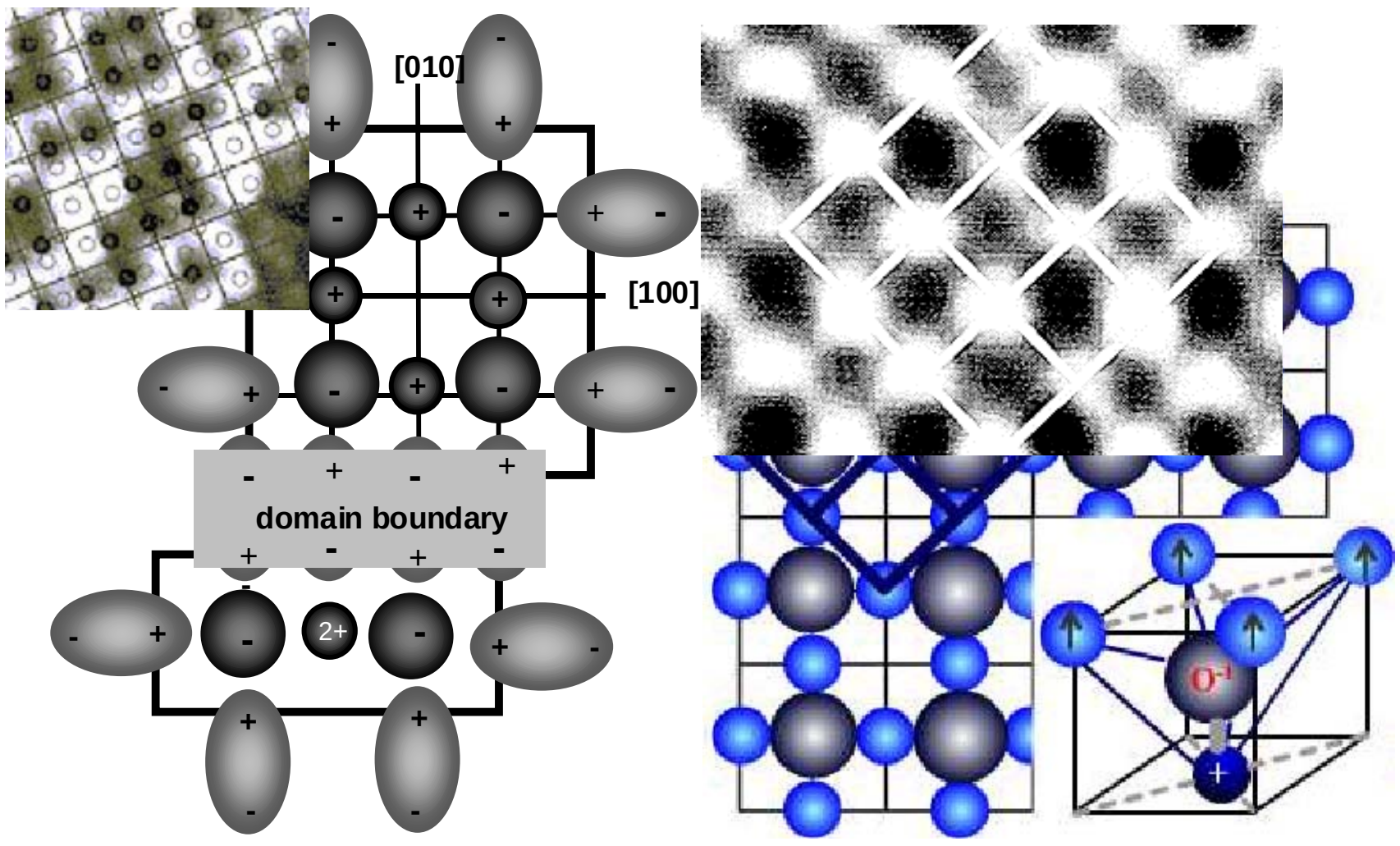

**(a) Cu(001)-c(2×2)-$O^{-1}$ (b) Rh(001)-c(2×2)-2$O^{-1}$**

Figure 12 STM images and the corresponding models illustrating $O^{1-}$-induced (a) tensile stress at the Cu(001) surface and (b) compressive stress at the Rh(001) surface. Link

o Cu(001)-(2×2√2)R45°-2$O^{2-}$ compressive stress

On further relaxation of the O-adsorbed Cu(001) surface, the O adsorbate varies its valence value from 1- to 2- at the same coverage of oxygen. As illustrated in Figure 13, $O^{1-}$ evolves into $O^{2-}$ accompanied with a Cu(001)-(2×2√2)R45°-2$O^{2-}$ "missing-row" type reconstruction. The sp-orbital hybridization of oxygen leads to a tetrahedron with involvement of two bonding and two non-bonding orbitals. The non-bonding lone pairs polarize the neighbouring Cu atoms to be $Cu^{dipole}$. At the Cu(001) surface, $Cu_3O_2$ clusters are formed with antibonding dipoles that are coupled oppositely, crossing over the missing row atoms. Because of the dipole-dipole repulsion, compressive stress becomes dominant at the Cu(001)-(2×2√2)R45°-2$O^{2-}$ surface. Therefore, the evolution of the $O^{1-}$ to the $O^{2-}$ turns the Cu(001) surface stress from tensile to compressive because of charge redistribution and polarization upon adsorbate bond making.

VLEED quantification reveals that the $Cu_3O_2$ formation occurs in four discrete stages and the bond lengths (1-$O^{2-}$ and 2-$O^{2-}$) are 0.163 and 0.175 nm, the bond angle is 104°; the non-bond distance ($Cu^{dipole}$ : $O^{2-}$) is 0.194 nm and the angle ($Cu^{dipole}$ : $O^{2-}$ : $Cu^{dipole}$) is around 140°. An expansion from 0.175 to 0.194 nm occurs to the separation between the top two Cu layers due to $O^{2-}$ insertion for tetrahedron formation. A pairing $O^{-2}$ : $Cu^{dipole}$ : $O^{-2}$ chains are coupled by the "dump-bell"-shaped quadruples which cross over the missing-row vacancies, where ":" represents the non-bonding lone pairs. The atoms "M" are missing because the bonds with its neighbouring atoms are broken and the "M" suffers from repulsion of the dipoles. Reconstruction and relaxation are consequences of bond formation and stress release. Similarly, $Ag_3O_2$ formation in the Ag(001) surface at temperature below 180 K proceeds in the same pattern of $Cu_3O_2$. When the temperature is raised, $O^{2-}$ reverts to $O^{1-}$. The inverse order of reconstructions of the Cu(001) and Ag(001) results from the difference in lattice size and electro-negativity between the two specimens.[213]

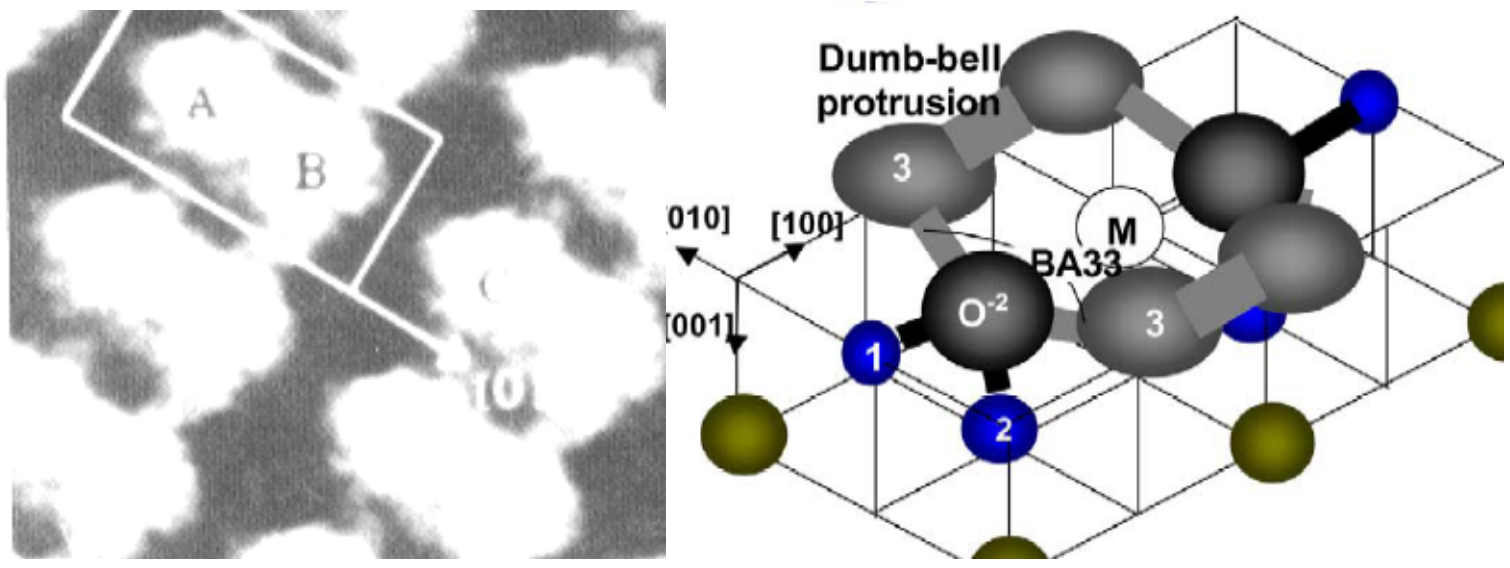

Figure 13 STM image [[228]] and the perspective of the $Cu_3O_2$ pairing tetrahedron in the Cu(001)-(2√2×√2)R45°-2$O^{2-}$ unit cell [[229]]. $O^{-2}$ prefers the center of a quasi tetrahedron. Atoms 1 and 2 are $Cu^{2+}$ and $Cu^{+}$. Atom 3 is $Cu^{dipole}$ and M is the vacancy of the missing Cu. This configuration also applies to Ag(001) low-temperature reconstruction. Link

- $C^{4-}$, $N^{3-}$, and $O^{2-}$ induced fcc(001) surface stress

C and N adsorption on a Ni(001) surface and O adsorption on an Rh(001) surface could derive the same pattern of c(2×2)p4g-2A (A = O, N, and C) "clock" reconstruction as identified using STM and a common rhombi-chain along the <11> direction, see Figure 14.[214] However, careful analysis from the perspective of tetrahedron bond making leads to a pattern of (4√2×4√2)R45°-16$A^{n-}$ reconstruction with clear identification of the atomic valences of the involved atoms, where $A^{n-}$ represents the $O^{2-}$, $N^{3-}$, and $C^{4-}$ ions. Atoms labeled 1 and 2 are $M^{+}$ and $M^{dipole}$, respectively. Atoms labeled with 1/2 are $M^{+/dipole}$. M refers to the metal atoms.

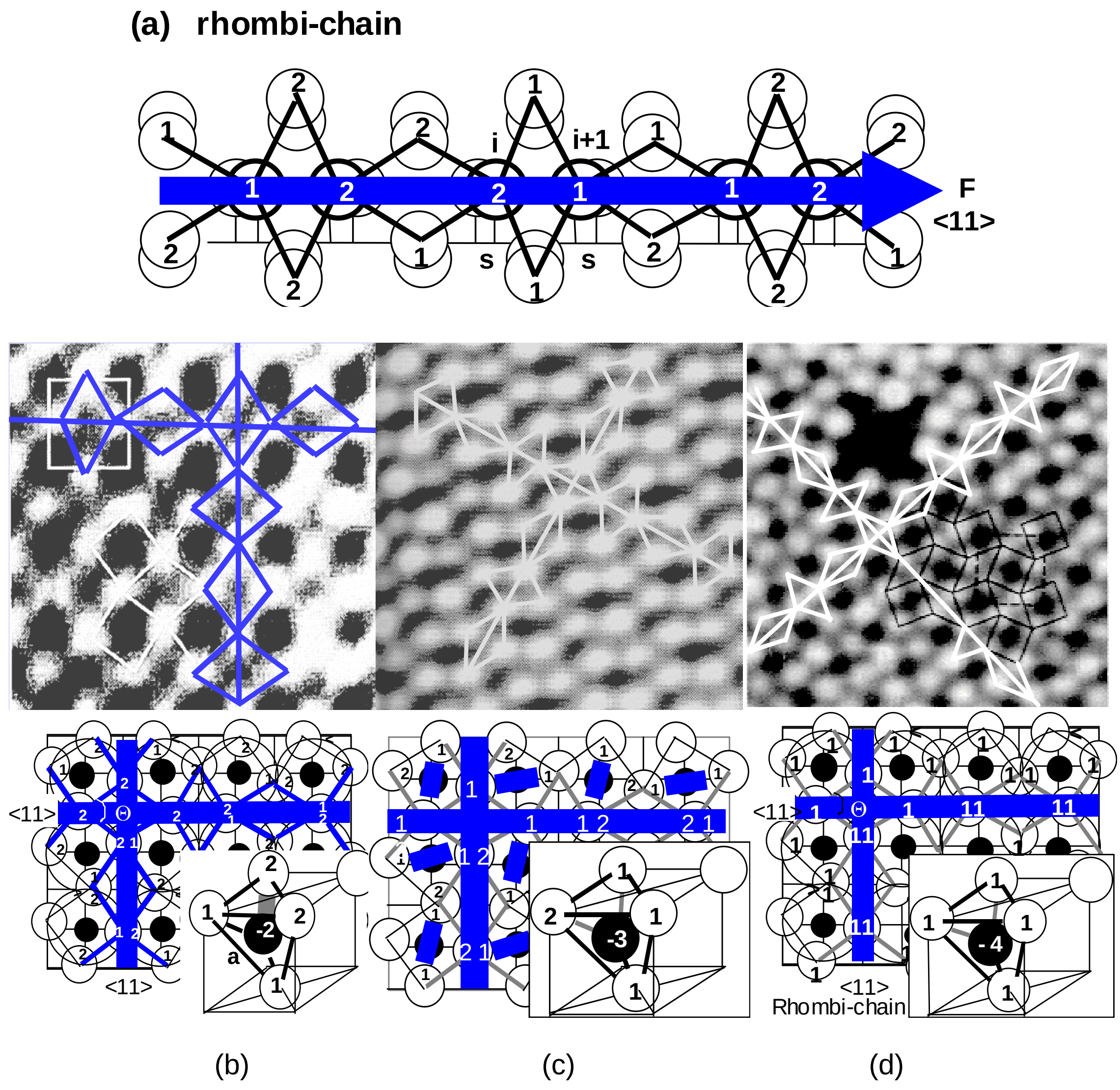

Figure 14 (a) The non-uniform '- 2 - 1 -- 1 - 2 -' rhombi-chain extracted from the STM images and the corresponding bond configurations for (b) the Rh(001)-(4√2×4√2)R45°-16$O^{-2}$ [230], (c) Ni(001) -(4√2×4√2)R45°-16$N^{-3}$ [231] and (d) Ni(001) -(4√2×4√2)R45°-16$C^{-4}$ [232] to estimate atomic dislocation, driving force, and bond stress. Labels 2 and 1 stand for different valences (refer to Table 6) for $O^{-2}$, $N^{-3}$ and $C^{-4}$ induced rhombi-chain chains. Individual atomic valences are indicated [214]. Link

Table 6 Summary of the unit cell rotation angle,Φ, derived from STM imaging, atomic dislocations along the <11> rhombus-chain, bond strain, driving force, $F_i$, bond tension, and T, of the (4√2×4√2)R45°-16$A^{-n}$ or 'p4g' clock rotation [214].

| (4√2×4√2)R45°-16$A^{-n}$ | $O^{2-}$/Rh(001) | $N^{3-}$/Ni(001) | $C^{4-}$/Ni(001) |
|---|---|---|---|
| Primary cell rotation Φ (°) | 0.0; 9.0 | 0.0; 12.0 | 0.0; 20.0 |
| Tangent shift <11> (Å) | 0.30 | 0.37 | 0.64 |
| Bond strain Δd/d (%) | 1.2 | 2.2 | 6.4 |
| Electrostatic forces $F_i$ (dyn) | 7.99; 11.09 | 8.21; 14.59 | -3.11; -91.17 |
| Bond strength $F_b$ (dyn) | 0; 35.16 | 0; 35.95 | 0; -133.00 |
| Strength of interaction along the rhombus chain | 2-2 ≈ 1-1 > 0 > 2-1 | 1-1 ≈ 2-2 > 0 > 1-2 | 2-2 > 2-1 > 1-1 > 0 |
| Surface atomic states | 1: dipole<br>2: +/dipole | 1: +<br>2: +/dipole | 1: +<br>2: 2+ |

We may take the O-Rh(001) surface as an example for detailed discussion of the bond stress. As the $O^{2-}$ has already bonded to one Rh host atom underneath, the tetrahedron defines one $Rh^+$ (labeled 1) and two lone-pair-induced $Rh^{dipole}$ (labeled 2) atoms at the unit cell surface. As can be seen from the primary unit cell containing the adsorbate, three of the four surface atomic neighbors are labeled with 1, 2, and 2, respectively. Because the surface atomic ratio O : Rh = 1 : 2, each oxygen bonds to one atom at the surface and needs two atoms to be polarized, and therefore, half of the overall surface atoms are defined as $Rh^{dipole}$ and another half as $Rh^{+/dipole}$ with eventually unlike charges. One can find that atoms labeled 1 change their positions in a clockwise fashion if one counts the O-occupied hollows along the <11> direction (gray thick lines). The adsorbate dislocates eccentrically in a periodic way. From this point of view, it would be essential and complete to consider a $c(4\sqrt{2}\times4\sqrt{2})R45°\text{-}16O^{2-}$ complex unit cell in practice due to the periodicity of the off-centered adsorbate positions. Strikingly, the 'rhombi' hollows without adsorbates form chains along the <11> direction with non-uniform forces of attraction between atoms of unlike charges or repulsion between the like ones. The alternative attraction (1-2) and repulsion (1-1 or 2-2) along the rhombi chain leads to a tensile response of the O-Rh bonds at the surface.

This process is also applicable to the ($N^{-3}$, $C^{-4}$)-Ni(001) surfaces. Sp-orbital hybridization of N and C gives one and none non-bonding lone pairs to the N and C centered tetrahedron, respectively, because of their different valences. The label "1" and "2" at the N-Ni(001) surface represents $Ni^+$ and $Ni^{+/dipole}$ and the "1" and "2" represent the $Ni^+$ and $Ni^{2+}$ at the C-Ni(001) surface. The repulsion between the like charges and the attraction between the unlike ones causes a tensile response of the N-Ni(001) surface bonds. The repulsion throughout the rhombus-chain leads to the compressive response of the bonds at the C-Ni(001) surface.[233] The bond stress response gives rise to the tensile or the compressive stress for the $O^{2-}$, $N^{3-}$, and $C^{4-}$ induced reconstruction systems, as summarized in Table 6. The understanding of C, N induced bond stress at Ni(001) surfaces has led to a novel approach of forming a TiCN graded buffer layer to neutralize the diamond-metal interfacial stress and hence strengthen the diamond-metal adhesion substantially.[234]

In summary, oxygen could turn the stress of O-Ag(001) and O-Cu(001) from tensile to compressive but turn the stress of O-Rh(001) from compressive to tensile when the $O^{1-}$ evolves into $O^{2-}$ in the specific surfaces. N turns the stress of N-Ni(001) from compressive to tensile when the $N^{1-}$ evolves to $N^{3-}$ but C retains the stress of the C-Ni(001) to be compressive when the $C^{1-}$ evolves to $C^{4-}$. Adsorbate induced stress is therefore very complicated and it is hard to learn without appropriate understanding of the process of bond making. These findings may provide improved understanding of the adsorbate-induced surface stress and the adsorbate-induced slope change of the temperature dependence of the surface tension of liquid metals.

### 3.6 Nitrogen enhanced elasticity and hardness

#### 3.6.1 Observations

Nitriding of metals has been a fully developed technology for hardness and elasticity enhancement of a surface.[235,236, 237,238,239,240,241] Even though the hardness and elasticity of nitrides has been intensively investigated and widely utilized, neither the local bonding structure nor the microstructure corresponding to the observed mechanical properties are well established.[222,242]

Recently, we found that the joint effect of surface bond contraction, bond nature alteration, and non-bonding lone-pair involvement is crucial to the hardness and elasticity of the nitride surfaces.[29] Nanoindentation measurements revealed that the elastic recovery of TiCrN and GaAlN surfaces could reach as high as 100% and the nitride films are harder under a relatively lower indentation load (< 1.0 mN) than the amorphous carbon (a-C) films that are slightly softer and less elastic at the same load scale of indentation. The same trend has been observed by Wang et al.[243] At a nanoindentation load of 1 mN and below, the hydrogenated CN:H film is harder and more elastic (75% elastic recovery) compared with that of a-C:H (60% elastic recovery). However, at higher indentation load, nitride films are softer than the a-C films or polycrystalline diamond. For C and Ti nitride films, the elastic recovery ranges from 65% to 85% at higher indentation load (5 mN).[242] The friction coefficient of the nitride films increases suddenly at critical loads for TiN and CN (5 N) indicating the breaking of the surface bonds. Similarly, sapphire ($\alpha$-$Al_2O_3$) with lone-pair presence[244] exhibits a pop-in critical load (1.0 mN) below which the elastic recovery of the specimen is 100% under nanoindentation.[167] The higher pop-in critical load of sapphire may indicate the breaking the high-density lone pairs in oxide.

#### 3.6.2 Atomistic understanding

We may consider a typical case in which the N reacts with metal atoms in a surface of $C_{3v}$ symmetry, such as the fcc(111) and hcp(0001) planes.[222,245] The $N^{3-}$ ion is located in between the top two atomic layers with the lone pair being directed into the substrate. The surface is hence networked with the smaller $M^+$ and the saturate bonded $N^{3-}$ ionic cores with densely packed electrons. The outermost shells of the $M^+$ ions are emptied due to charge transportation upon bond formation. Therefore, the top surface layer should be chemically inert as it is hard for one additional acceptor to catch electrons from the deeper energy levels of the $M^+$ ions. Electrons in the saturated bond should be more stable compared with the otherwise unbonded ones in the neutral host atoms.

The high intra-surface strength due to the ionic networking could be responsible for the hardness of the top atomic layer. On the other hand, the $N^{3-}$-$M^+$ network at the surface is connected to the substrate mainly through the non-bonding lone pairs. The non-bonding interaction is rather weak (~0.05 eV per bond) compared with the initially metallic bonds (~1.0 eV per bond) or the intra-surface ionic bond (2 ~ 3 eV per bond). The weak interlayer interaction due to lone-pair formation should be highly elastic within a certain range of loads, which makes the two adjacent surface layers more elastic at an indent load lower than a critical value at which the lone pairs will break. Therefore, the enhanced intra-layer strength makes a nitride usually harder, and the weakened inter-layer bonding makes the nitride highly elastic and self-lubricating. Results of indentation at various loads and the sliding friction measurements agree with the anticipated high elasticity and high hardness at lower indent load and the existence of the critical scratching load. [29]

The surface ionic layer determines the hardness of a nitride coating and the surface bond contraction further enhances the hardness at the surface. The lone pairs are responsible for the high elasticity and self-lubrication of the nitride. A recent first-principle calculation by Zhang, Sun, and Chen[246] suggested that the strength of the previously predicted hardest $C_3N_4$ phase is lower than diamond or cubic boron nitride by a surprisingly large amount while the $C_3N_4$ phase is highly elastic because of the excessive electrons on the N atoms, consistent with current understanding. The excessive electrons exist in the states of non-bonding lone pairs, play a key role in determining the high elasticity and low hardness, and hence, no nitride should be harder than a diamond. Nitride multilayers could be a different case because of the interface mixing where bond contraction, energy densification, and pinning effect may take place.

Briefly, nitrogen could enhance the hardness of a metal surface because of the bond nature alteration and surface bond contraction. An N-M bond is shorter than a C-M bond because of the ionic radius of $C^{4-}$ and $N^{3-}$. The involvement of lone pairs makes the nitride more elastic but readily broken under a critical load. Such an interpretation may provide a possible mechanism for the atomistic friction and self-lubrication of a nitride specimen.

## 3.7 Summary

An application of the T-BOLS correlation to the liquid and solid skins has led to a consistent insight into the mechanism for the surface energetics and its derivatives on the elasticity and strength. Major conclusions drawn from this section are as following:

(i) The broken bond may not contribute directly to the surface energetics but the broken bond causes the remaining ones to contract spontaneously associated with strength gain. The shortened and strengthened bonds between the under-coordinated surface atoms dictate the surface energetics and the mechanical behavior of liquid and solid skins. It is emphasized that the bond strain induces surface stress rather than the inverse: surface stress causes the surface bond to contract. The strain-induced stress is always tensile, giving ride to the pinning and trapping effect in the surface skins.

(ii) The concepts of the energy-density-gain and the cohesive-energy remnant per discrete atom are suggested essential to classify the origin and temperature dependence of surface energetics and their responsibilities for surface processes and phenomena. These new concepts may provide a complementary to the classical theories of continuum medium mechanics and statistic thermodynamics.

(iii) Functional dependence of the surface energetics on the bonding identities has been established to represent the fact that the variation of surface energetics from the bulk values arises from the shortened and strengthened bonds between the under-coordinated atoms.

(iv) The predicted volume average of the energy density gain and the residual atomic cohesive energy agrees with the measured size dependence of Young's modulus and the critical temperatures for evaporation, melting, and phase transition.

(v) The thermal weakening of surface energetics is dominated by thermal expansion and vibration through the internal energy, which follows the integration of the specific heat. This approach allows us to estimate the bond energy by reproducing the measured temperature dependence of surface tension and Young's modulus. The accuracy of estimation is strictly subject to the experimental data.

(vi) Adsorbate-induced inflection of the temperature coefficient of the surface tension and the adsorbate induced surface stress evolution of a solid surface can be consistently understood in terms of adsorbate bond making that causes charge redistribution and polarization and the corresponding patterns of reconstructions. The adsorbate-induced stress may change from situation to situation depending on the configuration of surface bonding networks. The involvement of non-bonding lone pairs and the bond nature alteration in nitridation makes a nitride surface highly elastic and robust at relatively lower indentation or scratching load.

# IV Single bond in monatomic chains

## 4.1 Observations

### 4.1.1 Temperature dependence of maximal strain

A metallic monatomic chain (MC) is an ideal prototype of a nanowire for extensibility and mechanical strength study as the extension of an MC involves only the process of bond stretching without other processes, such as bond unfolding or atomic gliding dislocations, as experienced by atoms in coarse-grained metallic chunks under external mechanical stimuli.[247] The intriguing phenomena appearing in MCs include the quantum conductance, higher chemical reactivity, lower thermal stability, and the unusually high mechanical strength and ductility. The quantum conductance has been well understood as arising from the enlarged sub-level separation, known as the Kubo gap, $\delta_k = 4E_F/3N$, where $E_F$ is the Fermi energy of the bulk solid and N is the total number of atoms of the cluster. [248,249] The metallic atomic chains of Au and Ag exhibit semiconductor features with a calculated band gap of 1.3 eV for Au and 0.8 eV for Ag.[250] However, knowledge about the equilibrium bond length, bond strength, extensibility, maximum strain, specific heat, and the thermal and chemical stability of the MC bond is still quite preliminary and the results are sometimes controversial.

For instance, the stretching limit of the Au-Au distance in the Au-MC has been measured using TEM at room temperature in the range of 0.29 nm,[251] 0.36 nm (± 30%),[252] 0.35 ~ 0.40 nm,[248] and even on a single occasion as 0.48 nm.[88] However, at 4.2 K, the breaking limit of the Au-Au bond is reduced to 0.23 ± 0.04 nm as measured using STM and to 0.26 ± 0.04 nm as measured using mechanically controllable break junctions.[87] The standard bulk value of the Au-Au distance is 0.288 nm. STM measurement[253] also revealed at 4.2 K that the Ir-MC and the Pt-MC breaking lengths are 0.22 ± 0.02 nm and 0.23 ± 0.02 nm, respectively, being substantially shorter than the corresponding bulk values of 0.271 and 0.277 nm. Low temperature measurements of the MCs show a commonly large extent of bond contraction with respect to the bulk values despite the applied tensile stress. An EXRAFS study[254] revealed that the covalent bond in the Tellurium monatomic chain (0. 2792 nm) is shorter and stronger than the bonds (0.2835 nm) in the trigonal Te bulk structure. The Debye-Waller factor (squae of the mean amplitude of lattice vinbration) of the Te chain is larger than that of the bulk but the thermal evolution of the Debye-Waller factor is slower than that of the bulk also suggesting hardening of the bond.

### 4.1.2 Theoretical approaches

Unfortunately, numerical calculations for the impurity-free Au-MC have yielded a maximum Au-Au distance of 0.31 nm under tension, [89, 255,256,257] which hardly match the values measured at the ambient or at the extremely low temperatures in ultra-high vacuum. In order to understand such unusual Au-MC elongation, several explanations have been proposed in terms of structural, chemical, or electrical effects:

- The fuzzy image mechanism [255] suggests that the measured MC elongation is an artifact linked to a rotatable zigzag structure of the atomic chain. For an MC with an odd number of atoms, every other one is at fixed position along the chain axis, while the atoms between the fixed ones rotate rapidly around the axis, offering a fuzzy image that could hardly be caught by TEM imaging. However, a deep analysis[258,259] of the high-resolution TEM images has ruled out this possibility.

- The impurity mediation mechanism[2] proposed that artificial mediation with impurities of light atoms such as X (= H, He, B, C, N, O, S, CH, $CH_2$, and CO) being inserted into the Au-Au chain in calculations are responsible for all the observed elongations because the light atoms cannot be seen in the TEM. The light atom insertion has indeed led to gold separation in the Au-X-Au chain that could match the observed Au-Au distances. For instance, the insertion of a carbon atom leads to the stretched Au-C-Au distance of 0.39 nm just before breaking; wires containing B, N, and O displayed even larger distances under tension. The Au-H-Au distance, 0.36 nm, matches one of the experimentally measured values, and the anomalously large distance of 0.48 nm matches the separation between gold atoms in an Au-S-Au chain.[260] Unfortunately, the shorter distance at 4.2 K could not be produced in calculations. Nevertheless, contamination appears unlikely under the ultrahigh vacuum conditions of the experiments.[248] Even though contamination may occur in the high-vacuum chamber it is unlikely that the contamination was always the cause as it disregards the bonding selectivity of elements such as H and He. A hydrogen atom can form only one bond and a He atom can form none with other atoms according to chemical bond theory.
- The charge mediation mechanism[90] assumed first that the enlarged electro-negativity and electro-affinity of the under-coordinated Au-MC atom catches selected number of electrons from the TEM radiation. Then, this charging effect modifies the shape of the pairing interatomic potential with the presence of a potential maximum or force zero (transition point at the second order differential of the potential curve), at a distance being attributed to the breaking limit. It was derived that the states of $Au_2^{1-}$, $Au_3^{2-}$, $Au_4^{2-}$, and $Au_4^{3-}$ could lead to the maximum Au-Au distances of 0.49, 0.41, 0.49 and 0.35 nm, respectively. Unfortunately, Au-Au distances shorter than 0.35 nm could not be approachable using this argument. In fact, charging effect exists only for thick insulating samples in the TEM measurement because of the non-conductive character of the specimen.[261] The energetic electrons ($E > 10^5$ eV) in the TEM readily transmit through insulating specimens thinner than 100 nm and the charging effect for conductors or thinner insulators becomes negligible, although an increase in the electro-affinity does occur for the under-coordinated Au atoms by 1.34 eV (25%).[203]

Nevertheless, the low temperature Au-Au, Pt-Pt, and Ir-Ir distances that are 15%~20% shorter than the bulk values appear to be beyond the scope of the models discussed. Therefore, mechanism for the MC elongation is still open for debate. Encouragingly, sophisticated DFT calculations[262] suggest that the pairing potential is valid and the Au-Au equilibrium distance (without the presence of an external stimulus) is between 0.232 and 0.262 nm and the cohesive energy per bond increases by 200% from –0.51 to –1.59 eV. Similar trends are also found for other metallic MC such as Pt, Cu, and Ag in theory although experimental observation of such MC formation at room temperature has been limited. According to calculations,[48] the mechanical strength of the Au-Au bond is about twice that of the bulk value and the metals that form chains exhibit pronounced many-atom interactions with strong bonding in the lower-coordinated systems. Results from a combination of high-resolution TEM and MD simulation[263] suggested that different initial crystallographic orientations lead to very differentiated linear Au atomic chain formations and suggested that kinetic aspects such as temperature values and elongation rates strongly affect the morphology and chance of MC formation. In addition to the DFT approaches, Jiang et al[264] correlated the formation tendency of a MC under tensile stress to the ratio between the Peierls stress of a bulk crystal having dislocations and the theoretical break shear breaking stress of the MC. They suggested that the metallic elements having the largest Poisson's ratio hold the largest MC forming ability since such metals have the smallest elastic energy storage within the crystals and can thus endure the largest plastic deformation.

## 4.2 T-BOLS derivatives

### 4.2.1 Chain energetics

It is known that the temperature of melting of an atom with $z_i$ coordinates, $T_{m,i}$, is proportional to the atomic cohesive energy, $T_{m,i} \propto z_i E_i$.[265,266] It is understandable that if one wants to melt or thermally rupture the bond, one has to provide thermal energy that is a certain portion of the entire binding energy. However, breaking a bond mechanically at a temperature T needs energy that equals to the net bond energy at T:

$$E_b(T)=\begin{cases} E_b(0)-\int_0^T \eta_1(t)dt & (T \le \theta_D) \\ \approx \eta_2+\eta_1(T_m-T) & (T>\theta_D) \end{cases}$$

$$E_i(T)=E_i(0)-\int_0^T \eta_{1i}(t)dt=\eta_{2i}+\int_T^{T_m}\eta_{1i}(t)dt=\begin{cases} C_i^{-m}E_b(0)-z_{bi}\int_0^T \eta_1(t)dt & (T \le \theta_D) \\ \approx C_i^{-m}\eta_2+z_{bi}\eta_1(T_m-T) & (T>\theta_D) \end{cases} \quad (23)$$

The constant $\eta_{2i}$ represents the $1/z_i$ fold energy that is required for evaporating a molten atom in the MC with $z_i = 2$. $\eta_{1i}$ and $\eta_{2i}$ can be determined with the known $C_i^{-m}$ and the known bulk values of $\eta_1$ and $\eta_2$ that vary with crystal structures.[136]

### 4.2.2 Elasticity and extensibility

Considering the contribution from heating, the strength and compressibility (under compressive stress) or extensibility (under tensile stress) at a given temperature can be expressed as:

$$P_i(z_i,T)=-\left.\frac{\partial u(r,T)}{\partial V}\right|_{d_i} \sim B_i(z_i,T)=-V\left.\frac{\partial^2 u(r,T)}{\partial V^2}\right|_{d_i} \propto \frac{N_i E_i(T)}{d_i^3(T)}$$

$$\beta_i(z_i,T)=-\left.\frac{\partial V}{V\partial P}\right|_T \propto [B_i(z_i,T)]^{-1}$$

(24)

Theoretically, β is in an inverse of bulk modulus in dimension. However, $\eta_{2i}$ does not contribute to the extensibility for the molten state as it approaches infinity at $T_m$. The $N_i$ is the total number of bonds in the $d_i^3$ volume. Calibrated with the bulk value at $T_0$, the reduced temperature dependence of the linear extensibility for the MC will be:

$$\frac{\beta_i(z_i,T)}{\beta_0(z,T_0)}=\frac{d_i\,(1+\alpha_i T)}{d(1+\alpha T_0)}\times\frac{\eta_1(T_m-T_0)}{\eta_{1i}(T_{mi}-T)}.$$

(25)

Note that the bond number density in the relaxed region does not change upon relaxation ($N_i \cong N_b$). For instance, bond relaxation never changes the bond number between the neighboring atoms in an MC (τ = 1) whether it is suspended or embedded in the bulk, nor does it change the number density between the circumferential atomic layers of a solid.

4.2.3 Maximum strain

Introducing the following effects: (i) atomic CN-imperfection-induced bond contraction, (ii) thermal expansion (with linear coefficient $\alpha_i$), and (iii) the temperature dependence of extensibility (with coefficient $\beta_i$), leads to an analytical expression for the distance between two nearest atoms in the interior of a MC, as a function of atomic CN, mechanical (P), and thermal (T) stimuli:

$$d_i(z_i,T,P)=d\times C(z_i)(1+\alpha_i T)[1+\beta_i(z_i,T)P]$$

or the maximum strain at constant T,

$$\frac{\Delta d_{iM}(z_i,T,P)}{d_i(z_i,T,0)}=\beta_i(z_i,T)\overline{P}$$

(26)

where $d_i(z_i,T,0)=d\times C(z_i)(1+\alpha_i T)$ is the bond length at T without mechanical stretching. At the bulk $T_m$, the linear thermal expansion ($\alpha_i\times T_m$) is around 3% for most metals, which is negligibly small compared with $C_i$. Considering the energy for mechanical rupture, as given in eq (23), we have the relation:

$$\begin{aligned}\int_{d_i(z_i,T,0)}^{d_{iM}(z_i,T,P)}Pdx &=\overline{P}[d_{Mi}(z_i,T,P)-d_i(z_i,T,0)]=\overline{P}\Delta d_{iM}(z_i,T,P)\\ &=E_i(T)=\eta_{2i}+\eta_{1i}(T_{mi}-T)\end{aligned}$$

(27)

The mechanical rupture energy has the same value as the thermal energy required for evaporating the same atom by warming it up from the initial T. One can approximate the P to the mean $\overline{P}$, if the $d_{iM}(z_i, T, P)$ represents the breaking limit, as the integral is a constant. Combining eqs (26) and

(27), one has,

$$\overline{P} = \pm \left\{ \frac{E_i(T)}{\beta(z_i, T) \times d_i(z_i, T, 0)} \right\}^{1/2} \tag{28}$$

For tensile stress, $\overline{P}$ takes a positive value; for compressive stress, $\overline{P}$ is negative. The combination of eqs (25) and (28) yields the maximum strain of a bond in the MC:

$$\begin{aligned}
\frac{\Delta d_{iM}(z_i, T, P)}{d_i(z_i, T, 0)} &= \beta_i(z_i, T)\overline{P} = \left\{ \frac{\beta(z_i, T) \times E_i(T)}{d_i(z_i, T, 0)} \right\}^{1/2} \\
&= \left[ \frac{\eta_1 d_i(z_i, T, 0) \times \beta_0(z, T_0)}{\eta_{1i} d \times d_i(z_i, T, 0)} \times \left( \frac{T_m - T_0}{T_{m,i} - T} \right) \times \left[ \eta_{1i}(T_{m,i} - T) + \eta_{2i} \right] \right]^{1/2} \\
&= \left[ \frac{\beta_0 \eta_1 (T_m - T_0)}{d} \left( 1 + \frac{\eta_{2i}}{\eta_{1i}(T_{m,i} - T)} \right) \right]^{1/2} \\
&\cong \left( \frac{\beta_0 \eta_1 (T_m - T_0)}{d} \right)^{1/2} \exp\left\{ \frac{\eta_2 / \eta_1}{2[T_m - T/(1+\Delta_i)]} \right\} = B \times \exp\left\{ \frac{A}{T_m - T/(1+\Delta_i)} \right\}
\end{aligned} \tag{29}$$

where $1+\Delta_i = z_{ib} C_i^{-m}$. For a metallic MC with $z_i = 2$ and $m = 1$, the melting point is, $T_{m,i} \propto z_{ib} c_i^{-1} T_m = T_m / 4.185 = 0.239 T_m$ . If the $\eta_{2i}$ is taken into consideration in the extensibility expression of eq (25), the strain will remain constant without features of temperature dependence:

$$\frac{\Delta d_{iM}(z_i, T, P)}{d_i(z_i, T, 0)} = \left[ \frac{z_{ib} \beta_0 (z, T_0)}{d} \times \frac{E_b(T_0)}{E_i(T)} \times E_i(T) \right]^{1/2} = const.$$

This derivative evidences the validity of Born's criterion applying to the extensibility.

The analytical expression of the maximum strain varies unapparent with the extrinsic parameter P or the strain rate but depends intrinsically on the inverse separation of $T_{m,i}$ - T in an exponential way. When the temperature approaches $T_{m,i}$, the strain will approach infinity. The constant A $= \eta_2 / 2\eta_1$ is crystal structure dependent. The factor B depends on the nature of the material and varies with the bulk extensibility (at $T_0$), bulk bond length, and $T_m$ as well as the specific heat per bond in the bulk.

### 4.3 Bonding identities quantification

#### 4.3.1 Atomic equilibrium distance

For a spherical dot at the lower end of the size limit, $K_j = 1.5$ ($R_j = K_j d = 0.43$ nm for an Au spherical dot), $\gamma_{1j} = 1$, $\gamma_{2j} = \gamma_{3j} = 0$, and $z_1 = 2$, which is identical in situation to an atom in the MC. At $R_j = 0.43$ nm, the Au-Au distance contracts by ~30% from 0.288 to ~ 0.201 nm. According to the BOLS correlation premise, the value of 0.23 ± 0.04 nm of Au-Au distance measured at 4.2 K under tension[87] is simply a ~15% strain response to the tensile stress of the original equilibrium bond length, 0.201 nm. Similarly, the values of 0.22 ± 0.02 nm and 0.23 ± 0.02 nm measured at 4.2 K for Ir and Pt monatomic chains,[253] respectively, are also 16% and 19% strain response to the tensile stress of the equilibrium MC bond length (0.190 and 0.194 nm, respectively) that are 30% shorter than the bulk values of 0.271 and 0.277 nm.

### 4.3.2 Binding energy and thermal stability

The core level shift measured using x-ray photoelectron spectroscopy (XPS) represents crystal binding energy and the melting point stands for the thermal stability. According to the BOLS correlation, the relative change of the core level energy and the melting point of a nanosolid are expressed as:[1]

$$
\begin{aligned}
\frac{\Delta T_m(z_i,m,K_j)}{T_m(z,m,\infty)} &= \sum_{i\le 3}\gamma_{ij}\left(z_{ib}C_i^{-m}-1\right)\\
\frac{E_c(z_i,m,K_j)-E_c(z,m,\infty)}{E_c(z,m,\infty)-E_c(1)} &= \sum_{i\le 3}\gamma_{ij}\left(C_i^{-m}-1\right)
\end{aligned}
\qquad \textbf{(30)}
$$

The term $E_c(z,m,\infty)-E_c(1)$ is the bulk core-level shift and $E_c(1)$ the reference energy level of an isolated atom of the specimen.

Figure 15(a) shows the reproduction of the measured size dependence of the Au-4f core-level shift $\Delta E_{4f}(K_j)$ for Au nanofilms deposited on octanedithiol,[267] $TiO_2$[268] and Pt[269] substrates, and thiol-capped Au particles.[270] Results indicate that the bond energy of an Au-MC is increased by $C_i^{-1} - 1 = 1/0.7 - 1 \sim 43\%$. Results also imply that the Au-4f core-level drops abruptly from that of an isolated atom by a maximum of 43% upon the MC (or a fcc unit cell) being formed and then the shift recovers in a $K_j^{-1}$ fashion to the bulk value when the solid grows from atomic scale to macroscopic size.[95] The core-level shift corresponds to an electro-affinity enlargement that can be estimated from the known bulk E-4f shift of 2.87 eV[94] to be 1.34 eV, which may provide quantization of the often-observed densely-and-deeply trapped “end states” of the gold chain on a Si substrate.[109]

Calibrated with $T_m(\infty) = 1337.33$ K and by use of the same m (= 1) value used in calculating the $E_{4f}(K_j)$, we obtained the theoretical $T_m$-suppression curves for different shapes, which are compared in Figure 15(b) with the measured size-dependent $T_m$ of Au on W[271] and C[272] substrates, and Au particles encapsulated in a Silica matrix.[273] The theoretically predicted melting curves merge at the lower end of the size limit, $K_j = 1.5$ with a ~76% $T_m$ suppression, being identical to the predicted temperature of the MC melting ($0.239T_m$). Therefore, thermal rupture of the Au-Au chain occurs just below the temperature for the MC melting of 320 K that is a 0.239 fold of the bulk $T_m$ of 1337.33 K.

At low temperatures, the shortened bond is twice as strong (strength = $E_i/d_i \propto C_i^{-2} = 0.7^{-2} \approx 2$), agreeing with that predicted in Ref 48. This result means that more force is required to stretch or compress a single bond in the MC by the same length compared to the force needed to stretch the same bond in the bulk at a very low temperature. This finding also explains why the thermal expansion coefficient of the mesoscopic system is smaller than the bulk value.[128-131] Table 7 lists the information derived from the reproduction of the size dependence of the Au-4f level shift and the melting point suppression of Au nanosolids.

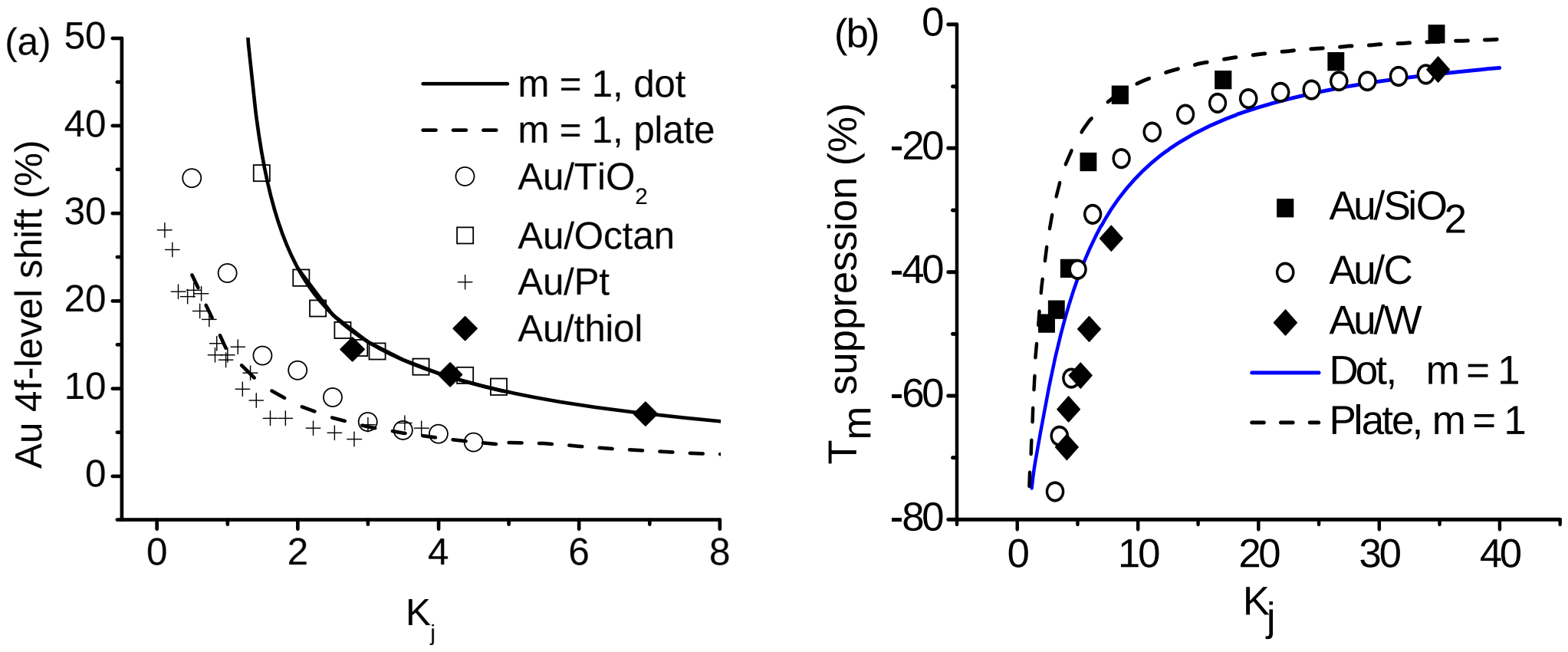

Figure 15 Link Theoretical reproduction of the measured size dependence of (a) $[E_{4f}(K_j) - E_{4f}(\infty)]/[E_{4f}(\infty) - E_{4f}(1)]$ of Au (nano-dot) on different substrates with derived information as given in Table 7, and, (b) $[T_m(K_j) - T_m(\infty)]/T_m(\infty)$ of Au nanosolids on various substrates, showing strongly interfacial effects and the dimensionality-based transition of the melting point of an Au nanosolid. Melting at the lower end of the size limit (K = 1.5, z = 2) corresponds to the situation of an Au-MC.

Table 7 The length and energy of the Au-Au bond in the monatomic chain and the core-level energy of an isolated Au atom obtained from decoding the size-dependent $E_{4f}(K_j)$ and $T_m(K_j)$ of nanosolid Au.

| | Au/Octan | $Au/TiO_2$ | Au/Pt | Au/Thiol |
|---|---|---|---|---|
| τ (dimensionality) | 3 | 1 | | 3 |
| $E_{4f}$(eV) | -81.504 | -81.506 | -81.504 | -81.505 |
| $\Delta E_{4f}(\infty)$(eV) | -2.866 | -2.864 | -2.866 | -2.865 |
| $d_{MC}/d$ | 0.2001/0.2878 = 0.7 | | | |
| $E_{MC}/E_b$ | 1.43 | | | |

| | |
|---|---|
| $P_{MC}/P$ (Strength) | 2.0 |
| $T_{m,MC}/T_m(\infty)$ | 1/4.185 = 0.239 |

4.3.3 The breaking limit and the specific heat

In order to examine the validity of the prediction, we employed the known values of thermal expansion coefficient $\alpha = 14.7\times10^{-6}$ $K^{-1}$, $T_m$ = 1337.33 K, and d = 0.2878 nm to predict the maximum strain of the Au-MC using eq (29). Using the measured breaking limits of $d_{iM}$(4.2 K) = 0.23 nm,[87] and the mean $d_{iM}$(300 K) = 0.35 nm,[248,251,252] leads to the quantification of the two unknown parameters of $\beta_0$ = 5.0 $TPa^{-1}$ and $\eta_{2i}/\eta_{1i}$ = 64 K.

Using the relation $E_i = c_i^{-1}E_b$, or, $\eta_{1i}T_{m,i} + \eta_{2i} = C_i^{-1}(\eta_1 T_m + \eta_2)$, and the given $\eta_1$ = 0.0005542 eV/K and $\eta_2$ = -0.24 eV for the fcc structures,[136] we obtained $\eta_{1i}$ = 0.0033325 eV/K and $\eta_{2i}$ = 0.2128 eV. The correct bulk value of $\eta_2 = C_i\eta_{2i}$ = 0.14897 eV, being compatible with that of diamond structures of $\eta_2$ = 0.24 eV. The $\eta_2 < 0$ in Ref [136] means that the actual energy for evaporating the molten atom is included in the term of $\eta_1 T_m$, and therefore, the $\eta_1$ may not represent the true value of the specific heat per coordinate. Accuracy of solutions gained herewith is subject strictly to the given $\eta_1$ and $\eta_2$ values and the precision of the measured $d_{iM}(T \neq 0)$ values used for calibration, as no freely adjustable parameters are involved in the iteration of calculations.

A refinement of eqs (27) and (29) by taking the pressure-induced deformation energy $-\int_0^P d(1+\beta p)dp = d(\overline{P} + \beta\overline{P}^2/2)$ into consideration, may lead to a high $\eta_{2i}/\eta_{1i}$ ratio without changing the trend of the predicted strain-temperature relation.

Figure 16(a) compares the calculated maximum strain versus $T/T_m$ with the measured values for the Au-MC at various temperatures. It is exciting to find that the theoretical curve covers all the divergent values measured at 4.2 K (0.23 ± 0.04 nm) and at the ambient temperatures (298 ± 6 K, 0.29 ~ 0.48 nm). The divergent data are actually centered at some 22 K below the melting point, 320 K, of the Au-MC with a 6 K fluctuation. The fluctuation may arise from differences in the temperature of testing, or the strain rate applied during measurement. Therefore, all reported values are correct but the measurements might have been conducted in different seasons or different locations. According to Egerton,[261] the temperature rise of a TEM specimen in micrometer size caused by electron beam in the TEM is less than 2 K. The energy released from bond stretching may contribute to the actual temperature but it is expected to be insignificant and is common to all measurements. Interestingly, the 16-19% elongation limits of Ir-MC and Pt-MC measured using an STM at 4.2 K are also within the prediction. The result shows clearly that the divergent values of the breaking limit measured at the ambient temperature is dominated by the extensibility factor that increases exponentially with temperature and reaches infinity at $T_{m,i}$ and by the thermal and mechanical fluctuations in the measurement. Therefore, the proposed mechanisms of fuzzy imaging, impurity mediation, or charge mediation may need reconsideration by taking the effect of bond order deficiency on the strength and thermal stability of the MC bond into consideration.

Figure 16(a) also suggests that the dominant factor $T_m$ has slight influences on the breaking mode of a MC. The bond of a low $T_m$ specimen breaks more readily at low temperature than the bond of a high $T_m$ specimen; the bond of the low $T_m$ specimen is more easily extended as T approaches $T_{m,i}$ than the ones with higher $T_m$ values.

The above analysis for metallic MC elongation also applies to the elongation of organic molecular chains. A curve shown in Figure 16(b) for the monomer persistence length versus temperature of a single polymer chain adsorbed on an Au(111) substrate showed a similar trend of temperature dependent extensibility to that for the Au-MC despite the complexity of polymer extension because of the involvement of worm-like extension and bond unfolding. [274] The rupture occurs at 315 K and the monomer increases its length from 0.18 to 0.38 nm in an exponential way when the temperature is increased from 300 to 315 K although the rupture event was a stochastic process and depends on many factors such as pulling speed, bond strength, and the temperature of operation.

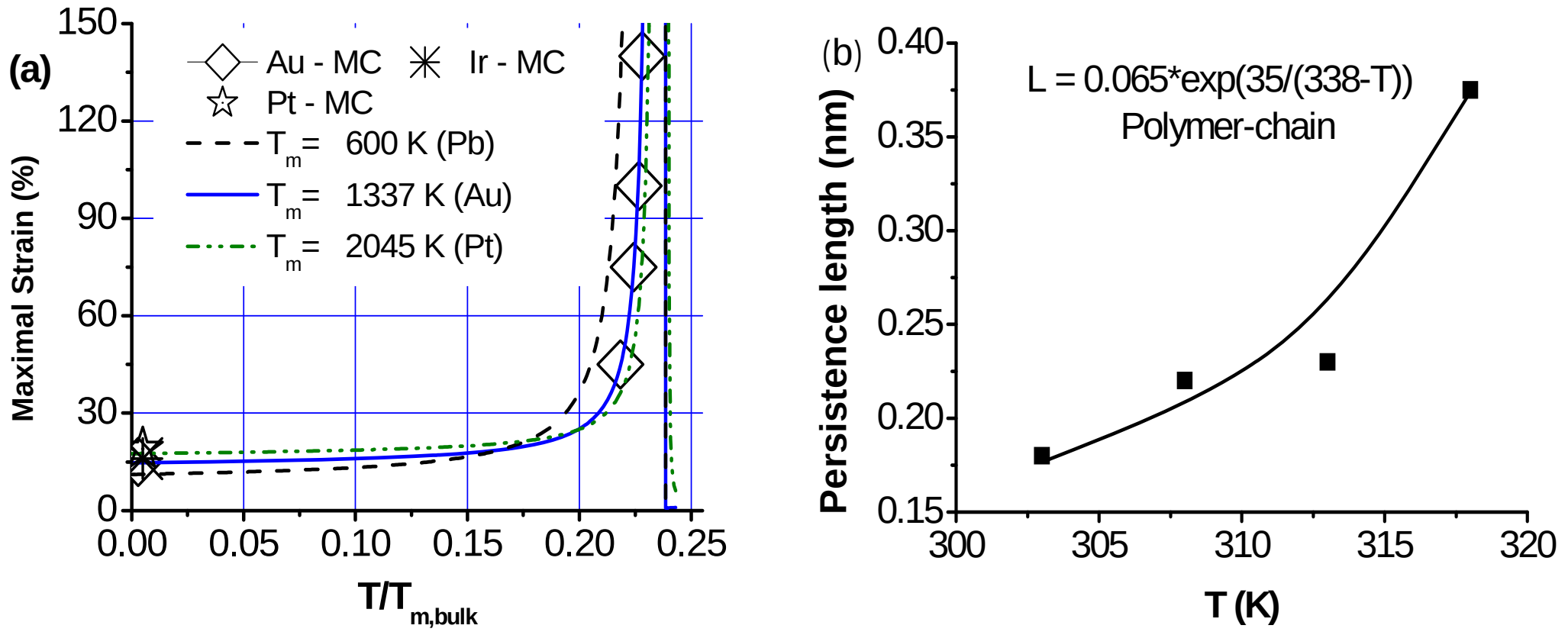

Figure 16 (link) (a) Temperature ($x = T/T_m$) dependence of an MC breaking limit in comparison with values for an Au-MC measured at 4.2 K (0.23 ± 0.04 nm) and at ambient (298 ± 6 K, 0.29 ~ 0.48 nm) indicates that the scattered data arise from temperature dependent extensibility and the thermal and mechanical fluctuations near the melting point of the Au-MC. Varying the $T_m$ changes slightly the ease of MC bond breaking at different temperatures. The 16-19% elongation of Ir and Pt measured using STM at 4.2 K[253] also within the prediction. (b) Temperature dependence of polymer extension also shows the exponential dependence of elongation before breaking. [274]

### 4.3.4 Criteria for MC formation

Equation (29) indicates that a metallic MC melts at a temperature that is a factor of 1/4.185 (0.239) of the bulk $T_m$. A metallic MC could be readily made at a temperature that is ~20 ± 6K lower than its melting point, $T_{m,i}$. Therefore, if one wants to make a MC of a certain specimen extendable at the ambient temperature (300 K), one has to work with the material whose $T_m$ satisfies (300 + 20)× 4.185 = 1343 K or a value near to this. However, an extendable MC can hardly form at room temperature or above if the bulk $T_m$ of a specific metal is below 300 × 4.185 = 1260 K, such as Sn (505.1 K), Pb (600.6 K), and Zn (692.7 K). An extendable Ti-MC (with $T_m$ = 1941 K) may form at ~440 K, slightly lower than its $T_{m,i} = T_m /4.185$ = 462 K. Therefore, it is quite possible to make a specific MC by operating at a carefully controlled temperature. It is not surprising that Au is favorable for such MC formation at the ambient temperature whereas Ag ($T_m$ = 1235 K), Al ($T_m$ = 933.5 K) and Cu ($T_m$ = 1356 K) are unlikely to form extendable MC's though they could form NWs with high extensibility at the ambient temperatures. Although the electronic structure may need to be considered in making an MC,[253] the operating temperature would be most critical. The high extensibility is apparent in the temperature range that corresponds to the quasi-solid state that is softer than the bulk. This finding may explain why some metallic extendable MCs can form at the ambient temperature and some cannot.

## 4.4 Summary

The T-BOLS correlation mechanism has enabled us to calibrate the length, the strength, the extensibility, and the thermal stability of the Au-MC bond under the conditions with and without thermal and mechanical stimuli. Major findings are summarized as follows:

(i) Without external stimuli, the metallic bond in an MC contracts by ~30%, associated with ~43% magnitude rise of the bond energy compared with the bulk standard situations. A metallic MC melts at 1/4.2 times the bulk melting point. The electro-affinity (separation between the vacuum level and the band edge of the 4f conduction band) of the Au-MC is 1.34 eV greater than the bulk, which is responsible for the high chemical reactivity of the under-coordinated MC.

(ii) The analytical solution shows that the strain limit of a metallic bond in a MC under tension does not apparently vary with mechanical stress or strain rate but apparently with temperature in the form $exp[A/(T_{mi}-T)]$, or $exp[A/(T_m- 4.2\times T)]$. This relation governs the tendency for a metallic MC to form or break, and therefore, an MC of other elements could be made by operating it at properly controlled range of temperature. However, as extrinsic factors, the stress field and the strain rate could be important in the experiments relating to MC elongation, which is beyond the modeling consideration.

(iii) Matching the calculated Au-Au distance to all the insofar-measured values indicates that the divergency in measurements originates from thermal and mechanical fluctuations and the extremely high extensibility near the melting point and no charge or atomic impurity mediation is necessary.

Findings provide a consistent insight into the coordination-imperfection-enhanced binding intensity, mechanical strength, the suppressed thermal stability, and the compressibility/extensibility of a MC, which could be extended to the thermal and mechanical behavior of other metallic nanowires. The developed approach provides an effective way of determining the bulk 0 K extensibility, $\beta_0$, the effective specific heat $\eta_{1i}$ per coordinate, and the energy ($\eta_{2i}$) required for evaporating an atom from the molten MC. Practical data would be helpful to give information on the MC bonding identities and the single electron energy level of an isolated atom; one of the challenging tasks for nanometrology.

# V Nanotubes and nanowires

## 5.1 C-C bond in single-walled carbon nanotubes

### 5.1.1 Modulus measurement

The strong bonds between adjacent carbon atoms make individual nanotubes one of the toughest materials ever known. A rather stiff CNT gel can be made by physically grinding up SWCNTs with ionic liquids. The stiffness results from a physical cross-linking of the tubes with the ionic liquid. The gels showed good thermal and dimensional stability and could be shaped into conductive sheets that also had enhanced mechanical properties. Using a coagulation-based CNT spinning technique, Dalton et al[50] spun surfactant-dispersed SWCNTs from a rotating bath of aqueous polyvinyl alcohol to produce CNT gel fibers that they then converted to solid nanotube composite fibers. The resulting 100 m long fibers were 50 μm thick and contained around 60% nanotubes by weight. The composite has a tensile strength of 1.8 GPa and an energy-to-break value of 570 J/g. The fibers, which are suitable for weaving into electronic cloth, are four times tougher than spider silk and 17 times tougher than the Kevlar fibers used in bulletproof vests. The fibers also have twice the stiffness and strength and 20 times the toughness of the same weight of a steel wire. As artificial muscles, nanotube fibers are some 100 times the force of natural muscle with the same diameter.[275] The toughness of spider silk that is dominated by hydrogen bonds[276] is due to chain extension in amorphous regions between relatively rigid crystalline protein blocks.

However, precision determination of the Young's modulus of the CNTs has been a challenging issue for a long time. [277] For instance, a TEM nanorobotic manipulator measured the Y values of 3.5,[278] 1.25,[279] and 1.23 TPa[280] for the SWCNT. It has been reported that the Y values vary over a range of 0.5 - 5.5 TPa being subject to the presumption of the wall thickness. The Y value for the bulk graphite or diamond is 1.02 TPa. In practice, one can only measure the product of the Young's modulus and the wall thickness, or called stiffness, Yt, of the SWCNT, rather than the individual component, Y or t. Therefore, the measured Y values are yet to be confirmed. If one assumes the equilibrium interlayer spacing of graphite sheet as $t_l$ = 0.34 nm, to represent the single-wall (bond) thickness, the derived $Y_l$ value is ~1.1 TPa.[49,281] If $t_l$ = 0.066 nm, which is close to the radius of a free carbon atom (0.0771 ~ 0.0914 nm), the $Y_l$ is derived as 5.5 TPa. [282,283,284,285] This uncertainty in bond thickness results in the large variety of the reported Y values.[286,287,288] Although the measured values of $t_l$ and Y are widely scattered, the product of Yt (stiffness) surprisingly approaches a constant value of 0.3685 ± 0.0055 TPa·nm, as documented in the literature.

For the SWCNTs, under a presumed $t_l$ value, the measured $Y_l$ varies insignificantly with the tube diameter or the tube helicity though the curvature-induced strain may contribute.[289] However, for the multi-walled (MW) CNTs, two typical trends of the change in Y values have been observed:

- The Y value remains almost constant for a given number of walls without changing with the tube diameter,[49] and,
- The Y value increases as the number of walls (λ) or wall thickness is reduced.[290]

5.1.2 Thermal stability

On the other hand, the CNTs are less chemically and thermally stable compared with the bulk graphite or diamond crystal. Atoms at the open edge of a SWCNT could even melt or coalesce prior to atoms in the tube body at temperature much lower than the melting point of the bulk graphite ($T_m$ = 3800 K). The coalescent temperature of the MWCNT increases as the number of walls increases. Coalescence of the SWCNT happens at 1073 K under energetic (1.25 MeV) electron beam irradiation and the coalescence starts at sites surrounding atomic vacancies via a zipper like mechanism,[291] indicating that the under-coordinated atoms melt more easily and the vacancy provides site for structure failure. The STM tip-end that is made of SWCNT starts to melt at 1593 K in ultrahigh vacuum.[292] The electron beam irradiation lowers at least the melting point by some 500 K due mainly to the impulse of the energetic electrons, according to Egerton.[261] Annealing at 1670 - 1770 K under medium-high vacuum, or in flowing Ar and $N_2$ atmospheres, 60% SWCNTs coalesce with their neighbors.[293] Heating under an Ar flow in the temperature range of 1873 - 2273 K results in a progressive destruction of the SWCNT bundle, and this is followed by the coalescence of the entire CNT bundle.[294] Coalescence starts at the edge of CNT bundles.[295] SWCNTs transform at 2473 K or higher to MWCNTs with external diameter of several nanometers. Fe-C impurity bonds can be completely removed from the CNTs at 2523 K.[296] MWCNTs are more thermally stable than the SWCNTs, and the stability of the MWCNTs increases with the number of walls.[286,297,298] On the other hand, an ordinary camera flash[299] could burn the SWCNT at the ambient conditions, showing the higher chemical reactivity for oxidation of the SWCNT. These observations evidence consistently the lowered chemical and thermal stability and at raised temperatures albeit the higher mechanical strength of the CNTs at the ambient temperature.

A recent work[300] suggested that a $T^{1/2}$-dependent dynamic thickness due to atomic vibration should be added to the effective thickness of the C-C bond in the SWCNTs and hence one should observe the sum of the static and the dynamic values. The dynamic thickness amounts at 0.0115 - 0.0304 nm depending on chiral index is suggested to be responsible for the Raman shift, Young's modulus and thermal conductivity reduction of the SWCNTs.

## 5.2 T-BOLS derivatives

### 5.2.1 C-C bond identities

The striking difference between the bulk carbon and a SWCNT is that the effective atomic CN (or $z_i$, i = 1 for a SWCNT) of a C atom reduces from a CN of 12 (diamond) to a CN of three upon SWCNT formation. For an atom surrounding a defect or being located at the open edge, the atomic CN is two. The effective atomic CN of a C atom in the diamond bulk is always 12 rather than 4 as the diamond complex unit cell is an interlock of two fcc primary unit cells. Comparing the covalent bond length of diamond (0.154 nm) with that of graphite (0.142 nm), the effective atomic CN of carbon in graphite is derived to be 5.5 according to the BOLS correlation (eq (2)).

The known values of $(Yt)_1$ = 0.3685 TPa·nm for z = 3, and the known $T_{m,1}$(for z = 2 at the tip end) value (= 1593 K) for a SWCNT and the known bulk values for carbon ($T_m$ = 3800 K, $Y_b$ = 1.02 TPa for z = 12) should satisfy a combination of the following relations:

$$\begin{cases} \dfrac{T_m(2)}{T_m(12)} = z_{2,12}C_i(2)^{-m} = C_i(2)^{-m}/6 & (tip-open-edge) \\ \dfrac{T_m(3)}{T_m(2)} = z_{3,2}\left(\dfrac{C_i(3)}{C_i(2)}\right)^{-m} = \dfrac{3}{2}\left(\dfrac{C_i(3)}{C_i(2)}\right)^{-m} & (wall-tip-relation) \\ \dfrac{(Yt)_1}{Y_b t_1} = [C_i(3)]^{-(2+m)} & (CNT-wall) \end{cases} \tag{31}$$

The value of $C_i(z)$ is given in eq (2). Solving these simple equations gives immediate solutions of m = 2.5585, $t_l$ = 0.142 nm and the tube-wall melting point, $T_m(3)$ = 1605 K. Furthermore, the activation energy for chemical reaction is also a portion of the atomic cohesive energy, $z_{ib}c_i^{-m}$. Therefore, the chemical stability of the under-coordinated atoms is lower than the bulk values, which may explain why the CNT could burn using an ordinary camera flash under the ambient conditions.

The accuracy of the numerical solutions is subject to the measured input of $T_m(2)$ and the product of $(Yt)_1$, as a whole. Errors in measurement or structural defects of the CNT may affect the accuracy of the solutions; however, they never determine the nature of the phenomena observed. Variation of any of the input (bold figures in Table 8) leads to no acceptable solutions. For example, replacing the quoted Yt value with Y = 0.8 TPa (disregarding the thickness $t_l$) gives m = -3.2. The corresponding values of $E_l/E_b = C_i(3)^{-m}$ = 0.52, and $T_m(3) = z_{1b}\times C_i(3)^{-m}\times T_m(12)$ = 493 K are unacceptable because the $T_m(3)$ value is much lower than any reported values. Assuming the SWCNT tip-end melting point to be $T_m(2) \geq 1620$ K gives an m value that is larger than 2.605 and $T_m(3) \leq 1620$ K, which is not in line with observation: the open edge melts first. If a value of m = 4 is taken, $T_m(2)$ = 2674 K and $T_m(3)$ = 2153 K, which are much higher than the measured values despite the order of tube edge and tube body melting. Therefore, the solution with the measured input values of Yt = 0.3685 and $T_m(2)$ = 1593 K is unique and, hence, the quoted Yt and $T_m(2)$ values are essentially true for the SWCNT.

Table 8 Comparison of the calculation results with various input parameters (bold figures) proving that the obtained solution is unique and the quoted data represent the true values.[120]

| Yt (TPa·nm)<br>or Y(TPa) | $(Yt)_l$= 0.3685<br>Y = 2.595 | **Y = 0.8** | $(Yt)_l$= 0.3685 | |
|---|---|---|---|---|
| Tip-end $T_m$ (2) (K) | 1593 | - | 2153 | **≥1620** |
| Tube wall $T_m$(3) (K) | **1605** | 493 | 2674 | ≤1620 |
| M | **2.5585** | 2.5585 | **4** | 2.6050 |
| Bond thickness $t_l$(nm) | 0.142 | - | - | - |
| Bond length $d_l$(nm) | 0.116 | - | - | - |
| Bond energy $E_l/E_b$ | 1.68 | 0.53 | - | - |
| Remarks | Acceptable | Forbidden | | |

### 5.2.2 Superplasticity

The theoretical maximum tensile strain, or elongation, of a SWCNT is almost 20%,[301,302] but in practice only 6% [303] has been achieved at room temperatures. However, at high temperatures, individual SWCNTs can undergo a superplastic deformation, becoming nearly 280% longer and 15 times thinner, from 12 to 0.8 nm, before tensile failure.[86]

It is significant to note that the temperature in the middle of the SWCNT is more than 2,270 K as estimated during deformation at a bias of 2.3 V with a current flow. Despite the apparent discrepency of this estimate compared with the STM measurements showing that the tube body melts at 1605 K, the nanotube appears to be completely ductile near the melting point, according to the findings for Au-MC (140% elongation near the $T_m(2)$). Compared to the Au-MC, kinks and point defects in the CNT are involved and fully activated, resulting in possible superplastic deformation of CNT at the elavated temperatures. The kink motion is evidence of kink-mediated plasticity at high temperatures. Experimental evidence shows that the processes of kink nucleation and motion and atom diffusion are important during superplastic deformation, helping to heal defects such as vacancies and to prevent the formation of large dislocation loops that might initiate cracks and lead to failure of the strained nanotubes.

Such large plastic strains in nanotubes demonstrate their ductile nature at high temperatures,[304,305,306] which concurs with BOLS predictions that the strain limit is exponentially proportional to the inverse of separation between the melting point and the temperatiure of operation. In contrast, tensile-pulling experiments at room temperature without any bias showed that almost all nanotubes failed at a tensile strain of less than 15%. It is expected that superplasticity of multi-walled CNT may be possible in vacuum at more elavated temperatures as the melting point of the MWCNT is higher than the SWCNT.

## 5.3 Multi-walled CNT

Compared with a nanosolid, the surface-to-volume ratio of a hollow tube includes both the inner and the outer shells, which can be expressed as:

$$\gamma_{ij} = \frac{2\pi\left(K_j + L_j\right)C_i}{\pi\left(K_j^2 - L_j^2\right)} = \frac{2C_i}{K_j - L_j} \propto \frac{C_i}{\lambda_j} \qquad (32)$$

$K_j$ and $L_j$ is the dimensionless form of the outer and the inner hollow radii for a MWCNT. $\lambda_j$ is the wall thickness. The ratio $\gamma_S = \sum_{i\le 3}\gamma_{ij}$ decreases in a $\lambda_j^{-1}$ fashion from unity to infinitely small, when $\lambda_j$ grows from unity to infinity. Therefore, it is not surprising that, for a solid rod or a MWCNT with $L_j << K_j$, the overall quantity change, $\Delta Q(\lambda_j)/Q$, varies with the inverse radius ($1/K_j$) and the $\Delta Q(\lambda_j)/Q$ value differs from the corresponding bulk value ($\Delta Q(\lambda_j)/Q = 0$). For a hollow MWCNT with a constant $\lambda_j$, the $\Delta Q(\lambda_j)/Q$ should vary with the diameter of the MWCNT insignificantly. These predictions agree well with the observed mechanical strength of nanobeams. The broad range of the reported Y values for MWCNTs should be due more to the fluctuation in the number of walls of the MWCNTs than to the error in measurement.

## 5.4 Nanowires

### 5.4.1 Elasticity and strength

The Young's modulus enhancement has been widely seen from nanorods of materials such as Ag, Pd,[51] and ZnO nanobelts[40] and nanowires.[42] The Young's modulus of Ag nanowires[45] of 20-100 nm diameter was measured to increase when the diameter is decreased, which was attributed to the effects of surface stress, the oxidation layers, and the surface roughness. ZnS nanobelts were measured to be 79% harder but 52% lower in elastic modulus compared to the bulk ZnS.[61] The ZnS nanobelts were also observed to exhibit significant creep under a constant indentation load at room temperature.

Using an AFM, Wong et al[49] measured the radius dependence of the Young's modulus of SiC nanorods and MWCNTs and found that the MWCNTs are about twice as stiff as the SiC nanorods. The modulus of MWCNTs varies insignificantly with the diameter of the CNTs. The strengths of the SiC nanorods are substantially greater than those found for large SiC structures (600 GPa). The Young's modulus is 610 and 660 GPa for SiC rods of 23.0 and 21.5 nm across, respectively, showing the trend of the Y value increasing with the inverse of the rod diameter.
However, an opposite trend has been measured using AFM bending methods in the (0001) oriented ZnO nanobelts/wires and showed a lower modulus than that of bulk ZnO (measured at 140 GPa) varying from 29±8,[55] 52,[56] to 38–100 GPa,[57]. The modulus of Cr[58] and Si[59] nanocantilevers is also measured to decrease sharply with decreasing diameter. In contrast, amorphous Si nanowires,[307] Au[62] and Ag[308] nanowires show no apparent change with size despite the scatter within the error bars in the measurement. Therefore, it appears quite confusing that, even for the same material such as Ni, Ag, ZnO, and Si, the Y value changes in different ways, depending on the experimental techniques and operation conditions. This is similar to the variation in the measured skin hardening and softening as discussed in section 3. For instance, the Y value of ZnO nanobelts of 50–140 nm in thickness and 270–700 nm in width were measured using an AFM three-point bending method to be 38.2 ± 1.8 GPa, which is about 20% higher than the Young's modulus of 31.1 ± 1.3 GPa obtained using nanoindentation.[39] A modeling study of the size effect on the elastic behavior of solid and hollow polymer nanofibers under uniaxial tension[309] shows that fiber radius has appreciable effect on the elastic response of polymer nanofibers. At nanoscale, solid nanofibers show less the effect of surface tension coupling. However, hollow nanofibers show greater axial stiffening effect with increasing axial stretch because of the coupling of surface tension depending upon the combination of the fiber exterior and interior radii and the material properties. The elastic modulus of single polymer nanofibers has been measured to increase exponentially as the diameter of the polymer nanofibers decrease to diameters of a few tens of nanometers.[310] The unusual behavior of nanobeams was attributed to the microstructure and confinement that play critical roles in engineering the mechanical properties of nanoscale materials.[308] The defect-free nanotubes are ideal cases of cylindrical nanocavities with shells that are much stronger than the bulk materials unless excessive defects are presented in the walls. Defects in the walls of nanotubes serve as centers initiating the failure in particular for the plastic deformation. For instances, the Young's modulus of a defected nanotube is reduced gradually with each atomic defect and the strength of the nanotubes is catastrophically influenced by the existence of just a few atomic defects.[93]

### 5.4.2 Nanowire superplasticity: bond unfolding

We may extend the extensibility of MC and CNT to the case of nanowires by replacing the $T_{m,i}$ and the $\eta_{2i}/\eta_{1i}$ for an MC in eqs (25) and (29) with the size dependent $T_m(K_j)$ and $\eta_2(K_j)/\eta_2(K_j)$. The T-dependent extensibility or compressibility and the strain limit of a defect-free nanosolid become:[311]

$$\frac{\Delta\beta_i(z_i,K_j,T)}{\beta_0(z_b,0)} = \frac{\eta_{1b}C(K_j)T_m}{\eta_1(K_j)[T_m(K_j)-T]}-1$$
$$\frac{\Delta d_M(K_j,T,P)}{d(K_j,T,0)} \cong B\exp\left\{\frac{\eta_2(K_j)}{2\eta_1(K_j)[T_m(K_j)-T]}\right\}$$
(33)

The counter plots in Figure **17** illustrate the size and $T/T_m$ dependent extensibility and the maximum strain of impurity-free Au-NWs with the parameters determined for the Au-MC. The ratio $\eta_1(K_j)/\eta_1 = 1+10/(1+\exp((K_j-1.5)/20))$ is assumed to change from $\eta_{1i}/\eta_1 = z_{bi} = 6$ to 1 (at $K_j = \infty$) gradually. We may note that the mean bond contraction coefficient $C(K_j)$ drops from 1 to 0.7, and the $T_m(K_j)/T_m(\infty)$ drops from 1 to 0.239 when a NW of infinite size shrinks into an MC. Eq (33) indicates that the extensibility enhancement happens ($\Delta\beta_i > 0$) when $[T_m(K_j) - T] < \eta_1 C(K_j)T_m/\eta_1(K_j)$, otherwise, the extensibility is lower than the bulk value. When T approaches $T_m(K_j)$, the extensibility increases exponentially up to infinity at $T \sim T_m(K_j)$.

Measurements have shown that the detectable maximum strain of a suspended impurity-free Au-MC bond is less than 140% ((0.48 - 0.20)/0.20) compared with the equilibrium Au-Au bond length in the MC, which is much lower than the detected strain ($10^3$) of nanograined Cu and Al NWs that could form at room temperature or at the sub-ambient temperatures. [312,313,314] Therefore, bond stretching discussed herewith is not the factor dominating the high extensibility of a NW. The factors dominating the NWs extensibility could be the bond unfolding, atomic gliding dislocations, creep and grain boundary movement.[247,312,313,314] However, the present understanding further confirms that the barrier or the activation energy for atomic dislocation and diffusion of the under-coordinated atoms at the grain boundaries is lower than that of the fully coordinated ones in the bulk, as these activities are subject to the atomic cohesion which, in turn, drops with atomic CN.

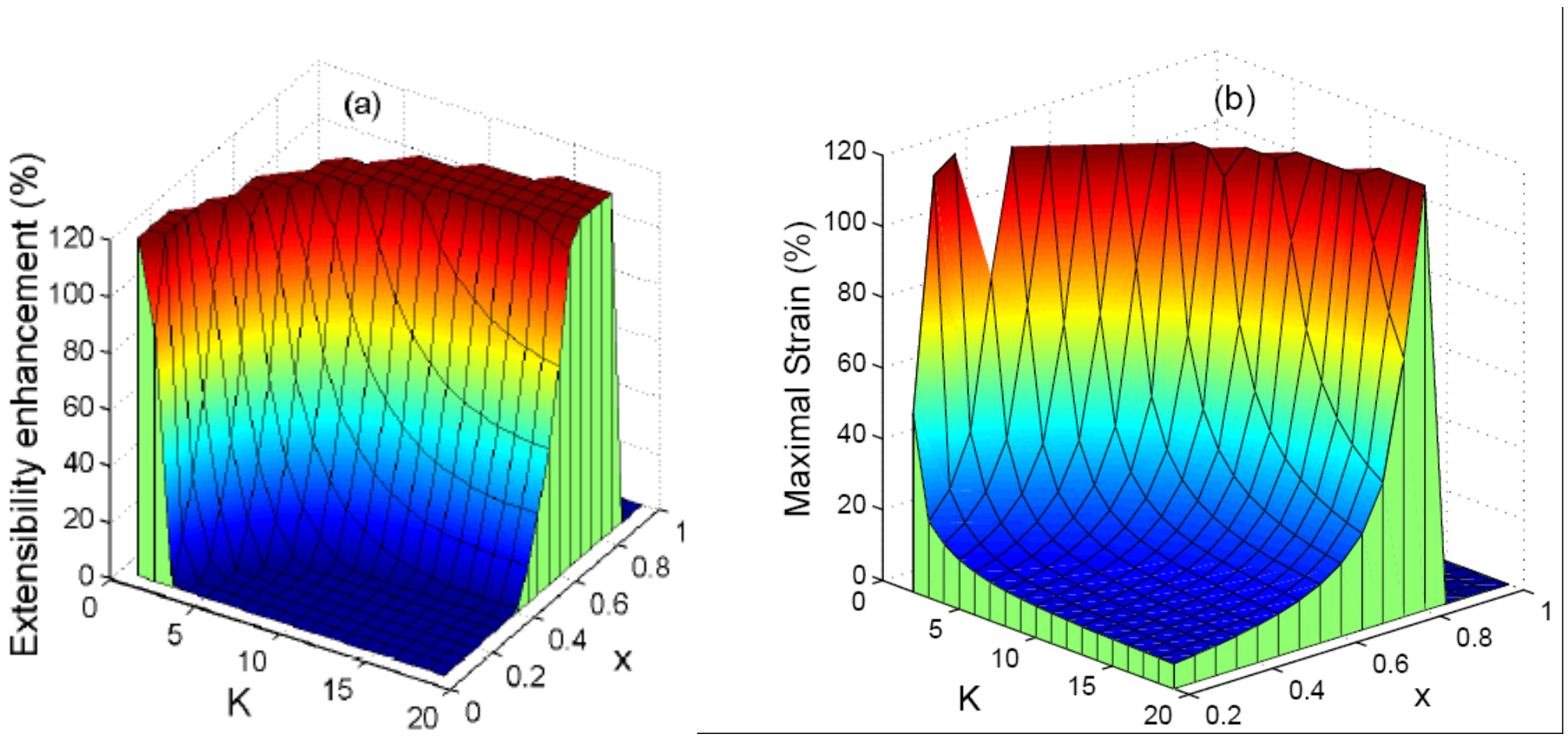

Figure 17 (linka; link b) Illustrative counter plots for the $K_j$ and x (=$T/T_m$) dependent (a) extensibility and (b) maximum strain of defect-free Au-NWs.

The extensibility and the maximum strain increase rapidly when T approaches to $T_m(K_j)$ which, in turn, drops with $K_j$.

5.4.3 Breaking modes of nanowires

Figure 18 showing the atomic CN and temperature dependence of maximum bond strain indicates that the melting of a nanosolid starts from the first atomic shell ($z_1 = 4$) and then the second ($z_2 = 6$) when the temperature is elevated, as observed in many cases. For instances, it has been confirmed that a flat or a curved surface melts at temperatures of 50 K [315] to 100 K[316] lower than the bulk interior. A quasi-liquid skin grows radially inward from the surface into the core center for both clusters and wires. The surface melting is followed by a breakdown of order in the remaining solid core. The melting of an impurity-free vanadium nanosolid proceeds in a stepwise way, i.e., the surface layer of 2 - 3 lattice-constant thick region melts first and then the abrupt overall melting of the entire cluster follows.[316] This surface pre-melting agrees with the 'liquid skin nucleation and growth' mode,[317,318,319,320,321] the 'surface phonon instability' models, [322] and the current BOLS predictions.[124]

At temperatures close to the $T_{m,1}$ of the surface, the maximum strain and the extensibility of the surface layer approach infinity whereas the strain limit of the core interior remains at the limited bulk values because of the higher bulk melting point. At temperatures near surface melting or above, bond breaking under tension should start from the NW interior, because the inner bonds firstly reach their strain limits, being lower than that of the surface skin bonds at the same temperature. However, at temperatures far below $T_{m,1}$, bond breaking may start from the outermost atomic shell and the nanosolid manifests brittle characteristics, as the shortened surface bonds break first. At very low temperatures, the theoretically allowed maximum strains for the entire NW should be constant. If one deforms the entire nanowire by xd per unit d length, the strain of the bonds in the respective shells will be $\varepsilon_i = x/C_i$. Because $C_1 < C_2 < C_3$, the actual applied strains are in this order, $\varepsilon_3 < \varepsilon_2 < \varepsilon_1$, which indicates that the surface shell bond breaks first at low T. Therefore, the breaking mode of a nanowire at low T is expected to be opposite to that at higher T. At very low T the surface bond breaks before the bulk ones while at $T \sim T_{m1}$ or higher the surface bond breaks after the ones in the core interior.

It has been measured using a "nano-stretching stage" located within a STM at room temperature that the MWCNTs break in the outermost shell.[323] This 'sword-in-sheath' failure mode agrees with the expectation of the current BOLS approach as the operating temperature is far below the tube melting temperature at 1600 K.[120] The helical multi-shell gold nanowires[324] become thinner and thinner without breaking the outer-shell atomic bond under tension at room temperature,[325] as the Au-NW breaking starts from the inner shell according to the current understanding.

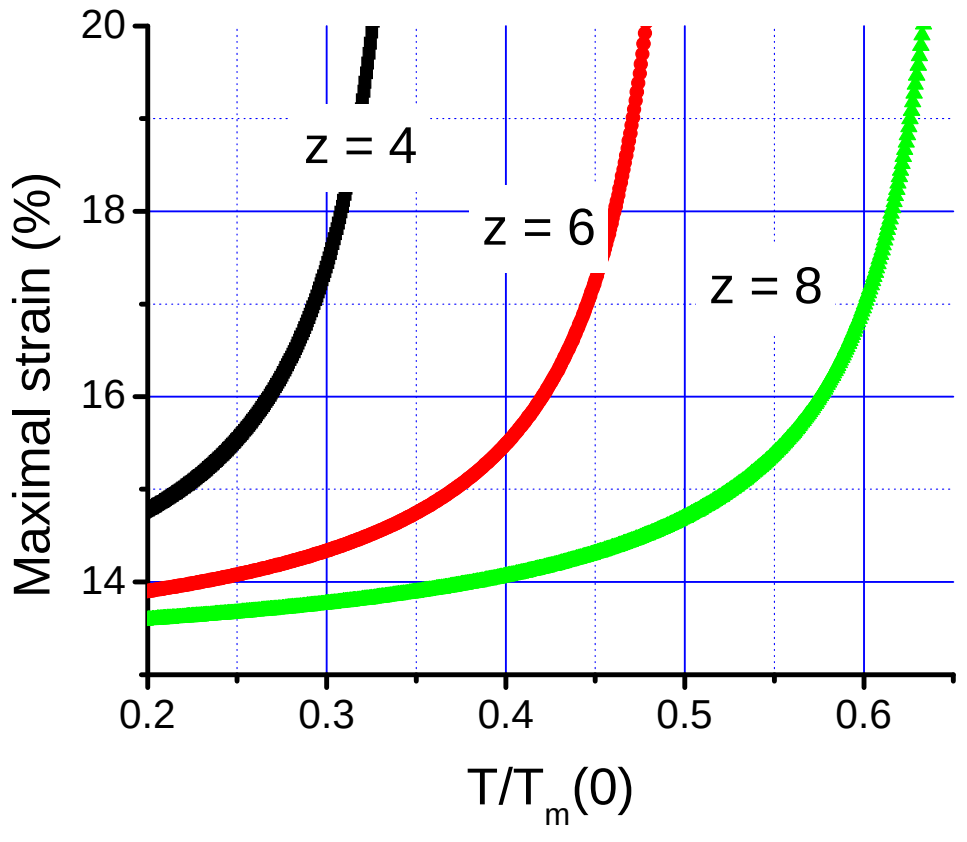

Figure 18 (link) Temperature and CN (z) -dependence of the Au-Au bond maximum strain shows the order of melting at a curved surface of a nanowire and infers the breaking mode of a nanowire at different temperature range.

5.5 Summary

Incorporating the T-BOLS correlation to the known Yt value and the known temperature of tip-end melting for a SWCNT, and their functional dependence on the bonding identities, the dimension and energy of a single C-C bond in SWCNT can be uniquely quantified, which, in turn, deepens our insight into the fascinating properties of CNTs. It has been clarified that the C-C bond of the SWCNT contracts by ~18.5% with an energy rise by ~68%. The effectively static thickness of the C-C bond is ~0.142 nm, which is the diameter of a C atom, rather than the graphite sheet separation (0.34 nm) or the radius of a free carbon atom (0.066 nm). The melting point of the tube-wall is slightly (~ 12 K) higher than that of the tube-end. The unique solution clarifies that the quoted $T_m(2)$ and the Yt values essentially represent the true situations of a SWCNT in which the Young's modulus is 2.5 times and the melting point is 0.42 times that of bulk graphite.

Predictions of the wall thickness dependence of the $T_m$ suppression and Y enhancement of the nanobeams agree well with the insofar-observed trends documented. It has been clarified that bond unfolding, atomic gliding dislocations, creep and kink formation dominates the superplasticity of nanograined crystals or the SWCNTS in quasi-solid states because the detectable maximum strain of an ideal bond is limited to 140% at melting. The findings provide a consistent insight into the unusual thermal, chemical, and mechanical behavior of nanobeams, as well as an effective approach toward bonding identities that are beyond the scope of direct measurement.

VI Nanograins: I. Elasticity and extensibility

## 6.1 Existing approaches

### 6.1.1 Size dependence

From the perspective of minimal strain energy at equilibrium, Ouyang *et al.* [326,327] developed a model for the size dependence of Young's modulus. The total strain energy of a nanocrystalline can be decomposed into the strain energy of the bulk $(U_b)$ and the surface, $(U_s)$, i.e., $U = U_b + U_s$. The deformation from the bulk crystal lattice to the self-equilibrium state will be achieved by minimizing the total strain energy of nanocrystals. The strain in a self-equilibrium state in nanocrystals can be calculated by $\partial U / V_0 \partial \varepsilon_{ij} = 0$, in which $V_0$ and $\varepsilon_{ij}$ $(\mathrm{i, j} = 1,2,3)$ are the volume and the elastic strain, respectively. The size dependent Y-modulus of spherical nanocrystals based on the size dependent surface free energy is derived as,[327]

$$\frac{Y(K_j)}{Y(\infty)} = \left(1 - K_j\right)^5 + \frac{5Y_s}{3Y(\infty)}\left(3K_j - 6K_j^2 + 4K_j^3\right),$$

where $Y_s$ denotes the surface Y-modulus and $K_j$ is the dimensionless form of the nanocrystal size.

### 6.1.2 Temperature dependence

The thermal depression of the Young's modulus $Y$ was attributed to anharmonic effects of the lattice vibrations.[328] Watchman *et al.* [329,330] suggested a semi-empirical formula, which is valid for silicon in the high-temperature limit:[331]

$$\frac{\Delta Y(T)}{Y_0} = -AT\exp(-T_0/T) \tag{34}$$

where $Y_0$ is the Young's modulus at 0 K. The constants $A$ and $T_0$ are temperature independent parameters for data fitting. Watchman *et al.* expected a correlation between the $T_0$ and the Debye temperature $\theta_D$ and between the $A$ and the Gruneisen parameter, $\gamma$. A complete theory for this problem was not available at that time. Later, Anderson[332] derived a similar expression for the adiabatic bulk modulus $B$:

$$\frac{\Delta B(T)}{B_0} = -\frac{3R\gamma\delta T}{B_0 V_0} F(T/\theta_D) \tag{35}$$

where

$$F(T/\theta_D) = 3\left(\frac{T}{\theta_D}\right)^3 \int_0^{\theta_D/T} \frac{x^3 dx}{e^x - 1}$$

$R$ is the ideal gas constant, $\delta$ is the Anderson-Gruneisen parameter, $B_0$ is the bulk modulus, and $V_0$ is the volume at 0 K. The product $\gamma\delta$ is assumed temperature independent. Note Eq (35) is an equivalent of,

$$\frac{\Delta B(T)}{B_0} \cong \frac{\Delta Y(T)}{Y_0} = -\frac{\gamma\delta R\theta_D}{B_0V_0}\int_0^T C_v dT = -\frac{\gamma\delta R\theta_D}{B_0V_0}U(T/\theta_D),$$

which shows the close relationship between the mechanical property $B$ and the specific heat $C_V$ or the internal vibration energy, $U(T/\theta_D)$. Eq. (35) satisfies Nernst's theorem, where the temperature derivatives of the elastic constants must vanish at 0 K. The B correlates to Y by $Y/B = 3\times(1-2\nu)$, where ν denoting the Poisson ratio is negligibly small, and therefore, Y ≈ 3B. Eq (35) could reproduce the measured Y(T) reasonably well at low temperatures by taking the $\gamma\delta R\theta_D/(B_0V_0)$ as an adjusting parameter.

The Young's modulus determines both the eigenfrequency ω(Y) and the spring constant $k_1$ of a homogeneous nanocantilever with uniform cross-section A = wt (width and thickness) in the following relations:[333,334]

$$\begin{aligned} \omega &= (1.875)^2 \frac{t}{L^2}\sqrt{\frac{Y}{12\rho}} \propto \sqrt{Y} \\ k_1 &= \frac{Yt^3w}{4L^3} \end{aligned}$$

where ρ is the density, L the length, w the width and t the thickness of the cantilever. Neglecting the thermally induced geometrical change of the cantilever, one can measure the vibration frequency and hence the Y-T relationships. Using this approach, Gysin et al[328] measured and fitted the vibration frequency of a silicon cantilever and derived $Y_0$ = 167.5 GPa and $\theta_D$ = 634 K. The derived values correspond to the documented Y value for the <110> direction and the documented $\theta_D$ = 645 K of Si.

## 6.2 T-BOLS considerations

### 6.2.1 Elasticity and extensibility

In order to solve the discrepancy in observations of size dependence, as summarized in Table 1, and the thermal depression of the Young's modulus, we developed expressions for the size, temperature, and bond nature dependence of the elasticity and extensibility of a solid based on the T-BOLS considerations, as discussed as follows.

At a specific *i*th atomic site at a given temperature, the temperature and bond nature dependence of the local stress, $P_i$, and the local Young's modulus, $Y_i$, can be expressed as:[140]

$$
\left.\begin{aligned}
P_i(z_i,m,T) &= -\left.\frac{\partial u(r,m,T)}{\partial V}\right|_{d_i,T} \\
Y_i(z_i,m,T) &= -V\left.\frac{\partial P(z_i,m,T)}{\partial V}\right|_{d_i,T}
\end{aligned}\right\} \propto \begin{cases} \eta_{1i}(T_{mi}-T)/d_i^3, & (Born's-criterion) \\ [\eta_{1i}(T_{mi}-T)+\eta_{2i}]/d_i^3 & (Energy-density) \end{cases}
$$

(36)

The Born's criterion applies only to the extensibility and the plastic deformation but not to the elastic modulus because of the non-zero sound velocity in gaseous and liquid phases. The sound velocity is proportional to the square root of the Young's modulus over mass density.

For a nanosolid, we may take the core-shell configuration of a nanosolid into consideration, then the bond nature (m), solid shape and size ($\tau$, $K_j$), and temperature (T) dependence of the relative change of the Y and the linear extensibility, $\beta$, can be obtained by summing contributions over the outermost three atomic layers. The relative change of the Y and the linear extensibility, $\beta$, of a solid measured at $T_0$ = 0 K can be expressed as,

$$
\begin{aligned}
\frac{\Delta Y(m,K_j,T)}{Y(m,\infty,0)} &= \sum_{i\le 3}\gamma_{ij}\delta_Y(m,z_i,T) \\
\delta_Y(m,z_i,T) &= \frac{C_i^{-3}}{(1+\alpha_i T)^3}\left(\frac{E_i(0)-\int_0^T \eta_{1i}(t)dt}{E_i(0)}\right)-1 \overset{T>>\theta_D}{=} \frac{C_i^{-(3+m)}}{(1+\alpha_i T)^3}\left(1-\frac{\eta_1 T}{z_{ib}C_i^{-m}E_b(0)}\right)-1 \\
\frac{\Delta\beta(m,K_j,T)}{\beta(m,\infty,0)} &= \sum_{i\le 3}\gamma_{ij}\delta_\beta(m,z_i,T) \\
\delta_\beta(m,z_i,T) &= C_i^{(1+m)}(1+\alpha_i T)\left\{\frac{E_i(0)}{E_i(0)-\int_0^T \eta_{1i}(t)dt}\right\}-1 \overset{T>\theta_D}{=} C_i^{(1+m)}(1+\alpha_i T)\left\{1+\frac{\eta_{1i}T}{\eta_{2i}+\eta_{1i}(T_{mi}-T)}\right\}-1
\end{aligned}
$$

**(37)**

Where the relations of $E_i = C_i^{-m}E_b = \eta_{1i}T_{mi} + \eta_{2i} = C_i^{-m}(\eta_1 T_m + \eta_2)$, and $T_{mi} = z_{ib}C_i^{-m}T_m = (1+\Delta_i)T_m$, and $\eta_{1i} = z_{bi}\eta_1$ and $\eta_{2i} = C_i^{-m}\eta_2$ are applied. The layer counting in Eq (37) indicates that the under-coordinated atoms in the surface skin dictate the relative change of the elasticity and extensibility of the entire nanosolid whereas atoms in the core interior retain their bulk features. The Y value drops nonlinearly with T at T << $\theta_D$ because of the involvement of the non-linear specific heat in Debye approximation (see figure 2b). At T > $\theta_D$, Y drops linearly with T in spite of the contribution from thermal expansion. Taking the approximation of $1+x \approx \exp(x)$, eq (37) also agrees with (34) but here we have included the effect of size, temperature, and bond nature.

For a bulk solid ($C_i$ = 1 and $\Delta_i$ = 0), the thermal softening exhibits the bulk feature and no summation over the surface layers is necessary. The T-suppressed Young's modulus can be simply expressed as,

$$\frac{Y(m,T)}{Y(m,0)} = \frac{1}{(1+\alpha_i T)^3}\left(1-\frac{\int_0^T \eta_1(t)dt}{E_b(0)}\right) = \frac{1}{(1+\alpha_i T)^3}\left(1-\frac{U(T/\theta_D)}{E_b(0)}\right)$$
$$\overset{T>>\theta_D}{=} \frac{1}{(1+\alpha_i T)^3}\left(1-\frac{\eta_1 T}{E_b(0)}\right)$$

**(38)**

Apart from the effect of thermal expansion, the present form agrees with Anderson's model of T-dependent Y with further identification of $\gamma\delta R\theta_D/(B_0V_0)$ = 1/$E_b$(0) and the internal vibration energy $3RT \times F(T/\theta_D) = U(T/\theta_D)$.

6.1.2 Debye temperature

The Debye temperature, which is defined as $\theta_D = \hbar\omega_D / k_B$ in Debye approximation of specific heat, is a key parameter that determines the thermal transport dynamics. The $\theta_D$ is actually not a constant but changes with the object size[335,336,337,338] and the temperature of testing,[339,340,341] as the $\theta_D$ depends functionally on Y. Earlier contributions[335] suggested that the size dependent $\theta_D$ result from the finite cut-off of frequency and the surface stress (effectively from a size dependent change of surface pressure), especially, if the size is smaller than 20 nm. Using x-ray-absorption spectra measurement and extended X-ray absorption fine structure spectroscopy, Balerna and Mobilio[336] confirmed the predicted size dependence of $\theta_D$. Calculations of the temperature dependent $\theta_D$ of some fcc and bcc metals[339] revealed that $\theta_D$ drops when the measuring temperature is increased because of the temperature dependence of elastic constants and the sound velocity of the solid. However, a discrepancy remains regarding the $T_m$ dependence of the T-independent $\theta_D$. One opinion is that $\theta_D$ varies linearly with $T_m$[339] and the other suggests a square root dependence of $\theta_D$ on $T_m$ according to Linderman's[342] criterion of melting.

The $\theta_D$ can be derived from the mathematical expression for the normalization of phonon density states: $\int_0^{\omega_D} g(\omega)d\omega = \kappa N_A$, with $N_A$ being the total number of vibration modes, giving rise to the relation, $\omega_D \propto v_s(n)^{1/\kappa} \propto v_s(d^{-\kappa})^{1/\kappa} = v_s / d$. The term $v_s = \sqrt{Y/\rho} \sim \sqrt{Yd^3} \sim \sqrt{E_i}$ is the sound velocity in the medium. The κ is the fractal and dynamic dimensions (κ = 1 for nanowire, κ = 2 for thin plate and κ = 3 for bulk material), which is different from the static dimensionality used for the surface-to-volume ratios. Because the Debye temperature is defined as $\theta_D = \hbar\omega_D / k_B$, the normalized expression for $\theta_D$ has the following form:

$$\frac{\theta_D(z_i,m,T)}{\theta_D(z_b,m,0)} = \frac{\omega_D(z_i,m,T)}{\omega_D(z_b,m,0)} = \frac{v_s(z_i,m,T)}{v_s(z_b,m,0)}\cdot\frac{d}{d_i} = \left(c_i\frac{Y(z_i,m,T)}{Y(z_b,m,0)}\right)^{1/2}$$

(39)

Combining eqs (36) and (39), leads to the immediate relation,

$$\begin{aligned}\frac{\Delta\theta_D(K_j,m,T)}{\theta_D(\infty,m,0)} &= \sum_{i\le 3}\frac{\tau C_i}{K_j}\left\{\frac{c_i^{-(1+m/2)}}{(1+\alpha T)^{3/2}}\left[1-\frac{\int_0^T \eta_1(t)dt}{z_{ib}C_i^{-m}E_b(0)}\right]^{1/2}-1\right\} \\ &\overset{T>>\theta_D}{\cong} \sum_{i\le 3}\frac{\tau C_i}{K_j}\left\{\frac{C_i^{-1}}{(1+\alpha T)^{3/2}}\left[\frac{\eta_{2i}+\eta_{1i}(T_{mi}-T)}{\eta_2+\eta_1 T_m}\right]^{1/2}-1\right\} \\ &\overset{T>>\theta_D}{\cong} \sum_{i\le 3}\frac{\tau C_i}{K_j}\left\{\frac{C_i^{-(1+m/2)}}{(1+\alpha T)^{3/2}}\left[1-\frac{\eta_1 T}{2z_{ib}C_i^{-m}E_b(0)}\right]-1\right\}\end{aligned} \tag{40}$$

Therefore, the current form of the $\theta_D(T)\propto\sqrt{Yd}=\sqrt{E_b(T)}/d$ agrees with the square root dependence of the T-independent $\theta_D$ on $T_m$: $\theta_D \propto T_m^{1/2}/d$.[342] A replacement of $E_b(T)$ with $E_b(T)\approx\eta_2+\eta_1(T_m-T)$ leads to the ($T_m$ – T) dependent $\theta_D=\sqrt{\eta_2+\eta_1(T_m-T)}/d\propto(T_m-T)^{1/2}/d$. For nanostructures, the $\theta_D$ drops with size because the $T_m$ depression.

### 6.2.3 Specific heat capacity

The specific heat capacity is a measurable physical quantity that characterizes the ability of a body to store the heat when the sample is being heated. Depending critically on the $\theta_D$ and hence the Y, the specific heat varies with both object size and the temperature of measurement. The effect of body size on the specific heat capacity has recently attracted considerable attention.[343,344,345,346] For instance, Novotny et al[343] measured the low temperature heat capacity of 2.2 and 3.7 nm sized lead particles and observed the enhancement of heat capacity below 5K. Lu[347] demonstrated that the specific heat of metallic or alloying nanosolids increases with the inverse of solid size. However, an ac micro-calorimeter measurement[344] showed the opposite trend for Al films where the specific heat drops with the thickness of the Al film from 370 to 13.5 nm. The decrease of specific heat was explained as the rise of the absorption and the loss of thermal waves with specific wave vectors in the small volumes. However, Lu et al[345] calculated the size effects on the specific heat of Al thin films employing the Prasher's approach[348] and derived that the reduction of phonon states was not the main reason causing the size effect on specific heat; but a thin layer of Al oxide was responsible for it. In Yu's measurement,[346] the heat capacity decreased with the film thickness; however, the specific heat increased as the film become thinner, in conflicting with the measured results of Song et al.[344] Therefore, discrepancies on the size and temperature dependence of the heat capacity and Debye temperature for metallic nanostructures remain unsolved theoretically.

The heat capacity per unit volume is defined as the ratio of an infinitely small amount of heat δE added to the body to the corresponding small increase in its temperature δT when the volume is kept unchanged. In the extended Debye model, the expression is given by:

$$C_v = \left(\frac{\partial E}{\partial T}\right)_V = \kappa^2 R(\frac{T}{\theta_D})^\kappa \int_0^{\theta_D/T} \frac{x^{\kappa+1}\exp(x)}{(\exp(x)-1)^2}dx$$

(41)

where $x = \hbar\omega/k_BT$. It can be shown that, when κ = 3, Eq. (41) is reduced to the standard form of the three-dimensional Debye model. For T >> $\theta_D$, the integration in Eq. (41) gives $(1/\kappa)(\theta_D/T)^\kappa$. The heat capacity $C_v$ is substituted with κR, in agreement with the Dulong-Petit law in the case of κ = 3. At higher temperatures, the $C_v$ approaches a constant.

At T << $\theta_D$, the upper limit of the integral of Eq. (41) approaches infinity and the integration gives $\int_0^\infty x^{\kappa+1}e^x/(e^x-1)^2 dx \approx$ 3.290, 7.212 and 25.976, for κ = 1, 2 and 3, respectively. Therefore, the heat capacity in the low temperature limit becomes:

$$C_v = A\kappa^2(\frac{T}{\theta_D})^\kappa \propto \kappa^2\theta_D^{-\kappa}$$

(42)

where $A$ is a fixed value. From Eq (41) and (42), we can see that $\theta_D$ has a strong effect on the heat capacity. Using the same core-shell structure for a nanosolid, we can obtain the expression for heat capacity per unit volume depending on the size, shape, and bond nature at very low temperatures (T ~ 0):

$$\begin{aligned}\frac{\Delta C_v(m,T,K_j)}{C_v(m.T_0,\infty)} &= \sum_{i\le 3}\frac{\tau c_i}{K_j}\left[c_i^{(1+m/2)\kappa}\left(\frac{1-T/T_m(1+\Delta_i)}{1-T_0/T_m}\right)^{-\kappa/2}-1\right] \\ &\cong \sum_{i\le 3}\frac{\tau c_i}{K_j}\left[c_i^{(1+m/2)\kappa}(1-T_0/T_m)^{\kappa/2}-1\right]<0 \qquad (T\sim 0)\end{aligned}$$

**(43)**

Since the coefficient of bond contraction $C_i$ is always smaller than unity, the heat capacity is always lower than the bulk value at lower temperatures. The heat capacity decreases with the inverse size ($K_j$). At temperatures close to $\theta_D$, the heat capacity should be evaluated using Eq. (41), where the $\theta_D$ is size, temperature, and bond nature dependent according to Eq. (40).

### 6.3 Predictions and applications

#### 6.3.1 Elasticity and extensibility

Using eq (37), we are able to predict the bond nature, solid shape and size, and the $T/T_m$ dependence of the Young's modulus and extensibility of a solid. For illustration purpose, we selected m = 1 (for metals), 3 (carbon, 2.56), and 5 (Si, 4.88), $x_m$ $(T/T_m)$ = 0.25, 0.5, and 0.75 and the dimensionality τ = 3 (for a sphere) in calculations. The $T_0$ was set at 0 K and T, respectively. For temperature dependence, we used $K_j$ = 10 and 50 sizes by fixing other parameters. $\eta_{21}$ was taken as zero for illustration purpose, otherwise a small offset would be achieved which could hardly be identified in the predicted trends of the relative changes.

Figure 19(a) shows that either elevation or depression of the Young's modulus with decreasing sizes may occur depending on the combination of the ($x_m$, m) values. For example, Y elevation occurs in the situations of ($x_m$, m) = (<0.25, >3). Y retention may happen at critical ($x_m$, m) combinations such as ($x_m$, m) = (~0.25, ~3). The critical combination of ($x_m$, m) can be obtained by allowing eq (37) to approach zero. If we select $T = T_0$, Y elevation also occurs in the situations of ($x_m$, m) = (<0.5, >5). Y value may also remain constant at ($T/T_m$, m) = (0.25, ~2) and (0.5, ~4). It is therefore not surprising that the modulus may rise, drop, or remain constant when the solid size is reduced, depending on the bond nature indicator m, the temperature ratio $x_m$, and the testing techniques as well. The predicted low-temperature stiffening agrees with the findings that the impact toughness of nanostructured Ti is enhanced at low temperatures of 200 and 77 K, a unique phenomenon that contradicts the observations in coarse-grained materials.[349]

According to the prediction, the Y values for pure metallic (m = 1) nanoparticle always drop with size at $T > 0.25\ T_m$. This prediction seems not to be in line with observations from pure metals such as Ag, Au, and Ni, as listed in Table 1. However, the surface chemical passivation, defects, and the artifacts in measurement could contribute to the measured data. For instance, surface adsorption alters the surface metallic bonds (m = 1) to new kinds of bonds with m >1. Surface compound formation or surface alloying alters the m value from 1 to a value around 4. Figure 20 shows the predicted temperature-induced relative change of (a) Y values and (b) the extensibility for $K_j = 10$ and m = 1, 3, and 5 samples. If $T_0 = T$, Y drops non-linearly with T until $T_m$ is reached. The insertion shows the case of $T_0 = 0$ in which the Y drops linearly with T. Detailed calculation with consideration of the thermal coefficient of expansion and the non-linear T dependence of the specific heat will give the nonlinear form at very low temperatures, as detailed in the next section. The extensibility approaches infinity at the corresponding $T_m(K_j)$. The insert in (b) shows singularities because of the shell-by-shell configuration. If we treat the outermost two atomic layers as the skin with a mean $z_i = (4+6)/2$ the singularity occurs at the melting point which drops in value with the characteristic size. On the other hand, a smaller nanosolid with lower m value is more easily extensible at elevated temperatures than the other cases. An MD computation[350] of the shell-resolved fluctuation of the root-mean-square bond length of a 147-atom Leonard-Jones cluster revealed that the process of surface melting starts from the migrating of the vertex atoms on the surface. Although the melting process of LJ147 cluster could be divided into stages of surface melting and general melting, the melting still exhibits a continuous process from the surface shell to the core interior. Therefore, the surface melting is not distinguished by a sharp layer-by-layer feature.

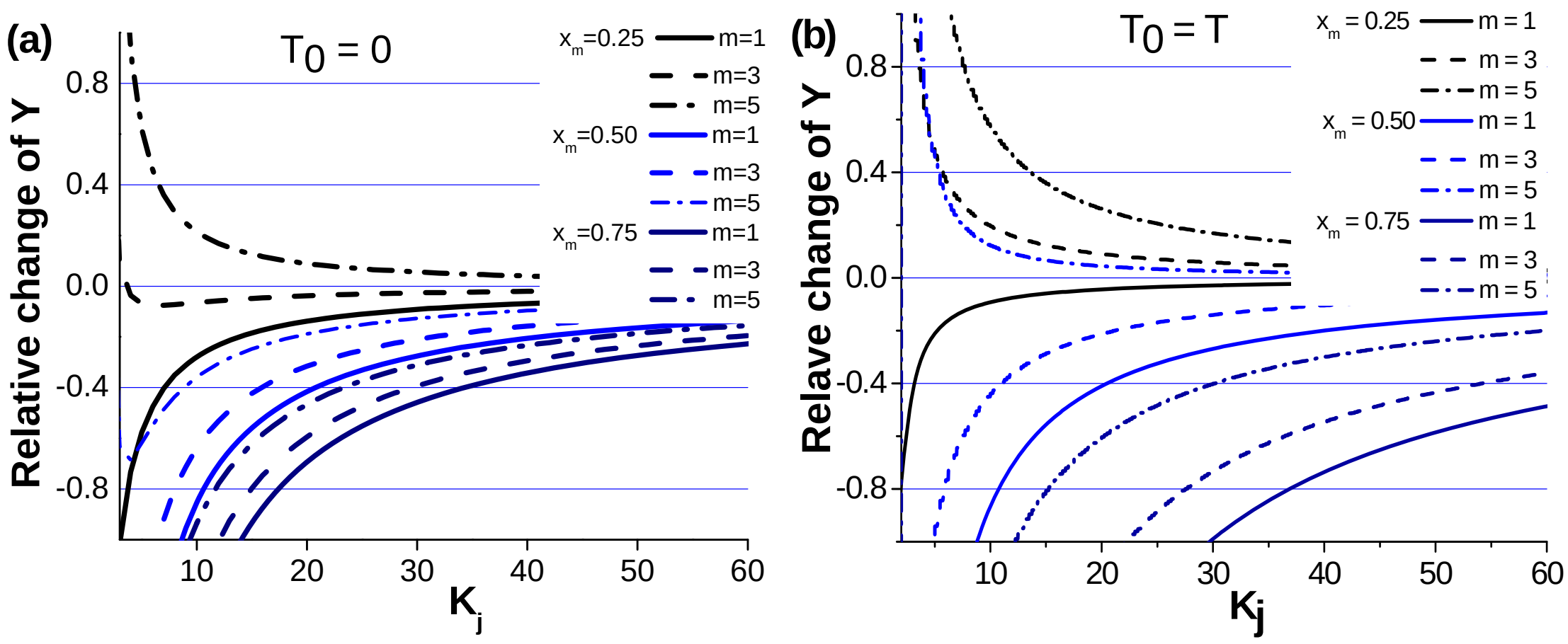

Figure 19 Prediction of size $K_j$ dependence of Young's modulus with (a) $T_0 = 0$ and (b) $T_0 = T$ of different bond nature and $x_m$ ($T/T_m$) values. Young's modulus enhancement occurs at the combinations of ($x_m$, m) = (< 0.25, > 3) for $T_0 = 0$ and ($x_m$, m) = (< 0.5, > 3) for $T_0 = T$. The Y retention may happen at critical ($T/T_m$, m) combinations. Link

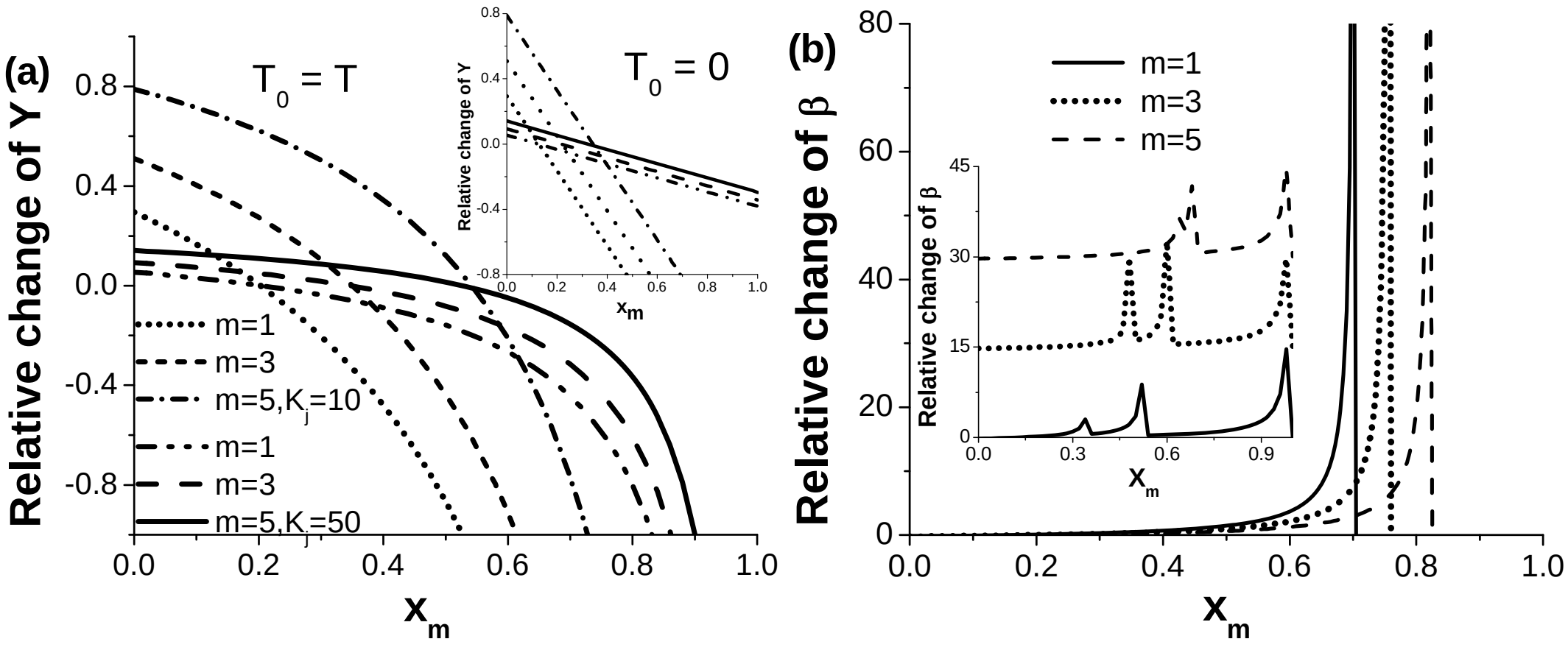

Figure 20 Prediction of $T/T_m$ dependence of (a) Young's modulus and (b) extensibility of a spherical nanosolid with different m and $T/T_m$ ($x_m$) ratios. For illustration purposes, we only present here the linear T dependence for the high temperature approximation. The insertion shows the cases of $T_0 = 0$. The Y values of a nanosolid of lower m values and smaller sizes drop faster when the test temperature is raised. The extensibility approaches infinity at the $T_m$. The singularities in the insert exhibit the shell-by-shell melting features. Link

6.3.2 Size and temperature dependence

Figure 21 compares the prediction with observations of the size dependence of the Y values for (a) ZnO[42] and (b) polymers measured at room temperature, [53] $T_0$ = 300 K. The prediction agrees exceedingly well with the measured data for ZnO nanowires (m = 4, $\tau$ = 2). For the polymer, the prediction agrees with the general trend of measurement (m = 4, $\tau$ = 1 (film)) with accuracy being subject to the precision of size determination.

The predicted m, $K_j$, and $T/T_m$ dependence of modulus and extensibility covers all the possible trends as observed. For example, the predicted Y-depression in Figure 19 agrees well with the measured trends of Al (m = 1, $T/T_m$ = 300/650 ~ 0.5),[74] and polymers ($T_g$ = 300/450 ~2/3).[76] Predictions agree well with the measured size dependence of compressibility, or the inverse of Young's modulus, for nanostructures. For instance, nanocrystalline $\alpha$-$Al_2O_3$ with particle sizes of 67, 37, 20 nm, and 6 nm up to 60 GPa shows a systematic decrease in the compressibility and transition pressure with an increase in particle size. In addition, a high-pressure phase above 51 and 56 GPa for $\alpha$-$Al_2O_3$ of 67 and 37 nm was reported.[127] The compressibility of both Ni and Mo also decreases with particle sizes.[351] Predictions also agree with the trends of temperature dependence of Young's modulus of CVD nanodiamond films,[75] the silicone resins,[76] and the yield stress (linearly proportional to modulus) of Mg nanosolid[77] of a given size.

The ductility increases exponentially with temperature until infinity at the $T_m$ value that drops with solid size. The extensibility of nanoscaled Al-Cu alloys in the quasi-solid state,[141] nanoscaled $Al_2O_3$[127] and PbS[352] at room temperature increases generally with grain refinement. The m values for compounds or alloys are around 4 or higher and their $T/T_m$ ratios are relatively lower. The increase of compressibility/extensibility of $Al_2O_3$ and PbS nanosolids is associated with a decrease of Young's modulus. The superplasticity of materials such as Cu wires (m = 1, $T/T_m$ ~ 1/2)[312] with grain size less than tens of nanometers in the temperature range 0.5 – 0.6 $T_m$[353] also agrees with the predictions. The Y-elevation Si nanosphere (m = 4.88, $T/T_m$ ~ 1/6) [354] is also within the prediction because of their high m values and low $T/T_m$ ratios.

However, the discrepancy for ZnO wires[42] and Si spheres[354] and Si belts[59] may arise from different $T/T_m$ of operation or different experimental conditions or methods. It is anticipated that modulus enhancement, as observed from TiCrN and GaAlN surfaces,[29] may not be observable at room temperature for the low-$T_m$ metals such as Sn, Pb, Al, Zn, Mg, and In. Figure 22 shows the match between predictions and the measured T-dependent Young's modulus for Au, Ag, AlN, $Al_2O_3$, $MgSO_4$, MgO, KCl, Si, Ge, and diamond with $T_m$ and $\theta_D$ as input and $E_B$ derivative of atomic cohesive energy in the bulk at 0 K. The accuracy of the derive $E_B(0)$ is subject to many factors such as surface finishing and measurement techniques. It is interesting to note that the $E_B(0)$ derived from the Ge(100) surface is different from that derived from Ge(111) surface. The anisotropy of binding energy may arise from the bond number density in different directions. Further examination of the anisotropy with more data would be interesting.

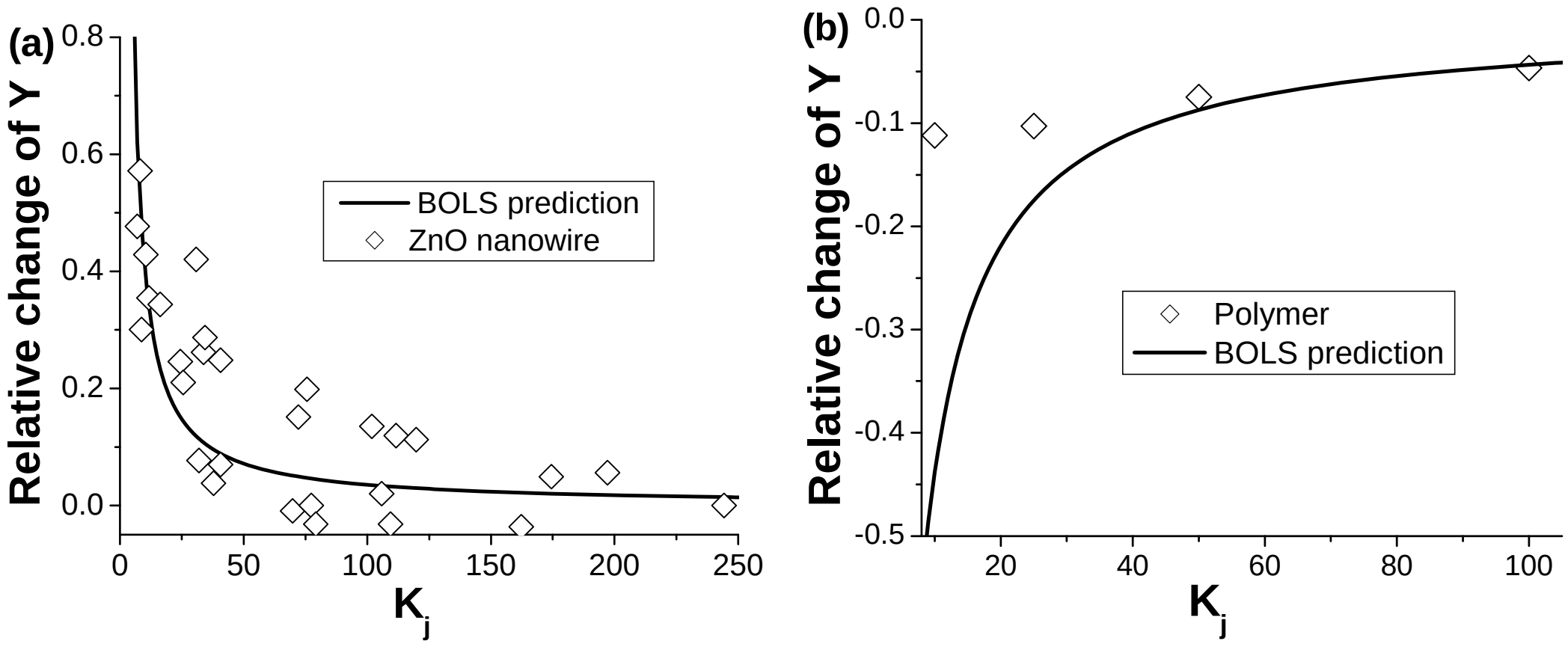

Figure 21 Matching predictions to the measurement of Y value (a) enhancement for ZnO nanowires [42] and (b) suppression for a polymer [53]. link.

Table 9 Summary of information derived from reproduction of temperature dependence of Young's modulus. The $T_m$ and $\theta_D$ for bulk are input and the atomic cohesive energy $E_b(0)$ are derived from the fitting [8,355].

| Sample | $T_m$(K) | $\theta_D$(K) | $E_B$(0)(eV) |
|---|---|---|---|
| AlN [356] | 3273 | 1150 | 5.19 |
| $MgSO_4$ | 1397 | 711 | 2.80 |
| $Al_2O_3$ [329] | 2303 | 986 | 3.90 |
| MgO[357] | 3100 | 885 | 1.29 |
| Ge [358] | 1210 | 360 | 2.57 (100); 3.47 (111) |
| Diamond [359] | 3820 | 1860 | 4.92 |
| Si [328] | 1687 | 647 | 4.80 |
| Ag[357] | 1235 | 160 | 1.24 |
| Au[357] | 1337 | 170 | 1.64 |
| KCl[357] | 1044 | 214 | 0.57 |

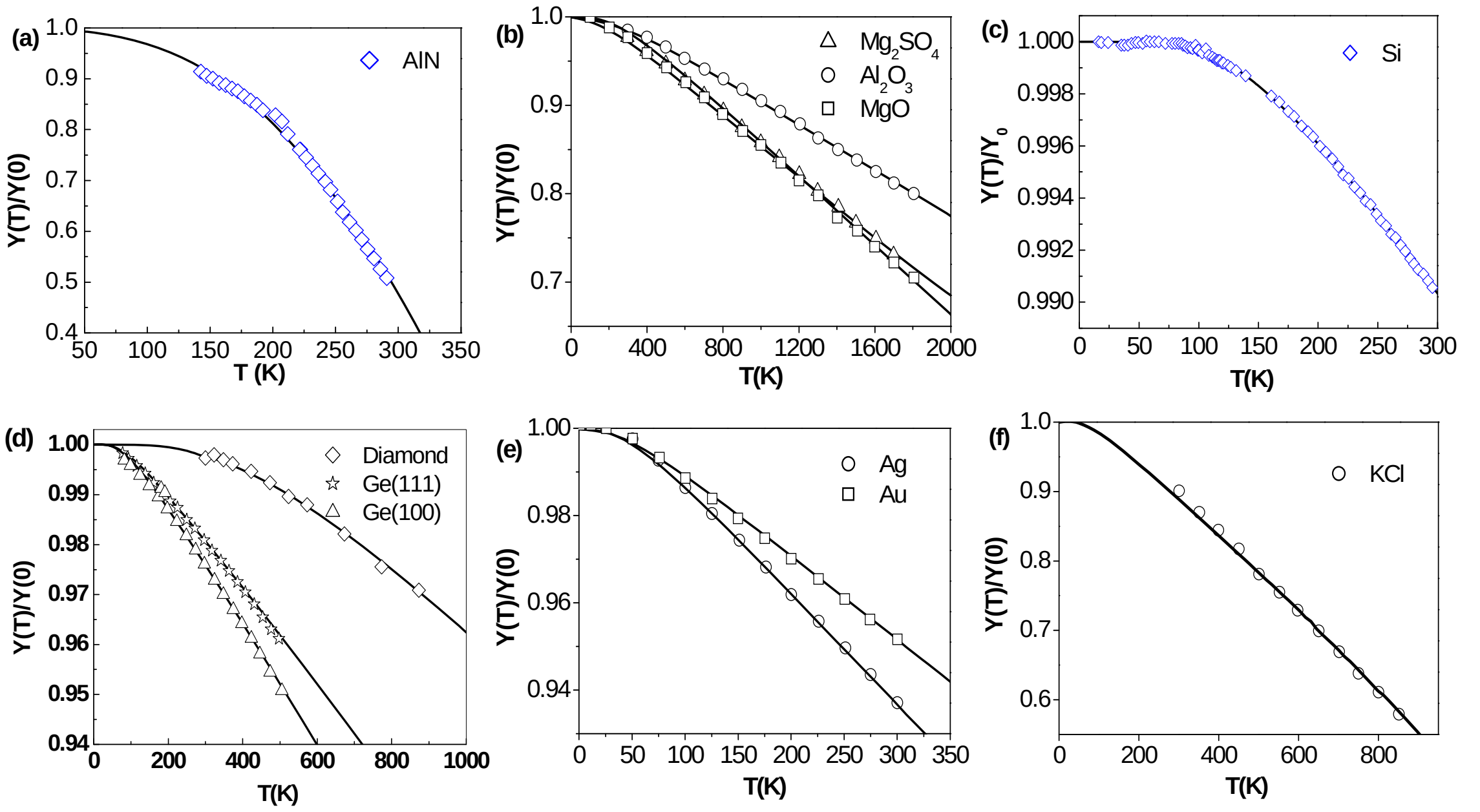

Figure 22 LBA reproduction (solid lines) of the measured (scattered data) T-dependent Young's modulus of (a - f) various specimens with derived information of mean atomic cohesive energy as listed in Table 9. Link

Numerical agreement of the temperature dependence of the *Y* at higher temperatures could be made by all the models, such as eqs (34) and (35), as discussed above despite the different physical mechanisms. However, we found that the constant B = $[E_b(0)\times(1+\alpha T)^3]^{-1}$ in Anderson's model and that the $T_0$ in Watchman's model corresponds to the turning point $T_0$ at which the *T*-*Y* curve transits from nonlinear to linear, which is governed by the Debye temperature. In addition, the present approach covers the contribution from thermal expansion and size induced bond contraction.

### 6.3.3 Debye temperature and specific heat

Using Eq. (40), we are able to predict the $\theta_D$(m, $K_j$, T). Figure 23 shows the relative change of $\theta_D$ for nanowires (τ = 2, κ = 1) with m = 1, 3, 5 and $x_m$ $(T/T_m)$ = 0, 0.25 and 0.50. When the measuring temperature is much lower than the melting point ($T/T_m << 1$), $\theta_D$ increases with the decrease of material dimension $K_j$; while $\theta_D$ increases faster at larger m. On the other hand, $\theta_D$ decreases with increasing operating temperature and the relative change of $\theta_D$ is greater for smaller m values. A close examination of Figure 23(a) and Eq (40) could lead to a conclusion that, for a certain ($T/T_m$, m) combination, $\theta_D$ may vary insignificantly with particle size, as for the insignificant change in Young's modulus. Figure 23 (b) shows the temperature-induced relative change of $\theta_D$ for nanowires (τ = 2, κ = 1) of size $K_j$ = 10 and 50. If we set $T_0$ = T, $\theta_D$ decreases non-linearly with T until the T approaches $T_{mi}$ (local melting temperature of the ith atomic site). The two transition points for each of (m, $K_j$) combinations arise from the loss of bonds, which happens only to the outermost two discrete atomic layers. Moreover, the variation of $\theta_D$ with the temperature and size is more pronounced for larger m and smaller $K_j$ values.

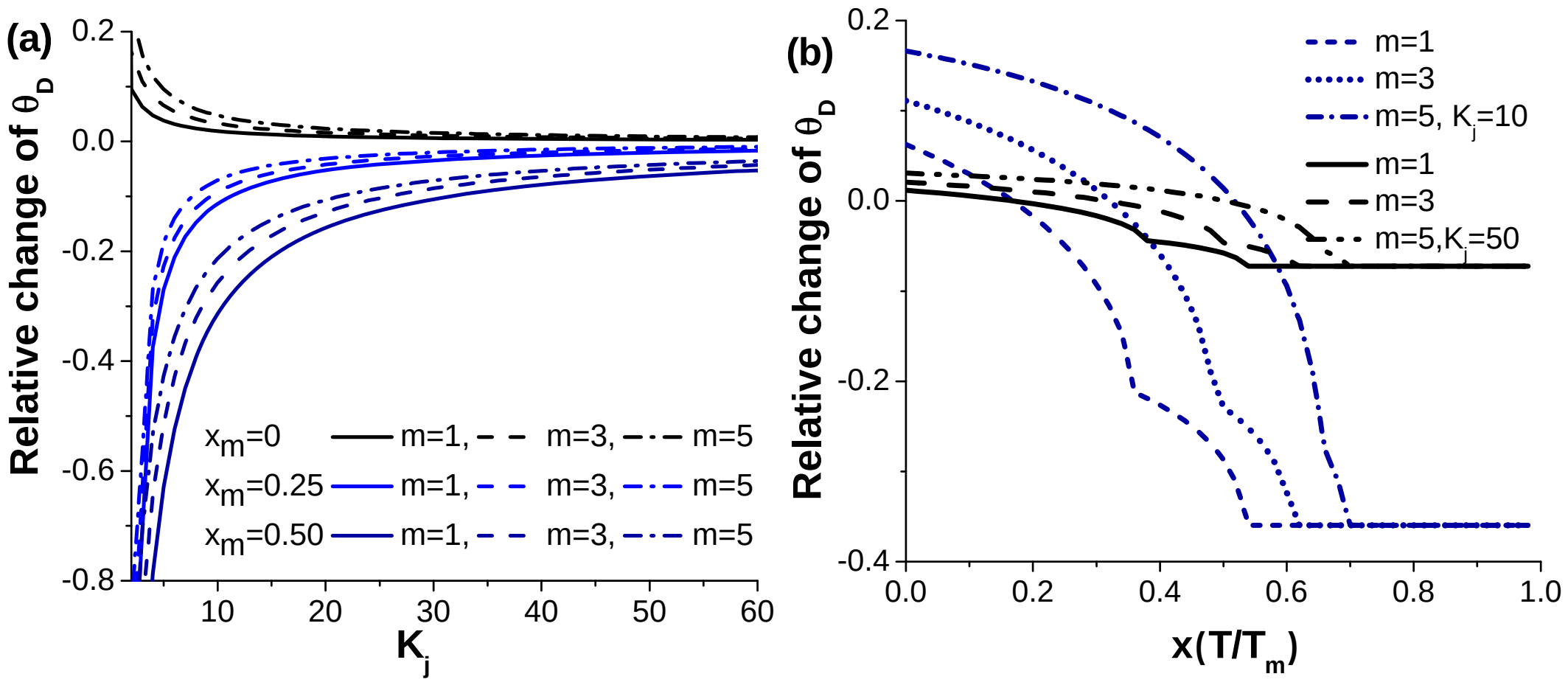

Figure 23 Prediction of (a) size ($K_j$) and (b) temperature $x(T/T_m)$ dependence of $\theta_D$ for different bond nature (m). The $\theta_D$ increases with the decreasing of size for very low T and decreases with decreasing the size for high temperature. The $\theta_D$ of nanowires with higher m values and smaller size drops faster when the temperature is elevated. The transition points correspond to local melting temperature of the outermost two atomic layers. Link

Figure 24 compares the predictions with various theoretical or experimental data (a) for Au particles and (b) $\theta_D$ from Debye-Waller parameter measurement for Se nanoclusters. Couchman and Karasz's approach[335] shows that the change of $\theta_D$ involves the particle size $K_j$ and cutoff acoustic wavevectors $K_0$ : $\Delta\theta_D/\theta_0 \approx -3\pi/(8K_jK_0)$, without temperature being involved. By applying eq (40), with $T_0 = 0.245T_m$ and $T = 0.16T_m$, agreement between the T-BOLS prediction and Couchman and Karasz's estimation has been reached. If we set $T_0 = 0.224T_m$ and $T = 0.204T_m$, The T-BOLS premise agrees well with the measured data of Balerna and Mobilio, as shown in Figure 24(a). In Figure 24 (b), the prediction shows the general trend of $\theta_D$ with respect to size although the precise agreement is not satisfied. However, since the measurement was conducted at T = 293K which is higher than the local melting temperature of the outermost two atomic layers, about $0.6T_m$ (for Se $T_m$ = 494K), the features measured are dominated by the core interior with an insignificant contribution from the temperature dependence.

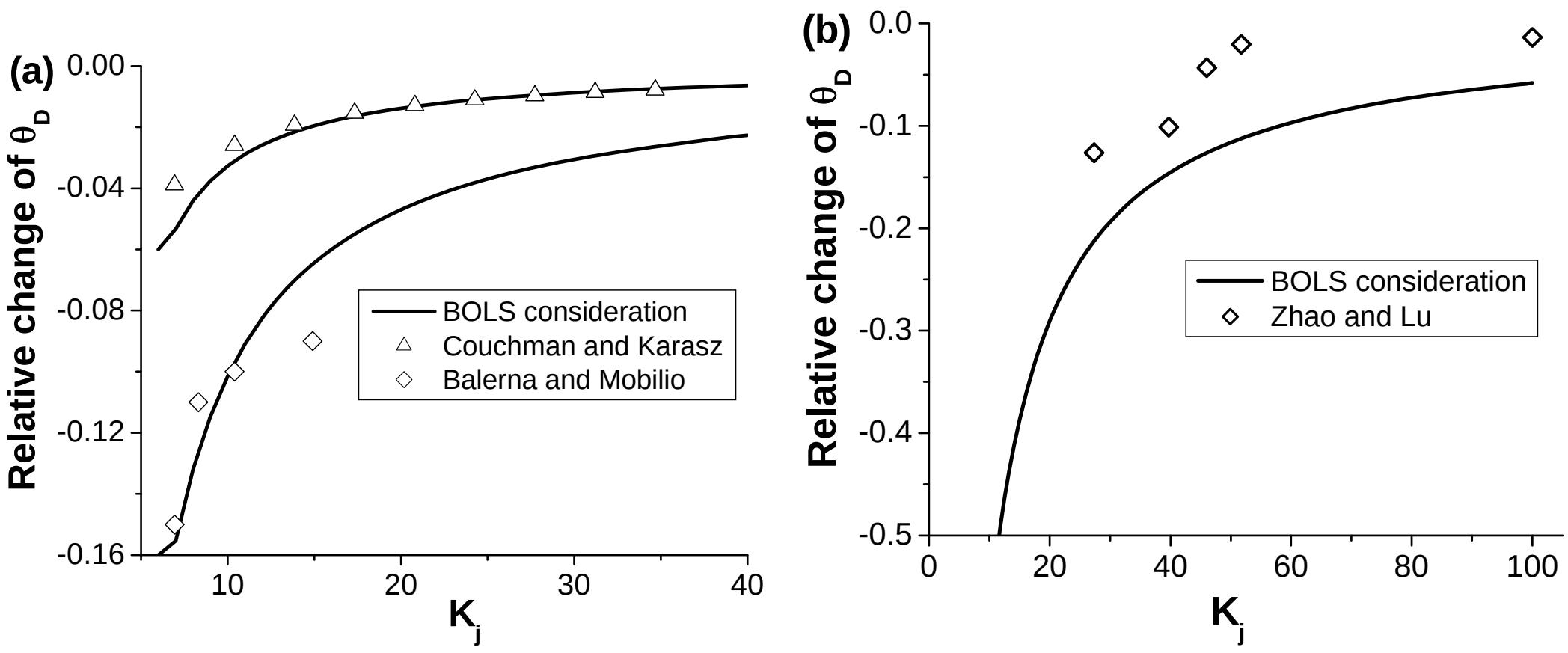

Figure 24 Comparison of the T-BOLS predictions with the observations based on (a) Couchman and Karasz's model (with $T_0 = 0.245\ T_m$, $T = 0.16\ T_m$, $T_m = 1337$ K) [335] and Balerna and Mobilio's measurement (with $T_0 = 0.224\ T_m$, $T = 0.204\ T_m$) [336] for Au particles and, (b) Zhao and Lu's measurement of Se particles (with $T_0 = T = 0.6\ T_m$, $T_m = 494$ K) [337]. link.

As indicated in eq (41), the specific heat capacity depends unambiguously on $\theta_D$ and hence on the size, temperature, and the bond nature involved. Figure 25(a) shows the reduced $C_v$ (in units of R, where R is the gas constant) versus temperature ($T/\theta_{D0}$) for Si nanowires (m = 4.88) and Al nanowires (m =1) of different diameters ($K_j$ = 5, 10 and 20). The shape of the $C_v$ curve is similar to that of the bulk but the size induces a depression over the whole temperature range. For the same $K_j$ at a given $T/\theta_{D0}$, the reduction of heat capacity is larger for greater m values. Figure 25(b) plots $C_v/C_{v0}$ (where $C_{v0}$ is the bulk heat capacity at a given temperature) versus $K_j$ at T = 100 K and 300 K for Al and Si nanowires. The heat capacity decreases with the size at fixed temperature (except for Al nanowires measured at room temperature, where the heat capacity is very close to the bulk $C_v$ value obtained when $K_j$ >15 and increases slightly with decreasing size). For a given size, the reduction of the heat capacity is more significant at lower temperatures or larger m values. In this analysis, we set $T_0 = T$. If $T_0$ is assumed 0 K, the general trend of heat capacity is preserved, but the reduction of heat capacity is more pronounced.

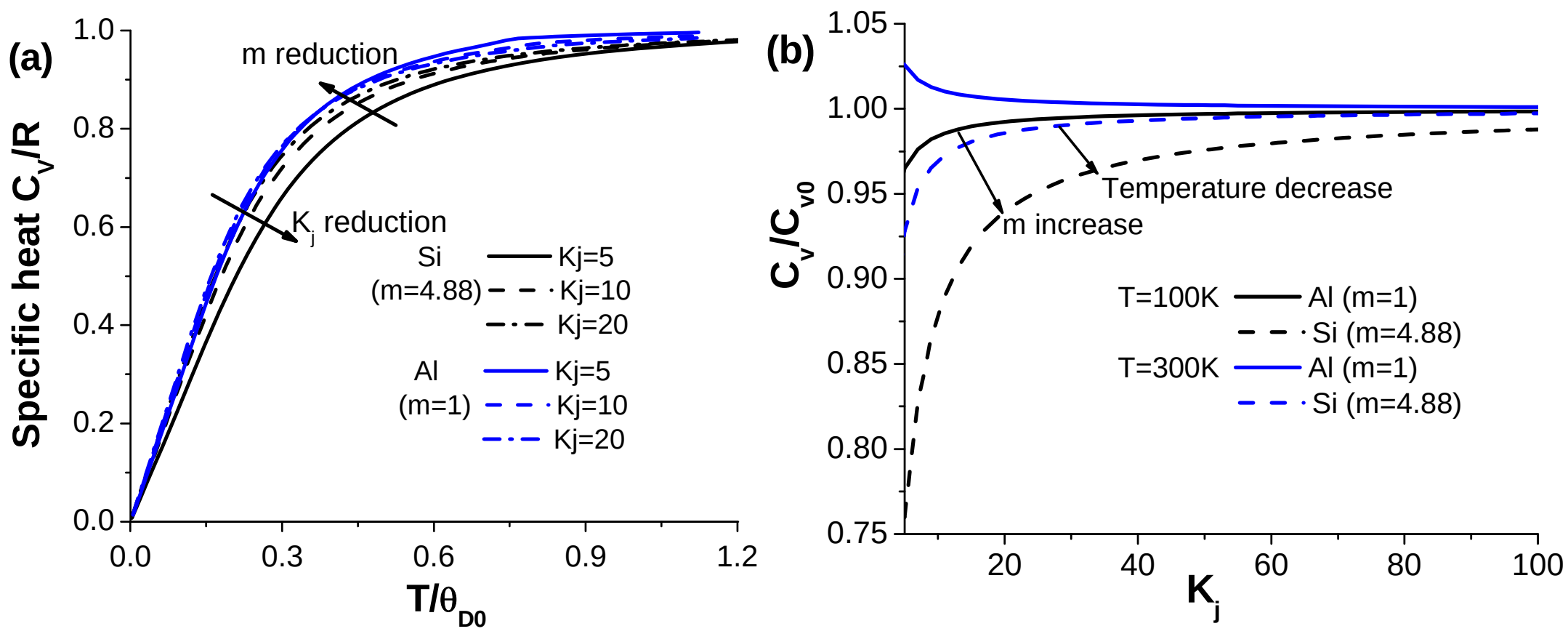

Figure 25 (a) Prediction of the temperature ($T/\theta_{D0}$) and size dependence of heat capacity (in units of R) for Si nanowires (m = 4.88) and Al nanowires (m = 1) with $K_j$ = 5, 10 and 20 and at T = 100K and 300K, in panels (a) and (b). The heat capacity approaches R at very high temperature, and decreases with solid size. The reduction of heat capacity is more pronounced for larger m values at a given $T/\theta_{D0}$. The heat capacity generally reduces with solid size and reduces faster at lower temperature and larger m values. link.

## 6.4 Summary

A set of analytical expressions for the size, temperature, and bond nature dependence of elastic modulus and its derivatives on the extensibility, Debye temperature, and specific heat of nanostructures has been established, which covers the essential parameters and their interdependence. It has been clarified why the Y values for some materials are elevated and why those of others are not increased upon size reduction. Conclusions can be drawn as follows:

(i) The Young's modulus of a nanosolid may drop, rise, or remain unchanged with size reduction, depending on the temperature of operation, and the nature of the bond involved, as well as experimental conditions. It is therefore not surprising to observe the elastic modulus change in different trends for different materials measured under different conditions.

(ii) The thermal softening arises from thermal expansion and vibration that weakens the bond through the increase of internal energy that follows Debye approximation.

(iii) One could not consider a single parameter on its own without addressing the rest when discussing the mechanical and thermal properties of a material, especially for a small object. One should not separate the mechanical performance from the thermal response in dealing with small objects, as they are interdependent.

(iv) The Debye temperature $\theta_D$ has a square root dependence on $(T_m-T)^{1/2}/d$, rather than a linear or square root dependence on $T_m$. The currently derived solution may provide complementary method to the T-independent form of $\theta_D$ given by Lindermann.

(v) The specific heat capacity generally decreases when the solid size is reduced. The reduction of the specific heat capacity is more pronounced for larger m values at lower temperatures.

## VII Nanograins: II. Plastic deformation and yield strength

### 7.1 Observations - Hall-Petch relation

In plastic deformation tests, the measured flow stress, or hardness, of nanograins changes with its grain size in a trend quite different from the predicted monotonic features of the elasticity response because of the involvement of extrinsic factors including artifacts. The flow stress of the solid grains increases under the mechanical stimulus when the grain size is reduced. For brittle materials, such as intermetallic compounds and ceramics, the ductility increases by grain refinement because of the increased grain boundary (GB) volume fraction and, hence, GB sliding and GB dislocation accumulation.[360] As suggested by Zhang and Zarudi,[361] the plastic deformation is the coupled result of mechanical deformation controlled by the stress field applied, chemical reaction determined by the external loading environment, and mechanical-chemical interaction governed by both loading type and environment. Temperature rise accelerates the penetration of oxygen into the grain interior and reduces the critical stress for plastic yielding. When the chemical effect is avoided, the initiation of plasticity is enabled by octahedral shear stress but the further development of plastic deformation is influenced by hydrostatic stress. Plasticity of silicon in the form of phase transitions, e.g., from the diamond to amorphous or from the amorphous to bcc structures, is determined by loading history. It has been found recently that the elastic deformation of the two-phase alloy obeys the rule of mixtures for a composite; the plastic deformation is governed by the softer matrix but enhanced, in part, by the deflection of shear bands by the globular harder phase.[362]

The size-induced mechanical hardening of nanograins is divided into three size regions, as illustrated in Figure 26a. Region I corresponds to the classical Hall-Petch relationship (HPR).[363,364,365] Regions II and III are known as the inverse Hall-Petch relationship (IHPR).[366,367,368, 369,370,371] Postulated explanations for this behavior include dislocation-based models, diffusion-based models, grain-boundary-shearing models, two-phase-based models and dislocation absorption based models.[372] The observations are also explained by considering two alternative and complementary rate mechanisms of plasticity, grain boundary shear and dislocation plasticity, each contributing to the overall strain rate in proportion to the volume fraction of the material in which they operate.[373]

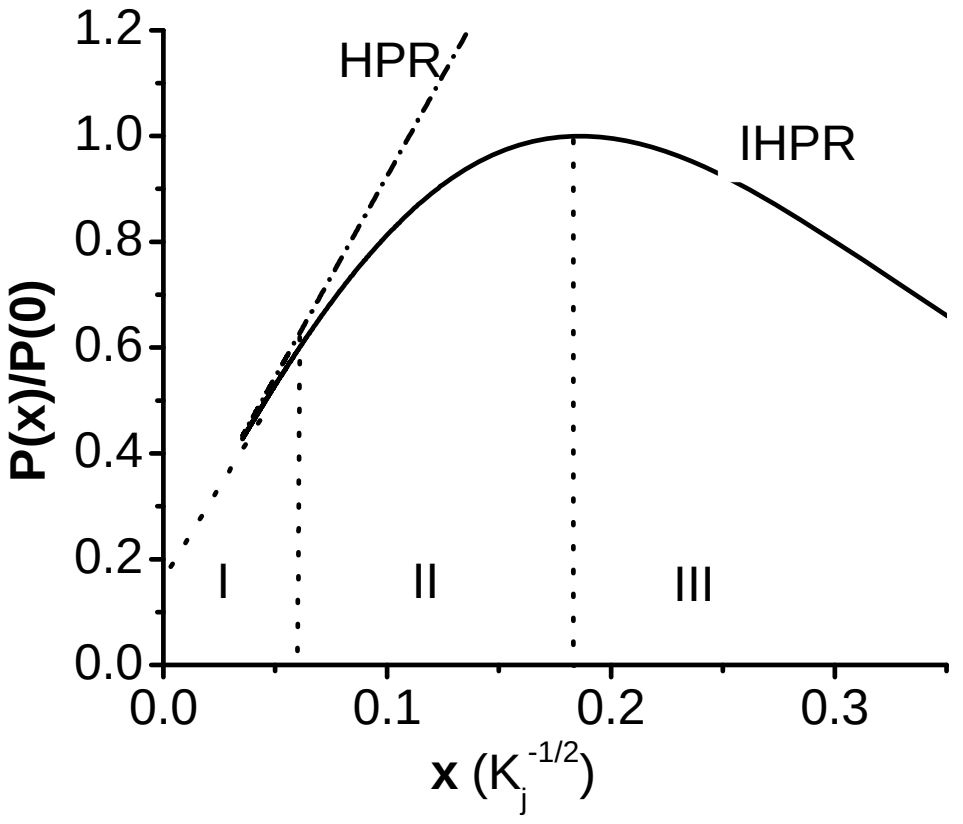

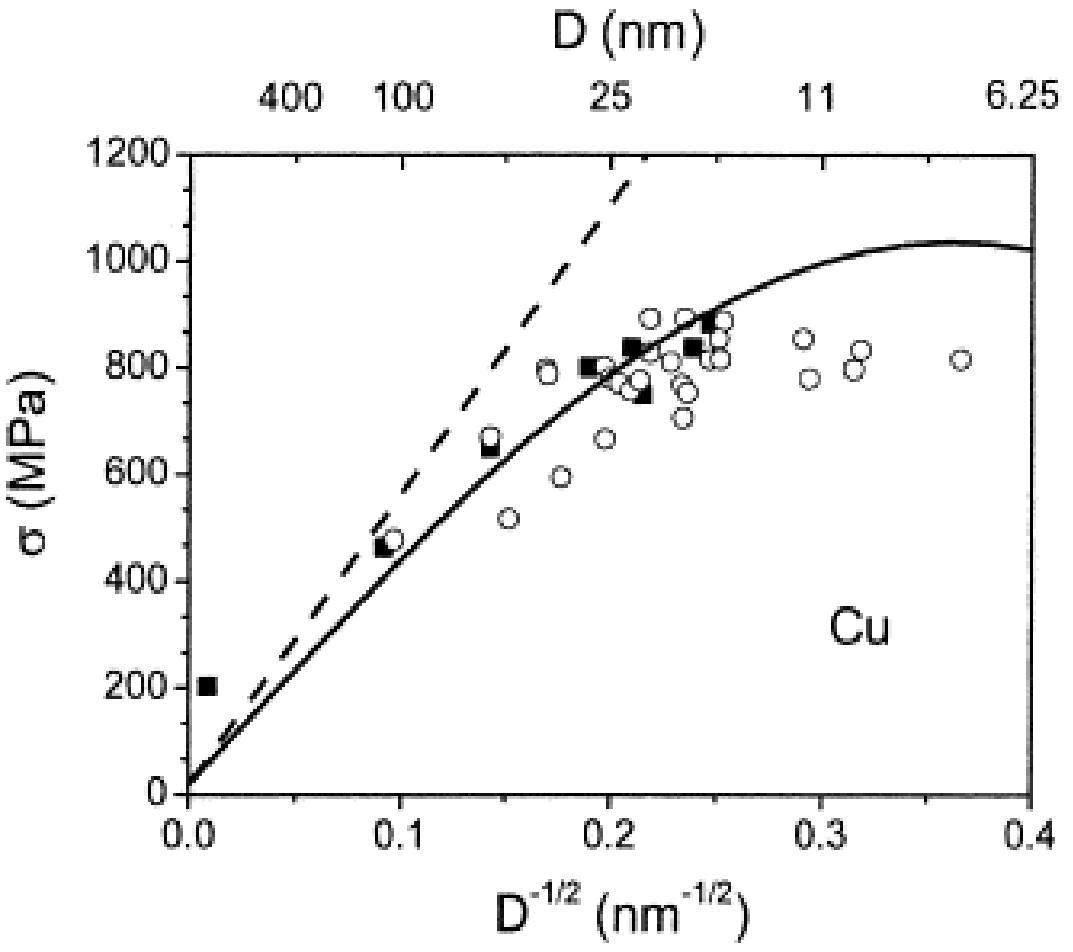

Figure 26 (a) Illustration of the three regions of hardness transition from the classical HPR to the IHPR and (b) the typical best fit of the IHPR for Cu nanograins by Zhao et al.[374] Link

The mechanical strengthening with grain refinement in the size range of 100 nm or larger (region I) has traditionally been rationalized with the so-called T-independent HPR that can be simplified in a dimensionless form being normalized by the bulk strength, P(∞), measured at the same temperature and under the same conditions:

$$P(K_j)/P(\infty) = 1 + AK_j^{-0.5} = 1 + A'\mu(b/K_j)^{0.5} \tag{44}$$

The slope A or A′ is an adjustable parameter for experimental data fitting, which represents both the intrinsic properties and the extrinsic artifacts such as defects, the pile-up of dislocations, shapes of indentation tips, strain rates, load scales and directions in the test. The μ and b correspond, respectively, to the shear modulus and the Burger's vector modulus reduced by atomic size, d. The bulk modulus B is related to the shear modulus μ and the Poisson ratio ν by: $\mu = B/[2(1+\nu)]$. Using the dimensionless form of the normalized strength is aimed to minimize the contribution from artifacts due to processing conditions, crystal orientations, and the purity of the specimens if the measurement is conducted under the identical conditions throughout the course of the experiment. For convenience, we use both $x_j = K_j^{-0.5}$ and $K_j$ as indicators of the dimensionless form of sizes.

As the crystal is refined from the micrometer regime into the nanometer regime (region II), the classical HPR process invariably breaks down and the yield strength versus grain size relationship departs markedly from that seen at larger grain sizes (region I). With further grain refinement, the yield stress peaks in many cases at a mean grain size in the order of 10 nm or so. A further decrease in grain size (region III) can cause softening of the solid, instead, and then the HPR slope turns from positive to negative at a critical size, or so-called the strongest grain size.[375]

There has been a huge database established on the HPR and IHPR effects. For example, by squeezing Si nanospheres of different sizes between a diamond-tipped probe and the sapphire surface, Gerberich et al.[354] determined at room temperature that a defect-free silicon nanoparticle with a diameter of 40 nm is ~3 times (50 GPa) harder than bulk silicon (12 GPa) while it was around 20 GPa with a diameter of 100 nm. The smaller the sphere, the harder it was. Tensile tests of the electrodeposited nanocrystalline Ni-W alloys with grain sizes of 20, 12, 8 and 5 nm revealed that the strongest grain size is at 8 ~ 10 nm that is an essential characteristics of the nanocrystalline Ni-W alloys. [376,377,378] The softening of nanograins is also attributed to the increase of intercrystalline volume fraction, especially the fraction associated with the triple junction.[379] The IHPR effect has also been experimentally observed in SiC,[380] Ni-P,[381] melt-spun AlMnCe, and AlCrCeCo alloys.[382] More examples are given in Table 10.

There has been a concerted global effort underway using a combination of novel processing routes, the state-of-the-art experimental measurements, and large-scale computations to develop deeper insight into the IHPR phenomena. Progress in fabrication, characterization, and the understanding of the HPR-IHPR relations of nanocrystalline materials have been reviewed by many specialists in the field.[383, 384,385,386,387,388,389,390,391,392,393]

## 7.2 Understanding at the mesoscopic level

Numerous models have been developed for the HPR and IHPR effects. The possible mechanisms for the plastic deformation of nanograins include the pile-up breakdown, GB sliding, core and mantle structures, GB rotation, grain coalescence, shear-band formation, gradient and twinning effects. The IHPR appears because of the increased porosity at small grain sizes, suppression of dislocation pile-ups, and dislocation motion through multiple grains, enhanced grain boundary diffusion, or absence of dislocations in the small grains without the mirror forces in the next grain being involved. It appears that the mechanism for the HPR and the IHPR is a topic of endless debate because of the numerous complicated factors. Typical mechanisms for the HPR, IHPR, and the transition from HPR to IHPR and the critical grain size are discussed as follows.

### 7.2.1 HPR-I – linear hardening

**GB barriers limited dislocations**. For large grain sizes, most of the models using a mechanism based on dislocation pile-up account for the grain size dependence of the macroscopic plastic flow stress. [394,395] In deriving the HPR, the role of GBs as a barrier to dislocation is considered in various models. One type of models[396] regards the GBs as barriers to the pile-up of dislocations. The GB stress of one grain concentrates the dislocations, which activates the dislocation sources in the neighboring grains, thus initiating slip from grain to grain. The other type of models[397] regards the GBs as dislocations barriers limiting the mean free path of the dislocations, thereby increasing strain hardening, resulting in the HPR.[398] The conventional dislocation theory for the HPR gives a linear dependence of P on $K_j^{-1/2}$ only when there are a large number of dislocations in a pile-up, which is equivalent to assuming that the grain sizes in the polycrystalline material under consideration is large. Deformation mechanism at larger scales is believed to occur through the production and motion of dislocations within the individual grains.[399]

### 7.2.2 IHPR-II – HPR deviation

**Transition of GB sourced dislocations**. The first theoretical attempt relied on dimensional arguments. It was proposed that dislocations in nanograins are nucleated at GBs, travel across the grain, and annihilate on the opposite GBs, and therefore, intragranular transition dominants the HPR deviation.[400]

**GB stacking fault drained dislocations**. As the grain size is decreased, dislocations are expected to pile up at the GBs and become less mobile because of the role of the stacking-fault energy, essentially a misfit energy caused by atomic planes stacked out of sequence, thus leading to increased strength.[399] Therefore, the IHPR is attributed to the GBs that serve as a dislocation drains rather than as dislocation sources.[401] In the nanometer regime, GBs are occupying a significant fraction of the material's volume, deformation proceeds by a mechanism that is intergranular rather than intragranular, instead, as argued above.

### 7.2.3 IHPR-III – softening

**GB migration and diffusion** mechanism suggests that the GB migration and diffusion dominate the size-induced softening. When the grain size is below the critical values of 10-20 nm, more than 50% of atoms are associated with GBs or interfacial boundaries. Therefore, GB atoms play the dominating roles in the softening of nanocrystalline materials. Nanocrystalline materials exhibit creep and super plasticity at lower temperatures than do conventional micro-grained counterparts. Similarly, plastic deformation of nanocrystalline coatings is considered to be associated with GB sliding assisted by GB diffusion or rotation. MD calculation using the embedded atom method[402] and the effective medium theory[403] suggested that GB migration and sliding are predominant instead of the mechanism of diffusion switching.[404] An intricate interplay between GB sliding and GB diffusion may occur, and therefore, the IHPR effect arises from sliding-accommodated GB diffusion creep.[385,405]

**Soft GBs embedded hard cores.** This composite core-shell structure model suggests that the IHPR deformed bulk metallic glass can be treated as a composite of hard amorphous grains surrounded by a shell of soft GBs.[406,407] The grain interior deforms elastically under external stresses, while the plastic deformation of the GB layer is governed by a Maxwell's equation. In such a microstructure, the deformation of the GB shell contributes significantly to the overall deformation. Nieh and Wang[408] indicated that the apparent IHPR in the BeN alloy is actually an artifact as it is caused by the presence of relatively soft amorphous Be-B phases when the grain size of Be is significantly refined by B alloying. When the grain size was less than 10 nm, the GBs were thicker than those of coarser grained materials. The interface region of the nanocrystallites, having a structure of non-periodic atomic array, expanded into the centre region. Factors such as residual porosity, impurities, residual stresses, minor surface flaws, and/or narrow shear bands are suggested to lower and scatter the tensile strength values.[409] Similarly, a phase mixture model[379,410] treats a polycrystalline material as a mixture of two phases: the grain interior material whose plastic deformation is governed by dislocation and diffusion mechanisms and the GB 'phase' whose plastic flow is controlled by a GB diffusion mechanism.[411] The size effect in this grain size range is governed by the reduction in dislocation mean-free path through GBs diffusion rather than dislocation nucleation.[412] The softening in plastic deformation was also ascribed as a large number of small 'sliding' events of atomic planes at the GBs, with only a minor part being caused by dislocation activity in the grains; the softening at small grain sizes is therefore due to the larger fraction of atoms at the GBs.[375] A recent acoustic emission spectroscopic study[413] suggested that a peculiar deformation behavior, due to the competition between different deformation mechanisms such as dislocation pileups in nanocrystalline grains and grain sliding-grain rotation within amorphous boundaries, plays a vital role in the deformation of superhard nanostructures.

### 7.2.4 HPR-IHPR transition

**Grain size triggers collective motion of dislocations**. Recent work[414] suggests that the classical HPR arises from the collective motion of interacting dislocations yet the breakdown for nanometric grains stems from the loss of such a collective behavior as the grains start deforming by successive motion of individual dislocations. Mohamed[415] interpreted the nanoscale softening in terms of dislocation-accommodated boundary sliding. The HPR breakdown may lead to three possible behaviors according to the patterns of dislocations: (i) If dislocations are nucleated at vertices, the IHPR will bend down showing softening characteristics. (ii) If dislocations are at the GB triple junctions, the IHPR curve will turn to be flat without softening or hardening taking place. (iii) If dislocations occur at the GBs, the IHPR will remain the classical HPR trend and there will be no IHPR.

**Dislocation absorption.** Carlton and Ferreira[372] revised the HPR with a statistical probability of dislocation absrption by grain boundaries, showing that the yield strength is dependent on strain rate and temperature and deviates from the HPR below a critical grain size. The HPR becomes IHPR in the following form,

$$P(K_j)/P(\infty) = 1 + A\left(\left(1 - P_{dis}\right)/K_j\right)^{0.5},$$

where $P_{dis}$ is the probability of a dislocation being absorbed by the grain boundary.

**Core-shell role exchange.** The IHPR slope transition is suggested to arise from the role-exchange of the grain interior and the GBs.[416] The thermally activated GB shear[368] and the excessive volume of the under-coordinated GB atoms[417] are suggested to be responsible for the IHPR softening.[418,419] A switch in the microscopic deformation mechanism from dislocation-mediated plasticity in the coarse-grain interior to the GB sliding in the nanocrystalline regime is suggested to be responsible for the maximum strength.[420] In the HPR regime, crystallographic slips in the grain interiors govern the plastic behavior of the polycrystallite; while in the IHPR regime, GBs dominate the plastic behavior. During the transition, both the grain interiors and the GBs contribute competitively. The slope in the HPR is suggested to be proportional to the work required to eject dislocations from GBs and the GB strain energy has to be taken into consideration.[421] The transition with decreasing grain size from a GB dislocation to a GB-based deformation mechanism in the nanocrystalline fcc metals provides also a possible mechanism.[422]

7.2.5 Strongest grain size estimation

The strongest grain size could be estimated by the following approaches. One approach is to assume that the plastic deformation of a nanocrystal switches abruptly from the pile-up of dislocations to the GB-relaxation mechanism at a grain size, $d = d_c$,[410]

$$P(K_c)/P(\infty) = 1 + AK_c^{-0.5} = 1 + g\left(\frac{d_c}{w} - 1\right)$$

where g is an adjustable parameter depending on grain morphology and w the GB width that is approximately three times the Burgers vector, b, i.e., $w \cong 3b$. According to this model, the strength of the IHPR material decreases linearly with its grain size below a certain threshold. The $d_c$ was estimated to be 25 nm for Cu, in comparison to the reported values of 14 nm,[423] 18 nm,[424] and 50 nm.[425]

Another approach for the flow stress of an A-B alloy suggests that the strongest grain size represents the emergence of GB diffusion that causes the HPR breakdown. For the A-B alloy, the critical grain size is given by:[378]

$$d_{AB}^{i} = \left[\frac{D_B}{D_A}(1-c) + c\right]^{2/7} d_A^{i}$$

where $d_A^i$ is the critical grain size for pure A, $D_B$ and $D_A$ are the diffusivity of atom B in A and the self-diffusivity of A, respectively. The c is the atomic ratio of B to A. A prediction of the strongest grain size for the W/Ni alloys (with 13 - 19.6 at% W and a presumption of $d_{Ni}^i$ = 12 nm) shows acceptable agreement to the measured critical size of around 8-10 nm.[376]

7.2.6 $T_m(K_j)$ dependent IHPR

Unfortunately, an atomistic analytical expression for reproducing the yield strength over the whole HPR and IHPR size range was lacking until recently when Zhao et al.[374] modified the T-independent HPR by introducing the activation energy for atomic dislocation to the HPR slope. The activation energy was related directly to the melting point suppression, $T_m(K_j) \propto E_A(K_j)$.[265] The suggested IHPR is in the following form,

$$P(K_j) = P_0 + \left(A_1 + A_2 K_j^{-1/2}\right)\exp\left(\frac{T_m(K_j)}{2T_0}\right)$$

where $P_0$, $A_1$ and $A_2$ are adjustable parameters for data fitting. $T_0$ is the reference temperature of measurement and $T_m(K_j)$ is the $K_j$ dependent melting point. The $T_m(K_j)$ drops with the inverse size of the nanoparticle, $K_j$. This approach could reproduce the IHPR observations quite reasonably for a number of specimens, as shown for a sample of such cases in Figure 26(b).

7.3 LBA and T-BOLS approach: intrinsic and extrinsic competition

The measured size and temperature dependence of the yield strength and compressibility of a nanosolid can be obtained by substituting the size and bond nature dependent $\eta_1(K_j)$, $d(K_j)$, and $T_m(K_j, m)$ for the $\eta_{1i}$, $d_i$, and $T_{m,i}$ in eq (24) that was derived for an atomic bond and a flat surface. The solution is expressed as:

$$\frac{P(K_j,T)}{P(\infty,T_0)} = \begin{cases} \dfrac{\eta_1(K_j)}{\eta_1(\infty)} \times \left(\dfrac{d}{d(K_j,T)}\right)^3 \times \dfrac{T_m(K_j,m) - T}{T_m - T_0} = \dfrac{P(K_j)}{P(\infty)}\varphi(K_j,m,T) & (Born-criterion) \\ \left(\dfrac{d}{d(K_j,T)}\right)^3 \times \dfrac{\eta_2(K_j) + \eta_1(K_j)\left[T_m(K_j,m) - T\right]}{\eta_2(\infty) + \eta_1(\infty)(T_m - T_0)} & (Full-energy) \end{cases} \quad (45)$$

In Born's criterion, the additional term $\varphi(K_j, m, T)$ includes contributions from bond nature, bond length, and the separation of $[T_m(m, K_j) - T]$ to the yield strength of a solid in the non T-dependent HPR treatment. The term $[T_m(m, K_j) - T]$ in the Born's criterion is replaced by the net energy difference in the full-energy approximation. Numerically, these approximations are substantially the same because the addition of $\eta_2$ leads to only an insignificant offset of the relative P values. However, according to the fact that the mechanical strength approaches zero at melting, the term of $\eta_2$ can be ruled out and the Born's criterion is applied. Therefore, the full-energy consideration can be ignored in the case of plastic deformation for the first order approximation.

By comparing the currently derived form of eq(45) with the traditional HPR of eq (44), one can readily find that the ratio of the size dependent specific heat per bond follows the traditional non T-dependent HPR, $\eta_1(K_j)/\eta_1(\infty) = P(K_j)/P(\infty) = 1 + AK_j^{-0.5}$. Incorporating the activation energy, $E_A(K_j) \propto T_m(K_j)$,[265,374, 426] for atomic dislocation into the pre-factor A, leads to an analytical expression for the size and temperature-dependent HPR:

$$\frac{P(K_j,T)}{P(\infty,T)} = \frac{\eta_1(K_j)}{\eta_1(\infty)}\left\{\left(\frac{d}{d(K_j)}\right)^3 \times \frac{T_m(K_j,m)-T}{T_m-T_0}\right\}$$
$$= \left[1 + AK_j^{-0.5}\right] \times \left\{\left[1+\Delta_d(K_j)\right]^{-3} \times \frac{T_m(K_j,m)-T}{T_m-T_0}\right\}$$
$$= \phi\left(A\left(T,m,T_m(K_j,m)\right),K_j\right) \times \varphi\left(K_j,m,T,T_m(K_j,m)\right)$$

(46a)

where,
$$A = A\left(T,m,T_m(K_j)\right) = f' \times \exp\left[\frac{T_m(K_j,m)}{T} - 1\right] = f \times \exp\left[\frac{T_m(K_j,m)}{T}\right]$$
$$\Delta_d(K_j) = \sum_{i\le 3}\gamma_{ij}(c_i - 1)$$

**(46**b)

The activation energy is proportional to the size and bond nature dependence of the atomic cohesive energy,

$$E_A(K_j,m,T) \propto T_m(K_j,m) - T = T_m(\infty,m)\left[1+\Delta_C(K_j)\right] - T$$
$$\Delta_C(K_j,m) = \sum_{i\le 3}\gamma_{ij}\left(z_{ib}C_i^{-m} - 1\right)$$

**(46**c)

The pre-factor $f$ is an adjustable parameter, which should cover both intrinsic and extrinsic contributions. The $\Delta_d(K_j)$ is the contraction of the mean bond length and $\Delta_C(K_j,m)$ the perturbation to the mean cohesive energy of a nanograin.

The difference of $T_m(K_j, m)$-T or the ratio of $T_m(K_j, m)/T$ is critical as it appears in both $\varphi\left(K_j,m,T,T_m(K_j)\right)$, and $\phi\left(A\left(T,m,T_m(K_j)\right),K_j\right)$. The derived form in eq **(46)** represents two types of competition in the IHPR:

(i) One is the intrinsic competition between the remaining atomic-cohesion-energy and the energy-density-gain in the GBs and their temperature dependence, in terms of the separation of $T_m(m, K_j)$ - T. If T remains unchanged during the measurement, a size reduction will decrease the $T_m(m, K_j)$ and hence the value of $T_m(m, K_j) - T$. Thus, the strength of the solid will drop with solid size. The energy-density-gain is represented by the term of $\eta_1(K_j)/\eta_1(\infty) = z/z(K_j)$, being inversely proportional to the average coordination, as discussed in the section dealing with MC elongation. In the nanograins, the deviations of both the energy density and atomic cohesive-energy arise from GB atoms as atoms in the grain interior retain their bulk nature.

(ii) The other is the competition between the activation and inhibition of dislocations. The exponential form of $T_m(m, K_j)/T$ represents the easiness of activating dislocations. Because $T_m(m, K_j)/T$ drops with solid size, the dislocation becomes easier to form at smaller sizes because of the under-coordinated GB atoms. The counterpart of competition comes from dislocation accumulation and strain gradient work hardening that inhibit further sliding dislocations, which complies with the traditionally established meaning of the T-independent HPR of $1+AK_j^{-0.5}$. The factors such as impurities or defects, shapes of indenter tips, strain rates, strain directions, and loading directions and scales contribute to the yield strength measured. [165]

Therefore, the intrinsic competition between the remaining atomic cohesive-energy and the binding energy-density-gain in the GBs and the extrinsic competition between the activation and the resistance of dislocations determine the entire process of IHPR, which depends on temperature of operation because of the thermally driven bond expansion and bond weakening. The analytical form also represents that the under-coordinated atoms in the surface skin dictate the IHPR whereas atoms in the core interior retain their bulk features and that both the bond length and bond strength are temperature dependent.

According to this solution, the grain boundary is harder at temperatures far below $T_m(K_j, m)$ because of the dominance of bond strength gain, whereas at temperatures close to $T_m(K_j, m)$, the GB is softer than the grain interior because of the dominance of bond order loss that lowers the barrier for atomic dislocation, concurring with the mechanism of core-shell role exchange. When operating at a given temperature, solid size reduction lowers the $T_m(K_j, m)$. When $K_j$ is sufficiently large, the analytical form degenerates into the traditional HPR, of which the slope is now clearly seen to be dominated by the term $f \times \exp(T_m(m, \infty)/T)$ that relates to the specific heat or activation energy for atomic dislocation.

The derived form supports the known mechanism of core-shell role exchange [406-421] and the composite or the mixed phase structure model[406,412] for the IHPR. The T-BOLS approximation further clarifies that the competition between the remaining atomic cohesive-energy ($T_m$ suppression) and energy-density-gain in the GB region dictates intrinsically the mechanical behaviour at GBs, which is sensitive to the temperature of operation. As the solid size is decreased, transition from the dominance of energy-density-gain to the dominance of remaining atomic cohesive-energy occurs at the critical size at which both mechanisms contribute competitively. It is emphasized that all the models discussed above give reasonable explanations of the HPR/IHPR transition but, with the currently proposed temperature dependence of the T-BOLS consideration as the atomistic and intrinsic origin, all the approaches would be correct and complete.

7.4 Applications

7.4.1 Strongest sizes in the IHPR

By equilibrating the differential of the natural logarithm of eq (46a) to zero, $d\left(Ln\left(P/P_0\right)\right)/dx_j = 0$, (with $x_j = K_j^{-0.5}$ and $\theta(0) = T/T_m(\infty)$ for calculation convenience), we can readily find the critical size $K_C(f, T_m(x_j)\ T, m)$, at which the slope of HPR transits from positive to negative, and hence, to distinguish factors dominating the $K_C$ value. For simplicity, we define $\theta(x_j) = T/T_m(x_j)$. P and $P_0$ represent $P(x_j, T)$ and $P(0, T_0)$, respectively. The numerical processing leads to the following relation:

$$\begin{aligned}\frac{d\left(Ln\left(P/P_0\right)\right)}{dLn(x_j)} &= A(x_j,\theta(x_j),m)\times\frac{\theta(0)+2\Delta_C(x_j,m)}{\theta(0)\left(1+A(x_j,\theta(x_j),m)\right)} \\ &-\frac{6\Delta_d(x_j)}{1+\Delta_d(x_j)}+\frac{2\Delta_C(x_j,m)}{1+\Delta_C(x_j,m)-\theta(0)}=0\end{aligned} \tag{47}$$

The solution indicates that the critical size depends on the bond nature indicator m, $T/T_m(x_j)$, and the pre-factor *f* in A. Solving this equation numerically gives rise to the critical sizes $x_C$ of different materials, agreeing with measurements, as listed in Table 10.

7.4.2 $T_C$ for phase transition

We may define the critical temperatures, $T_C$, and the corresponding critical size, $K_C$, for the transition from solid to quasi-solid and $T_m$ for transition from quasi-solid to liquid using the relations:

$$\frac{P(K_j,T_C)}{P(\infty,T_C)}=\frac{1+A(K_j,\theta(K_j),m)}{\left[1+\Delta_d(K_j)\right]^3}\times\frac{T_m(K_j,m)-T_C}{T_m-T_C}\begin{cases}\le 1 & (Quasi-solid)\\ =0 & (Liquid)\end{cases} \tag{48}$$

At temperature higher than $T_C(K_j, f, m)$, the solid of critical size $K_C$ is softer and easily compressible compared with the bulk counterpart at $T = T_C(K_j, f, m)$. At the melting point, $T_m(x_j, m)$, the quasi-solid becomes a liquid associated with zero hardness and infinite compressibility. This definition is much the same as in Born's criterion indicating the absence of shear modulus at melting.

7.4.3 Strongest grain sizes estimation

By incorporation of the value of $\eta_1(x_j)/\eta_1(0) = 0.00187/0.0005542 = 3.3742$ for an impurity-free gold MC into eq (46a), we may estimate the value of *f* for the gold MC with the parameters of $z_i = 2$, $K_j = 1.5$, $x_j = 1/\sqrt{K_j} = 1.223$, m = 1, T = 298 K, and $T_m(x_j) = 1337.33/4.1837 = 318$K:

$$f=\left[\frac{\eta_1(x_j)}{\eta_1(0)}-1\right]\Bigg/\left[x_j\times\exp\left[\frac{T_m(x_j)}{T}\right]\right]=2.3742\Bigg/\left\{1.223\times\exp\left[\frac{318}{298}\right]\right\}=0.663$$

(49)

The $f$ value is intrinsic for the elongation test of the impurity-free gold MC. However, in the plastic deformation test of nanograins, artefacts such as the external stress or strain rate or structural defects will contribute to the yield strength and hence the pre-factor $f$ value. The $f$ term is an adjustable parameter that may change value when the nature of the bond is altered, such as in the cases of Si and $TiO_2$ that will be shown shortly.

Further calculations using eq (46) were performed on some nanosolids with the known d and $T_m$ values as input and the $x_C$ as output as listed in Table 10. All the nanograins were taken as being spherical in shape. The prefactor $f$ is adjusted under the constraint that the HPR slope should match the observations and intercept at the positive side of the vertical axis. A negative intercept in measurement is physically unreasonable. The theoretical curves were normalized with the calculated peak values at the transition point, $x_C$, and the experimental data measured at room temperature were normalized with the measured maximum $P_M$.

Table 10

Comparison of the measured and predicted strongest critical grain sizes for IHPR transition with f = 0.5, 0.663 and m = 1 otherwise as indicated. The predicted critical sizes agree with measurement as derived from Figure 27 unless with given references. Changing the f value only affects the $D_C$ (= $2K_Cd$) of a material with $T_m$ below 1000 K.[170]

| Element | d/nm | $T_m$/K | $P_M$ | Measured $D_C$/nm | Predicted $D_C$/nm (f = 0.5; 0.663) |
|---|---|---|---|---|---|
| Mg | 0.32 | 923 | - | | 18.2; 17.3 |
| Ca | 0.394 | 1115 | - | | 22.41; 22.41 |
| Ba | 0.443 | 1000 | - | | 25.2; 23.9 |
| Ti | 0.293 | 1941 | - | | 23.6; 23.6 |
| Zr | 0.319 | 2182 | - | | 29.3; 29.3 |
| V | 0.2676 | 2183 | - | | 24.6;24.6 |
| Ta | 0.2914 | 2190 | - | | 26.8;26.8 |
| Fe | 0.252 | 1811 | 9765 | 18.2 | 19.1 |
| Co | 0.2552 | 1728 | - | | 19.3 |
| Ni | 0.2488 | 1728 | 7042 | 17.5;13[415] | 18.8 |
| Cu | 0.2555 | 1358 | 890 | 14[11]; 14.9; 25-30[427] | 16.1 |
| Zn | 0.2758 | 693 | 1034 | 17.2; 10;[428] 11[368] | 18.5; 16.5 |
| Pd | 0.2746 | 1825 | 3750 | 19.9 | 20.8 |
| Ag | 0.2884 | 1235 | - | | 17.3 |
| Pt | 0.277 | 1828 | - | | 21.0 |
| Au | 0.288 | 1337 | - | | 17.3 |
| Al | 0.286 | 993 | - | 18[11] | 16.3;15.4 |
| C | 0.154 | 3800 | - | | 20.5 (m = 2.56) |
| Si | 0.2632 | 1683 | 4 | 9.1 | 10.6 (m = 4.88, f = 0.1) |
| Ge | 0.2732 | 1211 | - | | 11.5 (m = 4.88, f = 0.1) |
| Sn | 0.384 | 505.1 | - | | 47.2; 40.6 |
| Pb | 0.3492 | 600.6 | - | | 28.0; 24.9 |
| Bi | 0.34 | 544.4 | - | | 36.0; 29.2 |
| $Ni_{80}P_{20}$ | 0.2429 | 1184 | 7063 | 7.9 (m = 4) | 8.9 |
| $NiZr_2$ | 0.4681 | 1393 | 7093 | 17.0 (m = 4) | 19.8 |
| $TiO_2$ | 0.3860 | 2098 | 7432 | 22.5 (m = 4) | 23.1(f = 0.01) |

Figure 27 compares the theoretically produced IHPR with the measured data of typical samples: (a) Cu,[429,430,431] (b) Fe,[429] (c) Ni,[369,432] (d) NiP,[433] and NiZr,[347] (e) $TiO_2$,[434] and (f) Si.[354] The straight lines are the traditional HPR and its slope is obtained by adjusting the *f* values in eq (46). The dashed lines only consider the extrinsic competition between activating and blocking dislocations represented by the activation energy without involving the contribution from the intrinsic competition, with φ(d, m, T) = 1. The solid lines are the current T-BOLS predictions covering both intrinsic and extrinsic competitions without using any other adjustable parameters. The scattered data are experimental results documented in the literature. All the panels were fitted at T = 300 K with f = 0.5 unless otherwise as indicated. Results indicate that the IHPR is dominated by the extrinsic competition whereas the intrinsic contribution plays an insignificant role in plastic deformation.

The T-BOLS predictions match reasonably well to all measurements. The perfect match of NiP alloy and $TiO_2$ may adequately support the evidence that the current T-BOLS and LBA approaches are close to the true situations of IHPR involving both intrinsic and extrinsic contributions. As can be seen from Table 10, changing the *f* values from 0.5 to 0.663 has no effect on the critical size for materials with $T_m(0) > 1000$ K, or $T/T_m(0) < 1/3$, and therefore, for the examined samples, using *f* = 0.5 or 0.663 makes no difference. The small *f* values for $TiO_2$ (0.01) and Si (0.1) may be dominated by the bond nature alteration that lowers the $T_m(x_j, m)$ insignificantly.

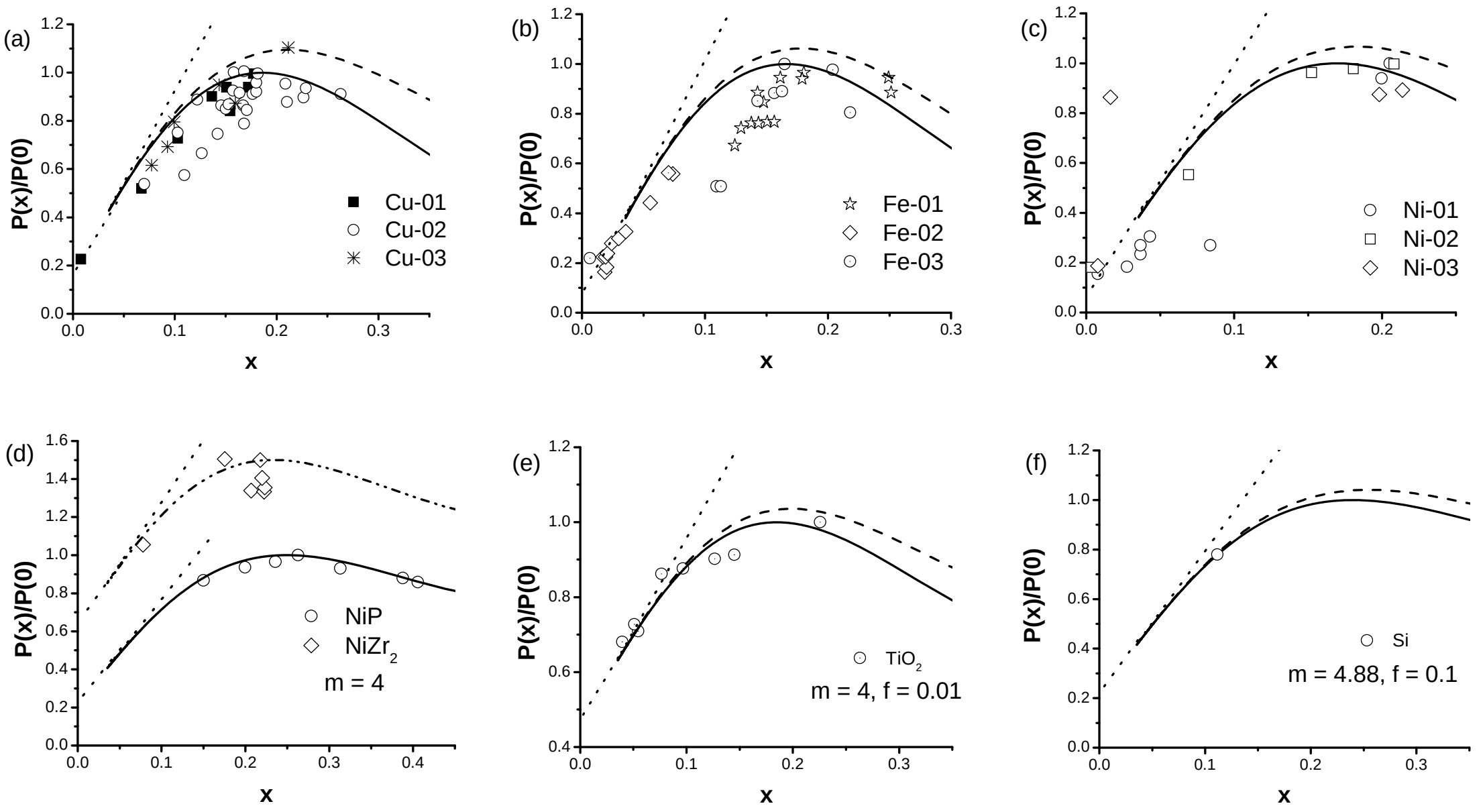

Figure 27 Link Comparison of the calculated (solid line) with the measured IHPR (scattered data) of (a) Cu,[429,430,431] (b) Fe,[429] (c) Ni [369],[432] (d) NiP[433] and $NiZr^{347,}$ (e) $TiO_2$[434] and (f) Si.[354] The solid curves represent the complete T-BOLS form, $\left[1+f\times x_j\times\exp\left(T_m\left(x_j\right)/T\right)\right]\varphi(d,m,T)$. The straight lines are traditional HPR representing $1+f\times x_j\times\exp\left(T_m(0)/T\right)$ and its intercept at the y-axis corresponds to the normalized hardness of the bulk counterparts. The dashed curves lines are the modified HPR, $\left[1+f\times x_j\times\exp\left(T_m\left(x_j\right)/T\right)\right]$. The slope f = 0.5 was optimised unless otherwise indicated for all the samples.

Strikingly, one datum point of measurement is sufficient to calibrate the IHPR of a specimen. For example, applying the measured relative hardness of the defect-free 40 nm silicon nanosphere to the IHPR equation results in the maximum hardness of Si nanosolid at room temperature being 5 times the bulk value and the IHPR critical size being at 9 ~ 10 nm. The calibration using one datum point and the known bulk value should be one of the advantages of the current approach in predicting and calibrating the IHPR of a solid. Applying m = 4.88 for Si to eq (46a) gives immediately $c_1 = (85 \pm 1)\%$ and the corresponding $z_1 \approx 3.60 \pm 0.25$. The $z_1$ for the spherical Si is slightly lower than $z_1 = 4$ for the flat surface and the $c_1$ is slightly higher than a flat surface (88 %) because of the positive curvature of the sphere.

### 7.4.4 Factors dominating the strongest size

Figure 28 shows that, as for the inverse of yield strength, the normalized compressibility drops first and then bends up at the IHPR critical point until it reaches the bulk value at $T_C(x)$ and then goes up to infinity at $T_m(x)$. This trend agrees with the breaking limits of metallic MC and polymer chains approaching $T_m$.

Results show that the Pb nanosolid of x = 0.34 (D = 6 nm) size becomes quasi-solid and of x = 0.52 (2.6 nm) becomes liquid at 300 K because of the low bulk melting point (600 K). In contrast, the Au nanosolid of x = 0.68 (D = 1.25 nm) becomes quasi-solid and x = 0.95 (D = 0.64 nm being smaller than one fcc unit cell) becomes liquid at 300 K. Therefore, the gold MC (or an fcc unit cell) is in the quasi-solid state at 300 K, which clarifies further the high extensibility of the Au-MC at the ambient temperature. However, the curve of IHPR-1 with φ(x, m, T) = 1 could hardly reach the quasi-solid or the liquid states in both cases.

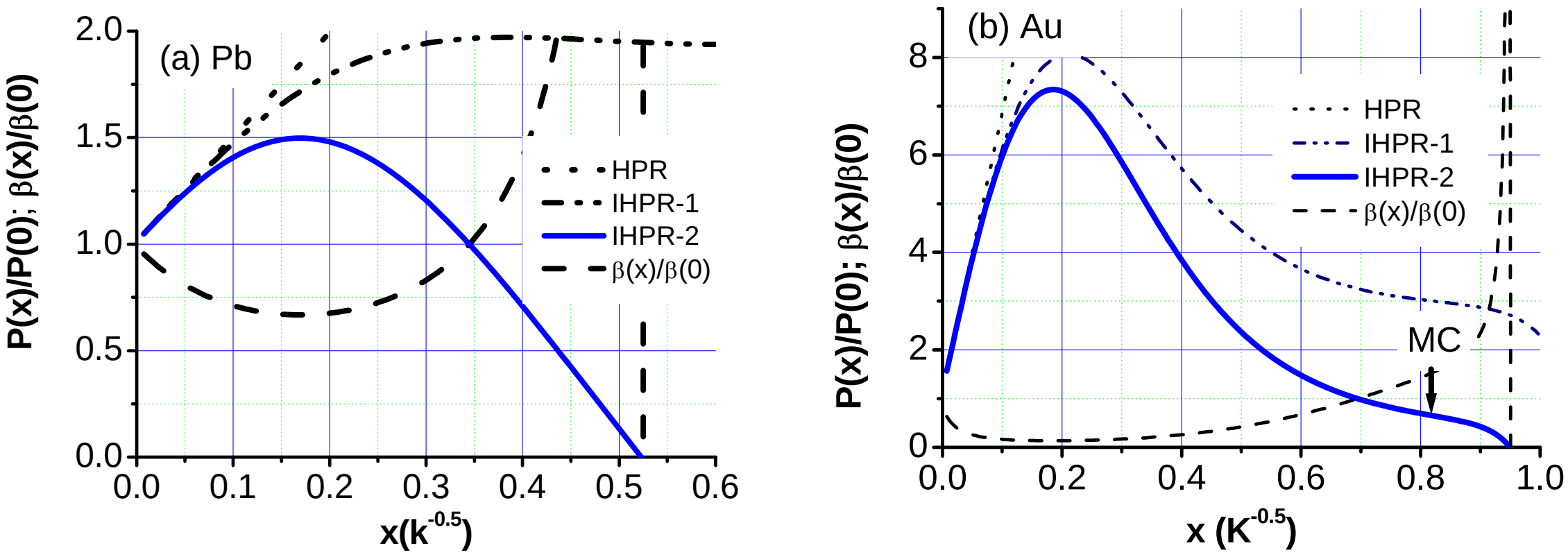

Figure 28 Link Comparison of the IHPR-2 (solid line) transition and critical sizes for solid-to-quasi-solid and quasi-solid-to-liquid transition at 300 K for (a) Pb (m = 1, $T_m$ = 600.6 K, f = 0.663) and (b) Au (m = 1, $T_m$ = 1337.33 K, f = 0.663). $P/P_0$ = 1 corresponds to quasi-solid state with critical temperature $T_C(x)$; $P/P_0$ = 0 to corresponds to quasi-solid-to-liquid transition $T_m(x)$. Broken line represents the reduced compressibility that drops first and then bends up until the bulk value at $T_C(x)$ and then goes up to infinity at $T_m(x)$. The gold MC is in quasi-solid state at 300K, which clarifies further the high extensibility of Au-MC at the ambient temperature. The curve of IHPR-1 could not reach the quasi-solid or the liquid states.

Figure 29 demonstrates the f, m, and $T/T_m(0)$ dependence of the $x_C$ values. The critical size varies significantly with the temperature of operation ($T/T_m(0)$) and the bond nature indicator (m). In the range of *f* = 0.1 and 1.0, $x_C$ depends less on the f values if the $T/T_m(0)$ ratio is smaller than 0.2. It is interesting to note that there exist two operating temperatures for one $x_C$ value in both panels (a) and (b). For the $x_C$ = 0.15 example, the temperature reads as T = 0.17 $T_m$ and 0.65 $T_m$ for the f = 1 curve in panel (a); and the T reads 0.1 and 0.7 $T_m$ for m = 5 in panel (b). This result means that the same critical size can be obtained by measuring the specimen at two different temperatures but the corresponding values of hardness are completely different; the strength at low T is much higher than the strength at the higher T, as demonstrated in Figure 30a.

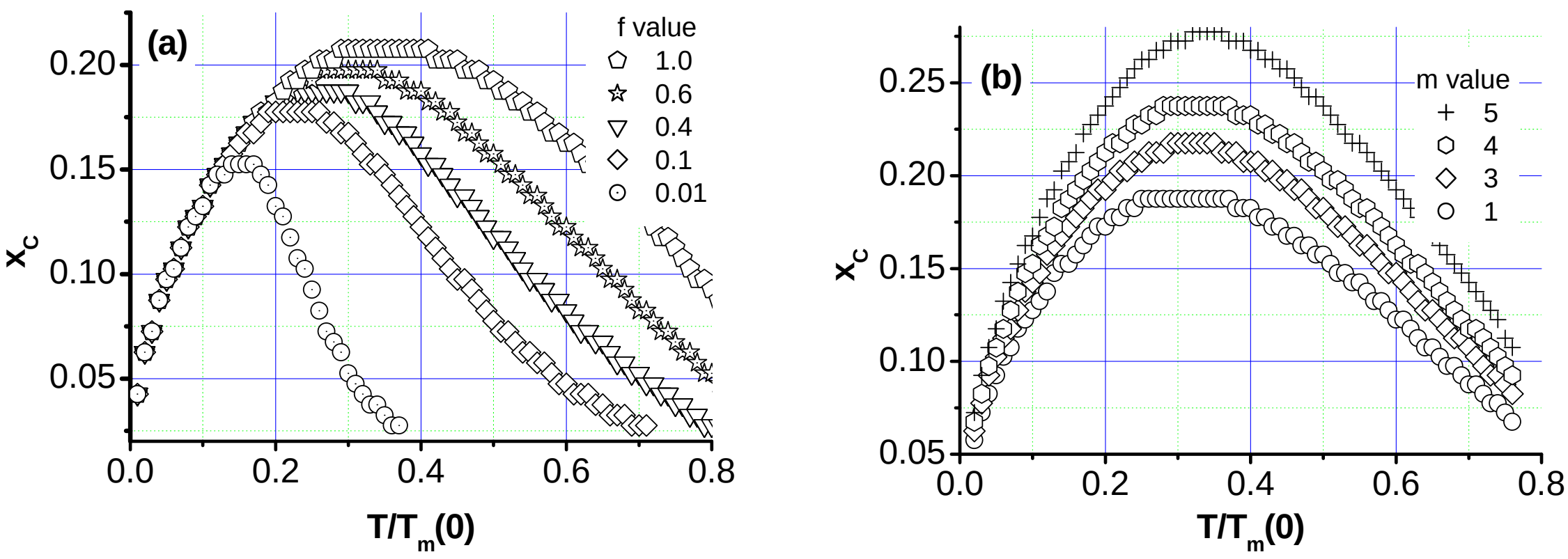

Figure 29 Link Dependence of $x_C$ (m, f, T/$T_m$(0)) on (a) extrinsic factor f and (b) bond nature indicator of m. At T/$T_m$(0) < 0.15, $x_C$ is independent of the *f* values but the bond nature has an influence over the whole range of temperature. One critical size $x_c$ can be obtained at two temperatures with different hardness.

7.4.5 Size effect on $T_C$(x) and $T_m$(x)

From the T/$T_m$ and size dependence of strength and ductility in Figure 30a, we may note the following trends in general:

(i) For a given material of a given size, the normalized mechanical strength drops from infinity to zero when T increases from zero to $T_m$(x, m), associated with an increase of ductility until a singularity is reached at $T_m$(x) that drops with size by 5% and 15%, respectively, for the $x_C$ = 0.15 and 0.21 examples.

(ii) Both P and β reach their bulk values (or transits into quasi-solid state) at $T_C$ = 0.75 $T_m$ and 0.65 $T_m$ for $x_C$ = 0.15 and 0.21, respectively. When T > $T_C$, P drops and β rises in an exponential way.

A comparison of the size dependence of the normalized atomic distance d($x_j$)/d(0), melting point, $T_m$($x_j$, m =1)/$T_m$(0), and the ratio of $T_C$($x_j$, f, m =1)/$T_m$(0) for solid-to-quasisolid transitions shown in Figure 30(b) indicates that the $T_C$($x_j$, f) drops more rapidly with size than $T_m$($x_j$, m). The bulk $T_C$ value is about 20% lower than the bulk $T_m$. It is interesting to note that the currently defined $T_m$($x_j$)/$T_m$(0)-BOLS curve overlaps the curve of $T_m$($x_j$)/$T_m$(0)-Born (derived from Born's criterion), indicating the consistency in the respective physical mechanisms of melting. The BOLS model suggests that the melting is governed by the atomic cohesive energy ($T_m \propto z_iE_i$) depression and the Born's criterion eq (48) requires that the modulus or mechanical strength approaches zero at $T_m$.[138] The trend of the $T_m(x_j)/T_m(0)$ curves also agree with the models of Shi[266] and Jiang et al[435] derived from Lindermann's criterion of atomic vibration, the model of surface lattice/phonon instability of Vallee et al,[436] the models of liquid drop[111] and the model of surface area difference determined cohesive energy.[437] Therefore, all the models are correct despite different perspectives.

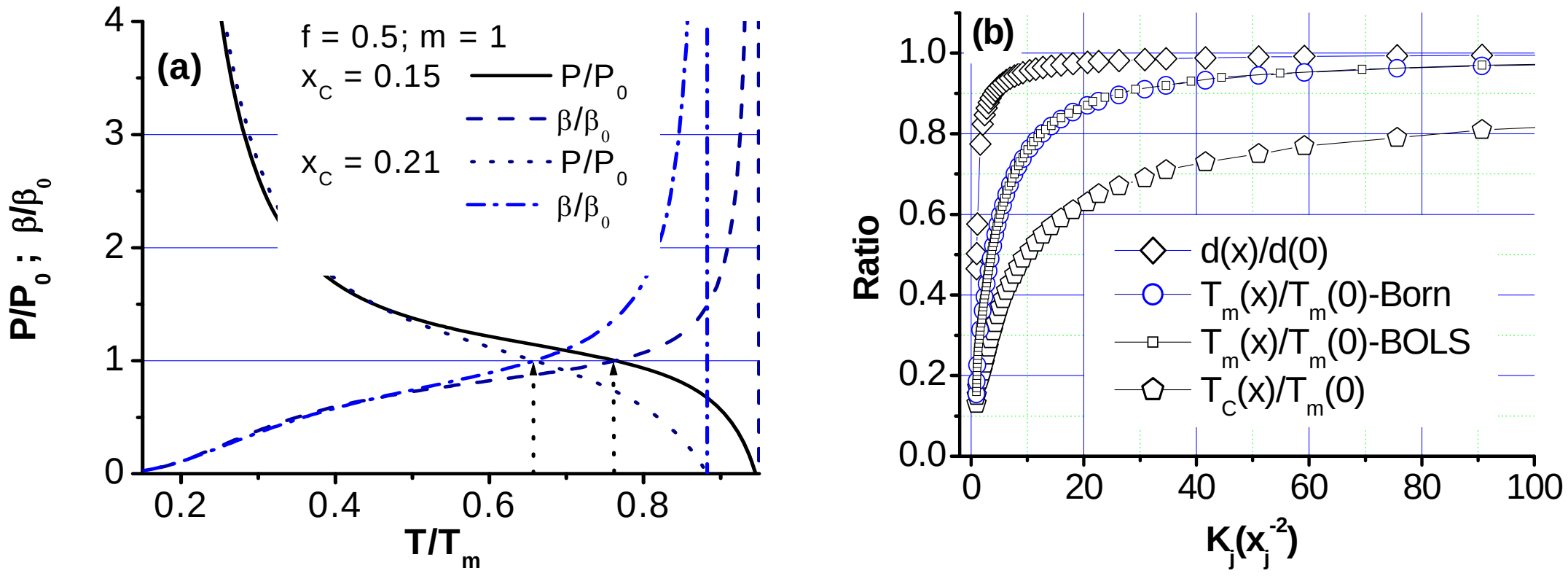

Figure 30 Link (a) The $T/T_m$ ratio dependent strength and ductility of size $x_C = 0.15$ and 0.21, and (b) comparison of size dependence of critical temperature for solid-quasi-solid transition, $T_C(x)/T_m(0)$, bond length contraction, $d(x)/d(0)$, and quasi-solid -liquid transition of metallic nanosolid ($m = 1$). $T_m(x)/T_m(0)$-BOLS curve overlaps the $T_m(x)/T_m(0)$-Born.

### 7.4.6 Quasi-solid state and superplasticity

The concept of quasi-solid may improve our understanding of the superplasticity of the nanograined solid. Superplasticity, an excessive strain of $10^3$ without any substantial necking region when loaded in tension, is generally observed in materials with grain size less than 10 mm in the temperature range of 0.5 – 0.6 $T_m$.[353] Because of the large GB volume fraction and the high self-diffusivity, superplasticity is achievable at lower temperatures and/or higher strain rates for some nanocrystalline materials. For a Cu nanosolid with size of $K_j = 10$ (5 nm in diameter) for example, the bond length contracts by a mean value of 5%, associated with a 25% drop of $T_m$ and a 50% drop of $T_C$ with respect to the bulk $T_m(0)$ value of 1358 K. The 5 nm-sized Cu solid is in a quasi-solid state at 680 K according to Figure 30(b). The self-heating in operation because of the energy released from the process of bond breaking and unfolding should raise the actual temperature of the specimen. Hence, the joint effect of size-induced $T_C$ suppression and the bond breaking/unfolding induced self-heating may provide an additional mechanism for the high ductility of Cu nanowires[312] at the ambient temperatures, which is dominated by bond unfolding and atomic gliding dislocations through creep and kink formation, instead of bond stretching, as discussed in the previous section.

The predicted m, f, and $T/T_m(0)$ dependence of $x_C$ and $T_C(x)$ and the trends of mechanical strength and compressibility/extensibility coincide exceedingly well with the cases as reported by Eskin, Suyitno, and Katgerman[141] on the grain size dependence of the tensile elongation (extensibility) of Al-Cu alloys in the quasi-solid state.[142] The ductility increases exponentially to infinity with temperature up to $T_m$ that drops with solid size. On the other hand, the ductility increases generally with grain refinement. This is also frequently observed in cases such as alumina[127] and PbS[352] in the nanometer range at room temperature.[438,439] The compressibility of alumina and PbS solid increases, whereas the Young's modulus decreases as the solid size is decreased. The predicted trends also agree with experimental observations[77] of the temperature dependence of the yield stress of Mg nanosolids of a given size showing that the yield strength drops as the operating temperature is elevated.[77] Temperature dependence of the tensile properties of ultrafine-grained ordered $FeCo_2V$ samples with grain sizes of 100, 150, and 290 nm revealed extremely high yield strengths (up to 2.1 GPa) present at room temperature with appreciable ductility of 3 -13%. The measured strengths declined gradually as the testing temperature was increased to 400 °C, while ductility was generally enhanced, up to 22%.[85] At $T > 0.7\ T_m$, the mechanical strength decreases rapidly with increasing temperature and with decreasing strain rate.[2]

## 7.5 Correlation between elasticity and hardness

Theoretically, the analytical expressions for elasticity, stress, and hardness, e.g., eq (37) and eq (46), should be identical in nature. The former represents the intrinsic property change without experimental artefacts being involved. However, the softening and slope transition in the IHPR plastic deformation arises from the extrinsic competition between activation of and resistance to glide dislocations, which is absent in the elastic deformation in particular using the non-contact measurement such as SWA techniques. Recent measurement of the size dependence of the hardness, shear stress and elastic modulus of copper nanoparticles, as show in Figure 31a, confirmed this expectation. The shear stress and elastic modulus of Cu reduce monotonically with the solid size but the hardness shows the strong IHPR. Therefore, the extrinsic factors become dominant in the plastic deformation of nanocrystals, which triggers the HPR and IHPR being actually a response to the detection. However, as compared in Figure 31b and c, the hardness and Young's modulus of Ni films are linearly interdependence. This observation indicates that extrinsic factors dominating the IHPR of nanocrystals contribute little to the nanoindentation test of film materials.

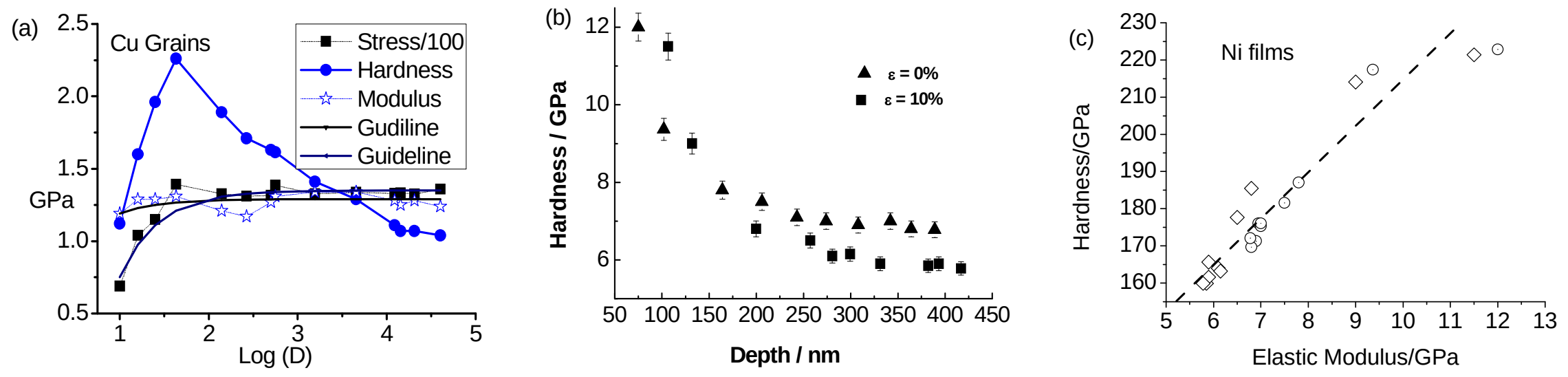

Figure 31 Link (a) Comparison of the measured size dependence of the elasticity, shear stress and the hardness of Cu nanostructures [427], and the nanoindentation depth dependence of (b) hardness and the correlation between the hardness and elastic modulus of Ni films [10], indicating the significance of extrinsic factors on the hardness measurement of nanocrystals but contribute little to films.

One may note that the Young's modulus is defined for elastic deformation, while its inverse, or the extensibility/compressibility, covers both elastic and plastic types of deformation. However, either the elastic or the plastic process is related to the process of bond distortion, including bond unfolding, stretching, or breaking, that consumes energy (obtained by integrating the stress with respect to strain) being a certain portion of the entire binding energy. No matter how complicated the actual process of bond deformation (with linear or nonlinear response) or recovery (reversible or irreversible) is, a specific process consumes a fixed portion of bond energy, and the exact portion for the specific process does not come into play in the numerical expressions for the relative change. Therefore, the relation of eq (37) is valid for describing the intrinsic properties for any substance of any scale of size and in any phase. However, for plastic deformation, the competition between dislocation activation and dislocation resistance becomes dominant, which presents a difference in the plastic deformation from the elastic response in terms of the IHPR features. It is important to note that, in a nanoindentation test, errors may arise because of the shapes and sizes of the tips. The stress-strain profiles of a nanosolid are not symmetrical when comparing the situation under tension to that under compression.[440] The flow stress is dependent of strain rate, loading mode and duration, and material compactness, as well as size distribution. These factors may influence measured data that are seen to be quite scattered and vary from source to source. As the Young's modulus and mechanical strength are quantities of intensity, they are volumeless. Therefore, it might not be appropriate to think about the stress or elasticity of a certain atom, but, instead, the stress and elasticity at a specific atomic site.

## 7.6 Summary

An analytical expression has been derived for the extensibility and plastic yield strength of a solid over the whole range of sizes based on the T-BOLS correlation mechanism, which has enabled us to reproduce the observed HPR and IHPR effect and identify the factors dominating the strongest sizes. Matching predictions to observations reveals the following:

(i) The IHPR originates from the intrinsic competition between the energy-density-gain in the surface skin and the remaining cohesive-energy of the under-coordinated surface atoms and the extrinsic competition between activation and prohibition of atomic dislocations. The activation energy is proportional to the atomic cohesion that drops with solid size whereas the prohibition of atomic dislocation arises from dislocation accumulation and strain gradient work hardening which increases with the indentation depth. As the solid size decreases, a transition from dominance of energy-density-gain to dominance of remaining cohesive-energy occurs at the IHPR critical size because of the increased portion of the under-coordinated atoms. During the transition, contributions from both processes are competitive.

(ii) The IHPR critical size is universally predictable, which can be calibrated with a few measured datum points for a specific system. The critical size is dominated intrinsically by the bond nature, the $T/T_m$ ratio, and extrinsically by experimental conditions or other factors such as size distribution and impurities that are represented by the factor *f*.

(iii) The IHPR at larger solid size converges to the normal HPR that holds its conventional meaning of the accumulation of atomic dislocations that resist further atomic displacements in plastic deformation. The slope in the traditional HPR is proportional to $\exp(T_m/T)$, which addresses the relationship between the hardness and the activation energy for atomic dislocations. The $K_j^{-0.5}$ in the conventional HPR should represent the accumulation of atomic dislocations that resists further dislocations.

(iv) This understanding of the process provides an additional explanation for the high ductility of a metallic nanosolid as the critical temperature for the solid-to-quasi-solid transition is much lower than the bulk melting point and the self-heating during detection should raise the actual temperature of the small specimen.

# VIII Atomic vacancy, nanocavity, and metallic foams

## 8.1 Observations

### 8.1.1 Atomic vacancies and point defects

It is expected that atomic vacancies or point defects reduce the number of chemical bonds of nearby atoms and hence the strength of the entire body of a material. However, the hardness of transition metal carbides and nitrides does not follow this simple picture of coordination counting. Vacancies or point defects not only provide sites initiating structure failure but also act as pinning centers inhibiting the motion of dislocations and hence enhancing the mechanical strength of the material.[104] It has been found that the hardness of iron aluminides (FeAlN) is proportional to the square root of vacancy concentration.[441] About 4% N vacancies are found to stabilize the rocksalt structure of MnN substantially.[442] Measurement has revealed that with the increase of C vacancies in the WAlC compounds, the mass density decreases whereas the hardness increases monotonically up to a maximum at 35% of C vacancies, as shown in Figure 32(a).[443] Fracture measurement and modeling analysis indicated that a small number of atomic defects govern the strength of the $WS_2$ nanotubes.[444] A study[93] using atomistic simulations and analytical continuum theory to describe the influence of the vacancy concentration related defects on the Young's modulus and tensile strength revealed the enormous influence of an atomic defect on the strength of the nanotubes because of the excessive voids presented in the wall of the nanotubes that serve as centers initiating structure failure in plastic deformation. Figure 32(b) shows the strength versus the number of missing atoms in the critical defect according to the quantized fracture mechanics (QFM) approximation.[445]

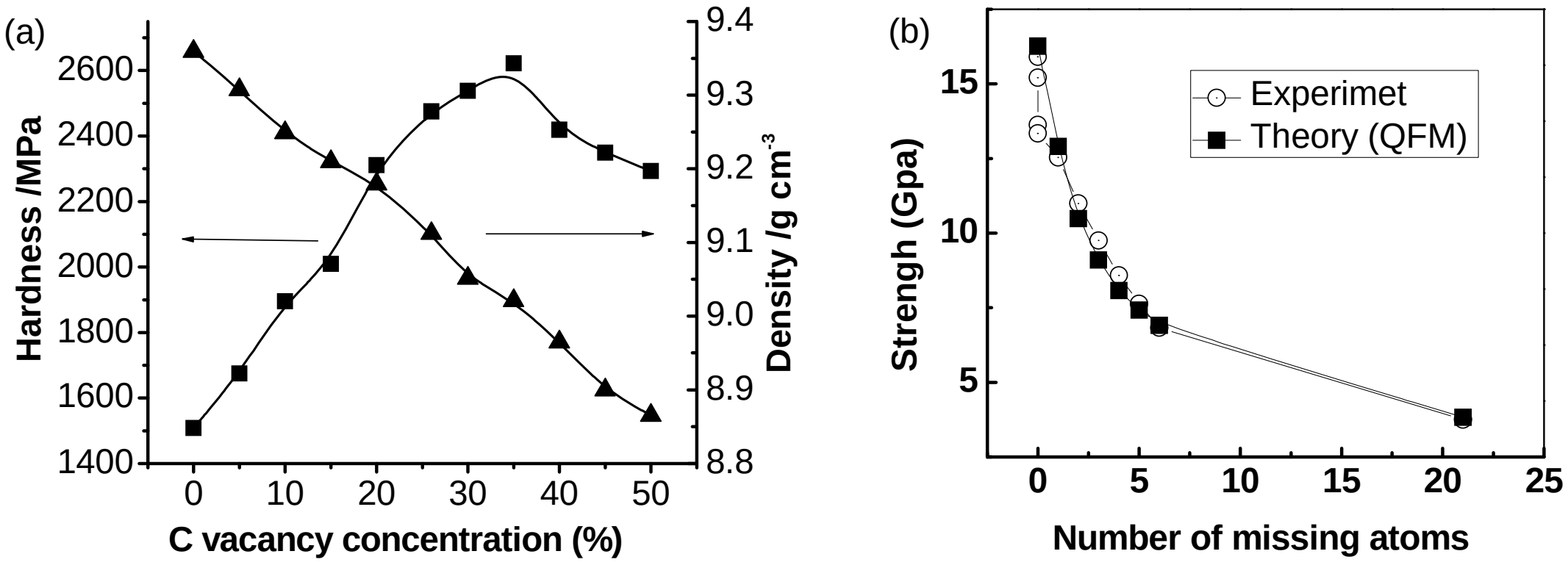

Figure 32 Mechanical hardening by (a) carbon vacancy density in $(W_{0.5}Al_{0.5})_xC_{1-x}$ compound [443] and (b) point defect in $WS_2$ nanotubes.[444] The squares represent the analysis of the experimental results, and the circles represent the calculated strength according to the QFM model. Link

An *ab initio* calculation of shear elastic stiffness and electronic structures suggested that the vacancy produces entirely different effects on the mechanical strength of group IVb nitrides and group Vb carbides. It is suggested that the occupation of shear-unstable metallic d-d bonding states changes essentially in an opposite way for the carbides and nitrides in the presence of vacancies, resulting in different responses to shear stress.[446] Experimentally, the hardness and the elastic modulus of group IVb nitrides such as $TiN_x$, $ZrN_x$, and $HfN_x$ decrease as the concentration of the non-metal vacancy increases.[447,448] In contrast, the hardness of the Vb carbides such as $NbC_x$ and $TaC_x$ increases consistently and reproducibly as the vacancy concentration increases up to a modest value and reaches a maximum at a vacancy concentration of about 12%.[449] It has long been puzzling how atomic vacancies affect the hardness of these two classes of compounds in such distinctive ways despite similar chemical bonding of these compounds at stoichiometry.

DFT calculations[450] suggested that the boron doping of $ScRh_3B_x$ could enhance the cohesive energy monotonically due to the strong covalent bonding between B-2p and Rh-4d states. However, at x = 0.5, a configuration is achieved in which each boron is surrounded by vacancies at the cube centers. This configuration reduces the strain in the structure and shortens the Rh-B bonds, leading to a maximum in the bulk modulus. The density of states at the Fermi energy is also minimum for x = 0.5 which adds further stability to the structure.

### 8.1.2 Nanocavity

Formation of nanometer-sized cavities could also enhance the mechanical properties of solid materials.[450,451,452] Amorphous carbon (a-C) films have an uniquely intrinsic stress (~12 GPa) which is almost one order in magnitude higher than those found in other amorphous materials such as a-Si, a-Ge, or metals (<1 GPa).[453] The internal stress of the a-C films can be modulated by changing the sizes of nanopores that are produced by the bombardment of noble gases (Ar, Kr, and Xe) during formation. [454,455] Using extended near-edge XAFS and XPS, Lacerda et al[454] investigated the effect of the trapping of noble gases in the a-C matrix on the internal stress of the a-C films and the energy states of the trapped gases. They found that the internal stress could be raised from 1 to 11 GPa by controlling the sizes of the pores within which noble gases are trapped. Meanwhile, they found an approximate ~1 eV lowering (smaller in magnitude) of the core level binding energy of the entrapped gases (Figure 33) associated with 0.03-0.05 nm expansion of the atomic distance of the noble gases trapped. The measured core-level shift is of the same order as those measured for noble gases implanted in Ge,[456] Al,[457] and Cu, Ag, and Au[458,459] and Xe implanted in Pd hosts.[460] The interatomic separation of Ar (Xe) increases from 0.24 (0.29) nm to 0.29 (0.32) nm when the stress of the host a-C is increased from 1 to 11 GPa.[461] Compressing the organic layers in organic light emitting diode (OLED) has been found to be an effective process for improving the performance of organic electroluminescent devices. [462] This process involves applying physical pressure to the organic layers of the device. The OLED fabricated by this method shows a notable increase in luminance intensity and current efficiency when compared with compression-free device, the current efficiency almost doubling.

Comparatively, an external hydrostatic pressure around 11 GPa could suppress the interplanar distance of microcrystalline graphite by ~15%,[463] gathering the core/valence electrons of carbon atoms closer together. The resistivity of a-C films decreases when the external hydrostatic pressure is increased.[464] These results are in agreement with the recent work of Umemoto et al[465] who proposed a dense, metallic, and rigid form of graphitic carbon with similar characteristics. The effect of hydrostatic pressure is very much the same as the pore-induced internal stress using noble gas sputtering and implanting.

The binding energy weakening and atomic distance expansion of the entrapped gases indicate clearly that the gas-entrapped pores expand in size and the interfacial C-C bonds contract because of the bond order loss of the interfacial C atoms, which contribute to the extraordinary mechanical strength of the entire a-C films. The pore-induced excessive stress is expected to play the same role as the external hydrostatic pressure causing densification, metallization, and strengthening of the graphite by lattice compression. It has been found that the surface interstitial nanovoids of 2~3 nm size induced by He ion implantation in crystalline Si could remarkably reduce B diffusion, whereas these nanovoids do not hinder the B electrical activation.[466] This finding may evidence the lowered activation energy for diffusion yet the densified charge for conductivity due to the strain and trapping effect at the void skins.

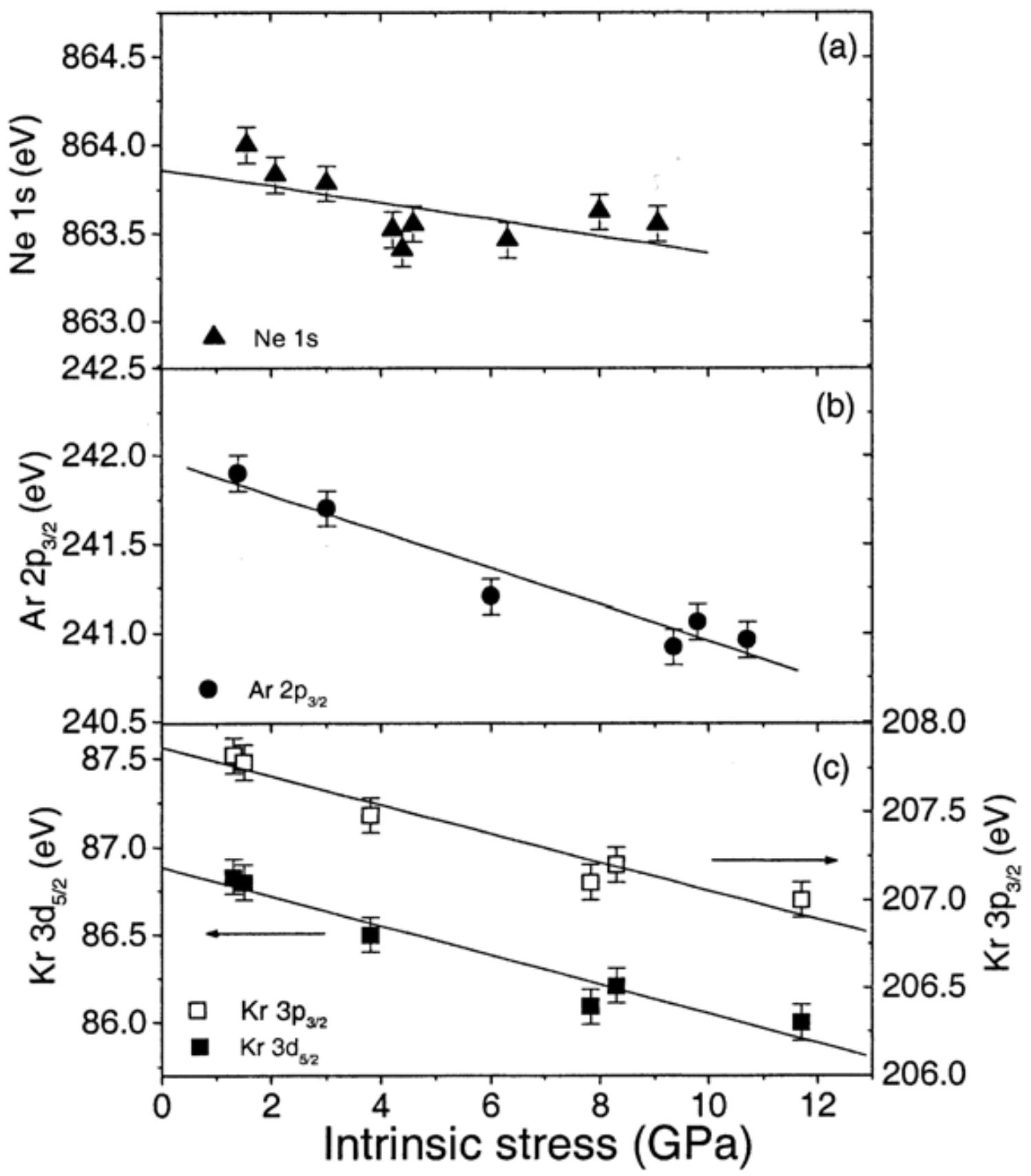

Figure 33 Implanted noble-gas binding energy shifts relative to the Fermi level as a function of the compressive stress of a-C films,[467] evidencing the weakening atomic interaction of the entrapped gases. Link

Ouyang et al have given an alternative explanation of the nanocavity hardening[468] from the perspective of surface energy. They suggested that the dangling bonds at the negatively curved surface could provide a large driving force for pore instabilities such as size shrinkage instead of pore size expansion. The surface free energy in the inner skin of nanocavities is suggested to be composed of two parts: one is chemical and the other is structural, i.e., $\gamma = \gamma^{chem} + \gamma^{stru}$, being similar to the core-shell nanostructures.[469] The chemical part of the surface energy originates from the dangling bond while the structural part is from elastic strain in the inner skin of one atomic layer thick of the nanocavity. Results showed that the surface free energy increases when the pore size is decreased. Importantly, the larger lattice deformation in the inner skin of nanocavity during shrinkage could raise the intrinsic modulus.

8.1.3 Nanoporous foams

Metal foams with excessive amount of discretely distributed nanocavities have formed a relatively new class of materials, which offer a variety of applications in fields such as lightweight construction or crash energy management.[470,471] Despite the versatility of geometrical forms of the pores,[472,473,474] the significance of the nanoporous foams is the large portion of under-coordinated atoms. The foams can be envisioned as a three-dimensional network of ultrahigh-strength nanowires or spherical holes in a matrix, thus bringing together two seemingly conflicting properties: high strength and high porosity. The foamed materials are expected stiffer at low temperatures and tougher at raised temperatures compared with bulk crystals. Figure 34 shows the typical open cell structure and the ligament size of Au nanofoams and the size and surface effect on the stiffness of Au foams of ~30% relative density samples.[475] Measurements revealed that foams of the smaller ligament sizes are stronger and that the surface is even stronger, agreeing with previously discussed surface mechanics and indentation-depth features of continuum thin films. A dramatic rise in the effective Young's modulus of nanoporous Au has also been detected with decreasing ligament size, especially below 10 nm.[476]

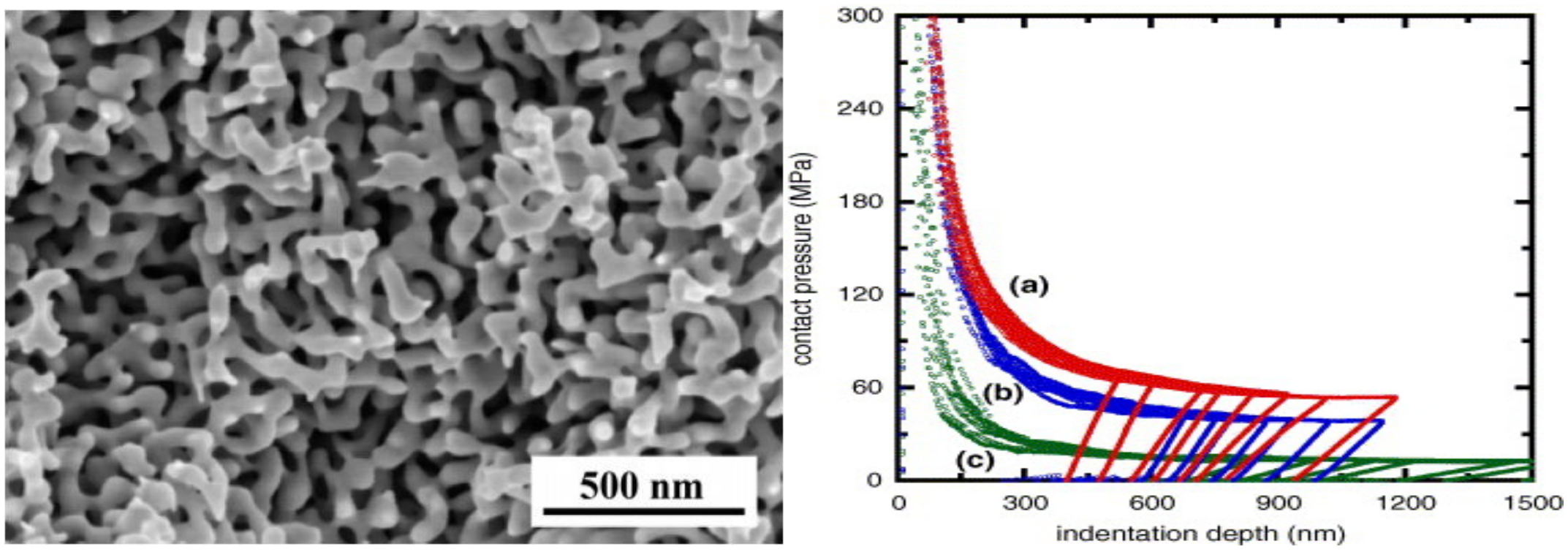

(a) (b)

Figure 34 (a) The open cell structure of Au nanofoams and (b) the size and surface effect on the hardness in the nanoindentation load-depth profiles for 30% relative density samples, given distinct ligament sizes in panel (b): a) 60 nm; b) 160 nm; and c) 480 nm [475].Link

Characterization[477] of the size-dependent mechanical properties of nanoporous Au using a combination of nanoindentation, column pillar microcompression, and MD simulations suggested that nanoporous gold could be as strong as bulk Au, and that the ligaments in nanoporous gold approach the theoretical yield strength of bulk gold, or even harder.[475] At a relative density of 42%, porous Au manifests a sponge like morphology of interconnecting ligaments on a length scale of similar to 100 nm. The material is polycrystalline with grain sizes of 10-60 nm. Microstructural characterization of the residual indentation reveals a localized densification via ductile (plastic) deformation under compressive stress. A mean hardness of 145 MPa and a Young's modulus of 11.1 GPa has been derived from the analysis of the load-displacement curves. The hardness of the investigated nanoporous Au has a value some 10 times higher than the hardness predicted by scaling laws of open-cell foams.[478] The compacted nanocrystalline Au ligaments exhibit an average grain size of < 50 nm and hardness values ranging from 1.4 to 2.0 GPa, which were up to 4.5 times higher than the hardness values obtained from polycrystalline Au.[479] A decrease in ligament size resulted in an increase in the yield stress, and the yield stress of the nanometer-sized Au ligament was much higher than that of bulk polycrystalline gold. It is suggested that the size and surface effects, such as a reduction in the number of defects in grains, are important to the strengthening of nanoporous Au.[480] Using scaling laws for foams, the yield strength of the 15nm diameter ligaments is estimated to be 1.5 GPa, close to the theoretical strength of Au. This value agrees well with extrapolations of the HPR yield strength at submicron scales.[481]

Similarly, the strength of Al foams can be increased by 60–75% upon thermal treatment and age hardening after foaming.[482] It was also found that the hardness of the Al foam is twice as high as pure Al, and the hardness decreases with increasing temperature.[483] Figure 35(a) shows the porosity dependence of the stress–strain curves of the aluminum foams with a relative density of 14.6–48.5%, where the pore size is between 1.6 and 2.0 mm. If the foam density was high, then the constant-stress plateau was short. On the other hand, foam of lower density exhibited a longer, flatter plateau because the structure affords more opportunity for cell walls to collapse and deform.[484] Foams with smaller pore sizes should follow the same trend of hardness change.

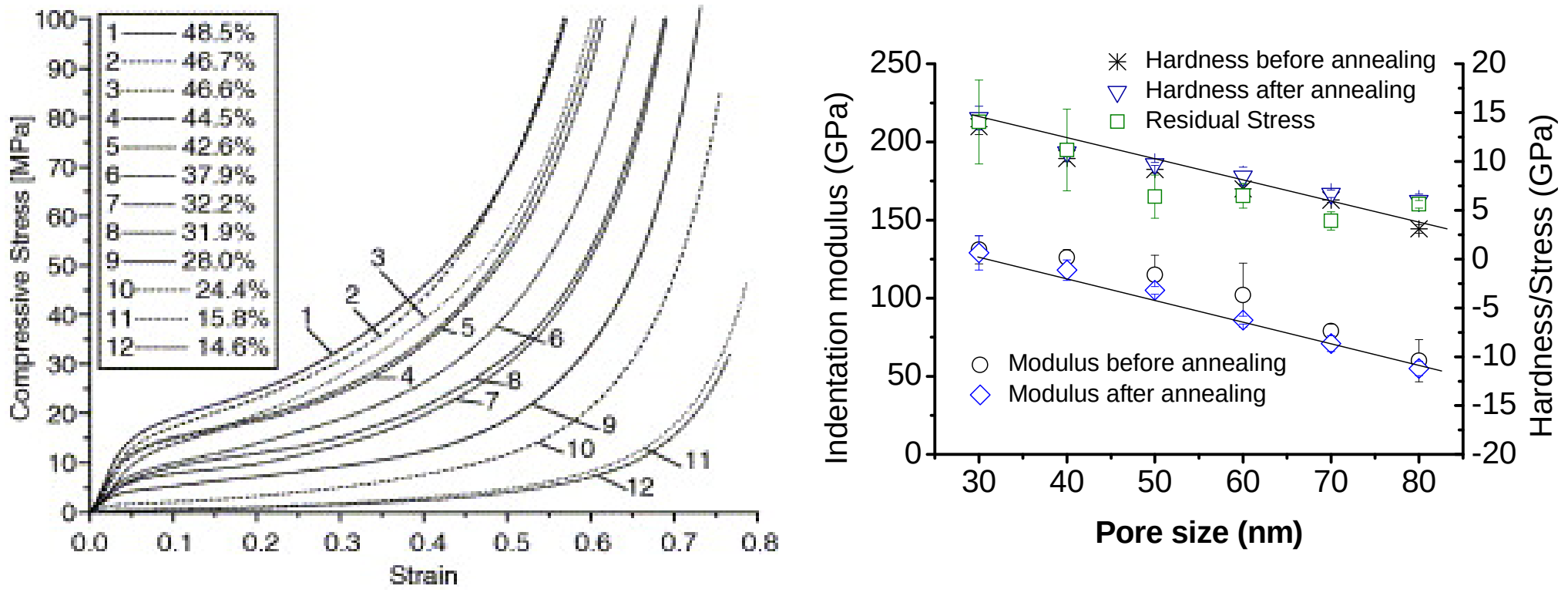

Figure 35 (a) Compressive stress–strain curves of the aluminium foams with different relative density (pore size: 1.60–2.00 mm) and (b) pore size dependence of anodic aluminium oxide foam modulus, hardness, and residual stress that follow the same trend of change [485], indicating the common origin of these seemingly different quantities from the perspective of energy density, according to the BOLS premise. linka linkb

However, nano-indentation measurement[485] revealed that the mechanical properties, such as the modulus, hardness, of anodic aluminum oxide structures decrease monotonically as the size of the hole is increased from 30 to 80 nm, see Figure 35(b). Although the elastic modulus, hardness, and viscosity are different mechanical properties, it is noteworthy that a similar behavior was observed for all properties, as observed from polymers.[486] The same trend of change of the seemingly different quantities indicates clearly their common origin and interdependence from the perspective of energy density, according to the current BOLS correlation premise.

On the other hand, the porous structure is thermally less stable. MD simulation[487] of the size effect on melting in solids containing nanovoids revealed four typical stages in void melting that are different from the melting of bulk materials and nanoparticles. Melting in each of the stages is governed by the interplay among different thermodynamic mechanisms arising from the changes in the interfacial free energies, the curvature of the interface, and the elastic energy induced by the density change at melting. As a result, the local melting temperatures show a strong dependence on the void size, which is the cause of the observed complex hierarchical melting sequence. Despite these exciting prospects, the understanding of the mechanical and thermal behavior of metal foams at the nanoscale is still very much in its infancy.[478,488]

### 8.1.4 Theoretical approaches

There have been several models regarding the cavity hardening of nanovoided systems. Quantize fracture mechanics (QFM) in terms of the classical continuum medium mechanics and the Gibbs free energy considers that a discrete number of defects arising from a few missing atoms in a nanostructure could contribute to the mechanical strength. Another theoretical approach considers the electronic structure around the Fermi energy. [489] Theoretical calculations suggested that the presence of two unsaturated electronic bands near the Fermi level responding oppositely to shear stress enhances the hardness of the voided systems behaving in an unusual way as the number of electrons in a unit cell changes. This finding agrees with the BOLS premise indicating that the density of states will shift to energy lower away from the $E_F$ because of the broken bond depressed potential well of trapping nearby. The deepened potential well of trapping provides also an atomistic understanding of effect of defect pinning that inhibits atomic dislocations.

It has been suggested that the macroscopic stresses are composed of two parts,[490] representing dynamic and quasi-static components. The dynamic part controls the movement of the dynamic yield surface in stress space, while the quasi-static part determines the shape of the dynamic yield surface. The matrix material is idealized as a rigid-perfect plastic. Two major effects of point defects are proposed to lead to significant anomalies in mechanical properties: spontaneous stress and stiffness. Amongst the two, one is the direct effects caused by non-interacting point defects, and the other is the collective effects induced by interacting point defects. [491] The first group includes: (i) changes in the linear dimensions of a solid in response to a change in defect concentration and, (ii) stress induced by an inhomogeneous distribution of point defects, a so-called chemical stress. The second group includes: (i) defect order–disorder transitions accompanied by self-strain and, (ii) deviations from linear elastic behavior because of the dissociation/association of point defects. All of the above effects become important if the concentration of point defects is high (above $10^{21}$ cm$^{-3}$).

For metallic foams, there are also a number of models. According to the standard empirical model of foam plasticity originated by Gibson and Ashby,[492] the relationship between the yield strength ($\sigma$) and the relative density ($\rho_f/\rho_b$) of a foam material follows the scaling laws,

$$\sigma_f = \sigma_b \times \begin{cases} (\rho_f/\rho_b)^{3/2} & (\textit{Gibson \& Ashby}) \\ C_b(\rho_f/\rho_b)^{3/2} & (\text{Hodge, et al}) \end{cases} \tag{50}$$

where the subscripts *f* and *b* denote foam and bulk properties, respectively. The $\rho_f$ = ($V_{total}$-$V_{void}$)/$V_{total}$. Substituting the $\sigma_b = \sigma_0\left(1 + AK_j^{-0.5}\right)$ in the HPR for the $\sigma_b$ in the modified scaling relation with a given porosity, Hodge et al[479] derived information of size dependence of ligament strength in Au foams, which follow the HPR relation with $C_b$ = 0.3 as a factor of correction as shown in Figure 39b.

The porosity-depressed Young's modulus of Pd and Cu specimens [429] was ever expressed in many mechanical simulations:[493,494,495]

$$Y = Y_0 \times \begin{cases} 1 - 1.9p + 0.9p^2 & (\text{Wachtman \& MacKenzie}) \\ (1 - p/p_0)^n & (Rice) \end{cases}$$

(51)

with $p$ the porosity being defined as $p = V_{void}/V_{total}$. The density relates to the porosity as $\rho_p = 1\text{-}p$. The decrease in Young's modulus and flow stress with density at larger pore sizes is attributed to the existing pores that provide sites for initiating failures. Rice[495] proposed a normalized porosity dependence of the Young's modulus in the form, where $P_0$ is the value of $P$ for which the porosity dependent properties go to zero. The index n and $P_0$ are adjustable parameters. A linear fit with n = 1 to the measured data of various pores has been realized using this model. Figure 36 compares the model predictions of density dependence of mechanical strength of metallic foams.

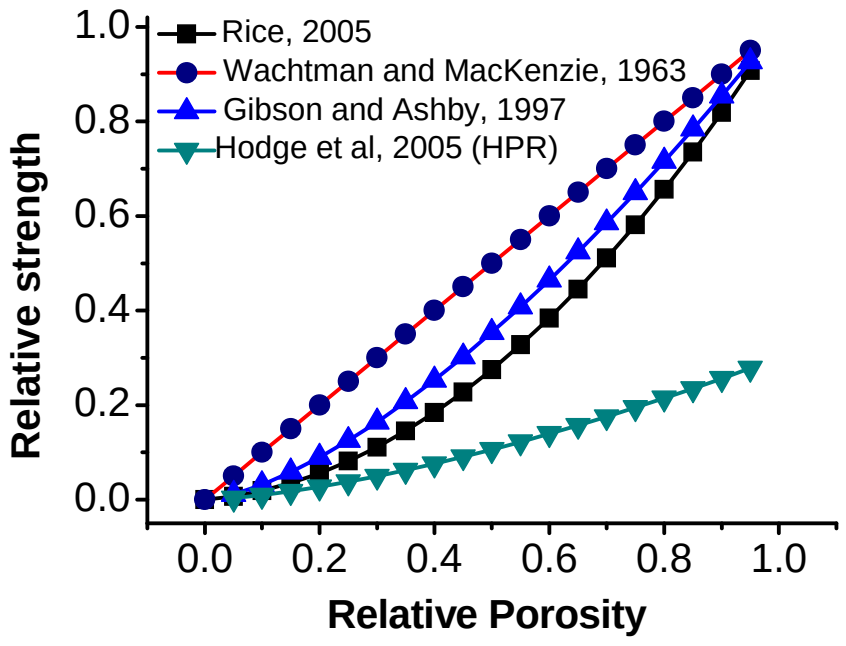

Figure 36 Model comparison of the density dependence of mechanical strength of metallic foams. Link

The theory of linear elasticity has been successfully used to describe the deformation behavior of multiphase materials of larger pore sizes showing that the strength of foam materials always decreases with increasing porosity. The mechanical properties of foams are related to the porosity or the relative density of the material as the dominating parameter and neither the effect of sizes of pores nor the effect of bond nature of the matrix are involved. In fact, nanomechanical measurements of nanoporous foams on a submicron scale [62,496] revealed close resemblance of the nanosized ligaments in foams showing a dramatic increase in strength with decreasing sample size.[478,481] Therefore, the effects of pore size, bond nature, and temperature must be considered in practice because the high portion of the under-coordinated atoms become predominant in controlling the physical behavior of nanocomposites.

Because the mechanical behavior of the surface is different from the bulk interior of the material, it would be an effective approach to consider the effective elastic constants of nanocomposites in terms of a three-phase structure, i.e., the bulk matrix, the voids, and the interfacial skins.[497] The effect of surface energy and surface stress has been considered in analyzing the deformation of microscale structures by many researchers. For examples, Sader[498] analyzed the effects of homogeneous surface stress on the deflection of cantilever plates used in AFM. Cammarata and Sieradzki[499] evaluated the effect of surface stress on the elastic modulus of thin films and superlattices. Yang and Li[500] discussed the effect of chemical stress on the bending of a single-layer and bilayer micromechanical beam. Sharma *et al.*[501] investigated the effect of surface energy and stress on the size-dependent elastic state of embedded inhomogeneities and found that surface elasticity can significantly alter the stress state of materials at the length scale of nanometers. They suggested that the dependence of the stress state on the size of the embedded inhomogeneities could be significant in determining the effective elastic modulus of composites. Yang[502] investigated the effect of surface energy on the effective elastic properties for elastic composite materials containing spherical nanocavities at dilute concentration and proposed solutions of the effective shear modulus and bulk modulus, which turned out to be a function of the surface energy and the size of the nanocavity. It is found that the dependence of the elastic response on size of the nanocavity in composite materials differs from the classic results obtained in the linear elasticity theory, suggesting the importance of the surface energy of the nanocavity in the deformation of nanoscale structures. In order to apply the scaling relation to nanoporous metal foams, the yield strength should be considered as a variable of the ligament or void size. An atomistic analysis of the effective elastic modulus of the porous systems from the perspective of bond breaking and the associated nearby strain and trapping is necessary.

## 8.2 BOLS consideration: defects induced strain and trapping

Strikingly, atomic vacancies, point defects, nanocavities, and nanoporous foams perform very much the same to nanostructures in the enhancement of stiffness (elastic strength) and toughness (plastic strength) and in the depression of thermal stability because of the large portion of the under-coordinated atoms. Nanovoids increase the portion of under-coordinated atoms at the negatively curved pore surfaces whereas the nanodots or rods provide such less-coordinated atoms in the positively curved particle surfaces. These two kinds of nanostructures differ one from another mainly in the exact portion of the lower-coordinated atoms and the slight difference in the coordination environment, given the same surface chemical finishing conditions.

Naturally, the T-BOLS correlation applies directly to the porous structures, which allows us to predict the elasticity, thermal stability, and the yield strength in plastic deformation by repeating the practice in previous sections with the derived surface-to-volume ratio, $r_{ij}(n, L_j, K_j)$, represented with eq (11) in section 2.3.2, and with the constraint of,

$$2(L_j+1)(n+1/2) \le K_j - 2 \tag{52}$$

with n being the number of spherical voids of $L_j$ radius aligned along the radius of a spherical dot of $K_j$ size, for instance. With the known expressions for the size, bond nature, and temperature dependence of the melting point, elasticity, extensibility, and the yield strength in the IHPR premise, as given in Table 11, we are ready to predict the performance of a porous system.

Table 11 Functional dependence of mechanical and thermal properties of porous structures on the atomic CN, bond nature, and temperature of testing.

| Q <br> $[\Delta Q(K_j)/Q(\infty) = \Delta_{qj} = \sum_{i\le 3}(\Delta q_i/q)]$ | $q_i(z, m, d(T), E_b(T))$ | Ref |
|---|---|---|
| Melting point ($T_{mi}$) | $\propto z_i \times E_i(0)$ | Sec 4.3.2 |
| Young's modulus ($Y_i$) | $\propto E_i(T) \times [d_i(T)]^{-3}$ | Sec 6.2.1 |
| Extensibility ($\beta_i$) | $\propto d_i(T) \times [\eta_i(T_{mi}-T)]^{-1}$ | Sec 6.2.1 |
| Yield strength ($P_i$, IHPR) | $\propto \{1+ f\times K_j^{-0.5} \times \exp[T_m(K_j)/T)]\} \times [d_i(T)]^{-3} \times (T_{mi}-T)$ | Sec 7.3 |

8.3 Results and discussion

8.3.1 Critical hollow-sphere size: total energy storage

Assuming a hollow sphere of K exterior radius with a shell of $C_{1L} + C_{1K}$ thick, or three contracted atomic layers, we have the total energy stored in the shell skin at 0 K in comparison to that stored in an ideal solid sphere of the same size without surface effect being considered,

$$\frac{\Delta E_{shell}}{E_{sphere}} = \frac{C_{1L}^{-(m+3)}\int_{K-(C_{1L}+C_{1K})}^{K-C_{1K}} 4\pi R^2 dR + C_{1K}^{-(m+3)}\int_{K-C_{1K}}^{K} 4\pi R^2 dR}{\int_0^K 4\pi R^2 dR} - 1$$
$$= C_{1L}^{-(m+3)}\left[(1-C_{1K}/K)^3 - (1-(C_{1L}+C_{1K})/K)^3\right] + C_{1K}^{-(m+3)}\left[1-(1-C_{1K}/K)^3\right] - 1 \quad (53)$$

Calculations were conducted based on the given $C_i$ and the curvature dependent $z_i$ values. From the results shown in Figure 37, we can find the critical size below which the energy stored in the shell of the hollow sphere is greater than that in the ideal bulk of the same volume without considering the surface and temperature effects. The estimation indicates that the critical size is bond nature dependent. The critical size is K = 7, 10, and 14 for m = 1, 3, and 5, respectively. Similarly, for hollow tubes, the corresponding critical K values are estimated as 5, 7, and 10. For the single walled hollow structure, the integration crosses only the diameter of the wall atom. For the elasticity, the shell is always higher than the bulk because the elasticity is proportional to the energy density. For plastic deformation, the hollow sphere could be tougher than the ideal bulk because of the long distance effect in the indentation measurement. On the other hand, the thermal stability of the hollow sphere is always lower than the solid sphere.

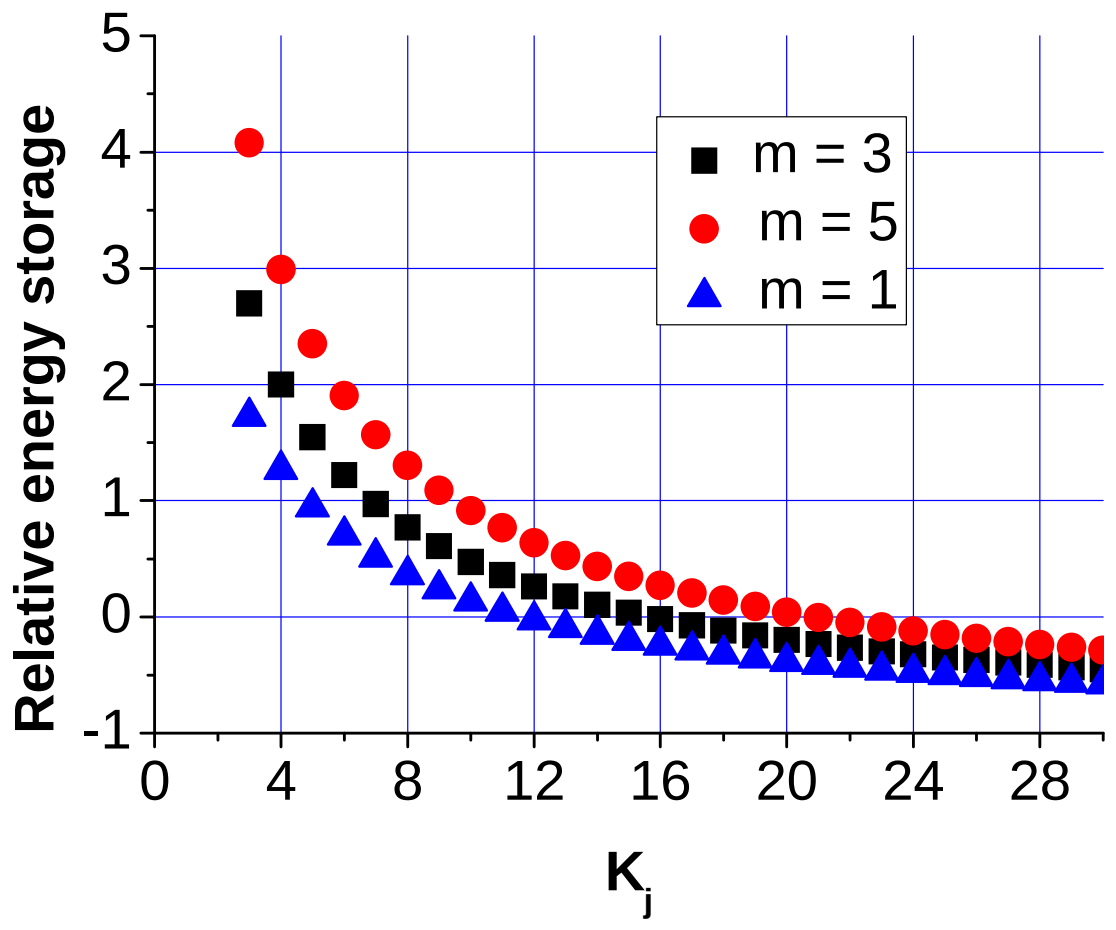

Figure 37 Bond nature dependence of the critical size below which the total energy stored in the shell of the hollow sphere of three atomic layers is greater than the energy stored in an ideal bulk of the same size. Link

8.3.2 Elasticity and thermal stability

With the derived $r_{ij}\left(n, L_j, K_j\right)$ relation and the given expressions for the q($z_i$, $d_i$(t), $E_i$(t)), one can readily predict the size, cavity density, and temperature dependence of the melting point, elasticity, and the flow stress of a system with large portion of under-coordinated atoms without involving hypothetic parameters.

Calculations of the elasticity and melting point were conducted by using a fixed value of sphere radius $K_j$ = 600 with different pore sizes and pore numbers. Correlation between pore size and pore number follows the constraint, eq (52), with a given $K_j$ value. Figure 38 shows the predicted porosity dependence of $T_m$ and Young's modulus of Au foams (m = 1) with different pore sizes. Generally, the $T_m$ drops when the porosity is increased; at the same porosity, the specimen with smaller pore size is less stable than the ones with larger pores. The Young's modulus increases with the porosity and the Young's modulus of the specimen with smaller pores increases faster. The predicted trends of thermal stability and strength agree well with the experiment observations for the size-dependent mechanical properties of nanoporous Au characterized by nanoindentation and column microcompression.[479] It is important to note that there exists porosity limit for the specimens with small pore sizes. This discussion applies to the small pores but the conclusion may be subject to further verification for specimen of larger pores that are expected to drop in strength with pore size depending on the surface-to-volume ratio. For $T_m$ consideration, the surface-to-volume ratio should not include the volume of pores as given in eq **(11)**; for Y consideration, the pore volume contributes to eq **(11)**. Unfortunately, experimental data on pore size and density dependence $T_m$ and Y are to be collected.

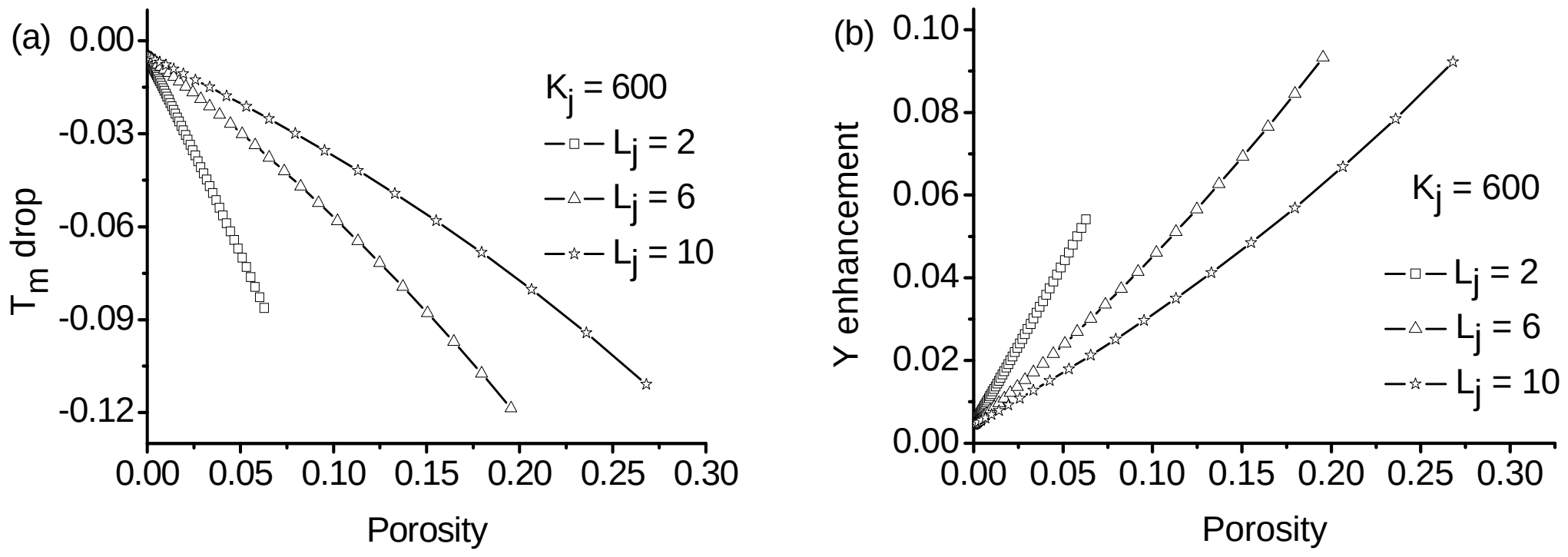

Figure 38 Prediction of the porosity dependence of (a) $T_m$ and Y of porous Au foams with different pore sizes of a $K_j = 600$ specimen. **Link**

8.3.3 Plasticity and IHPR

In dealing with the plastic deformation using IHPR, we may use the following relation to find the effective volume by excluding the pore volume in the specimen:

$$\frac{4\pi}{3}K'^3_j = \frac{4\pi}{3}\left[K^3_j - \frac{4\pi}{3}(n+1/2)^3 L^3_j\right]$$

$$x = K'^{-0.5}_j$$

**(54)**

Figure 39(a) shows the predicted IHPR as a function of $L_j$ for $10< K_j < 600$ specimens. Compared with the situation of single nanoparticle, the strongest size is significantly reduced for the foams. Figure 39(b) compares the predicted IHPR of Au with experimental results. The ligament size $x(K_j^{-1/2})$ is derived from Au foams with the modified scaling relation of eq (50). IHPR 2 and IHPR 1 are the IHPR with and without involving the intrinsic competition of energy density and atomic cohesive energy as discussed for the nanoparticles. The extremely high strength of ligaments smaller than 5 nm deviates from the expected IHPR is beyond the expectation of IHPR. One possibility for the extremely high strength of 5 nm ligands is the surface chemical conditions of the ligands because the higher chemical reactivity of small particles. A combination of the present IHPR with the scaling relation of (50) may describe the observed trends at larger porosities, and further investigation is in progress.

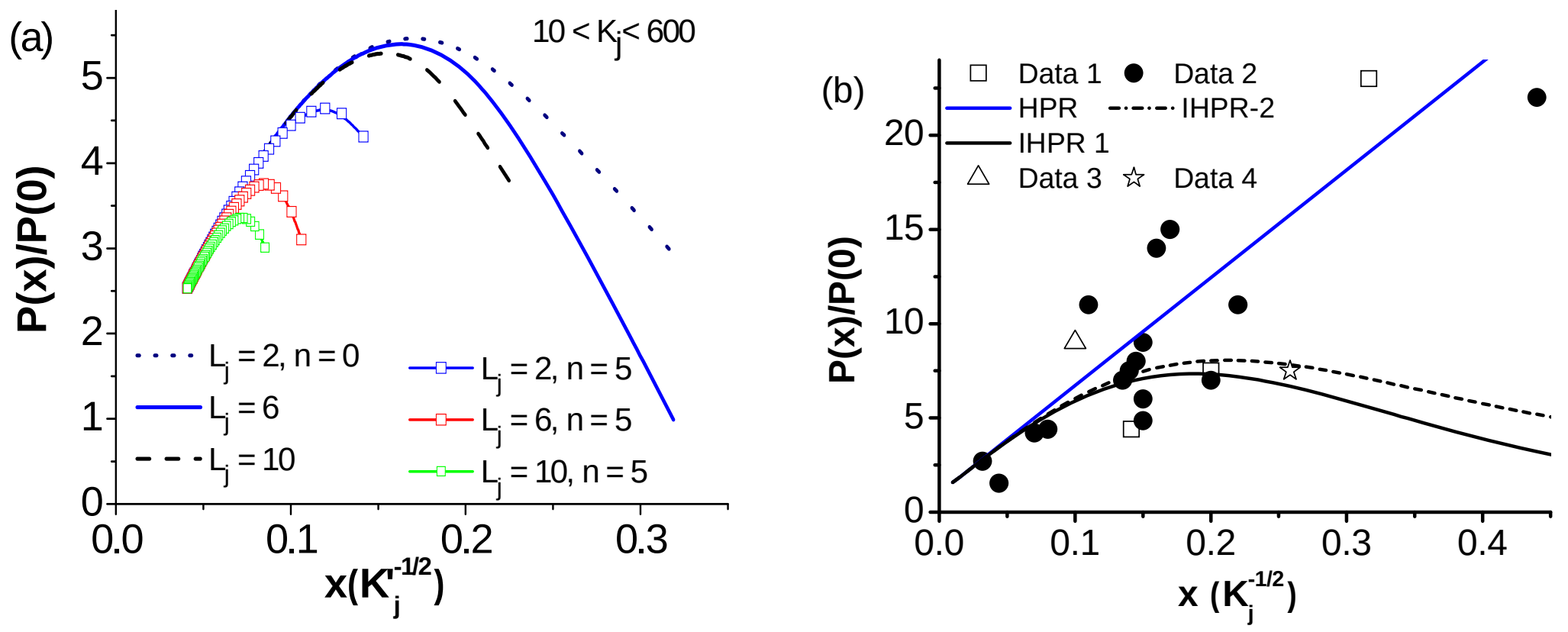

Figure 39 Prediction of (a) the IHPR for nanoporous Au sphere with $10 < K_j < 600$ and different pore sizes $L_j$ and pore numbers. (b) Comparison of the predicted IHPR of Au with measurement, Data 1 [477], Data 2 [475], Data 3 [478], and data 4 [481]. The ligament size $x(K_j^{-1/2})$ is derived from Au foams with the modified scaling relation of Ashby. Link

According to the currently developed understanding, the magnitude of $T_m - T$, or the ratio $T/T_m$, plays a key role in determining the relative strength. The $T_m$ of Al (933.5 K) is lower than that of Au (1337 K), which explains why the relative strength of Al foam to Al bulk is lower than that of Au.

Unfortunately, no immediate mechanism is apparent to explain the difference in vacancy-induced hardening between the IVb-nitride and the Vb-carbide compound. The excessive electron lone pair of the nitrogen atom and the shorter ionic radius of nitrogen (0.17 nm) compared to that of carbon (0.26 nm)[1] could be possible reasons for this difference. Increasing the number of N vacancies or reducing the number of the shorter ionic bonds of nitrides may lower the strength. We may leave this question open for further investigation.

## 8.4 Summary

The under-coordinated atoms in the negatively curved surfaces of atomic vacancies, point defects, nanocavities, and the synctatic foams are responsible for the strain hardening and thermal stability depression of the negatively curved systems, being the same by nature to those positively curved systems such as nanorods, nanograins and flat surfaces. Numerically, the negatively curved systems differ from the zero or the positively curved systems only by the fraction of the under-coordinated atoms and the coordination environment that determines the extent of the BOLS induced property change. Therefore, all derivatives and conclusions for the flat surface and the positively curved surface apply to the negatively curved ones without needing any modifications. It is emphasized that the nanopores play dual roles in mechanical strength. The shorter and stronger bonds near the pores act as pinning centers inhibiting motion of atomic dislocations because of the strain and the trapping; the pores provide sites for initiating structure failure under plastic deformation.

# IX Compounds and nanocomposites

## 9.1 Observations

### 9.1.1 Multilayers and nanocomposites

The multilayered structures such as nc-TiN/a-$Si_3N_4$, nc-TiN/a-$Si_3N_4$/ and nc-$TiSi_2$, nc-$(Ti_{1-x}Al_x)N$/a-$Si_3N_4$, nc-TiN/$TiB_2$, nc-TiN/BN, [503,504] TiN/CrN,,[505] CrN/AlN, TiN/AlN[506] and Cu/Au[507] manifest enhanced mechanical strength and thermal stability, because of the interfacial bond strengthening. The nanocrystalline or amorphous composites show hardness approaching that of diamond,[504] yet oxygen interface impurities cause a significant reduction of the interface strength.[508] Similarly, high density twins in pure Cu,[509, 510,511] and Al,[512] for example, leads to enhanced mechanical properties because of the pinning effect that prevents atoms from gliding dislocations. The presence of the grown-in twins in nanocrystalline-Al is found to enhance plastic deformation via twin-migration in which partial dislocations, emitted at the intersection of the twin boundary and the GB, travel through the entire grain. It has been found that generally the microhardness of multilayer nitride films increased with the number of layers, except if the two layers mutually dissolved. A microhardness of 78 GPa was achieved in a 180-layer TiN/NbN film.[513] Figure 40 shows the microhardness of TiN/ZrN, TiN/NbN, and TiN/CrN multilayer films as a function of the layer number for films with a similar total thickness of nearly 2 nm. For TiN/NbN and TiN/ZrN films with ~180 layers (with monolayer thickness of ~10 nm) the hardness is about 70–80 GPa, i.e. approximately that of diamond. The different behavior of TiN/CrN is connected with (Ti, Cr)N solution formation for more than 100 layers, which has been shown by XRD analysis. This observation indicates the significance of interlayer mixing to the mechanical strength that is proportional to the local energy density, and therefore, the presence of interface strain and trapping.

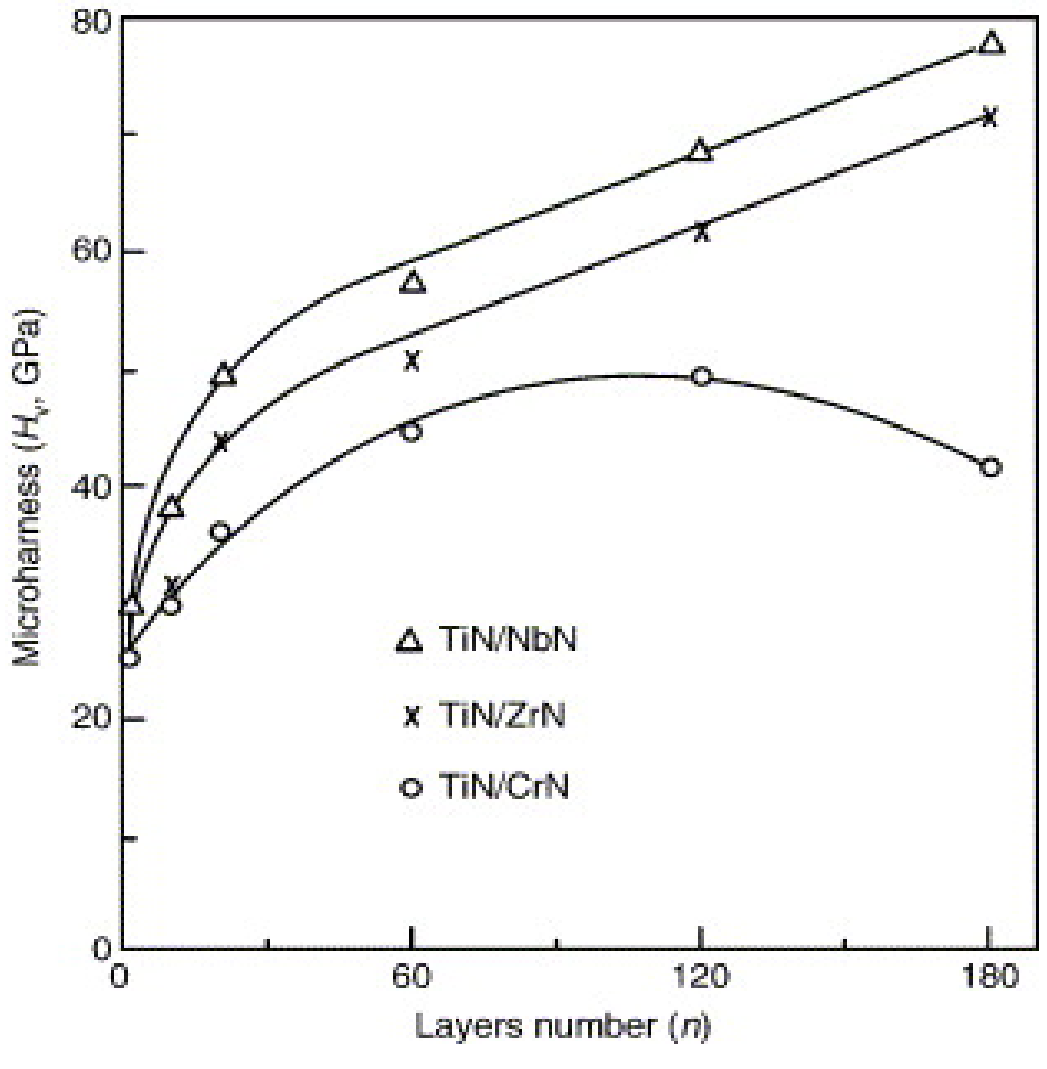

Figure 40 The layer number dependence of the microhardness of nitride films,[513] indicates the significance of interlayer mixing to the mechanical strength that is proportional to the local energy density, and therefore, the presence of interface strain and trapping. Link

Polymer nanocomposite foams filled with hard infillers have received increasingly attention in both scientific and industrial communities. The combination of functional nanoparticles such as SiC, CNT, clay, glass fibers, rubbers, and supercritical fluid foaming technology has a high potential to generate a new class of materials that are lightweight, high strength, and multifunctional.[514,515,516,517,518,519,520] For a 60% clay nanocomposite example, its elastic modulus increases up to 21.4 GPa, which is five times higher than that of the agarose matrix.[521] A small amount of well-dispersed nanoparticles in the polymer domain may serve as the nucleation sites to facilitate the bubble nucleation process. Moreover, the nano-scaled particles are suitable for micrometer-scaled reinforcement because of the large surface area for interface chemical bonding between the infiller and the matrix, thus achieving the macroscopic mechanical enhancement.

Haraguchi *et al.*[522] used gel formation in an aqueous medium to create a composite of hydrophobic poly (2-methoxyethyl acrylate) and hydrophilic hectorite clay. During the polymerization, the clay platelets are excluded from the polymer particles. Once dried, the clay shells comprised a three-dimensional network. A surprising feature of the composites was the ability to undergo huge elongations when being subject to a stress. After an initial irreversible necking deformation, subsequent applied large strains were shown to be reversible, with good shape recovery observed on release. The remarkable mechanical properties of nanocomposite coatings, such as superhardness, high elastic modulus, high elastic recovery, excellent resistance against cracking, low wear rate, and high thermal stability, are due to their unique structures and deformation mechanisms at the nanometer scale.

Theoretical progress in understanding the role of interface in the nanocomposite is still limited. Jiang et al[523] investigated the cohesive interfaces between the CNT fillers and the polymer epoxy that are not well bonded with the van der Waals force in terms of the tensile cohesive strength and the interface cohesive energy. The interaction is described by the area density of CNT and the volume density of polymer as well as the parameters in the van der Waals force. For a CNT in an infinite polymer, the shear cohesive stress vanishes, and the tensile cohesive stress dominates depending only on the opening displacement. For a CNT in a finite polymer matrix, the tensile cohesive stress remains unchanged, but the shear cohesive stress depends on both opening and sliding displacements, i.e., the tension/shear coupling. Analytical expressions of the cohesive behavior have been used to study the CNT-reinforced composites to give an improved understanding of the interfacial effect. However, challenges remain yet regarding the roles of cavity surfaces, interfaces, or the strong infiller particles in the reinforcement of nanocomposites.

A recent review by Lu and co-workers[524] provides the latest understanding of the origin of superhardness in nanocomposite coatings with models based on analyses and simulations at different levels from continuum to atomistic scales. Useful and timely overviews on the mechanical hardening of composite coatings are also available in Refs [[92, 525,526,527,528]].

### 9.1.2 Compounds and alloys

Blending different types of atoms in a solid could enhance the hardness of the solid preferably in an amorphous state, as so called high entropy materials.[529] A number of models for the mechanical strength have been developed based on the concept of ionicity to predict the hardness of several compounds. [530,531, 532,533] It is anticipated that covalent bonds take the responsibility for increasing the hardness. The hardness, or the activation energy required for plastic gliding, was related to the band gap $E_G$ which is proportional to the bond length in a $d^{-n}$ fashion with the power index n varying from 2.5 to 5.0. The typical relation proposed by Liu and Cohen is,[534]

$$B = N_c/4\times(19.71-2.20f_i)d^{-3.5}(Mbar)$$

$N_c$ is the nearest atomic neighbors. The parameter $f_i$ accounts for the reduction in the bulk modulus $B$ arising from increased charge transfer. The value of $f_i$ = 0, 1 and 2 for groups IV, III-V, and II-VI solids in the periodic table. For a tetrahedral system, $N_c$ = 4, otherwise, the $N_c$ is an average of atomic CN. For diamond, $f_i$ = 0, d = 1.54 Å, and hence $B$ = 4.35 Mbar, compared with an average experimental value of 4.43 Mbar. This relationship was applied to BN and β-$Si_3N_4$ with corresponding prediction of $B$ = 3.69 and 2.68 Mbar. These predictions stimulated tremendous interest of experimental search for a superhard carbon-nitride phase worldwide,[535,536,537] as the diameter of an N atom is 0.140 ~ 0.148 nm shorter than the C-C bond in a diamond. In spite of the difference in the power index, -2.5, -3.5, -5, and –(m+3) in the current approach, all the expressions indicate that shorter and denser chemical bonds as well as smaller ionicity should favor hardness.

In addition to the high-entropy superhard materials, intermetallic superalloys, such as NiAl, FeAl, and TiAl systems, form another kind of important materials of tougher than ever the individual constituent elemental solid. In these alloys, raising the ductility is an important goal when metals are added into the alloy.[538] The significant characteristics of the superalloys is the mixture of at least two kinds of elements with different bulk melting points. The mixture as such may make the alloy both ductile and thermally stable. Mechanism for the superalloy is of importance, and further exploration is under consideration.

## 9.2 BOLS considerations: interface effects

Compared with the surface strengthening of nanosolids and nanocavities, the mechanical strengthening of twin GBs, multilayers, and blending composites arises from the coordination environment of the interfacial atomic bonding and the alteration of bond nature of the interatomic bonding upon the compound or alloy formation at the interface. The bond contracts at the interface because of the bond order (length and angle) distortion that also stores energy locally. Therefore, bond contraction and bond nature alteration occurs at the interface, generate local strain and energy trapping, which is suggested here to be possible mechanism for the enhanced mechanical strength of these kinds of materials. At a given temperature, the local energy density dictate the mechanical strength in the mixed interface region.

### 9.2.1 Interface trapping, charge and energy desification

The currently available database allows us to propose models for the interface trapping, as illustrated in Figure 41 for the compact and dissociated interface, agreeing with the recent finding of Popovic and Satpathy[539] in calculating oxide superlattices and microstructures. They found the essentiality to introduce a wedge-shaped potential well for the monolayer structure sandwiched between the $SrTiO_3$ and $LaTiO_3$ superlattices. The potential well of trapping is originating from the Coulomb potential of a two-dimensional charged La sheet, which in turn confines the electrons in the Airy-function-localized states. For the compct interaface, there is a monotrap at the interface. Localization and densification of charge end energy occurs to the interface. The energy levels of atoms in the interface region also shit positively unless interface dipoles are formed. For the dissociated interface, double potential traps will present associated with one barrier between. The Z is the coordinate directing into the bulk. Cutting off at Z = 0, one side of the double well converges to the case of free surface or atomic vacancy defect- A bbarier followed by an immeadiate trap in the surface skin.

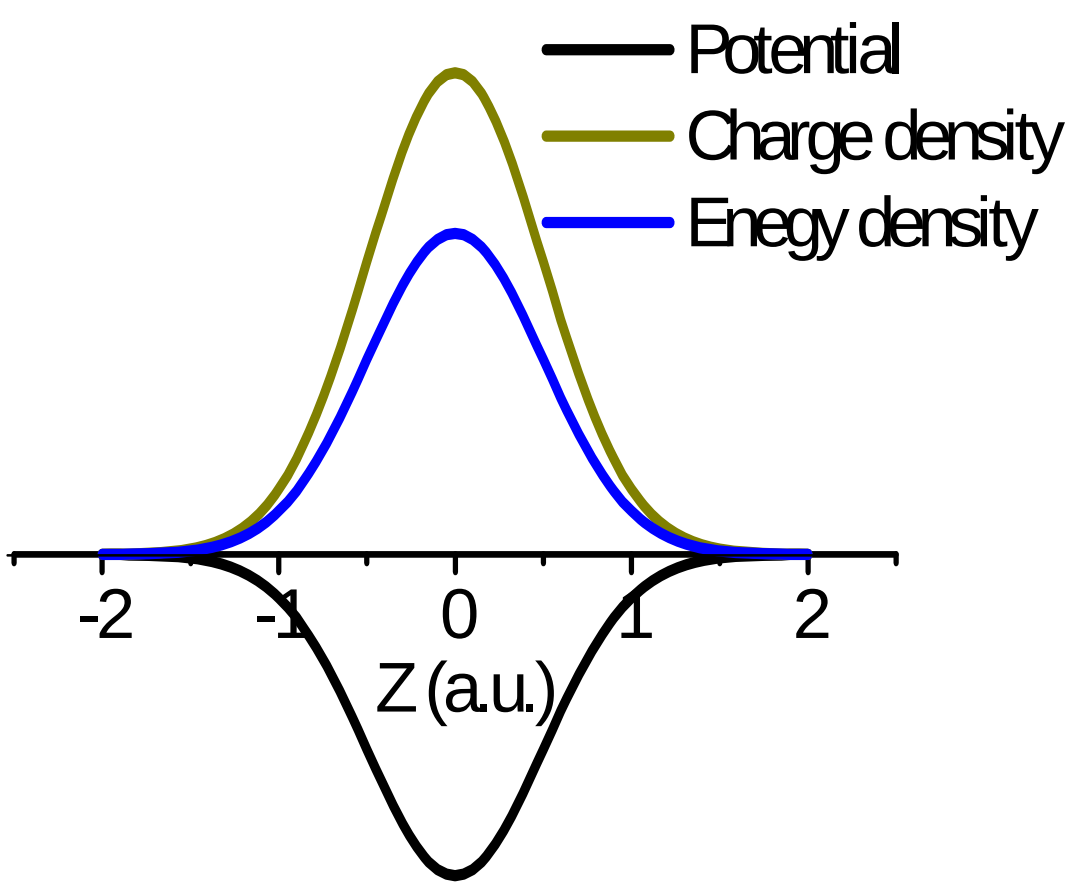

(a)

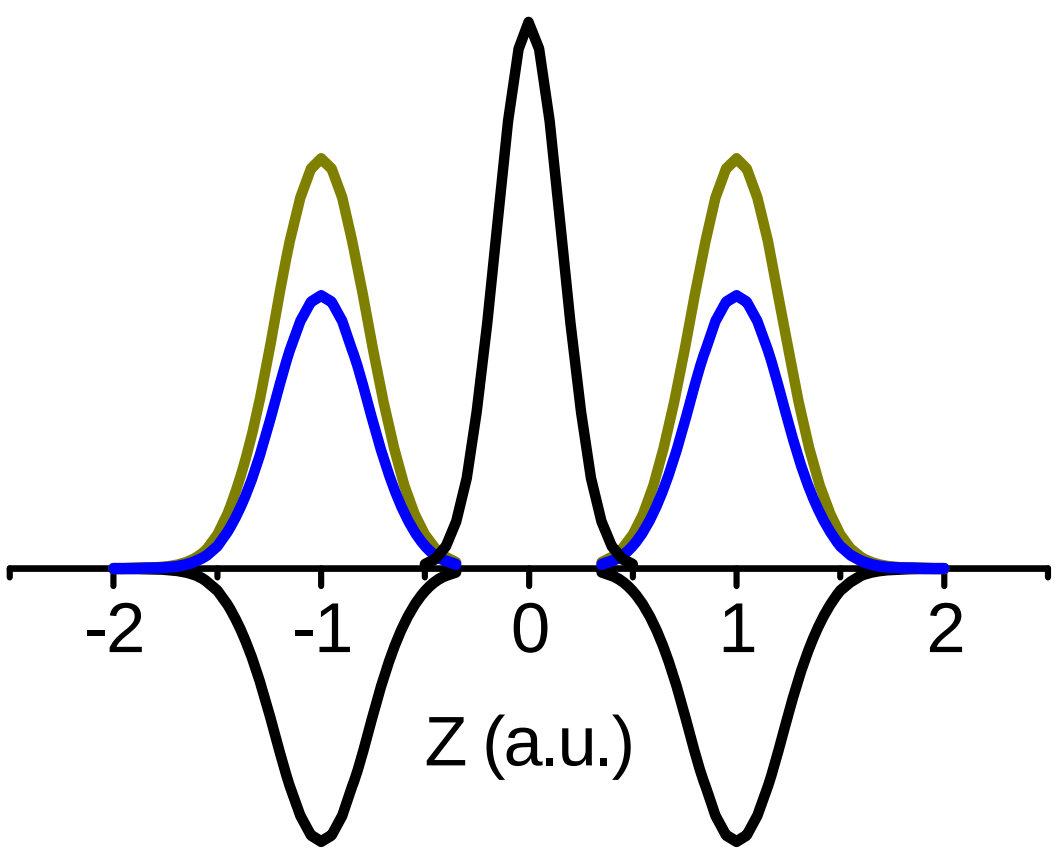

(b)

Figure 41 Models for the atomic coordination imperfection induced surface and interface trapping and the associated charge and energy density gain in the (a) compacted

and the (b) dissociated interfaces. One addition barrier is located at the dissociated interface. Z is the coordination directing into one medium. For defect edge or a free surface, Z > 0 in (b) applies.

### 9.2.2 Evidence for interface bond contraction

Evidence has shown that bond contraction takes place at sites surrounding impurities and at the interfaces. For instances, a substitutional As dopant impurity has induced an 8% bond contraction around the impurity (acceptor dopant As) at the Te sublattice in CdTe, as observed using EXAFS (extended X-ray absorption fine structure) and XANES (X-ray absorption near edge spectroscopy).[540] A fluorescence x-ray-absorption fine structure and x-ray diffraction has recently revealed that a 3% Ti-N bond contraction occurs at the $TiN/Si_3N_4$ interface,[541] being responsible for the hardening of the crystalline/amorphous $TiN/Si_3N_4$ multilayer films. EXAFS investigation[542] has revealed size-dependent inter-atomic distance contraction in thiol-capped gold nanoparticles. A slight nearest-neighbor distance reduction was observed as a function of particle diameter (2-4 nm range) but, for all samples, it was less than 1%. This value is smaller compared with the larger effect expected and found in other systems, especially for the smallest particles (2 nm). The Au-Au bond of Au nanocrystals embedded in $SiO_2$ was found to contract by 0.04~0.03 nm.[543] Raman scattering spectroscopy, XRD, EXAFS, and XANES revealed clearly that the Sb-In bond in the first shell of the SbIn embedded in a $SiO_2$ matrix contracted slightly by about 0.002 nm compared with that of the bulk SbIn,[544] which partly weakens the phonon confinement effect. XANES and EXAFS analysis[545] also revealed that Mn ion implantation on a heated Si substrate to form clusters with 6–8 atoms located in the first coordination sphere in three subshells. The first subshell has one atom at a distance of 2.31 Å, the second subshell has three at 2.40 Å, the third subshell has three atoms at 2.54 Å, and finally the fourth subshell contains six Mn atoms at a distance of 2.80 Å. The finding of dopant-induced bond contraction and the interface bond contraction could provide an atomic scale understanding of the bond in a junction interface that has been a puzzle for decades. According to the recent review of Veprek and Veprek-Heijman [546] on the a-XN interface monolayer strengthening of nc-MN/a-XN nanocomposites (M = Ti, W, V, (TiAl), X= Si, B), the enhanced strength of the one-monolayer interface is a general nature that can be understood in terms of an extension of the BOLS correlation to the effect of monolayer interface mixing. The bond contraction and bond nature evolution upon interface formation do cause local strain and the associated trapping and pinning, which should be responsible for the monolayer strengthening as observed.

Interestingly, recent theoretical calculations, confirmed by electron microscopy measurement,[547] revealed that homo-junction dislocations in aluminum have either compact or dissociated core interlayers. The calculated minimum stress ($\sigma_P$) required for moving an edge dislocation is approximately 20 times higher for the compact dislocations than for the equivalent dissociated dislocations. As compared with the theoretical tensile strength in the direction [001] or [111] of an Al single crystal, the Al-GB is still strong due to the interface reconstruction, which indicates the special strength of the reconstructed and less-coordinated GB bonds, as suggested by Lu et al [23].

Encouragingly, a grazing incidence X-ray reflectivity (GIXR) study revealed that the interface of $SiO_2$/Si has a higher electron density than the Si and $SiO_2$ constituents in separate forms. The higher density persists disregarding the thickness of the layered films.[548] The electron density in the Pt/$BaTiO_3$ interface dead layer is about 10% higher than the bulk because of the lattice strain.[549] A dead layer of 3 nm thick has also been found in the $Ba_{0.7}Sr_{0.3}TiO_3$/$SrRuO_3$ interface.[550] The dead layer formation could lower the dielectric permittivity and hence the microcapacitance.[551] These observations provide robust evidence for the expectation for compact interfaces to have deeper trapping potential wells and higher bond strength, resulting in localization and densification of charge and energy in the surface/interface skins, which makes the interface region an insulator dead layer because of the potential well depression.

### 9.2.3 Bond nature alteration

It is understandable that an atom performs differently at a free surface compared to an atom at the interface. Although the coordination ratio at the interfaces undergoes little change ($z_{ib} \sim 1$), formation of an interfacial compound or alloy alters the nature of the interatomic bond which should be stronger. Energy storage due to bond geometry distortion also contributes to the bond energy. Investigations[203] revealed that overheating occurs in substances covered by relatively higher $T_m$ substances, or stronger binding systems, as the $T_m$ relates directly to the atomic cohesive energy.

At the mixed interface, the $z_i$ may not change substantially, so we can introduce the interfacial bond energy as $E_{int} = \lambda E_b$ and the interfacial atomic cohesive energy as $E_{C,int} = \lambda z E_b$, and then all the equations for the surface effect are transferable to the interface properties. A numerical fit of the size dependence of overheating for In/Al,[552] Ag/Ni,[553] and Pb/Al and Pb/Zn[554] core-shell nanostructure has led to a $\lambda$ value of 1.8, indicating that an interfacial bond is 80% stronger than a bond in the bulk of the core material.[42] If we take the bond contraction to be 0.90 ~ 0.92 as determined from the As and Bi doped CdTe compound[540] into consideration, it is readily found that the m value is around 5.5 ~7.0. The high m value indicates that the bond nature indeed evolves when a compound is formed. The m value increases from 1 for the initially metallic to 4 or higher for the interfacial compound, which indicates the covalent interfacial bond nature. The electro-affinity and the interfacial DOS are expected to shift positively by 80% of the corresponding bulk $\Delta E_C(\infty)$ value. Therefore, the deformed and shortened interfacial bond is much stronger, meaning that electrons at an interface are deeply trapped; giving the observable interfacial local DOS. From this perspective, twins of nanograins[511] and the interfaces of multilayered structures[555] should be stronger and thermally more stable because of the interfacial strain and trapping. Experiments also show that the interfacial fcc-SiN can strengthen the TiN/SiN heterostructures only when its thickness is about 1–2 ML.[556] A recent work[557] on the size-dependent melting points of the silica embedded crystalline germanium nanowires with mean diameters ranging from 2.2 to 8.5 nm revealed strong interactions at the interface between the nanowire and the matrix. The bond enthalpy between Ge–O and Ge–Ge is significantly larger, i.e. 385 kJ $mol^{-1}$ compared to 188 kJ $mol^{-1}$, leading to the observed superheating by some 60 K compared to the bulk Ge (1200 K). This finding also evidences that the interface is stronger.

In order to obtain a compound with large bulk modulus, one must find such a covalent compound that has both a shorter bond and smaller ionicity, and high compactness in internal atomic arrangement. Thus, the atomic CN-imperfection induced bond contraction should contribute directly to the hardness at the surface or sites surrounding defects. Therefore, a nanometer sized diamond is expected to be 100% ($0.88^{-5.56}$ - 1) harder than the bulk natural diamond but the hardness should be subject to the inverse Hall-Petch relationship.[370]

The detrimental effect of oxygen in the interface may arise from the introduction of non-bonding lone pairs and the lone pair induced dipoles, according to the understanding of O-induced stress.[181] It is expected that the excessive lone pair of nitrogen should weaken the local non-bond strength and hence the strength of the entire sample. However, the measured strength depends on the sum of binding energy per unit volume. If the energy density of nitride in the interface region is higher than that of a diamond, the nitride interface is then stronger than the diamond. In contrast, oxygen involvement doubles the number of lone pairs and lowers the local binding energy density. This understanding may explain why the nitride interface is stronger and why oxygen addition could lower the strength. It is interesting to note that nitriding multilayers show the high strength compatible to diamond while the strength of carbide multilayers are seldom reported. The current understanding suggests that carbon induced compressive stress may prevent interlayer mixing, as being in the case of diamond/Ti poor adhesion.[234] Neutralizing the interfacial stress by introducing the graded metallic carbon-nitride buffer layers such as TiCN could have overcome the difficulty in nitride/carbide multilayer formation.

## 9.4 Summary

The cohesive interface, or the interfacial mixing, can also cause spontaneous bond strain associated with excessive energy storage. The stored energy and the associated effect of trapping serve as pinning in the interface region to inhibit atomic dislocations and hence interfacial hardening. On the other hand, alloy or compound formation at the interface alters the originally metallic bonds of the components to be ionic or covalent and hence the bond nature indicator, m, evolves from the original value for each constituent element to higher values. The bond nature alteration provides an additional mechanism of interfacial strengthening. As the mesoscopic mechanics depends functionally on atomic CN, bond length, and bond energy, and their temperature dependence, the T-BOLS correlation can readily be extended to the interface by introducing a parameter, λ, to correlate the interfacial bond energy to the bulk standard, $E_{int}(T) = \lambda E_b(T)$, and then the T-BOLS correlation mechanism applies to the cohesive interface, if the atomic CN change insignificantly and the bond strain is not reliably determined.

## X Concluding remarks

### 10.1 Attainment

A set of analytical expressions has been developed from the perspective of local band average for the elasticity, extensibility, and mechanical strength of low dimensional systems in terms of bond order, bond length, bond strength, and their coordination environment, temperature, and stress field dependence. The effect of a broken bond on the identities of the remaining bonds between the under-coordinated atoms is shown to dominate the mechanical performance and thermal stability of the mesoscopic systems. The competition between the residual atomic cohesive-energy of the less-coordinated atoms and the energy-density-gain per local unit volume dictates the effect intrinsically and the competition between the activation and the prohibition of atomic glide dislocations determine extrinsically the plastic deformation and yield strength of the mesoscopic systems. The described approaches connect the macroscopic properties to the atomistic factors by developing the functional dependence of the measurable quantities on the bonding identities and the response of the bonding identities to external stimulus, which may provide complementary to the classical theories of continuum medium mechanics and statistic thermodynamics that have demonstrated the limitation to mesoscopic systems.

The following facts have been taken into consideration as physical constraints in developing the atomistic solutions:

(i) For a given specimen, the nature and the total number of bonds do not change under external stimulus unless phase transition occurs, whereas the length and strength of all the bonds involved will response to the stimulus. This fact enables one to implementing the LBA approach to focus on the behavior of the representative bonds and their average for the mechanical behavior of the entire specimen.

(ii) The mechanical properties of a substance are proportional to the sum of binding energy per unit volume. The broken bond triggers the BOLS correlation and the temperature rise causes volume expansion and atomic vibration that weakens the bond energy through the increase of internal energy in Debye approximation. Therefore, the joint effect of bond breaking and the associated local strain and trapping as well as the bond vibrating under thermal stimulus represented by the T-BOLS correlation is of key importance to the mesoscopic mechanics.

(iii) The molten phase is extremely soft and highly extensible, following Born's criterion, yet the elastic modulus remains non-zero at $T_m$ or above because of the detectable sound velocity in liquid and gaseous phases.

The major progress made by using the LBA approach and the T-BOLS mechanism can be summarized as follows:

(i) Concepts of the energy-density-gain per unit volume in the surface skin and the residual cohesive-energy for the under-coordinated surface atoms are proven essential and more effective to describe the behavior of atoms and processes occurring at a surface. The deepened potential well of trapping at sites of the under-coordinated atoms provides a mechanism for the pinning effect as observed in mechanical deformation tests because of the locally densified and lowered energy states though the broken bonds provide sites initiating structure failure.

(ii) Consistent insight has been obtained into the atomistic mechanism of surface tension and surface stress and their adsorption and temperature dependence from the perspective of charge redistribution and polarization upon adsorbate bond making. A solid surface is harder at temperatures far below $T_m$ but the surface melts more easily compared to the bulk interior. The elastic sheet of a liquid surface tends to solidify more easily than the liquid interior, which may provide a mechanism for liquid drop and bubble formation.

(iii) It has been clarified that the observed HPR-IHPR transition arises from the intrinsic competition between the atomic residual cohesive-energy of the less-coordinated atoms and energy-density-gain near the grain boundaries and the extrinsic competition between the activation and prohibition of atomic dislocations. The under-coordinated atoms in the surface skin dominate such transition yet atoms in the grain interior retain their bulk nature. For thin film samples, the hardness shows no IHPR and the hardness and elastic modulus correlate linearly. Therefore, the HPR and IHPR effects in nanograins are dominated by the extrinsic competition between dislocation activation and dislocation inhibiting.

(iv) It is understood that the mechanical strengthening and thermal weakening induced by vacancies, cavities, and pores arise from the increased portion of the under-coordinated atoms at the negatively curved surfaces, which is naturally the same as that of the flat surfaces or the positively curved surfaces of nanograins. It is expected that nanoporous foams should be stiffer with smaller pore sizes and lower porosities whereas it would be tougher with higher porosities and larger pores, being similar to the IHPR effect.

(v) Interfacial bond contraction and the associated bond strengthening, and the bond nature alteration upon alloy and compound formation at the junction interfaces are responsible for the hardening and overheating of twin gains, interfaces, and nanocomposites.

(vi) The Debye temperature, $\theta_D$, has a square root dependence on $(T_m\text{-}T)^{1/2}$, rather than the linear or square root dependence on $T_m$. The currently derived solution may provide a complementary one to the T-independent form of $\theta_D$ given by Lindermann. The specific heat capacity generally decreases when the solid size is reduced. The reduction of the specific heat capacity is more pronounced for larger m values at lower temperatures.

Reproduction of observations reveals the following:

(i) Surface relaxation and reconstruction of an elemental solid arises from the effect of bond breaking, which originates the surface energetics. The surface stress results from the surface strain rather than result in the strain, as one often believes. The surface and size induced hardening arises from bond breaking and the associated nearby strain and trapping and the thermally-induced mechanical softening results from bond expanding and vibrating upon being heated.

(ii) The adsorbate induced surface stress arises from charge repopulation and polarization upon adsorbate bond making, which is very complicated and varies from situation to situation. Knowing the bonding kinetics and dynamics is crucial to understanding the adsorbate induced surface stress and the process of reconstruction.

(iii) The equilibrium bond length, strength, thermal stability, and the strain limit of monatomic chains have been quantified without needing involvement of charge or atomic mediation. Without external stimuli, the metallic bond in a MC contracts by ~30%, associated with ~43% magnitude rise of the bond energy compared with the bulk standard cases. A metallic MC melts at 0.24 fold of the bulk $T_m$. The strain limit of a bond in a metallic MC under tension does not vary apparently with mechanical stress or strain rate but apparently with temperature difference $T_{mi}$ - T. It is anticipated that extendable MCs of other elements could be made at an appropriate temperature of operation.

(iv) The developed approach has allowed us to determine the actual values of the Young's modulus and the wall (C-C bond) thickness by solving a set of equations, to advance a consistent understanding of the mechanical strength and the chemical and thermal stability of CNTs. The C-C bond in the SWCNT contracts by ~18.5% with an energy rise by ~68%. The effective thickness of the C-C bond is ~0.142 nm, which is the diameter of an isolated C atom. The melting point of the tube-wall is slightly (~ 12 K) higher than that of the open edge of the tube-end. The literature-documented values of the tip-end $T_m$ and the product of Yt essentially represent the true situations of a SWCNT in which the Young's modulus is 2.5 times and the $T_m$ is 0.42 times that of bulk graphite. Predictions of the wall thickness dependence agree well with the insofar-observed trends in $T_m$ suppression and Y enhancement of the multi-walled hollow tubes and nanowires. It is anticipated that at temperature far below the surface $T_m$, a nansolid should be fragile with lower extensibility, whereas when T approaches the surface $T_m$ or higher, the nanosolid should be ductile.

(v) The Young's modulus of a nanosolid may be depressed, increased, or remain unchanged when the solid size is decreased, depending on the nature of the bond involved, temperature of operation, surface passivation, and experimental techniques. This understanding clarifies why the Y values for some materials are elevated and why those of others are not upon size reduction. It is therefore not surprising to observe the elastic modulus change in different trends of different materials measured under different conditions.

(vi) Reproduction of the temperature dependence of surface tension and Young's modulus has led to quantitative information regarding the mean bond energy of a specimen in the bulk at 0K though the accuracy is subject to the involvement of artifacts in measurement.

(vii) For a nanograined nanowire, the bond unfolding and atomic sliding dislocations of the under-coordinated atoms at grain boundaries are the dominant mechanism for the super plasticity, as the detectable bond strain at a temperature close to the melting point is limited to within 140%. On the other hand, the self-heating during bond unfolding and breaking should raise the actual temperature of the small samples. A metallic nanosolid at $T << T_m$ is in quasi-solid state as the critical temperature for the solid-to-quasi-solid transition is much lower than the temperature of melting.

(viii) The under-coordinated atoms in the negatively curved surfaces of atomic vacancies, point defects, nanocavities, and the syntactic foams are responsible for the strain hardening and thermal instability of the negatively curved systems, being the same by nature to the atoms at the positively curved and flat surfaces. It is emphasized that the nanopores play dual roles in mechanical strength. The shorter and stronger bonds near the pores act as pinning centers inhibiting motion of atomic dislocations because of the strain and the trapping; the pores provide sites for initiating structure failure under plastic deformation.

(ix) The IHPR originates intrinsically from the competition between the bond order loss and the associated bond strength gain of atoms in the GBs. The artifacts in plastic deformation play dominant roles in observations, which can be described with introduction of activation energy for atomic dislocation. As the solid size shrinks, a transition from dominance of energy density gain to dominance of cohesive energy loss occurs at the IHPR critical size because of the increased fraction of lower coordinated atoms. During the transition, both bond order loss and bond strength gain contribute competitively.

(x) The IHPR critical size is universally predictable, which can be calibrated with a limited number of measurements for a specific system. The critical size is dominated intrinsically by bond nature and the $T/T_m$ ratio and extrinsically by experimental conditions or other factors such as size distribution and impurities. The IHPR at larger solid size converges to the normal

HPR that maintains its conventional meaning of the accumulation of atomic dislocations that resist further atomic displacements in plastic deformation. The slope in the traditional HPR is proportional to $\exp(T_m/T)$, which represents the relationship between the hardness and the activation energy for atomic dislocations. Understanding these events provides a consistent insight into the factors dominating the critical size for HPR transition and the critical temperature for solid-quasi-solid and quasisolid-liquid transitions at a given temperature.

(xi) Finally, achievement evidence that the proposed LBA approach and the T-BOLS correlation could provide complementary to the classical and quantum approaches for the size and temperature dependence of low-dimensional solid mechanics.

## 10.2 Limitations

One may wonder that there is often competition between various factors for a specific phenomenon to occur. However, the broken bond affects almost all the intrinsic aspects of concern, and therefore, the atomic CN imperfection should dominate the performance of a nanosolid through the competition factors. In the nanoindentation test, errors may arise because of the shapes and sizes of the tips, such as in the cases described in Ref 33. In practice, the stress-strain profiles of a nanosolid are not symmetrical when comparing the situation under tension to the situation under compression, and the flow stress is dependent of the strain rate, loading mode and time, and materials compactness, as well as particle size distribution. Fortunately, the effect of tip shape and loading mode never affects the origin and the hardness peaking at the surface in detection. By taking the relative change of the measured quantity into account, the present approach of seeking for the change relative to the bulk values can minimize contributions from the artefacts. On the other hand, the extrinsic factors could be modelled by changing the prefactor in the IHPR modelling procedure. The relative change of intensity, the peak position, and the trend of change could reflect the intrinsic physical characteristics.

Furthermore, the thermal energy released from bond breaking and bond unfolding should raise the actual temperature of the system. The fluctuation in grain size distribution and surface passivation also affects the actual melting point of the individual atoms at boundaries of grains of different sizes. These factors make the measurements deviate from the predicted results. Nevertheless, one could not expect to cover fluctuations due to mechanical (strain rate, stress direction, loading mode and time, etc.), thermal (self-heating during process), crystal structure orientation, impurity density, or grain-size distribution effects in a theoretical model, as these fluctuations are extrinsic, ad-hoc, and hardly controllable. [558,559]

In dealing with the effect of temperature, we used the Debye specific heat for constant volume. In fact, most of the measurements were conducted under constant pressure. The specific heat for constant pressure should be preferable. However, for solid state, the ration of $(C_p-C_v)/C_p$ is 3% or less.[103] Therefore, assumer the specific heat follows the Debye approximation is acceptable.

It is known that the internal stress signature of nano and microstructures plays a critical role in determining the appropriate rate limiting process. When dislocations mediate the plasticity, the defect structure (over many length scales) through the internal stress, plays a critical role in determining the final macroscopic plasticity. The strength of a material is therefore a function of many temporal and spatial length scales. Plastic deformation is dominantly related to the mechanism of the atomic movements of long scale for strain accumulation, accompanied with the atomic interaction among many atoms. On the other hand, in non-crystalline regions, the local stress at an atom site arises from all the bonds that the atom is involved in. However, the average of local bonds over the measured size could represent the measurement collecting statistic information from the given volume. The disordered structures or system with large amount of defects will results in the deviation of the derived bond energy from the true value in ideal case.

For compounds, different kinds of bonds are involved and each kind of bond is suitably described using one kind of interatomic potential, the description using a representative bond herewith seems inappropriate. However, the average of the local bond is substantially the same to the DFT or MD calculations solving the Schrödinger equation or the motion of equations with the average interatomic potential as a key factor. The long-range atomic dislocation and the effect of pressure induced strengthening would be focus of future study. Nevertheless, we should focus more on the nature and trend of the unusual behaviour in mechanics, as accurate detection of the absolute values remains problematic.[165] As the LBA approach and the T-BOLS correlation deals with only the joint effects of temperature and bond order loss, none of the particularities of the elements, crystal or phase structures or the form of pair potential is involved. What we need to consider are the nature of the bond and the equilibrium atomic distance with and without external stimulus, which could be one of the advantages of the reported approach. It is emphasized that it is impractical for one theory model to cover all the factors simultaneously in particular those extrinsic, kinetic, and random that contribute little to the nature.

## 10.3 Prospects

Further extension of the LBA approach and the T-BOLS premise should open up new ways of thinking about the mesoscopic systems from the perspective of a broken bond that seems not to contribute directly to the physical properties of the mesoscopic systems. Consistent understanding and consistency in the numerical match give evidence of the validity of the approaches that may represent the true situation of the less-coordinated systems. More attention is needed to be paid to the following continuing challenges:

(i) Miniaturization of dimensionality not only allows us to tune the physical properties of a solid but also provides us with opportunity to elucidate information including the energy levels of an isolated atom, the vibration frequency of an isolated dimer, and the specific heat per bond, as well as discriminating the contribution of bond order loss from the effect of chemical passivation to the detectable quantities. Further exploitation in these areas may provide profound information that is beyond the scope of conventional approaches.

(ii) Extension of the LBA and T-BOLS correlation to domains such as pressure and external electric field could lead to new knowledge that would be even more fascinating and useful. For instance, the bending strength of PZT-841 ceramics was measured to decrease by 25% when the temperature was increased from 300 K to the Curie point $T_c$ (543K) with a valley at temperatures around 498 K. A positive or negative electric field larger than 3 kV/cm reduces the bending strength of the specimen significantly,[560] indicating the joint effect of Joule heat and electric field. The mesoscopic systems show unusual performance under an external stress field such as the critical temperature for phase transition, spectroscopic features of lattice vibration and photoluminescence.

(iii) Transport dynamics in thermal conductivity and electric conductivity play an important role in the performance of nanostructured devices. Introduction of the deepened potential well of trapping near the defects would be a topic that leads to new knowledge and phenomenon. In addition to the known mechanism of surface scattering, the surface/interface trapping may play a dominant role in determining the intrinsic transport dynamics of nanometer-sized electronic and photonic devices. Employing the BOLS-induced barrier and trap in the crystal potential for a nanosolid and for an assembly of nanosolids could improve the understanding on the kinetic and dynamic performance of a nanosolid under external stimuli.

(iv) The effect of bond order loss may correlate the nanostructures and amorphous states in their physical behavior. The bond order loss occurs orderly at the skins of nanostructures yet it presents randomly in the bulk amorphous. Exploration of the similarities and differences between the nanophous state and the amorphous state would be interesting.

(v) The new degree of the freedom of size, and its combination with temperature, pressure, chemical, and other domains could amplify the parameter space for functional materials design. For example, a comparative study of thermo mechanical properties of nano-polycrystalline nickel (nano-Ni) and micrometer-polycrystalline nickel (micron-Ni) by in situ high pressure-temperature (P - T) diffraction experiments[561] revealed that the yield strength of 2.35 GPa for the nano-Ni measured under high-pressure triaxial compression is more than three times that of the micron-Ni. Significant work hardening for the nano-Ni in high-pressure plastic deformation stage was observed, whereas the micron-Ni experiences minor high-pressure work-softening and considerable energy dissipation into heat. The significantly reduced energy dissipation for the nano-Ni during the loading-unloading cycle indicates that the nanostructured materials can endure much greater mechanical fatigue in cyclic loadings. The nano-Ni exhibits steady grain growth during bulk plastic deformation at high-pressure loading, and drastic stress reduction and grain growth occur during the high P-T cycle. The results should be of considerable interest for further study.

(vi) It is interesting as noted in a recent non-linear constitutive modeling[562] and first principle calculation[563] that in the indentation test the pressure enhancement of both elastic modulus and yield strength is significant. A comparison with classical linear elasticity, that uses constant, zero-pressure, values of the modulus and constant yield strength, shows that the enhancement of the elastic modulus and of the yield strength due to extreme high pressures that develop in material under the indenter has a significant effect in hard and superhard materials. This enhancement has to be accounted for if accurate modeling of the mechanical response of such materials with extreme properties is to be achieved. The pressure effect on the elasticity enhancement could be described by the extension of the BOLS to pressure domain that causes bond deformation associated with storage of the deformation energy.

(vii) Incorporating the concepts of BOLS correlation and temperature dependence to the established computation methodologies could not only refine the parameters used in practice but also expand the capability of computation tools to reveal true situation and minimize the gap between measurement and calculation.

These topics would form challenging branches of study towards profound knowledge and practical applications. The progresses in the current and previous reports may evidence the validity of the LBA approach and the BOLS correlation, which enable us to touch the surface of the vast field of mesoscopic studies.

Acknowledgment

The author would like to express his sincere gratitude to Professors S. Veprek, J.S. Colligon, P.J. Jennings, C.L. Bai, S.Y. Tong, R. A. Andrievski, Y.W. Mai, F.G. Shi, C. Humphreys, M. Wautelet, M.F. Ashby, R.H. Baughman, A. Zakhidov, E. Roduner, S. Suzer, S.X. Dou, W. Gao, L.D. Zhang, E.K. Wang, B.R. Mehta, D.P. Yu, T.Y. Zhang, H.C. Huang, J. Zhou, S.Y. Fu, W.T. Zheng, Q. Jiang, D.H. Feng, and E.Y. Jiang for their encouragement and valuable communications. The author also would like to thank J.S. Colligon and Y.W. Mai for critical reading and valuable input. The author is indebted to S. Li, H. Huang, L.K. Pan, Y.Q. Fu, Z. Sun, M.X. Gu and Y. Ding for their support and input in one way or another. A fascinating and encouraging discussion in person with Dr Alan G. MacDiarmid at UT Dallas is particularly memorable. Permission of reprinting diagrams from Elsevier, IOP, APS, ACS, and AIP is also acknowledged. The project is supported by NNSF of China (Nos. 10525211 and 10772157) and MOE (RG14/06), Singapore.

[561] Y. Zhao, J. Zhang, B. Clausen, T.D. Shen, G.T. Gray III, and L. Wang, Thermomechanics of Nanocrystalline Nickel under High Pressure-Temperature Conditions, Nano Lett 7 (2007) 426-32.

[562] R.G. Veprek, D.M. Parks, A.S. Argon, S. Veprek, Non-linear finite element constitutive modeling of mechanical properties hard and superhard materials studied by indentation, Mater Sci Eng A 422 (2006) 205–17.

[563] W.M. Mook, J.D. Nowak, C.R. Perrey, C.B. Carter, R. Mukherjee, S.L. Girshick, P.H. McMurry, and W.W. Gerberich, Compressive stress effects on nanoparticle modulus and fracture, PHYS REV B **75** (2007) 214112.